%% file: Manuscrit.tex
\documentclass[a4paper,11pt,twoside]{StyleThese}
\include{formatAndDefs}

\begin{document}

\include{TitlePage}

\pagenumbering{roman}

\setcounter{page}{0}
\cleardoublepage

\section*{Remerciements}
Je tiens d'abord à remercier David Dean et Julien Randon-Furling, qui ont accepté de rapporter mon manuscrit, ainsi que les membres de mon jury, Jean-François Joanny, Laure Dumaz, et Kirone Mallick. 

Un immense merci à Olivier Bénichou et Raphaël Voituriez pour m’avoir accompagné tout au long de cette thèse. J’ai énormément appris à vos côtés, et je reste profondément admiratif de votre manière d’aborder les problèmes complexes : avec des outils concrets, sans détour, en sachant toujours isoler les mécanismes essentiels sans jamais négliger l’exigence technique. Merci pour votre bonne humeur constante, pour nos discussions au-delà des sciences, et pour la liberté que vous m’avez laissée dans les méthodes et les questions explorées au fil de cette thèse. Vous m’avez appris qu’avec suffisamment de ténacité et de curiosité, même les problèmes les plus ardus finissent par céder quand on en comprend intimement chaque recoin.

Merci à Giulio Biroli et à Grégory Schehr pour m'avoir accepté dans leurs superbes écoles d'été respectives et pour m'avoir donné le goût de la physique statistique.

Merci à l’équipe originale de la salle du couloir 12-13 — Jérémie, Brieuc et Léo —, à Anna, Louise, Tim et Marc, pour avoir été des compagnons de travail et de détente aussi précieux qu’enthousiastes. Les discussions au tableau, les pauses café ou bière, et les sorties ciné ont ponctué ces années avec un équilibre parfait entre recherche et convivialité. Vous avez donné au LPTMC cette atmosphère chaleureuse et vivante qui en fait un lieu où il fait bon vivre et travailler. Merci tout particulièrement à Léo pour notre collaboration scientifique aussi rapide que féconde : en à peine deux mois, on a compris un sacré paquet de trucs pas triviaux ! 

Merci à Aurélien pour les bières, les discussions, le ciné, ma compréhension approximative du monde merveilleux de Tracy-Widom et de Painlevé VI, et notre collaboration scientifique – certes moins productive, mais menée avec enthousiasme (on y a vraiment cru à cette histoire de fBM !). Les futurs doctorants que tu encadreras auront une chance inouïe : en plus de ta bienveillance et de ton incroyable compétence technique, ils profiteront de l’expertise du grand maître des intégrales gaussiennes, capable d’évaluer \[
\int \! \mathrm{erfc}^5(x) \, dx
\] en la fittant numériquement, une bière à la main.

Merci aux anciens, aux nouveaux, et aux futurs anciens : Thibaut (mon co-bureau et mon co-vélo adoré), Arthur (le monde est juste pas prêt pour la transformée de Laplace non-linéaire), Paul (si si je te promets il reste des trucs à faire), Pierre-Louis (fourmisses), Nicolas, Théotime, Vincent, Sankarshan, Louis-Martin, Davide — ainsi qu’aux très sympathiques Pascal, Bertrand, Maria, Nicolas, Maxim, Liliane, Jean-Noël, Hadrien, Rémy (un jour on reparlera de groupes de tresses !), Laura et Marco. Grâce à vous, je suis convaincu que le LPTMC restera (voire deviendra encore plus) un endroit aussi agréable et vivant après mon départ.

Un mot maintenant pour les polytechniciens, qui font partie de ma vie depuis mon intégration — et qui, je crois, y resteront jusqu’à ma désintégration. Merci à Simon d'avoir été mon éternel coloc depuis maintenant plus de 3 ans, merci d'être toujours de bonne humeur et chaud pour n'importe quoi même quand ca implique des randos improvisées avec des pneus crevés, de cuisiner des journées entières, de passer des concours d'oenologie en sachant à peine ce qu'est un Beaujolais ou de faire une thèse en physique expérimentale (oups). Essaie de pas mourir en Géorgie.

Merci infiniment à la team Maxence, Antoine, Henri, et plus spécialement à mon cher coloc Arthur, pour tous les cinés, les plats, les soirées, les cuites, le sport, les voyages, passés ou à venir (je te regarde Max). 

Merci à mes très chers Benoît et Juliette pour les Sherlock, pour les bières, pour vos personnalités incroyables et complémentaires, et pour les remarques douteuses fenêtre ouverte, au bon Thomas (depuis la Courtine !) pour ses talents de MJ et de cuisinier, et à l'incroyable Brice (une fois j'ai fait du vélo avec lui, la thèse c'était rien à côté), ainsi qu'à la team de Saint-Malo, Axel, Jeanne, Louis, Marie, Léa, Hugo, Alice.

Merci infiniment à mon co-thésard Mathis et à la si talentueuse Margaux pour la musique, nécessaire échappatoire pendant ces longues semaines de thèse parfois arides. Merci à Youenn et Guillaume, également partenaires du 2bis et du ouzo à 2e.

Merci aux Beynois Manu (Vidauban vie), Alex, PA, Audrey, Paul, Mathieu, Inès, Félicien, Baptiste pour avoir été là depuis le début. Mention spéciale à Vincent mon bro de toujours, j'espère que la peaulisse c'est pas trop nul, faut qu'on revienne à Rome manger des pizze et faire du vélo.

Merci à ma famille, Axelle, Romain et en particulier à mon père pour m'avoir donné le goût des sciences et du travail bien fait (même si je sais pas utiliser un tournevis) et à ma mère pour une grande partie du reste.

Merci enfin à ma star Alexandra, que j’admire pour son talent et sa discipline et que j'aime infiniment pour sa bienveillance, son envie de tout essayer... et tant d'autres choses encore. Tout ce que tu fais m’inspire à me dépasser, et je suis convaincu que tu réussiras tout ce que tu entreprendras. Je serai là pour t’accompagner tout au long du chemin — à l’étranger pour t’écouter en concert, ou partout ailleurs. 

\dominitoc
\tableofcontents

\mainmatter

\include{Intro}

\include{LARW}
\include{SIRW}

\include{UniversalExploration}

\appendix

\cleardoublepage
\mtcaddpart[Annexes]
\chapter*{Annexes}

\include{Annexe1}

\bibliographystyle{StyleThese}
\bibliography{These}

% Et j'ajoute ici un modèle d'abstract extrait de ma thèse (les thèses ont souvent besoin de ca à la fin donc j'inclus pour montrer comment formater ça)

% \cleardoublepage
% \begin{vcenterpage}
% \noindent\rule[2pt]{\textwidth}{0.5pt}
% \\
% {\large\textbf{Résumé :}}
% blabla
% \\
% {\large\textbf{Mots clés :}}
% Non-Markovien 
% \\
% \noindent\rule[2pt]{\textwidth}{0.5pt}
% \end{vcenterpage}

\end{document}

%% file: formatAndDefs.tex
\usepackage{amsmath,amssymb}
\usepackage[T1]{fontenc}
\usepackage[utf8]{inputenc}
\usepackage[french]{babel}
\usepackage{ dsfont }
\usepackage{physics}
\usepackage{tikz}
\usepackage{mathtools}
\usepackage{subcaption}
\usepackage{diagbox}
\usepackage{amssymb}  

\usepackage[left=1in,right=.75in,top=1in,bottom=1in,includefoot,includehead,headheight=13.6pt]{geometry}

\usepackage{silence}

\usepackage{aecompl}
\usepackage{url}

\usepackage[printonlyused,withpage]{acronym}

\graphicspath{{.}{images/}}

\usepackage{color}
\definecolor{linkcol}{rgb}{0,0,0.4}
\definecolor{citecol}{rgb}{0.5,0,0}

\usepackage[nottoc, notlof, notlot]{tocbibind}
\usepackage{minitoc}
\usepackage{pdfpages}
\newcommand{\dw}{d_\text{w}}

\newcommand{\Lim}[1]{\lim\limits_{#1}}
\newcommand{\erfc}[1]{\text{erfc}\left({#1}\right)}
\newcommand{\Sim}[1]{\mathrel{\mathop{\kern 0pt\sim}\limits_{#1}}}
\newcommand{\Propto}[1]{\mathrel{\mathop{\kern 0pt\propto}\limits_{#1}}}
\newcommand{\To}[1]{\mathrel{\mathop{\kern 0pt\to}\limits_{#1}}}

\usepackage{rotating}                    % Sideways of figures & tables
\usepackage{fancyhdr}                    % Fancy Header and Footer

\let\headruleORIG\headrule
\renewcommand{\headrule}{\color{black} \headruleORIG}

\usepackage{colortbl}
\arrayrulecolor{black}

\fancypagestyle{plain}{
  \fancyhead{}
  \fancyfoot{}
  
}

\usepackage{FrAlgorithm}
\usepackage[noend]{FrAlgorithmic}
\usepackage{scrextend}

\makeatletter

\def\cleardoublepage{\clearpage\if@twoside \ifodd\c@page\else%
  \hbox{}%
  \thispagestyle{empty}%              % Empty header styles
  \newpage%
  \if@twocolumn\hbox{}\newpage\fi\fi\fi}

\makeatother

{%

\hrulefill
\vspace*{0.5cm}%
\end{minipage}
}

\let\minitocORIG\minitoc
\renewcommand{\minitoc}{\minitocORIG \vspace{1.5em}}

\usepackage{multirow}
{ \begin{list}%
	{$\bullet$}%
	{\setlength{\labelwidth}{25pt}%
	 \setlength{\leftmargin}{30pt}%
	 \setlength{\itemsep}{\parsep}}}%
{ \end{list} }

\renewcommand{\epsilon}{\varepsilon}

\ifpdf
  \usepackage[pagebackref,hyperindex=true]{hyperref}
\else
  \usepackage[dvipdfm,pagebackref,hyperindex=true]{hyperref}
\fi

\renewcommand*{\backref}[1]{}
\renewcommand*{\backrefalt}[4]{%
\ifcase #1 %
(Non cité.)%
\or
(Cité en page~#2.)%
\else
(Cité en pages~#2.)%
\fi}

\hypersetup
{
bookmarksopen=true,
pdftitle="Titre du manuscript",
pdfauthor="Votre nom", %auteur du document
pdfsubject="Description rapide du sujet du manuscrit", %sujet du document
pdfmenubar=true, %barre de menu visible
pdfhighlight=/O, %effet d'un clic sur un lien hypertexte
colorlinks=true, %couleurs sur les liens hypertextes
pdfpagemode=UseNone, %aucun mode de page
pdfpagelayout=SinglePage, %ouverture en simple page
pdffitwindow=true, %pages ouvertes entierement dans toute la fenetre
linkcolor=linkcol, %couleur des liens hypertextes internes
citecolor=citecol, %couleur des liens pour les citations
urlcolor=linkcol %couleur des liens pour les url
}

\usepackage[most]{tcolorbox}

\newtcolorbox{boxedeq}{
  enhanced,
  colback=blue!5,
  colframe=blue!80,
  boxrule=0.6pt,
  sharp corners,
  left=1mm,
  right=1mm,
  top=1mm,
  bottom=1mm,
  before skip=10pt,
  after skip=10pt,
  breakable
}

\newtcolorbox{keyboxedeq}[1][]{
  enhanced,
  sharp corners,
  colback=red!5,
  colframe=red!75!black,
  boxrule=1pt,
  left=1.5mm,
  right=1.5mm,
  top=1.2mm,
  bottom=1.2mm,
  before skip=12pt,
  after skip=12pt,
  fonttitle=\bfseries,
  coltitle=black,
  title={#1},
  breakable
}

%% file: TitlePage.tex
%!TEX root = Manuscrit.tex

\pagenumbering{Alph}

\begin{titlepage}
\begin{center}
\noindent {\large \textbf{SORBONNE UNIVERSITE}} \\
\vspace*{0.3cm}
\noindent {\LARGE \textbf{ÉCOLE DOCTORALE PHYSIQUE EN ÎLE-DE-FRANCE}} \\
\noindent \textbf{PHYSIQUE THÉORIQUE} \\
\vspace*{0.5cm}
\noindent \Huge \textbf{T H È S E} \\
\vspace*{0.3cm}
\noindent \large {pour obtenir le titre de} \\
\vspace*{0.3cm}
\noindent \LARGE \textbf{Docteur en Sciences} \\
\vspace*{0.3cm}
\noindent \Large de Sorbonne Université \\
\noindent \Large \textbf{Mention : \textsc{Physique Théorique}}\\
\vspace*{0.4cm}
\noindent \large {Présentée et soutenue par\\}
\noindent \LARGE Julien \textsc{Brémont} \\
\vspace*{0.8cm}
\noindent {\Huge \textbf{Exploration et vieillissement de processus stochastiques non-Markoviens}} \\
\vspace*{0.8cm}
\noindent \Large Thèse dirigée par Raphaël \textsc{Voituriez} et Olivier \textsc{Bénichou} \\
\vspace*{0.2cm}
\noindent \Large préparée au Laboratoire de Physique Théorique de la Matière Condensée (LPTMC) \\
\vspace*{0.2cm}
\noindent \large soutenue le  \\
\vspace*{0.2cm}
03/07/2025 
\vspace*{0.5cm}
\end{center}
\noindent \large \textbf{Jury :} \\
\begin{center}
\noindent \large 
\begin{tabular}{llcl}
      \textit{Rapporteurs :}	& David \textsc{Dean}		& - & CNRS (LOMA)\\
				& Julien \textsc{Randon-Furling}		& - & ENS Paris-Saclay\\
      \textit{Co-Directeur :}	& Raphaël \textsc{Voituriez}		& - & CNRS (LJP) \\
      \textit{Co-Directeur :}	& Olivier \textsc{Bénichou}		& - & CNRS (LPTMC) \\
      \textit{Président :}	& Jean-François \textsc{Joanny}		& - & Collège de France\\
      \textit{Examinateurs :}   & Laure \textsc{Dumaz}          & - & CNRS (DMA-ENS)\\
      				& Kirone \textsc{Mallick}			& - & CEA (IPhT)\\
\end{tabular}
\end{center}
\end{titlepage}

\pagenumbering{arabic}
\sloppy

\titlepage

%% file: Intro.tex
\chapter{Introduction générale}
\addcontentsline{toc}{chapter}{Introduction générale}

\section{Contexte général}
Le mouvement brownien et ses variantes discrètes, les marches aléatoires (RWs), forment depuis plus d'un siècle un cadre fondamental pour la modélisation de phénomènes aléatoires, tant en physique statistique qu'en probabilité \cite{pearson1905,einstein1905,spitzer1964,kacRandomWalk}. Sous leur forme minimale, ces processus reposent sur une dynamique locale élémentaire — chaque déplacement étant indépendant de l'historique de la marche — mais donnent lieu à des comportements collectifs riches et universels à grande échelle.

Ce paradoxe apparent entre simplicité microscopique et complexité macroscopique explique leur large succès. Les marches aléatoires et le mouvement brownien décrivent efficacement une vaste gamme de phénomènes : diffusion thermique \cite{vankampenDiffusionInhomogeneous}, transport de particules dans les milieux homogènes \cite{hughes1995random}, fluctuations des cours financiers à court terme \cite{mandelbrot1963variation}, ou encore des algorithmes de recherche \cite{benichou2011intermittent}.

De tels processus sont dits \emph{markoviens} : leur évolution future est conditionnée uniquement par leur état présent, sans dépendance explicite au passé. Cette propriété confère au formalisme des chaînes de Markov et des processus de diffusion une puissance d’analyse remarquable, permettant des résultats analytiques précis sur une large variété d’observables : les probabilités de premier passage \cite{redner2001guide}, les temps de retour et les lois de récurrence \cite{feller1971introduction}, les fonctions de corrélation temporelle \cite{kubo1966fluctuation}, ou encore les distributions extrêmes et les statistiques de l’occupation spatiale \cite{majumdar2005brownian, godreche2001statistics}.

De nombreux systèmes naturels, biologiques ou artificiels s'écartent pourtant de ce cadre idéal. Des effets de mémoire, d'adaptation ou d'interaction avec l'environnement génèrent des dynamiques \emph{non-Markoviennes}, dans lesquelles l'évolution future dépend explicitement de la trajectoire passée. Ce constat a conduit à l'émergence de nombreux modèles à comportement atypique, tels que les marches renforcées \cite{pemantleSurveyRandom}, les processus à accroissements corrélés tels que le mouvement Brownien fractionnaire \cite{mandelbrotFractionalBrownian,molchanMaximumFractionalBrownian1999}, les marches aléatoires avec temps d’attente à queue lourde \cite{metzlerAnomalousDiffusionModels2014}, ou encore les marches en milieu désordonné corrélé \cite{bouchaudAnomalousDiffusion}. 

Ces processus présentent souvent plusieurs propriétés remarquables communes\footnote{Le mouvement brownien fractionnaire constitue une exception notable, car il possède un propagateur gaussien et des incréments stationnaires.} : le vieillissement, c’est-à-dire la perte d’invariance temporelle dans les observables dynamiques \cite{barbier-chebbahSelfInteractingRandom,bremontAgingDynamics,schulzAgingRenewalTheory2014}; la diffusion anormale, marquée par des lois d’échelle non classiques \cite{metzlerAnomalousDiffusionModels2014,barbier-chebbahAnomalousPersistence,metzlerRandomWalkGuide2000}; des propagateurs non-Gaussiens, révélant des dynamiques asymétriques ou non locales \cite{dasilvaNonGaussianPropagator,bremontExactPropagators,dumazMarginalDensities,ghoshAnomalousNonGaussian}; et enfin, une altération profonde des modes d’exploration de l’espace, résultant de l’influence persistante de la trajectoire passée \cite{barbier-chebbahSelfInteractingRandom,regnierRecordAges,regnierCompleteVisitationStatistics2022}.

Toutefois, la diversité structurelle des dynamiques non-Markoviennes rend difficile l’élaboration d’un cadre théorique unificateur. Un tel cadre demeure, à ce jour, hors de portée. Cette difficulté reflète une vérité fondamentale : il n’existe qu’une seule manière d’être Markovien, mais une infinité de façons de s’en écarter.
Ce constat motive le développement de modèles minimalistes. Ces modèles doivent être assez simples pour permettre un traitement analytique explicite. Dans le même temps, ils doivent capturer des mécanismes fondamentaux : mémoire locale ou globale, interaction avec un environnement statique ou évolutif, ou encore modification irréversible du paysage visité. Ils doivent enfin conserver une interprétation physique claire et robuste.

L’intérêt de tels modèles réside aussi dans les liens qu’ils entretiennent avec des outils puissants issus de la théorie des probabilités \cite{tothGeneralizedRayKnightTheory1996,tothTrueSelfrepelling} et de la physique statistique hors équilibre \cite{derridaExactSolution,bertiniMacroscopicFluctuation}. Leur étude permet non seulement de produire des résultats analytiques rares pour des processus non-Markoviens, mais aussi d’établir des ponts vers des classes de phénomènes plus larges, tels que les modèles de croissance cinétique \cite{kardarDynamicScaling} ou les dynamiques de transport dans des milieux encombrés \cite{grabschExactSpatial}.

C’est dans cette perspective que s’inscrit ce travail, qui étudie comment l’introduction de mémoire ou de rétroactions locales dans des marches aléatoires affecte leurs propriétés d’exploration. Une attention particulière est portée aux effets de vieillissement — c’est-à-dire à la manière dont l’âge du système influence les lois statistiques d’observables temporelles.

\section{Objectifs de cette thèse}
Au fil de cette thèse, nous avons exploré sous diverses questions comment la mémoire modifie profondément les propriétés d’exploration spatiale d’un marcheur aléatoire non-Markovien. Plusieurs questions centrales ont guidé notre démarche. Comment la mémoire — sous forme d’une mémoire du temps passé dans des régions de l'espace fixes, ou d’une altération durable de l'environnement du marcheur — altère-t-elle les lois de diffusion et la structure des observables statistiques ? Pouvons-nous quantifier rigoureusement l’impact de cette mémoire sur la dynamique d’exploration, les temps de premier passage, la géométrie du territoire visité, ou encore sa taille ? Dans quelle mesure l’âge du système — le temps écoulé depuis le début de la marche — influence-t-il ces observables, révélant des phénomènes de vieillissement dynamiques ? À ces différentes questions, nous avons apporté des réponses analytiques précises, souvent exactes et nouvelles, reposant sur des outils issus de la théorie des processus stochastiques, des probabilités, de la physique statistique et usant d'arguments physiques, en étroite comparaison avec des simulations numériques. Nos résultats révèlent des phénomènes aussi exotiques et complexes que des lois non-gaussiennes, des comportements de multiscaling, des transitions dynamiques et des signatures universelles du vieillissement, élargissant considérablement le champ des modèles non-Markoviens analytiquement accessibles. \par 
\vspace{3ex}
Exposons succinctement le plan de cette thèse. La première partie est consacrée aux Locally Activated Random Walks (LARWs), un modèle simple mais riche de marche aléatoire non markovienne, où la dynamique du marcheur dépend de son état d’activation. Celui-ci est défini par le temps passé sur certains sites particuliers de l’espace, appelés hotspots. D’un point de vue physique, ce paramètre d’activation représente un degré de liberté interne du marcheur — comme, par exemple, sa température interne ou son inertie directionnelle — qui évolue dynamiquement en fonction des interactions locales avec l’environnement. Ce couplage entre mémoire spatiale et état interne donne lieu à des comportements dynamiques nouveaux et non triviaux, que cette première partie explore en détail. Notons qu'une version minimale de ce modèle, unidimensionnelle et comportant un unique hotspot, a été introduite dans \cite{benichouNonGaussianityDynamical}. À notre connaissance, il s'agit à ce jour de la seule référence disponible dans la littérature sur ce sujet. Nous en proposons ici une généralisation substantielle, en l’étendant à des configurations à plusieurs dimensions et à plusieurs hotspots, tout en explorant une variété de mécanismes d’influence sur la dynamique du marcheur.
Par exemple, le marcheur peut accélérer, ralentir, ou encore renforcer sa persistance directionnelle à mesure que l'activation augmente. Nous analysons en détail les mécanismes de piégeage dynamique induits par décélération, la distribution jointe de l’activation et de la position du marcheur, les statistiques de premier passage, les effets de vieillissement, ainsi que l’influence de la répartition spatiale des hotspots sur la dynamique du système.
Nous apportons également une réponse quantitative au cas physiquement pertinent mais techniquement plus difficile où l’activation décroît en dehors des hotspots : ce cadre révèle une transition de phase dynamique, accompagnée d’une densité de position fortement non gaussienne, au comportement inattendu. Ce modèle, à la fois naturel et analytiquement accessible, constitue un terrain d’étude privilégié pour explorer les propriétés fondamentales des marches non-Markoviennes. Il présente un intérêt particulier en physique statistique hors équilibre, en donnant lieu à des phénomènes émergents riches malgré sa simplicité apparente. Par ailleurs, il trouve des applications potentielles en biologie, notamment dans la modélisation des dynamiques de cellules dendritiques \cite{alraiesCellShape,moreauIntegratingPhysicalMolecular2018}. \par 
\vspace{3ex}
La deuxième partie de cette thèse est consacrée à une autre classe de marches aléatoires non-Markoviennes : les \emph{Self-Interacting Random Walks} (SIRWs), dans lesquelles la dynamique dépend explicitement du passé du marcheur — notamment des sites déjà visités. Ce type de mémoire introduit des rétroactions entre la trajectoire du marcheur et l’environnement qu’il explore, modifiant en profondeur les mécanismes d’exploration spatiale. De tels modèles trouvent des motivations physiques concrètes, par exemple dans le comportement d’organismes vivants ou d’agents diffusants qui évitent ou revisitent certaines régions selon leur histoire de navigation. Ces processus donnent lieu à des dynamiques individuelles complexes, souvent contre-intuitives, qui ne peuvent être comprises à l’aide des outils classiques des marches aléatoires Markoviennes. 

Ces modèles ont d’abord été introduits par les physiciens, à commencer par l’article séminal \cite{amitAsymptoticBehaviorTrue1983}, puis principalement étudiés à l’aide de simulations numériques et d’arguments heuristiques \cite{barbier-chebbahAnomalousPersistence,grassbergerSelfTrappingSelfRepellingRandom2017,barbier-chebbahSelfInteractingRandom,fosterReinforcedWalks}. Du côté mathématique, un grand nombre de travaux théoriques ont été menés sur ces processus \cite{pemantleSurveyRandom,tothTrueSelfAvoiding,tarresDiffusivityBounds1D2012,tothGeneralizedRayKnightTheory1996}. Malgré l’abondance des travaux mathématiques sur les SIRWs, peu de résultats explicites sont disponibles concernant des observables physiques, à quelques exceptions notables près \cite{carmonaBetaVariables,dumazMarginalDensities}. Cette rareté s’explique en grande partie par la technicité du formalisme de type \emph{Ray–Knight}, développé principalement par B. Tóth \cite{tothGeneralizedRayKnightTheory1996} et W. Werner \cite{tothTrueSelfrepelling}, qui demeure à ce jour le seul cadre analytique capable de traiter explicitement les SIRWs unidimensionnelles.

Ce constat a motivé l’approche originale menée dans cette partie : en adaptant ces outils issus de la probabilité théorique, nous montrons qu’il est possible d’obtenir des expressions explicites pour plusieurs observables fondamentales caractérisant l’exploration spatiale des SIRWs. Nous dérivons en particulier la distribution exacte de la position du marcheur, la densité de premier passage, ainsi que la taille et l’asymétrie du territoire visité. Ces résultats établissent un lien inédit entre les méthodes rigoureuses de la théorie des processus stochastiques et les grandes questions de la physique statistique hors équilibre, liées à la mémoire, à l’irréversibilité et à la complexité émergente.

Nous prolongeons ensuite cette approche en deux directions majeures. La première explore l’application de ces méthodes sur des structures plus complexes, en particulier les SIRWs sur des arbres. Cette généralisation conserve une certaine simplicité topologique tout en s’écartant du cadre unidimensionnel, et nous permet d’obtenir des résultats exacts sur des processus non-Markoviens dans des géométries non triviales. La seconde consiste à étendre le formalisme de Ray–Knight, historiquement limité à des marches unidimensionnelles et à des observables à un temps, à des contextes vieillis où l’âge du système, ici le temps écoulé depuis le début de la marche, entre en jeu. Ce prolongement est d’un grand intérêt théorique, car il permet d’accéder à des informations sur des observables à deux temps dans un cadre où les outils analytiques sont traditionnellement inopérants. Cette avancée permet, pour la première fois à notre connaissance, de quantifier de manière exacte l’effet de l’âge du système sur l’exploration spatiale dans des processus non-Markoviens. Elle met en évidence des dynamiques restées jusqu’ici inaccessibles — parfois anticipées par des arguments poussés d’analyse dimensionnelle et d'invariance d'échelle \cite{regnierCompleteVisitationStatistics2022,regnierRecordAges}, mais jamais quantifiées — et ouvre une voie nouvelle pour l’étude du vieillissement dans les systèmes stochastiques hors d’équilibre. \par 
\vspace{3ex}
Après l’étude approfondie des deux grandes classes de marches non-Markoviennes que sont les LARWs et les SIRWs, la troisième partie de cette thèse aborde une question transversale, présente en filigrane dans les chapitres précédents : comment quantifier l’efficacité de l’exploration spatiale lorsqu’on s’intéresse à une dynamique non-Markovienne \emph{arbitraire} ? Dans ce cadre général, la mémoire du système induit des corrélations temporelles complexes, qui altèrent profondément les mécanismes d’exploration de l’espace. 

Pour répondre à cette question, nous introduisons une observable centrale : la \emph{probabilité de flip}, qui mesure la tendance d’un marcheur à inverser la direction de son exploration après avoir atteint une extrémité du domaine visité. Dans le cas unidimensionnel, elle correspond à la probabilité d’incrémenter le minimum de sa trajectoire juste après avoir augmenté le maximum, plutôt que de continuer à explorer dans la même direction. Cette observable met en évidence un comportement asymptotique universel remarquable : elle décroît en $1/n$, où $n$ représente la taille du territoire exploré, indépendamment du processus non-Markovien considéré. Ce résultat souligne une idée centrale et prometteuse : conditionner un processus à mémoire sur la taille de l’espace qu’il a exploré semble faire émerger des lois universelles, au-delà des particularités dynamiques du système. \par 
Dans cette même perspective, nous quantifions également l’influence de l’asymétrie du domaine visité sur la probabilité de flip, pour un large éventail de processus non-Markoviens, incluant notamment les SIRWs et les processus gaussiens corrélés. Nous montrons également, de manière exacte et explicite, comment l’asymétrie du territoire exploré influence la probabilité de flip, pour un large éventail de dynamiques à mémoire, incluant notamment les SIRWs et les processus gaussiens corrélés. Ce résultat inédit met en lumière comment une information géométrique globale — l’asymétrie du domaine visité — peut gouverner une observable dynamique comme les flips. Il révèle un couplage profond, universel et, surtout, analytiquement accessible entre mémoire, géométrie et directionnalité de l’exploration, qui subsiste même dans des dynamiques fortement corrélées.

Notre cadre théorique, basé sur le concept de flip, fournit un socle conceptuel pour définir d’autres observables liées à la persistance de l’exploration et à la structure du domaine visité. L’approche que nous développons dépasse le cadre unidimensionnel, et peut être généralisée à des géométries plus complexes, suggérant l’existence de lois d’échelle robustes dans une large classe de dynamiques stochastiques à mémoire. Cette partie établit ainsi un cadre unificateur pour quantifier la persistance exploratoire, et met en lumière l’impact universel de la mémoire dans les systèmes hors d’équilibre.
\par \vspace{3ex}
La suite de cette thèse est rédigée en anglais. Elle se compose de trois parties, largement indépendantes les unes des autres.

%% file: LARW.tex
%!TEX root = Manuscrit.tex

\part{Locally Activated Random Walks (LARWs)}
% In this chapter, we develop a general theoretical framework for analyzing locally activated random walks (LARWs), which are random walks where the dynamical parameters depend on the time \( a(t) \) spent by the RWer in specific regions of space, referred to as \textit{hotspots}, up to a given observation time \( t \). We refer to \( a(t) \) as the \textit{activation}. This framework, defined on lattices, encompasses different classes of LARWs in which the RWer's dynamics—such as speed or persistence—may increase or decrease with growing activation.  
% We begin by studying the case of single-hotspot LARWs, where we derive exact results for the propagator and demonstrate that it exhibits non-Gaussian behavior. The mean square displacement \( \langle x^2(t) \rangle \) follows an anomalous scaling law, and the system displays aging dynamics, where the statistical properties depend explicitly on the time since the process started.  
% We then extend the analysis to the case of multiple hotspots, which we solve exactly in one dimension. We identify sufficient conditions on the spatial distribution of hotspots under which the RWer behaves as a time-changed Markov random walk at long times.  
% Finally, we investigate the effect of relaxation on the activated dynamics, where the activation \( a(t) \) decreases when the RWer is outside of the hotspots. For relaxing LARWs with multiple hotspots, we uncover a universal dynamical phase transition between a non-Gaussian and a Gaussian regime, which we characterize quantitatively in the one-dimensional case.
\chapter{Introduction}

\section{Localised activation and memory-driven dynamics}

In this chapter, we study a class of non-Markovian random walks known as \emph{Locally Actived Random Walks} (LARWs), in which the walker’s memory is encoded in a scalar \emph{activation} variable $a(t)$ that evolves jointly with its spatial position $x(t)$. Activation increases upon visits to predefined \emph{hotspots}—fixed locations in space—and in turn modulates the walker’s dynamics, such as its persistence or jump frequency. Typically, the instantaneous hopping rate is given by $1/\tau(a)$, where $\tau(a)$ is an activation-dependent waiting time function.

This coupling between spatial structure and internal memory gives rise to feedback loops: the RWer’s trajectory shapes its present dynamical state, which in turn conditions future interactions with the environment. Despite its apparent simplicity, this model deviates significantly from classical RW behavior, leading to robust and nontrivial phenomena across parameter regimes. Depending on the hopping rate $\tau(a)$, the LARW can localize, spread anomalously, or exhibit aging, and generally shows non-Gaussian statistics. In some limits, its dynamics approximate those of time-changed CTRWs~\cite{metzlerRandomWalkGuide2000}; in others, they lead to fundamentally new behavior.

Such locally activated dynamics, where visits to specific sites either accelerate or hinder motion, are relevant to a variety of biological systems. For instance, cells navigating through tissues often respond to local environmental cues, modulating their motility upon encountering specific microenvironments~\cite{moreauIntegratingPhysicalMolecular2018,naderCompromisedNuclear}. On larger scales, animals may adapt their motion in response to localized resources or hazardous zones~\cite{benichouIntermittentSearch,viswanathanPhysicsForaging}. A striking example involves dendritic immune cells, which switch from a slow, non-persistent state to a fast, persistent one once they accumulate sufficient antigenic signals — typically through repeated visits to chemically enriched regions~\cite{moreauIntegratingPhysicalMolecular2018}. Interestingly, similar transitions can be triggered mechanically, even in the absence of external signals, as observed when cells migrate through densely packed pores~\cite{alraiesCellShape}.

To our knowledge, the only prior work introducing this activation mechanism is~\cite{benichouNonGaussianityDynamical}, which studied a continuous-space LARW in a one-dimensional setting. The present chapter develops the theory of LARWs in full generality — in both discrete and continuous space, with and without relaxation, and under various hotspot configurations. Our approach blends physical arguments, exact calculations, and asymptotic analysis.

\section{Summary of the results}
Our main contributions are summarized below, according to the structural organization of the chapter. The activation-dependent hopping rate is defined as $1/\tau(a)$.

\begin{itemize}
    \item \textbf{General formalism.} We formulate the LARW model on discrete and continuous lattices, introducing the joint process $(x(t), a(t))$ and deriving its exact evolution equations. 
    
    \item \textbf{Trapping criterion.} We identify a \emph{trapping criterion} that separates free ($\tau(a)\ll a$ as $a \to \infty$) and trapped ($\tau(a)\gg a$) LARWs. In the trapped regime, the LARW gets permanently trapped due to diverging waiting times.

    \item \textbf{Single-hotspot dynamics.} We study the dynamics of a LARW with a single hotspot at the origin: activation $a(t)$ is the accumulated time the RWer spent at the hotspot. We derive the joint distribution of position and activation $P(x,a,t)$ and show that the marginal $P(x,t)$ is both non-Gaussian and either bimodal (for an accelerating LARW), peaked away from the origin, or peaked at the origin (for a decelerating LARW). Aging properties and scaling exponents are computed explicitly, revealing a deviation from classical scaling laws.

    \item \textbf{Persistent LARWs.}
    We extend the LARW framework to include persistence, whereby the RWer retains a memory of its direction of motion between jumps. This modification introduces velocity correlations and breaks detailed balance at the microscopic level. However, our analysis shows that persistence does not qualitatively alter the main features of the non-persistent LARW, as persistence can be taken into account in a renormalization of the activation-dependent waiting time.

    \item \textbf{Multiple periodic hotspots.} For a free LARW with a periodic hotspot distribution, we show that the long-time behavior is captured by a time-changed CTRW through the renormalized time $\tilde{t}(t) = \int_0^t (\tau(a(t')))^{-1} dt'$, leading to Gaussian spreading with anomalous time scaling.

    \item \textbf{Trapped regime.} In the regime $\tau(a) \gg a$, the LARW becomes eventually trapped at a hotspot. We characterize the distribution of the final trapping site and the trapping time. In the $1d$ periodic case, we show that the trapping hotspot follows a discrete Skellam distribution. In dimension $d=1$, $2$, and $3$, the large-time behavior of the activation distribution is computed, with distinct decay forms reflecting recurrence or transience of the underlying RW.

    \item \textbf{Disordered hotspot distributions.} We extend the model to quenched disorder, where hotspots are distributed randomly (e.g., Bernoulli with density $p$). In the trapped regime, the LARW reduces to a CTRW with random, spatially-dependent waiting times. Averaging over disorder, we derive expressions for the trapping probability at the origin and the typical spreading of the LARW before trapping.

    \item \textbf{Relaxation dynamics.} We introduce activation decay at a rate $\gamma$ via a relaxation term $-\gamma a(t)$, and investigate its effect in the $L$-periodic hotspot setting. A dynamical phase transition is identified by the parameter $\nu = \sqrt{\gamma} L$ at the critical value $\nu_c = \pi/\sqrt{2}$, separating two regimes:
    \begin{itemize}
        \item The \emph{Gaussian phase} ($\nu < \nu_c$), where activation spreads over the whole system, and the RWer retains memory between hotspot visits. The position distribution is Gaussian with renormalized diffusion.
        \item The \emph{non-Gaussian phase} ($\nu > \nu_c$), where activation is localized near hotspots, and the RWer loses memory between visits. In this regime, the position distribution develops a non-Gaussian profile, which we compute by approximating the LARW dynamics as a CTRW with spatially-dependent waiting times. We finally provide an accurate approximation to describe the activation profile, and reconstruct the full propagator with excellent agreement with simulations.
    \end{itemize}
\end{itemize}

This work forms the basis of a first publication \cite{bremontAgingDynamics} as well as several future publications.

\chapter{The single-hotspot LARW}
\begin{figure}
    \centering
        \begin{subfigure}{0.35\textwidth}
            \includegraphics[width=\textwidth]{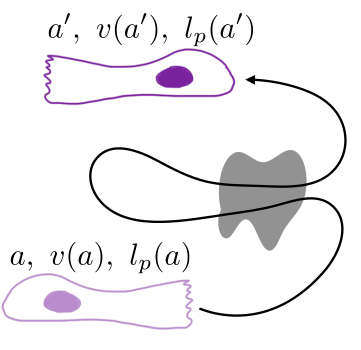}
            \vspace*{-20pt}
        \end{subfigure}
        \begin{subfigure}{0.55\textwidth}
            \includegraphics[width=\textwidth]{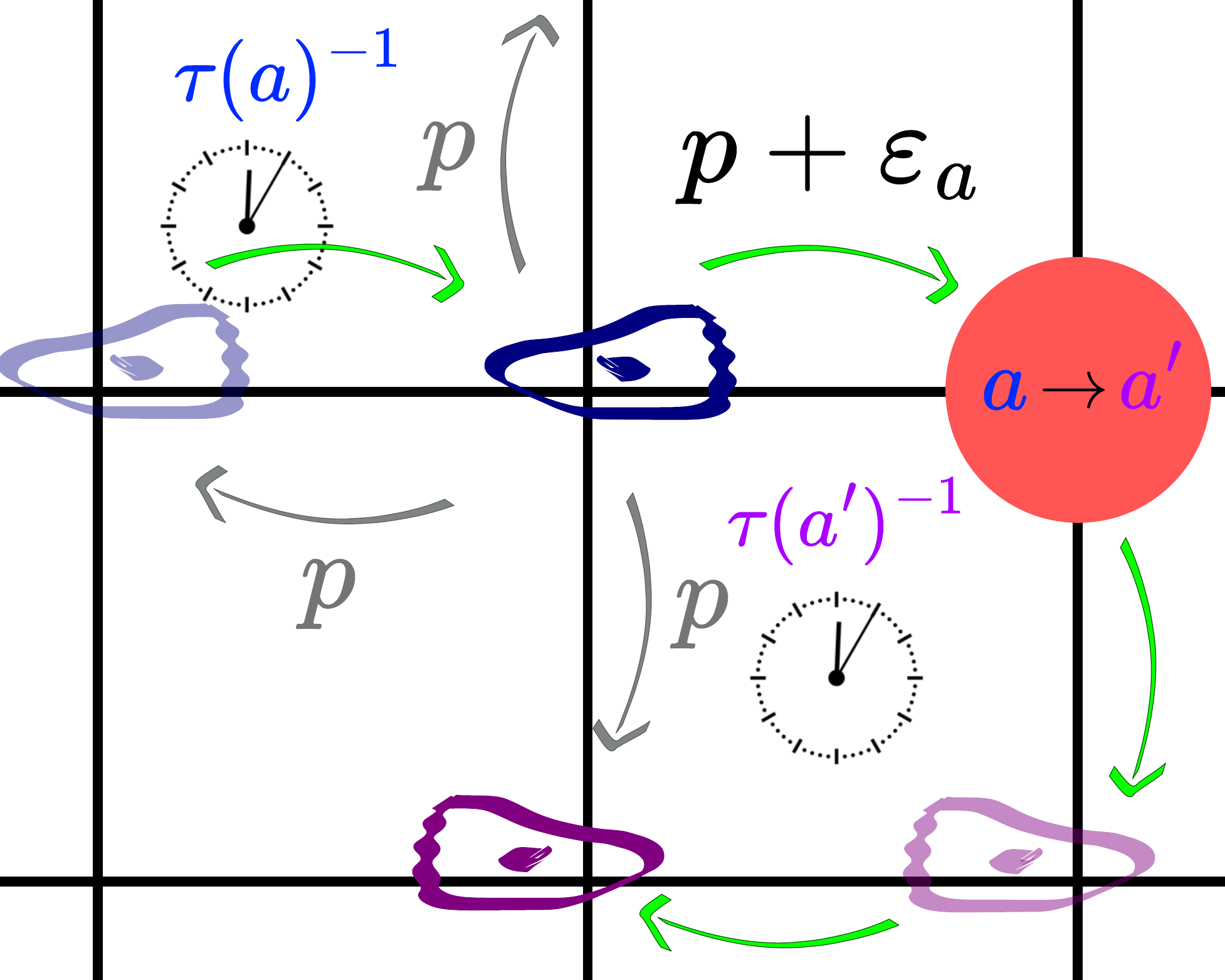}
            \vspace*{-20pt}
        \end{subfigure}
        \caption{Left: Sketch of a LARWer (e.g. immune cell), whose dynamic parameters (speed $v$ and persistence length $l_p$) depend on the activation $a$, which increases upon each visit to the activation site (e.g. antigen carrying site). Right: A persistent LARW on a 2D square lattice, with persistence parameter $\varepsilon_a$ ($p=\frac{1-\varepsilon_a}{4}$) and waiting time $\tau(a)$, which are the  counterparts  of the speed $v$ and persistence length $l_p$ for lattice models that we consider in this article. The trajectory of the RWer is represented by the green arrows, and  the hotspot is represented by the  red disk. The activation increases from $a$ to $a'$ upon the visit of the hotspot. }
        \label{fig:dessin}    
\end{figure}
We begin by introducing the minimal Locally Actived Random Walk (LARW), defined as a $d$-dimensional random walk whose dynamics are modulated by the cumulative time spent in a designated region of space, referred to as the \emph{hotspot}. 

We consider a RWer that performs a generic, continuous-time RW on a general graph $\mathcal{G}$. More precisely, the RWer, if at site $s$ with activation $a$, performs a jump with rate $1/\tau(a)$ to a site $s'$ drawn from a given distribution. Note that $s'$ does not need to be a neighbor of $s$. The RWer starts with activation $a=0$ from the origin $x=0$, which we assume here to be the only hotspot of the lattice.  

For this minimal model, the activation $a$ of the RWer is non-decreasing, and is given by the cumulative time spent on the hotspot up to time $t$, so that
\begin{equation}
    a(t) = \int_0^t \mathds{1}\left(x(t')\right)dt',
\end{equation}
where $\mathds{1}\left(s\right) = 1$ if $s$ is the hotspot $s=0$, and $0$ otherwise.  

In turn, the jump rate $\tau(a)$ is model-dependent and can capture both accelerated (decreasing $\tau(a)$) and decelerated (increasing $\tau(a)$) processes.  

Importantly, the aging dynamics of $a$ makes the position process $x(t)$ non-Markovian, because the jump rate of the RW depends explicitly on the activation $a$ that is controlled by the full history of visits of the RWer to the hotspot. Nonetheless, the process $(x(t),a(t))$ is Markovian, and the joint distribution $P(x,a,t)$ fully characterizes the process; we denote  
\begin{equation}
    \hat{P}(x,a,s) \equiv \int_0^{\infty} e^{-st} P(x,a,t) dt
\end{equation}
its Laplace transform.

\section{General results on LARWs}
\subsection{The trapping criterion}
Before presenting a detailed analysis of LARWs, we emphasize that, like time-changed CTRWs defined earlier,  LARWs can either become trapped at the hotspot or remain unbounded ('free'), depending on the growth of the waiting time $\tau(a)$ with the activation $a$. Specifically, we show that the LARW is trapped when $\tau(a) \gg a$, and free when $\tau(a) \ll a$.  
To see this, let $P_s(t)$ denote the probability that a RWer, initially at the hotspot at time $t = t_0$ with activation $a_0$, is still on the hotspot at time $t + t_0$. This quantity satisfies the equation:
\begin{equation}
    \frac{dP_s}{dt} = -\frac{P_s(t)}{\tau(a_0 + t)}.
\end{equation}
which integrates to:
\begin{equation}
    P_s(t) = e^{-\int_0^t \frac{dt'}{\tau(a_0 + t')}}.
\end{equation}
It follows that the RWer always leaves the hotspot (i.e., $P_s(\infty) = 0$) only if $\tau(a)^{-1}$ is not integrable, which we weaken to the condition $\tau(a) \ll a$. \footnote{We do not consider the case $\tau(a) \approx a$ in our study.} Our exact analysis below does not explicitly distinguish between trapped and free LARWs, but the asymptotic behavior derived in later sections will clarify the distinction between these two regimes.
\subsection{Master equation for $P(0,a,t)$}

In the following, we first show that even though $a(t)$ alone is not a Markovian process, an explicit evolution (master) equation for $P(0, a, t)$ can still be obtained. The importance of this observable $P(0, a, t)$ is twofold.  

\bigskip

First, since $a(t)$ can only increase, a RWer with activation $a(t) \geq a$ must have necessarily been at the hotspot $x = 0$ with activation $a$ at some earlier time $t'\leq t$. This leads to the relation:
\begin{equation}
    \label{a-from-p0a}
    P(a(t) \geq a) = \int_0^t P(0, a, t') dt',
\end{equation}
which fully characterizes the stochastic process $a(t)$ in terms of $P(0, a, t)$.  

\bigskip

Second, as shown below, $P(0, a, t)$ yields the full joint distribution $P(x, a, t)$.  

We now proceed to derive the master equation for $P(0,a,t)$. Writing $P(0, t + dt, a + dt)$ as a partition over the last time the hotspot was visited gives two possible scenarios:  
\begin{enumerate}
    \item[(i)] The RWer was at $x = 0$ at time $t$ with activation $a$ and did not jump during the interval $dt$, or  
    \item[(ii)] The last visit to the hotspot before the visit at $t + dt$ ended at some earlier time $t' < t$ with activation $a + dt$.  
\end{enumerate}
The key point is that between two consecutive visits to the hotspot, the activation $a$, and thus the jump rate $1/\tau(a)$, remain constant. Therefore, the probability of events involved in scenario (ii) can be written in terms of the first-passage time (FPT) density $F_a(0|{0^+}, t)$ to site $x = 0$ for a non-activated RW with constant jump rate $1/\tau(a)$ starting from site $x = 0$ and jumping at time $t = 0^+$. This yields
\begin{equation}
    \begin{gathered}
        P(0,a+dt, t+dt) = \mathbb{P}(\text{being at $0$ with $a$ at $t$, stay at $0$ for $dt$}) \\ + \mathbb{P}(\text{jump from $0$ at $t'\leq t$ with $a+dt$ and return to $0$ exactly at $t+dt$}) \\
        = \left(1-\frac{dt}{\tau(a)} \right) P(0,a,t) + \int_0^{t+dt} \frac{dt'}{\tau(a)} P(0,a+dt,t') \left(F_{a+dt}(0|{0^+}, t+dt-t')dt\right)
    \end{gathered}
\end{equation}
Taylor-expanding both sides of the equation and neglecting $O(dt^2)$ terms gives us the following master equation
\begin{boxedeq}
\begin{equation}
    \label{single-hotspot-transport-realtime}
    \partial_t P(0,a,t) + \partial_a P(0,a,t) = 
    -\frac{P(0,a,t)}{\tau(a)} + 
    \int_0^t \frac{dt'}{\tau(a)} P(0,a,t') F_a(0|{0^+}, t-t').
\end{equation}
\end{boxedeq}
Next, we define 
\begin{equation}
    \xi_a(s) \equiv (1+s\tau(a))^{-1}
\end{equation}
and Laplace-transform \eqref{single-hotspot-transport-realtime} to obtain a first important result
\begin{equation}
    \label{single-hotspot-transport-laplace}
    \partial_a \hat{P}(0,a,s) = \frac{[\mathcal{F}\big(0|0,\xi_a(s)\big)-1]\hat{P}(0,a,s)}{\tau(a)\xi_a(s)}
\end{equation}
where we introduced the discrete-time, non-activated first-passage generating function $\mathcal{F}(0|0,\xi)$, related to its continuous-time, activated counterpart $\hat{F}_a(0|{0^+},s)$ by $\hat{F}_a(0|{0^+},s) = \mathcal{F}(0|0,\xi_a(s))/\xi_a(s)$ \cite{hughes1995random}.
Integrating \eqref{single-hotspot-transport-laplace} is straightforward and yields an explicit expression for $\hat{P}(0,a,s)$, provided that $\mathcal{F}(0|0,\xi)$ is known.

\bigskip

We now show how to obtain the full joint distribution $P(x,a,t)$ from \eqref{single-hotspot-transport-laplace}. For the RWer to be on site $x$ with activation  $a$ at $t$, it must jump away from  $0$ at an earlier time $t'$ with activation $a$, and next reach site $x$ without hitting $0$ in the remaining time $t-t'$. This analysis yields the following renewal equation
\begin{equation}
\label{convolution_fulljoint}
    P(x,a,t) = \int_0^t \frac{dt'}{\tau(a)} P(0,a,t')P^a_{\text{surv}}(x|{0^+},t-t')
\end{equation}
where we define $P^a_{\text{surv}}(x|{0^+},t)$ to be the (survival) probability for a RWer with activation $a$, starting from site $0$ and jumping at time $0^+$, to be at site $x$ at time $t$, all this without visiting site $0$ again. This quantity is related \cite{hughes1995random} to its non-activated, discrete-time counterpart $\mathcal{P}_{\text{surv}}(x|0,\xi)$ by $\hat{P}^a_{\text{surv}}(x|{0^+},s) = \tau(a) \xi_a(s) \frac{\mathcal{P}_{\text{surv}}(x|0,\xi_a(s))}{\xi_a(s)}$. The Laplace transform of \eqref{convolution_fulljoint} thus yields 
\begin{equation}
\label{fulljoint-laplace}
    \hat{P}(x,a,s) = \hat{P}(0,a,s) \mathcal{P}_{\text{surv}}(x|0,\xi_a(s)).
\end{equation} 
We now recall how all the quantities entering \eqref{single-hotspot-transport-laplace}, \eqref{fulljoint-laplace} can be deduced from the generating function  $\mathcal{P}(x|y,\xi)$ of the propagator of the corresponding non activated RW. Standard results \cite{hughes1995random} yield the discrete-time generating function $\mathcal{P}_{\text{surv}}(x|0,\xi)$
\begin{equation}
\label{p_surv_lattice}
    \mathcal{P}_{\text{surv}}(x|0,\xi) = \mathcal{F}(x|0,\xi) = \mathcal{P}(x|0,\xi)/\mathcal{P}(0|0,\xi)
\end{equation}
as well as the first-return time to $0$:
$   \mathcal{F}(0|0,\xi) = 1-1/\mathcal{P}(0|0,\xi).$
Using these results, one finds finally the exact expression of the Laplace transformed  joint law
\begin{keyboxedeq}[Joint distribution of activation and position for the LARW]
\begin{equation}
    \label{jointlaw_final}
    \hat{P}(x,a,s) = \frac{\mathcal{P}(x|0,\xi_a)}{\mathcal{P}(0|0,\xi_a)} \exp\left(-\int_0^a \frac{db}{\tau(b)\xi_b \mathcal{P}(0|0,\xi_b)}\right).
\end{equation}
\end{keyboxedeq}
To our knowledge, even for non-activated RWs, such an explicit and general expression for the joint distribution of position and local time at the origin is not known in the literature. We emphasize that this determination of the joint law is fully explicit for all processes for which the propagator of the underlying, non-activated random walk is known. Explicit examples will be presented in the next sections.  

Moreover, the generality of \eqref{jointlaw_final} can be exploited to derive exact results for the FPT density of LARWs.

\subsection{Survival probability and FPT density of LARWs}
Suppose we are interested in the FPT density of the LARW to a given target $\Omega$. A natural starting point is to consider the auxiliary, non-activated random walk that becomes trapped at $\Omega$ upon reaching this target. Let $\mathcal{P}^{\dag}(x|0,t)$ denote the propagator of this trapped random walk, with $\mathcal{P}^{\dag}(x|0,\xi)$ its generating function. We can use the general expression \eqref{jointlaw_final} to obtain the Laplace transform of the joint probability for the LARW to be at $x$ at time $t$ with activation $a$, while having never hit the site $\Omega$, as:
\begin{equation}
    \label{dag-propag}
    \hat{P}^\dag(x,a,s) = \frac{\mathcal{P}^\dag(x|0, \xi_a)}{\mathcal{P}^\dag(0|0, \xi_a)} \exp\left(- \int_0^a \frac{db}{\tau(b) \xi_b \mathcal{P}^\dag(0|0, \xi_b)}\right).
\end{equation}
According to classical results of RW theory \cite{hughes1995random}, the generating function of the trapped propagator can be written in terms of the non-trapped propagator $\mathcal{P}(x|0,\xi)$ as:
\begin{equation}
    \label{dag-propag-rw}
    \mathcal{P}^{\dag}(x|0,\xi) = \begin{cases}
        \mathcal{P}(x|0,\xi) - \frac{\mathcal{P}(x|\Omega,\xi)\mathcal{P}(\Omega|0,\xi)}{\mathcal{P}(0|0,\xi)}, & x\neq \Omega, \\
        \frac{\mathcal{P}(\Omega|0,\xi)}{(1-\xi)\mathcal{P}(0|0,\xi)}, & x=\Omega.
     \end{cases}
\end{equation}
Hence, \eqref{dag-propag} is fully explicit whenever $\mathcal{P}$ is known.  \par 
The propagator $\mathcal{P}^\dag$ is normalized, meaning that $\sum_{x} \mathcal{P}^\dag(x|0,\xi) = \frac{1}{1 - \xi}$, because the non-activated RWer is not absorbed when it reaches $\Omega$; instead, it remains trapped at $\Omega$. Therefore, the quantity $\mathcal{P}^\dag(\Omega|0,t)$ corresponds exactly to the probability that the non-activated RWer reached $\Omega$ at some earlier time $t' \leq t$ and has remained trapped there since. From there, we deduce the FPT density of the LARW to site $\Omega$
\begin{equation}
    F(\Omega|0,t) = \partial_t \int_0^\infty  P^\dag(\Omega|0,a,t) da.
\end{equation}
In Laplace space, this becomes:
\begin{keyboxedeq}[FPT density of the LARW]
\begin{equation}
    \label{fpt_larw_laplace}
    \hat{F}(\Omega|0,s) = -\int_0^\infty \frac{\mathcal{P}(\Omega|0, \xi_a)}{\mathcal{P}(0|0, \xi_a)} \partial_a \left[ \exp\left(- \int_0^a \frac{db}{\tau(b) \xi_b \mathcal{P}^\dag(0|0, \xi_b)}\right) \right] da.
\end{equation}
\end{keyboxedeq}
This expression admits a natural interpretation: the quantity
\begin{equation}
    \label{a-at-fpt}
    \hat{F}(\Omega, a | 0, s) = - \frac{\mathcal{P}(\Omega | 0, \xi_a)}{\mathcal{P}(0 | 0, \xi_a)} \, \partial_a \left[ \exp\left( - \int_0^a \frac{db}{\tau(b)\, \xi_b\, \mathcal{P}^\dag(0 | 0, \xi_b)} \right) \right]
\end{equation}
is the Laplace transform (with respect to time) of the joint density of the pair $(t, a)$ corresponding to the LARW reaching the target $\Omega$ for the first time at time $t$ with activation $a$.

\section{$d$-dimensional nearest-neighbor LARWs}
In this section, we consider the paradigmatic example of a symmetric, nearest-neighbor LARW on the hypercubic lattice $\mathbb{Z}^d$. The generating function of the (non-activated) propagator $\mathcal{P}(x|y,\xi)$ admits an explicit integral representation \cite{hughes1995random}
\begin{equation}
    \label{propagator_hypercubic}
    \mathcal{P}\left (x = (x_k)_{k=1\dots d},\xi \right) = \int_0^\infty e^{-u} \prod_{k=1}^d I_{x_k}\left(\frac{u\xi}{d} \right) du.
\end{equation}
Together with \eqref{jointlaw_final}, Eq. \eqref{propagator_hypercubic} allows, up to Laplace inversion, the complete and exact determination of the distribution of $(x(t), a(t))$. Explicit computations follow when specifying the dimension $d$. We first focus on the distribution of $a(t)$, both in the free and trapped regimes.

\subsection{Distribution of the activation $a(t)$}
\subsubsection{Free regime $\tau(a) \ll a$}
Before computing the full joint distribution of $x(t), a(t)$, we first show how to obtain the distribution of the activation $a(t)$ from Eqs. \eqref{a-from-p0a}, \eqref{jointlaw_final} and \eqref{propagator_hypercubic}. First, from \eqref{a-from-p0a} and \eqref{jointlaw_final}, the following identity gives the Laplace transform of this distribution:
\begin{equation}
    \label{cdf-a-laplace}
    \mathcal{L}_{t\to s}P(a(t) \geq a) = \frac{\hat{P}(0,a,s)}{s} = \frac{\exp\left(-\int_0^a \frac{db}{\tau(b)\xi_b \mathcal{P}(0|0,\xi_b)}\right)}{s}.
\end{equation}
Using \eqref{propagator_hypercubic} and known results \cite{hughes1995random}, we obtain
\begin{equation}
    \label{asymptotic-p0-sm}
    \mathcal{P}(0|0,\xi_a) = \begin{cases}
        \frac{\xi_a}{\sqrt{1 - \xi_a^2}} & \Sim{as \ll 1} \frac{1}{\sqrt{2s \tau(a)}}, \quad d=1, \\[10pt]
        \frac{2}{\pi K(\xi_a^2)} & \Sim{as \ll 1} \frac{1}{\pi} \log \frac{8}{s\tau(a)}, \quad d=2, \\[10pt]
        & \To{as \ll 1} \frac{1}{1 - R_d}, \quad d \geq 3,
    \end{cases}
\end{equation}
where $R_d$ is the return probability of the non-activated RW on the $d$-dimensional hypercubic lattice.

Applying a Tauberian theorem, we obtain the exact asymptotics of the distribution of $a$ in the regime $a \ll t$, which is the relevant one for free LARWs:
\begin{keyboxedeq}[Distribution of the activation $a(t)$]
\begin{equation}
    \label{cdf_a_time}
    \mathbb{P}(a(t) \geq a) \Sim{t \to \infty, a \ll t} 
    \begin{cases}
        \text{erfc} \bigg( \frac{\int_0^a \frac{db}{\sqrt{\tau(b)}}}{\sqrt{2t}} \bigg), & d = 1, \\[10pt]
        \exp\bigg(-\frac{\pi \int_0^a \frac{db}{\tau(b)}}{\log 8t} \bigg), & d = 2, \\[10pt]
        \exp\bigg(-(1 - R_d) \int_0^a \frac{db}{\tau(b)}\bigg), & d \geq 3.
    \end{cases}
\end{equation}
\end{keyboxedeq}
Note that choosing $\tau(a)=1$ in \eqref{cdf_a_time} returns the well-known distribution of the local time at the origin of a simple RW. It is distributed as positive truncation of a Gaussian random variable with mean $0$ and standard deviation $\sqrt{t}$ for $d=1$ \cite{borodinHandbookBrownianMotion2002}. For $d=2$, it is exponential with mean $\frac{\log (8t)}{\pi}$ \cite{erdosProblemsConcerning}.

\bigskip 

Eq. \eqref{cdf_a_time} allows us to compute the typical scaling with time of $a(t)$, e.g for the choice $\tau(a) \sim a^{-\alpha}, \alpha>-1$. We obtain
\begin{equation}
    \label{typical_a}
    a(t) \propto
    \begin{cases}
        t^{\frac{1}{2+\alpha}}, & d = 1, \\[10pt]
        \log^{\frac{1}{1+\alpha}} t, & d=2 \\[10pt]
        O(1), & d \geq 3.
    \end{cases}
\end{equation}

\subsubsection{Trapped regime $\tau(a) \gg a$}
\begin{figure}
\centering
    \includegraphics[width=.6\textwidth]{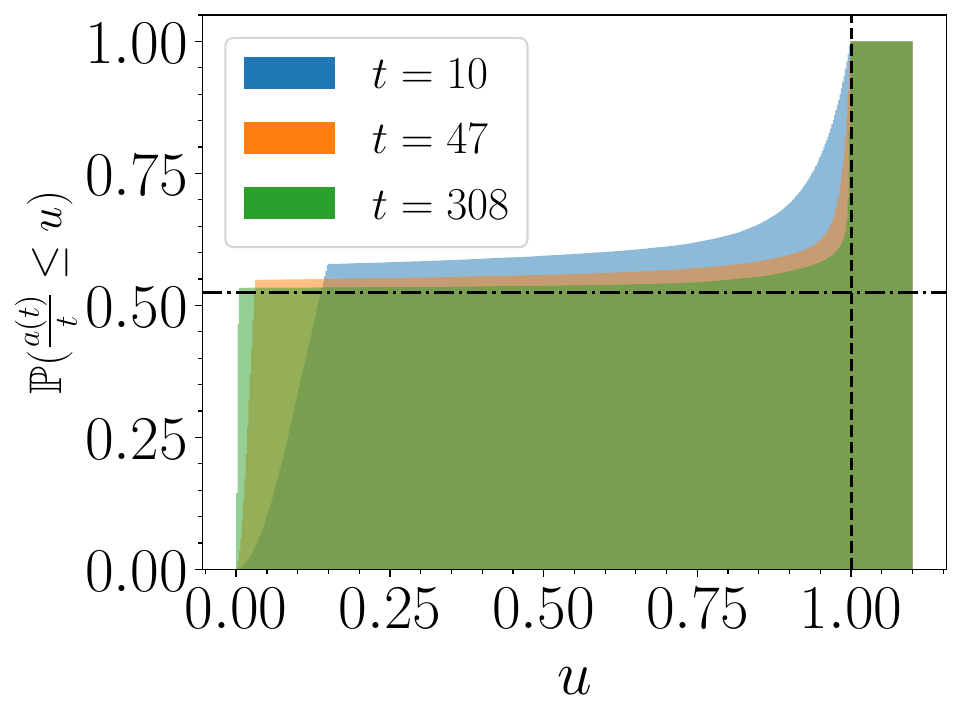}
    \caption{Cumulative distribution of $\frac{a(t)}{t}$ for a $3d$ accelerating LARW with $\tau(a) = a^{-1}$, which we choose to trap at the hotspot when $a>a_f=1.5$. The black dash-dot line represents the probability $\mathbb{P}(a(t=\infty)<a_f) = \exp(-\frac{(1-R)a_f^2}{2})$ as given by \eqref{cdf-a-3d}.}
    \label{fig:bimodal-3d}
\end{figure}
We now focus on the \emph{trapped regime} defined by $\tau(a) \gg a$. As established earlier, in this regime the RWer has a nonzero probability of becoming trapped at a hotspot with each visit. In this section, we aim to compute the distribution of the \emph{trapping time}—that is, the time required for the RWer to become permanently localized at the origin.

Importantly, in this regime the assumption $a \ll t$ used in earlier asymptotic expansions no longer holds. To quantify the fraction of trajectories that have become trapped by time $t$, we consider the probability $\mathbb{P}(a(t) \geq \gamma t)$ for some fixed $\gamma > 0$. Indeed, once the RWer is trapped, its activation $a(t)$ grows linearly with $t$, so this probability serves as an indicator of trapping. We define the trapping time $T_{\text{trap}}$ as the time at which half of the trajectories have become trapped.

To compute $\mathbb{P}(a(t) \geq \gamma t)$, we refer to the Laplace-transformed distribution from \eqref{cdf-a-laplace}, which requires evaluating $\hat{P}(0,a,s)$, which is given by
\begin{equation}
    \label{p0as-2}
    \hat{P}(0,a,s) = \exp\left( -\int_0^a \frac{db}{\tau(b)\, \xi_b\, \mathcal{P}(0 | 0, \xi_b)} \right),
\end{equation}
where $\xi_b = (1 + s \tau(b))^{-1}$ as before. We are interested in the regime where $a s \sim 1$, corresponding to $a \sim t$. In this regime, the integral in \eqref{p0as-2} contains two distinct contributions, depending on whether $s \tau(b) \ll 1$ or $s \tau(b) \gg 1$. Assuming that $\tau(b)$ is monotonically increasing, these two regions are separated at the crossover point $b_* = \tau^{-1}(1/s)$.
Since we are in the trapped regime where $\tau(b) \gg b$ and $a s \sim 1$, we have $b_* \ll a$. We therefore split the integral as
\begin{equation}
    \int_0^{a} \frac{db}{\xi_b \tau(b) \mathcal{P}(0 | 0, \xi_b)} 
    = \int_0^{\tau^{-1}(1/s)} \frac{db}{\xi_b \tau(b) \mathcal{P}(0 | 0, \xi_b)} 
    + \int_{\tau^{-1}(1/s)}^{a} \frac{db}{\xi_b \tau(b) \mathcal{P}(0 | 0, \xi_b)} 
    \equiv A + B.
\end{equation}
We now evaluate the contributions $A$ and $B$ in their respective asymptotic regimes:
\begin{itemize}
    \item For $A$, where $s \tau(b) \ll 1$, we use $\xi_b \sim 1 - s \tau(b)$.
    \item For $B$, where $s \tau(b) \gg 1$, we use $\xi_b \sim \frac{1}{s \tau(b)}$.
\end{itemize}
This yields the estimates:
\begin{equation}
    A \sim \int_0^{\tau^{-1}(1/s)} \frac{db}{\tau(b)\, \mathcal{P}(0 | 0, 1 - s \tau(b))}, 
    \qquad 
    B \sim s \int_{\tau^{-1}(1/s)}^{a} \frac{db}{\mathcal{P}(0 | 0, \frac{1}{s \tau(b)})} \sim a s.
\end{equation}
Thus, in the regime $s \to 0$ with $a s \sim 1$, we obtain the asymptotic form
\begin{equation}
    \label{p_0_a_s_nonergodic}
    \hat{P}(0, a, s) \Sim{s \to 0,\ as \sim 1} 
    \exp\left( -a s - \int_0^{\tau^{-1}(1/s)} \frac{db}{\tau(b)\, \mathcal{P}(0 | 0, 1 - s \tau(b))} \right).
\end{equation}
In real time, the factor $e^{-a s}$ in Laplace space corresponds to a time shift $t \mapsto t - a$. Hence, the probability $\mathbb{P}(a(t) \geq a)$ is governed by the small-$s$ behavior of the integral
\[
\int_0^{\tau^{-1}(1/s)} \frac{db}{\tau(b)\, \mathcal{P}(0 | 0, 1 - s \tau(b))}
\]
which we investigate for LARWs in dimensions $d=1,2,3$.

\subsubsection{$1d$ LARW}

We consider here the case of a $1d$ trapped LARW. Using the asymptotic form of \eqref{p_0_a_s_nonergodic} together with $P(0|0,1 - s \tau(b))^{-1} \sim \sqrt{2s \tau(b)}$, we obtain
\begin{equation}
    \hat{P}(0,a,s) \Sim{s \to 0,\ as \sim 1} \exp\left( -a s - \sqrt{2s} \int_0^{\tau^{-1}(1/s)} \frac{db}{\sqrt{\tau(b)}} \right).
\end{equation}

Two regimes arise depending on whether the integral $\int_0^\infty \frac{db}{\sqrt{\tau(b)}}$ is finite or divergent. If $\int_0^\infty \frac{db}{\sqrt{\tau(b)}} < \infty$, then the large-$a$ asymptotics simplify to
\begin{equation}
    \hat{P}(0,a,s) \Sim{s \to 0,\ as \sim 1} \exp\left( -a s - \sqrt{2s} \int_0^\infty \frac{db}{\sqrt{\tau(b)}} \right).
\end{equation}
Performing a Laplace inversion with $a = \gamma t$, we obtain
\begin{equation}
    P(a(t) \geq \gamma t) \Sim{t \to \infty} \erfc{\frac{I}{\sqrt{2t(1 - \gamma)}}}, \qquad
    I = \int_0^\infty \frac{db}{\sqrt{\tau(b)}}.
\end{equation}
This implies that the trapping time's dependence on the waiting time function $\tau(a)$ is through the integral $I$, namely $T_{\text{trap}} \propto I^2$. \par 
If instead $\int_0^\infty \frac{db}{\sqrt{\tau(b)}} = \infty$, the Laplace inversion cannot be carried out explicitly, but numerical Laplace inversion is possible.

\subsubsection{$2d$ LARW}
We now turn to the case of a trapped $2d$ LARW. Using \eqref{p_0_a_s_nonergodic} along with known results for the Green's function in two dimensions, we obtain the asymptotic form
\begin{equation}
    \hat{P}(0,a,s) \Sim{s \to 0,\ as \sim 1} \exp\left( -a s - \pi \int_0^{\tau^{-1}(1/s)} \frac{db}{\tau(b) \log \frac{8}{s \tau(b)}} \right).
\end{equation}
Because $I = \int_0^\infty \frac{db}{\tau(b)} < \infty$ in the trapped regime, then, for small $s$, we have
\begin{equation}
    \int_0^{\tau^{-1}(1/s)} \frac{db}{\tau(b) \log \frac{8}{s \tau(b)}} \Sim{s \to 0} \frac{I}{\log \frac{8}{s}}.
\end{equation}
Applying the Tauberian theorem then yields the large-time asymptotics
\begin{equation}
    P(a(t) \geq \gamma t) \Sim{t \to \infty} \exp\left( -\frac{\pi I}{\log(8(1 - \gamma) t)} \right).
\end{equation}
This implies that the trapping time's dependence on the waiting time function $\tau(a)$ is through $I$ via $\log T_{\text{trap}} \propto I$. 

\subsubsection{$3d$ LARW}
Finally, we consider the case of a $3d$ trapped LARW. Using again \eqref{p_0_a_s_nonergodic}, and the fact that $\tau(a) \gg a$ implies $\int_0^\infty \frac{db}{\tau(b)} < \infty$, we find
\begin{equation}
    \hat{P}(0,a,s) \Sim{s \to 0,\ as \sim 1} \exp\left( -a s - (1 - R) I \right), \qquad I = \int_0^\infty \frac{db}{\tau(b)},
\end{equation}
where $R\approx 0.34$ is the return probability of the simple RW in $3d$. Laplace inversion then yields
\begin{equation}
    \label{cdf-a-3d}
    P(a(t) \geq \gamma t) \Sim{t \to \infty} 1 - e^{-(1 - R) I}.
\end{equation}
In contrast to the recurrent cases ($1d$, $2d$), the weight of non-trapped trajectories,
\[
w_S = e^{-(1 - R) I},
\]
remains strictly positive at long times. See Fig.~\ref{fig:bimodal-3d} for a numerical verification of \eqref{cdf-a-3d} in the case of a waiting time $\tau(a)$ that blows up at finite activation $a_f$.

\subsection{Propagators and FPT densities of $d$-dimensional nearest-neighbor LARWs}
\subsubsection{The $1d$ LARW}
Because the $1d$ LARW is the most analytically tractable case, we compute several of its key properties:  
(i) the joint distribution of position $x(t)$ and activation $a(t)$,  
(ii) the marginal distribution of the position $x(t)$,  
(iii) the first-passage time density to a target site $\Omega > 0$,  
and (iv) the aging properties of the $1d$ LARW.

\paragraph{(i) Joint distribution of position and activation.}  
In one dimension, the generating function \eqref{propagator_hypercubic} yields the explicit expression:
\begin{equation}
    \label{1d-propag-nonac}
    \mathcal{P}(x, \xi) = \left( \frac{1 - \sqrt{1 - \xi^2}}{\xi} \right)^{|x|} (1 - \xi^2)^{-1/2}.
\end{equation}
From this, using \eqref{jointlaw_final}, we obtain the exact expression for the Laplace-transformed joint distribution:
\begin{equation}
    \label{eq:fulljoint_Z}
    \hat{P}_{1d}(x,a,s) = \left( \frac{1 - \sqrt{1 - \xi_a^2}}{\xi_a} \right)^{|x|} \exp\left( -\int_0^a \frac{db}{\tau(b)} \frac{\sqrt{1 - \xi_b^2}}{\xi_b} \right).
\end{equation}

For free LARWs, it is always the case that $a \ll t$ at long times, corresponding to the Laplace regime $s a \ll 1$, which implies $s \tau(a) \ll 1$ under the assumption $\tau(a) \ll a$.

Using the expansion $\xi_a \Sim{a s \ll 1} 1 - s \tau(a)$, we find:
\begin{equation}
    \left( \frac{1 - \sqrt{1 - \xi_a^2}}{\xi_a} \right)^{|x|} \Sim{a s \ll 1} e^{ -|x| \sqrt{2s \tau(a)} }.
\end{equation}
Hence,
\begin{equation}
    \label{eq:fulljoint_Z_laplace}
    \hat{P}_{1d}(x,a,s) \Sim{a s \ll 1} \exp\left( -\sqrt{2s} Z \right), \quad Z = \sqrt{\tau(a)} |x| + \int_0^a \frac{db}{\sqrt{\tau(b)}}.
\end{equation}
Laplace inversion then yields the long-time behavior:
\begin{equation}
    \label{eq:fulljoint_Z_time}
    P_{1d}(x,a,t) \sim \frac{Z}{\sqrt{2\pi} t^{3/2}} \exp\left( -\frac{Z^2}{2t} \right),
\end{equation}
with $Z = \sqrt{\tau(a)} |x| + F_1(a)$ and $F_1(a) = \int_0^a \frac{db}{\sqrt{\tau(b)}}$. This is exactly the joint distribution of position and activation found in the seminal article \cite{benichouNonGaussianityDynamical}.

\paragraph{(ii) Marginal distribution of the position.}  
To allow for explicit computation, we now choose the waiting time $\tau(a) \sim a^{-\alpha}$ with $\alpha > -1$ to ensure $\tau(a) \ll a$. Under this choice, the $1d$ LARW becomes self-similar with walk dimension $d_w$, such that the rescaled variable $x(t)/t^{1/d_w}$ becomes independent of $t$ at long times. The propagator then acquires the scaling form:
\begin{equation}
    \label{scaling-propag}
    P_{1d}(x, t) = \frac{1}{t^{1/d_w}} f_\alpha\left( \frac{x}{t^{1/d_w}} \right).
\end{equation}

We emphasize that this scaling behavior holds only for the one-point function $P_{1d}(x,t)$ and only for the specific class of waiting times $\tau(a) = a^{-\alpha}$. The process $(x(t))_{t \geq 0}$ itself is not invariant under standard rescaling, reflecting a breakdown of scale invariance at the level of the full stochastic process. This phenomenon, discussed further in the following section, is analogous to the behavior observed in continuous-time random walks (CTRWs), another class of time-changed stochastic processes that also break standard scaling relations.

We now compute the one-point scaling function for the $1d$ LARW. Starting from the asymptotic form \eqref{eq:fulljoint_Z_laplace}, and using the specific choice $\tau(a) = a^{-\alpha}$, we obtain
\begin{equation}
    \hat{P}_{1d}(x,a,s) \Sim{a s \ll 1} \exp\left( -\sqrt{2s} Z \right), \quad Z = a^{-\alpha/2}|x| + \frac{a^{1+\alpha/2}}{1 + \alpha/2}.
\end{equation}

To compute the marginal distribution of the position, we integrate over $a$. Changing variables via $a \to u\, x^{\frac{1}{1+\alpha}}$, we obtain:
\begin{equation}
    \label{laplace-marginal-scaling}
    \hat{P}_{1d}(x,s) = x^{\frac{1}{1+\alpha}} \int_0^\infty \exp\left( -\sqrt{2s} x^{\frac{2+\alpha}{2+2\alpha}} \phi(u) \right) du,
\end{equation}
with
\begin{equation}
    \phi(u) = \frac{u^{\alpha/2 + 1}}{1 + \alpha/2} + u^{-\alpha/2}.
\end{equation}

Laplace inversion yields the exact expression for the propagator:
\begin{equation}
    P_{1d}(x,t) = \int_0^\infty \frac{\phi(u) |x|^{\frac{4+\alpha}{2(1+\alpha)}}}{\sqrt{2\pi} t^{3/2}} \exp\left( -\frac{\phi(u)^2 |x|^{\frac{2+\alpha}{1+\alpha}}}{2t} \right) du.
\end{equation}

Matching this with the scaling form \eqref{scaling-propag}, we identify the walk dimension:
\begin{boxedeq}
\begin{equation}
    \label{dw-1d}
    d_w = \frac{2 + \alpha}{1 + \alpha}
\end{equation}
\end{boxedeq}
and extract the explicit form of the scaling function:
\begin{boxedeq}
\begin{equation}
    \label{scalingfunction-final-1d}
    f_\alpha(y) = \frac{y^{\frac{4 + \alpha}{2 + 2\alpha}}}{\sqrt{2\pi}} \int_0^\infty \phi(u) \exp\left( -\frac{y^{\frac{2 + \alpha}{1 + \alpha}}}{2} \phi(u)^2 \right) du.
\end{equation}
\end{boxedeq}

The function $f_\alpha(y)$ is properly normalized, as can be verified using the identity:
\begin{equation}
    \int_{-\infty}^\infty \frac{y^{\frac{4+\alpha}{2+2\alpha}}}{\sqrt{2\pi}} \phi(u) \exp\left( -\frac{y^{\frac{2+\alpha}{1+\alpha}}}{2} \phi(u)^2 \right) dy = \frac{(1 + \alpha)(2 + \alpha)\, u^\alpha}{\left( \alpha + 2 u^{\alpha + 1} + 2 \right)^2},
\end{equation}
which integrates to $1$ over $u$.

To better understand the behavior of the scaling function, we consider its asymptotic limits. For large values of the rescaled position $y = x / t^{1/d_w}$, we obtain:
\begin{boxedeq}
\begin{equation}
    \label{eq:marginal1d}
    f_\alpha(y) \Sim{y \gg 1}
    \begin{cases}
    A\, y^{\frac{1}{1+\alpha}} \exp\left( -B\, y^{\frac{2+\alpha}{1+\alpha}} \right), & \alpha > 0 \quad (\text{accelerating}), \\
    \dfrac{2^{\frac{1}{2} - \frac{1}{\alpha}} \Gamma\left( \frac{1}{2} - \frac{1}{\alpha} \right)}{\sqrt{2\pi} |\alpha|} y^{-\frac{2}{|\alpha|}}, & -1 < \alpha < 0 \quad (\text{decelerating}),
    \end{cases}
\end{equation}
\end{boxedeq}
where the constants $A$ and $B$ depend on the function
\begin{equation}
    g(u) = \frac{u^{-\alpha/2} \left( \alpha + 2 u^{\alpha + 1} + 2 \right)}{1 + \alpha/2},
\end{equation}
and its minimum $g(u_*)$ through:
\begin{equation}
    A = \frac{g(u_*)}{\sqrt{g''(u_*)}}, \quad B = g(u_*)^2.
\end{equation}

\begin{figure}
    \centering
    \includegraphics[width=0.6\textwidth]{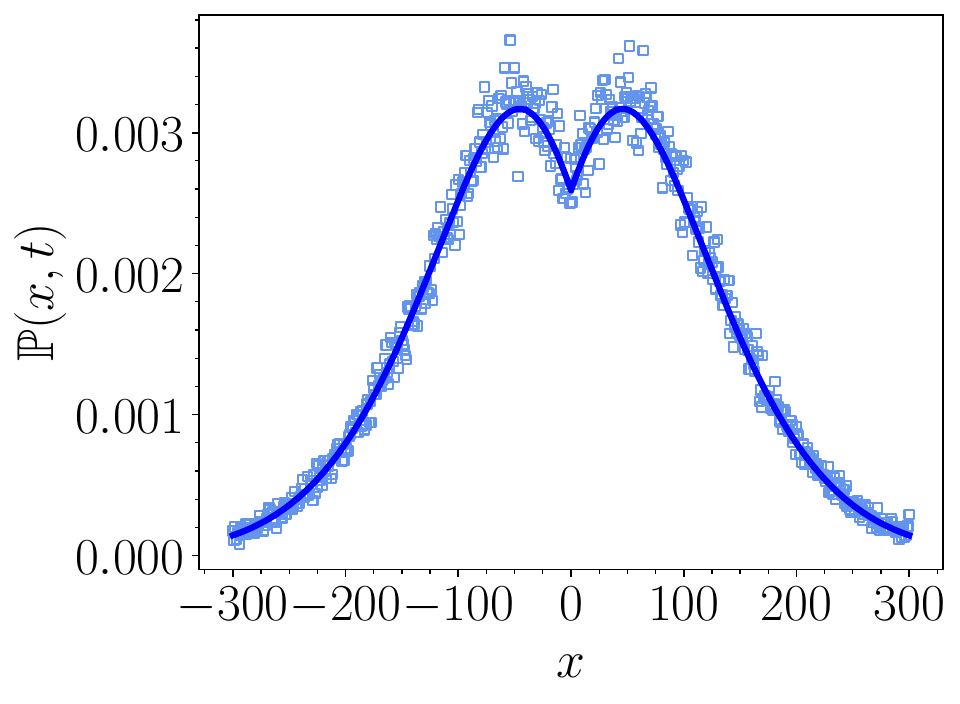}
    \caption{Propagator of the accelerating $1d$ LARW with $\tau(a) = a^{-\alpha}$, here with $\alpha = 1$. Symbols are obtained from numerical simulations; the thick line shows the asymptotic prediction from \eqref{eq:marginal1d}.}
    \label{fig:pxt-1d}
\end{figure}

Equation \eqref{eq:marginal1d} recovers and extends the results of \cite{benichouNonGaussianityDynamical}, originally derived for a Brownian LARW in continuous space. In particular, the marginal distribution of the position in the accelerating regime is not only non-Gaussian, but also non-monotonic in $x$. The derivation of the asymptotic forms follows two distinct methods: saddle-point analysis for $\alpha > 0$, and boundary-layer analysis for $-1 < \alpha < 0$. This distinction reflects the location of the minimum of the function $\phi(u)$: in the accelerating regime, the minimum occurs at a finite $u_* > 0$, while in the decelerating regime it is attained at the boundary $u_* = 0$.

The power-law tails observed in the decelerating case arise because the waiting time at hotspots is negligible at $t=0$ for the choice $\tau(a)=a^{-\alpha}, \alpha<0$. Thus, the RWer quickly reaches abnormally large distances, resulting in a broad, heavy-tailed distribution of the position.
 
\begin{figure}
    \centering
    \includegraphics[width=.4\textwidth]{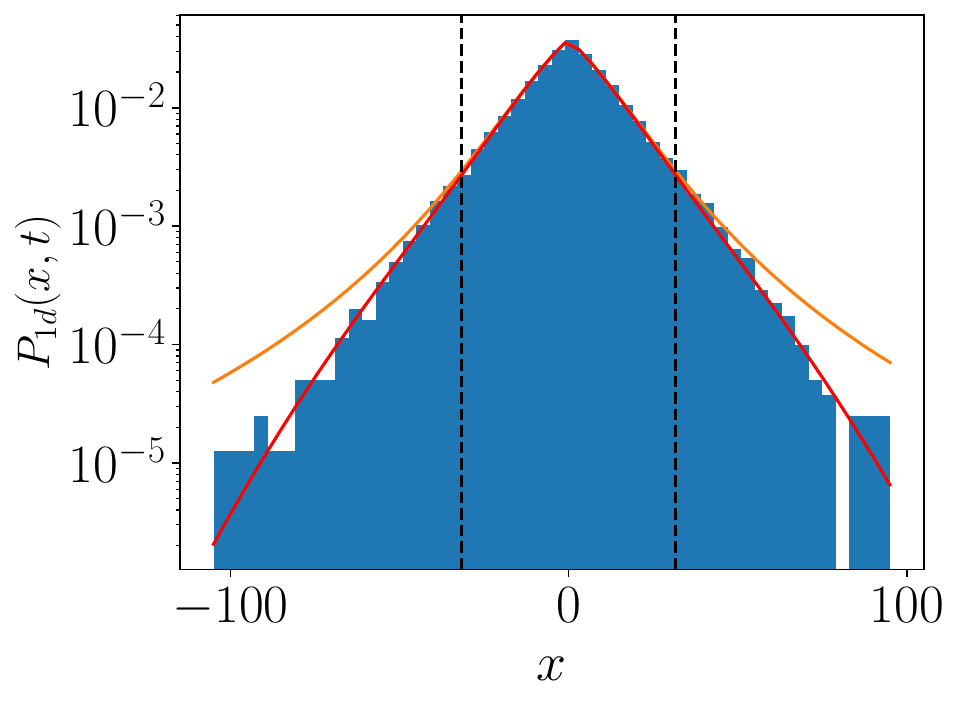}
    \includegraphics[width=.4\textwidth]{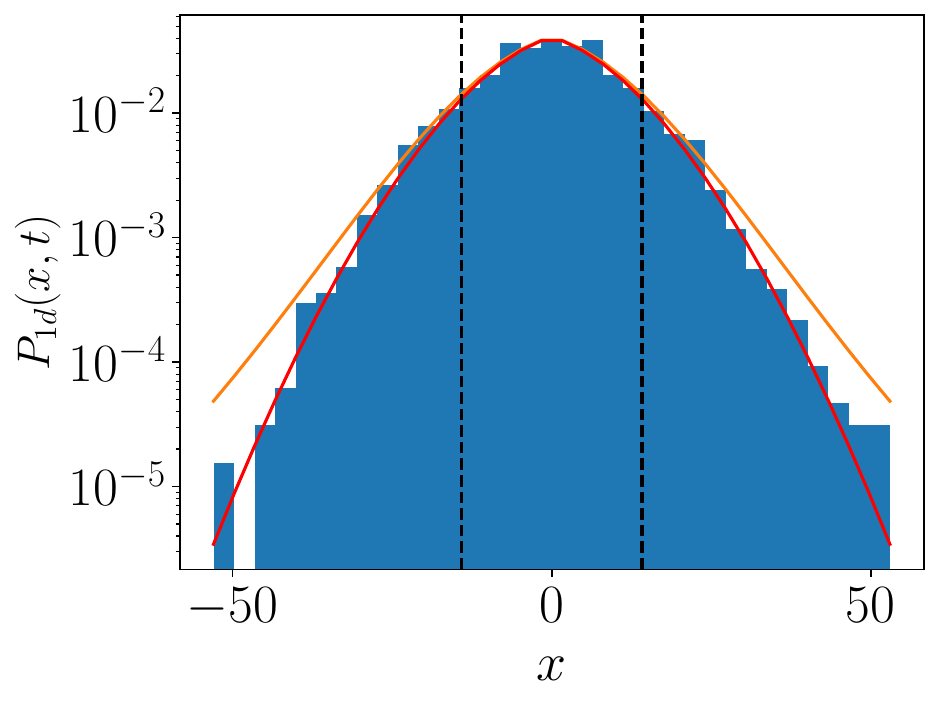}
    \caption{Propagator of decelerating $1d$ LARWs with $\tau(a) = (1+a)^{-\alpha}$. Left: $\alpha=-1/2$, right: $\alpha=-1/4$. The red curves show the exact result \eqref{propag-1d-choice1pa} integrated numerically, while the orange curves show the propagator of the LARW with $\tau(a)=a^{-\alpha}$ for comparison. The latter overestimates the real tails of the propagator. We see that the two propagators differ after a lengthscale $\sqrt{t}$ indicated by the dashed black lines, as predicted in \eqref{propag-1pa-decelerating}.}
    \label{fig:tau1pa-simus}
\end{figure}

This slightly singular behavior can be regularized by modifying the waiting time at $a=0$, for instance by choosing $\tau(a) = (1+a)^{-\alpha}$. We adopt this form to build physical intuition for the behavior of decelerating LARWs. With this choice, the $Z$ term defined in \eqref{eq:fulljoint_Z_laplace} becomes
\[
Z = (1+a)^{-\alpha/2}|x| + \frac{(1+a)^{1+\alpha/2}-1}{1+\alpha/2}.
\]
When integrating $P_{1d}(x,a,t)$ over $a$, we may neglect the constant $-\frac{1}{1+\alpha/2}$ in the term $Z$ at large times, since $a(t) \gg 1$ (a fact supported by numerical evidence). Introducing the change of variable $(1+a) = u x^{\frac{1}{1+\alpha}}$, we obtain the expression
\begin{align}
\label{propag-1d-choice1pa}
P_{1d}(x,t) =  \int_{0}^\infty \frac{Z e^{-\frac{Z^2}{2t}}}{\sqrt{2\pi } t^{3/2}} da \Sim{t \gg 1} \int_{x^{-\frac{1}{1+\alpha}}}^\infty \frac{\phi (u) | x| ^{\frac{\alpha +4}{2 \alpha +2}} e^{-\frac{\phi (u)^2 | x| ^{\frac{2+\alpha}{1+\alpha}}}{2t}}}{\sqrt{2\pi } t^{3/2}} du.
\end{align}
This representation reveals the loss of scale invariance due to the $x$-dependent lower bound $x^{-\frac{1}{1+\alpha}}$. See Fig.~\ref{fig:tau1pa-simus} for a numerical confirmation and a comparison with the unregularized choice $\tau(a)=a^{-\alpha}$. We now derive the explicit asymptotics of \eqref{propag-1d-choice1pa}. Importantly, a saddle-point approach is not applicable here, as the function \( Z \) reaches its minimum at the lower boundary \( a = 0 \) of the integration domain. Expanding \( Z \) near \( a = 0 \) gives:
\[
Z^2 = |x|^2 + |x|(2 - \alpha |x|)a + O(a^2),
\]
indicating that we need to use Laplace's method at the boundary. In the regime \( x \gg \sqrt{t} \), this yields:
\[
P(x,t) \Sim{x \gg \sqrt{t}} \frac{e^{-x^2/2t}}{\sqrt{2\pi t^3}} \int_0^\infty e^{-x(2 - \alpha |x|)a / 2t} |x|\, da = \frac{e^{-x^2 / 2t}}{\sqrt{2\pi t} \left(1 - \frac{\alpha}{2} |x| \right)}.
\]

On the other hand, in the regime \( x \ll \sqrt{t} \), the minimum at \( a = 0 \) becomes irrelevant. The lower bound in \eqref{propag-1d-choice1pa} can then be extended to zero, effectively recovering the result for the unregularized waiting time \( \tau(a) = a^{-\alpha} \). Physically, in this regime, the initial value \( \tau(0) = 0 \) or \( \tau(0) = 1 \) has negligible impact, since unphysical trajectories enabled by vanishing waiting times contribute only to the tails of the distribution, and not to the bulk where \( x \ll \sqrt{t} \).

As a result, the full asymptotic behavior of the propagator reads:
\begin{keyboxedeq}[Propagator of the decelerating LARW]
\begin{equation}
\label{propag-1pa-decelerating}
P_{1d}(x,t) \sim \begin{cases}
    \dfrac{1}{t^{1/d_w}} f_\alpha\left( \dfrac{x}{t^{1/d_w}} \right), & x \ll \sqrt{t}, \quad d_w = \dfrac{2+\alpha}{1+\alpha}, \\[0.8em]
    \sqrt{\dfrac{2}{\pi t}}\, \dfrac{e^{-x^2 / 2t}}{|\alpha| x}, & x \gg \sqrt{t},
\end{cases}
\end{equation}
\end{keyboxedeq}
where \( f_\alpha \) is the scaling function defined in \eqref{scalingfunction-final-1d}.

\begin{figure}
    \centering
    \includegraphics[width=.8\textwidth]{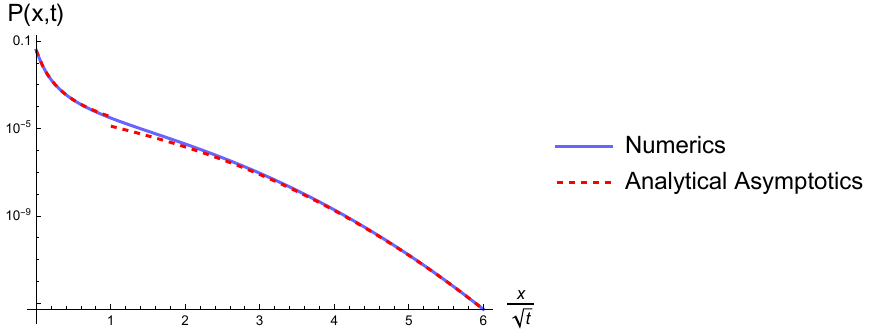}
    \caption{Comparison between the exact expression \eqref{propag-1d-choice1pa}, integrated numerically, and the asymptotics \eqref{propag-1pa-decelerating}, for $\alpha = -3/4$. The apparent discontinuity in the orange curve is a consequence of the change of regime, when $x \approx \sqrt{t}$.}
    \label{fig:comp-numerics-asymp-1d}
\end{figure}

This result reveals a clear multiscaling structure in the decelerating regime, governed by two competing timescales: \( t^{1/d_w} \) and \( \sqrt{t} \), which coincide only in the non-activated case \( \alpha = 0 \). Notably, the strong decay $\frac{e^{-x^2 / 2t}}{x}$ in the far tail \( x \gg \sqrt{t} \) is a physically consistent feature that corrects the unphysical power-law tail obtained from the unregularized waiting time \( \tau(a) = a^{-\alpha} \). This highlights the importance of regularization at small activation. The asymptotics in \eqref{propag-1pa-decelerating} are supported by numerical simulations, as shown in Fig.~\ref{fig:comp-numerics-asymp-1d}.

We now evaluate the mean-square displacement (MSD) of the decelerating LARW in one dimension, with $\tau(a) = (1+a)^{-\alpha}$ and $\alpha<0$, using the expression \eqref{propag-1pa-decelerating}. The MSD reads:
\begin{align}
\label{msd-decel-1d}
\langle x^2(t) \rangle &= \int_0^{\sqrt{t}} x^2 P_{1d}(x,t) \, dx + \int_{\sqrt{t}}^\infty x^2 P_{1d}(x,t) \, dx \notag \\[0.5em]
&= \int_0^{\sqrt{t}} \frac{1}{t^{1/\dw}} f_\alpha\left(\frac{x}{t^{1/\dw}} \right) x^2 \, dx + \int_{\sqrt{t}}^\infty \sqrt{\frac{2}{\pi t}} \frac{e^{-\frac{x^2}{2t}}}{|\alpha| x} x^2 \, dx \\[0.5em]
&= t^{2/\dw} \int_0^{t^{1/2 - 1/\dw}} f_\alpha(u) u^2 \, du + \frac{\sqrt{\frac{2}{e\pi}} \sqrt{t}}{|\alpha|}, \notag
\end{align}
where we used the change of variable $u = x/t^{1/\dw}$ and the walk dimension $\dw = \frac{2 + \alpha}{1 + \alpha}$. For $\alpha < 0$, we have $\dw > 2$, implying $t^{1/2 - 1/\dw} \to \infty$ as $t \to \infty$. 

The behavior of the first term in \eqref{msd-decel-1d} depends on the asymptotic decay of $f_\alpha(u) \sim u^{2/\alpha}$ for large $u$. Thus, the convergence of the integral $\int_0^\infty u^{2/\alpha + 2} \, du$ determines the scaling. This integral converges if and only if $\alpha > -\frac{2}{3}$. 

When $\alpha \leq -\frac{2}{3}$, the integral diverges and $\dw \geq 4$, so that the leading contribution of the first term behaves as
\begin{equation}
    t^{2/\dw} \int_0^{t^{1/2 - 1/\dw}} f_\alpha(u) u^2 \, du \propto \frac{\alpha  \left(\sqrt{t} - t^{2/\dw}\right)}{3 \alpha + 2} \propto \sqrt{t}.
\end{equation}
In this case, both terms in \eqref{msd-decel-1d} scale as $\sqrt{t}$, and we conclude:
\begin{equation}
    \langle x^2(t) \rangle \propto \sqrt{t}, \qquad \text{for } \alpha \leq -\frac{2}{3}.
\end{equation}

In contrast, for $\alpha > -\frac{2}{3}$, the integral in the first term converges and yields the dominant behavior:
\begin{equation}
    \langle x^2(t) \rangle \propto t^{2/\dw}, \qquad \text{for } -\frac{2}{3} < \alpha < 0.
\end{equation}

\begin{figure}
    \centering
    \includegraphics[width=.48\textwidth]{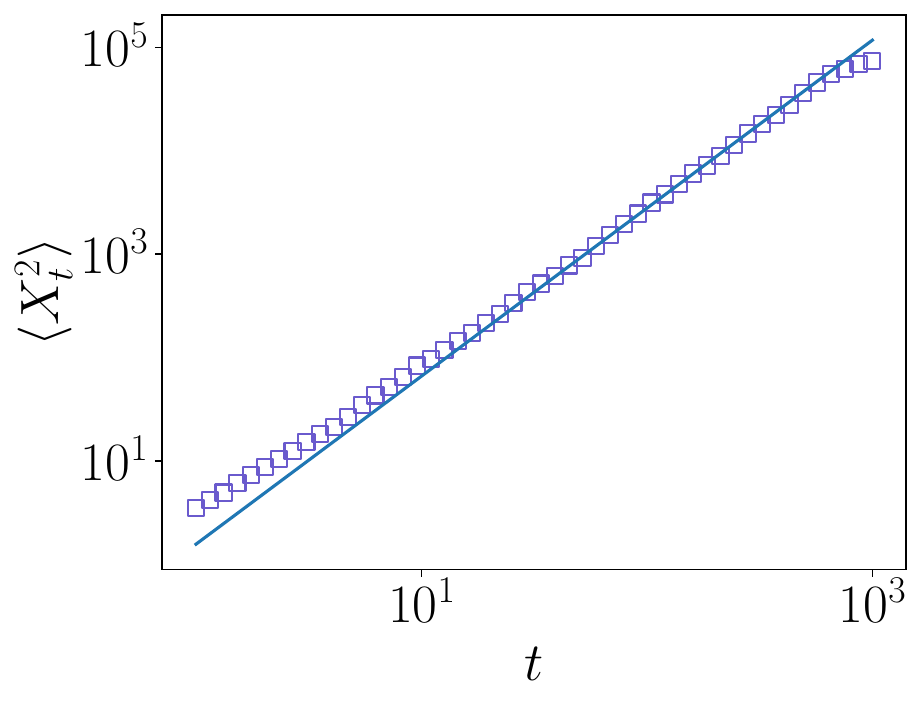}
    \includegraphics[width=.48\textwidth]{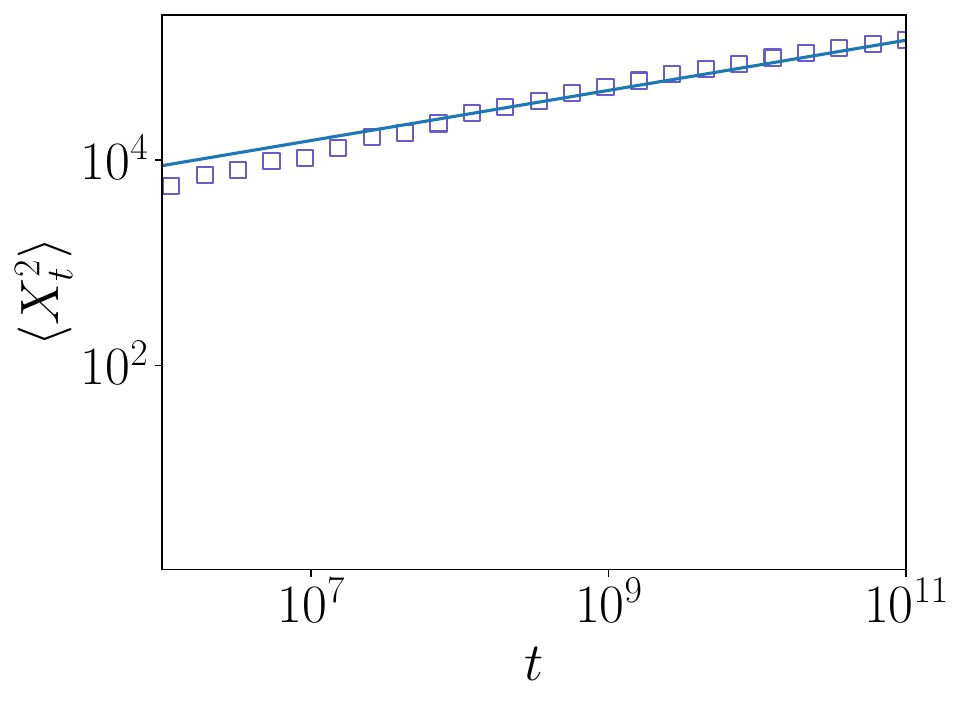}
    \caption{MSD of a LARW with $\tau(a)=(1+a)^{-\alpha}$. Left: decelerating case $\alpha=-3/4<-2/3$ with $\langle x^2(t) \rangle \sim \sqrt{t}$; right: accelerating case $\alpha=3$ with $\langle x^2(t) \rangle \sim t^{2/\dw}$. Symbols denote simulation data, while full lines represent theoretical predictions from \eqref{msd-decel-recap}.}
    \label{fig:msd-1d}
\end{figure}

Summarizing, the MSD for decelerating LARWs exhibits two different regimes depending on the value of $\alpha$:
\begin{boxedeq}
\begin{equation}
    \label{msd-decel-recap}
    \langle x^2(t) \rangle \propto \begin{cases}
        t^{2/\dw} = t^{\frac{2+2\alpha}{2+\alpha}}, & -\frac{2}{3}<\alpha, \\[0.5em]
        \sqrt{t}, & -1 < \alpha \leq -\frac{2}{3}.
    \end{cases}
\end{equation}
\end{boxedeq}
We note that the result \eqref{msd-decel-recap} also holds in the accelerating case $\alpha > 0$. Numerical verification is presented in Fig.~\ref{fig:msd-1d}.
\paragraph{(iii) First-passage properties.}  
From Eqs.~\eqref{dag-propag} and \eqref{dag-propag-rw}, we obtain for a free LARW (i.e., in the regime $\tau(a) \ll a$):
\begin{equation}
\label{eq:Pdag_xi_a}
    \mathcal{P}^\dag(0|0,\xi_a) \To{a s \ll 1} 2 \Omega,
\end{equation}
which we interpret as the expected number of returns to the hotspot before the RWer is absorbed at the target site $\Omega$. 

Using Eq.~\eqref{a-at-fpt}, the joint Laplace-transformed density of first-passage time $t$ and activation $a$ is given by
\begin{equation}
\label{eq:FPT_laplace}
    \hat{F}(\Omega,a|0,s) \Sim{a s \ll 1} -e^{-\sqrt{2\tau(a) s}|\Omega|} \partial_a \exp\left( -\int_0^a \frac{db}{2\Omega \tau(b)} \right).
\end{equation}
Inverting this expression with respect to $s$ yields
\begin{keyboxedeq}[FPT density of the 1d LARW]
\begin{equation}
\label{eq:FPT_real}
    F(\Omega,a|0,t) \Sim{a \ll t} \frac{e^{-\frac{\tau(a) \Omega^2}{2t}}}{\sqrt{8\pi \tau(a) t^3}} \exp\left( -\int_0^a \frac{db}{2\Omega \tau(b)} \right).
\end{equation}
\end{keyboxedeq}

For the physical choice $\tau(a) = (1+a)^{-\alpha}$, the tail of the FPT density can be obtained by integrating \eqref{eq:FPT_real} over $a$, while neglecting the term $e^{-\frac{\tau(a) \Omega^2}{2t}}$ for large $t$. This is justified since only values $a \ll t$ contribute significantly, and $\tau(a) \ll t$ in this regime.%
\footnote{This approximation is validated numerically: the neglected term has no effect on the exponent $\theta = 1/2$ nor on the prefactor.}

We then obtain the exact asymptotic form of the FPT density to site $\Omega$:
\begin{equation}
\label{eq:FPT_density}
    F(\Omega|0,t) \Sim{t \to \infty} \frac{e^{\frac{1}{2\alpha\Omega + 2\Omega}} E_{\frac{\alpha}{2\alpha + 2}}\left( \frac{1}{2\alpha\Omega + 2\Omega} \right)}{\sqrt{2\pi} (2\alpha + 2) t^{3/2}},
\end{equation}
\begin{figure}
    \centering
    \includegraphics[width=.48\textwidth]{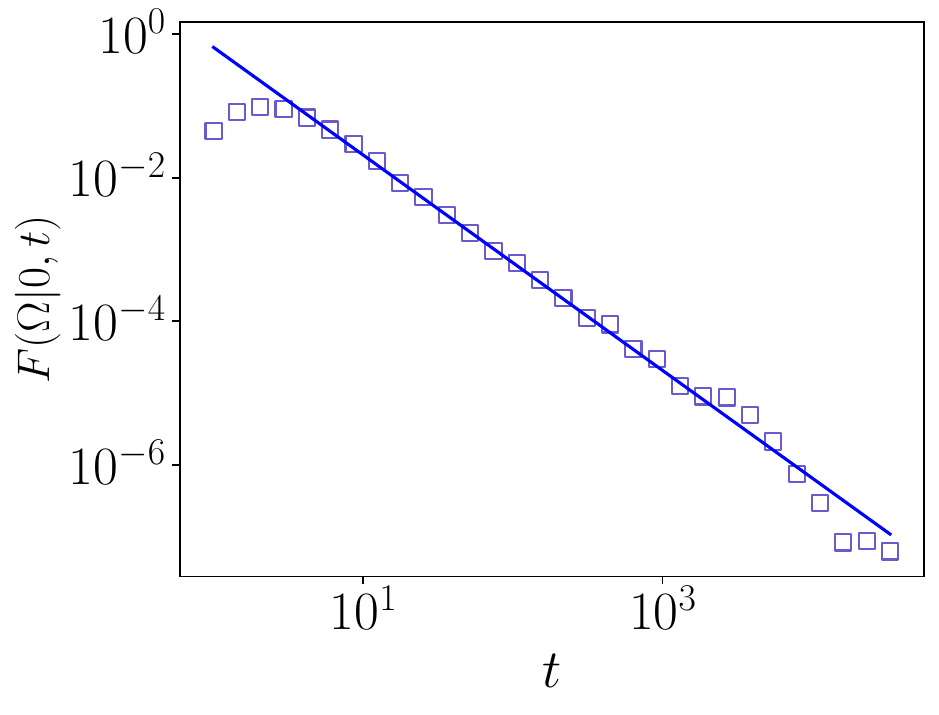}
    \includegraphics[width=.48\textwidth]{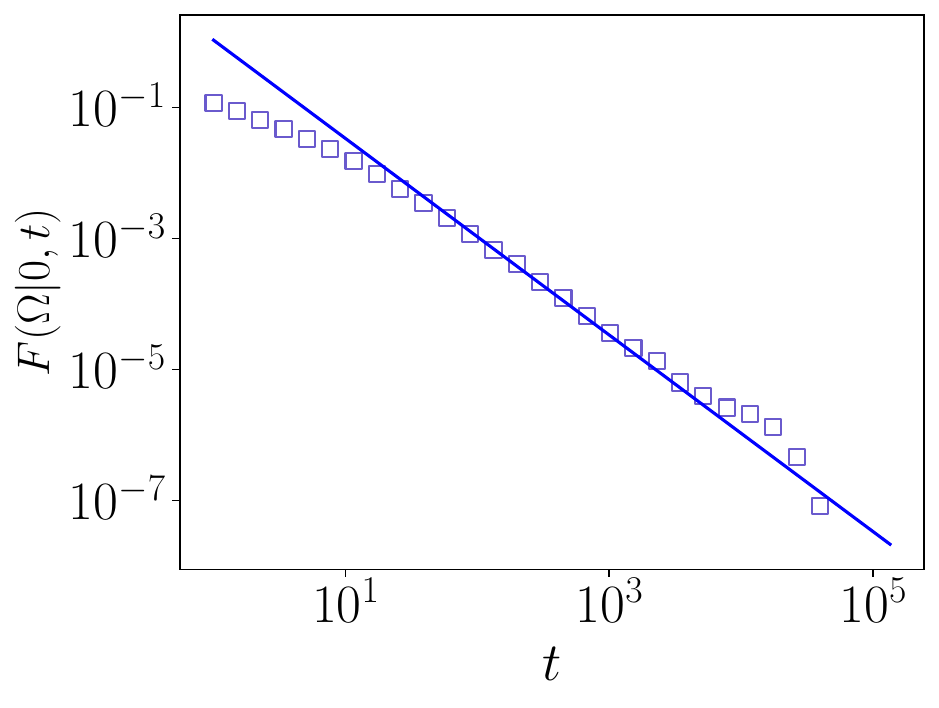}
    \caption{FPT density of a LARW to site $\Omega$ with $\tau(a)=(1+a)^{-\alpha}$. Left: $\alpha=3, \Omega=5$, right: $\alpha=-1/3,\Omega=1$. Symbols come from numerical simulations while the full line comes from the asymptotics \eqref{eq:FPT_density}.}
    \label{fig:fpt-1d}
\end{figure}
where $E_\nu$ is the exponential integral function. This expression is confirmed numerically in Fig.~\ref{fig:fpt-1d}. Remarkably, the persistence exponent remains $\theta = 1/2$, as in the case of the non-activated RWer. However, the prefactor in \eqref{eq:FPT_density} exhibits a nontrivial dependence on the target distance $\Omega$.

In the small-$\Omega$ regime, the tail simplifies to
\begin{equation}
\label{fpt-small-omega}
    F(\Omega|0,t) \sim \frac{\Omega}{\sqrt{2\pi t^3}},
\end{equation}
which matches the exact result for the standard RW. This is physically expected: when the target is close to the hotspot at $x = 0$, the activation remains small and the LARW behaves as a simple RW.

Interestingly, \eqref{fpt-small-omega} violates the scaling law $F(\Omega|0,t) \propto \Omega^{d_w \theta}$ predicted in \cite{levernierUniversalFirstpassage} for strictly scale-invariant RWs. This indicates that LARWs are not scale-invariant as full stochastic processes. Indeed, although the one-point function of an accelerating LARW with $\tau(a) = (1+a)^{-\alpha}$ and $\alpha > 0$ satisfies a scaling form:
\begin{equation}
    P_{1d}(x,t) \sim \frac{1}{t^{1/d_w}} f\left( \frac{x}{t^{1/d_w}} \right), \quad d_w = \frac{2 + \alpha}{1 + \alpha},
\end{equation}
this does not imply scale invariance of the full process.

If the LARW were truly scale-invariant in the sense that $(X_t)_{t \geq 0} \overset{d}{=} (\lambda^{1/d_w} X_{t/\lambda})_{t \geq 0}$ for all $\lambda > 0$, we would expect the FPT density to scale as $F(\Omega|0,t) \propto \Omega^{d_w \theta} = \Omega^{\frac{2 + \alpha}{2 + 2\alpha}}$, which is clearly not observed. This breakdown in scaling behavior reflects the existence of multiple competing timescales in the LARW dynamics, beyond the dominant scale $t^{1/d_w}$—a phenomenon also evident in the multiscaling form \eqref{propag-1pa-decelerating}.

Moreover, the classical relation $\theta = 1 - \frac{1}{d_w}$—which holds for scale-invariant random walks with stationary increments \cite{krugPersistenceExponents,levernierUniversalFirstpassage}—does not apply in this context. This breakdown motivates a closer examination of the increments of LARWs, with the aim of showing their non-stationarity and aging behavior.

\paragraph{(iv) Aging properties.}  
We now examine the aging behavior of the $1d$ LARW by analyzing the aged mean squared increments:
\begin{equation}
    \langle (X_{T+t} - X_T)^2 \rangle.
\end{equation}
For a scale-invariant process—such as the $1d$ LARW with $\tau(a) = a^{-\alpha}$—this quantity is expected to follow the scaling form
\begin{equation}
    \label{aged-incr}
    \langle (X_{T+t} - X_T)^2 \rangle \propto T^{\beta} t^{2/d_w^0},
\end{equation}
in the aging regime $1 \ll t \ll T$, where $d_w^0$ and $\beta$ are exponents characterizing the non-stationary dynamics of the LARW.

These exponents are related to the walk dimension $d_w = \frac{2+\alpha}{1+\alpha}$ (see Eq.~\eqref{dw-1d}) through the relation \cite{levernierUniversalFirstpassage,regnierRecordAges}
\begin{equation}
    \label{rel-dw-dw0}
    d_w^0 = \frac{1}{\frac{1}{d_w} - \frac{\beta}{2}}.
\end{equation}
During the short time window $t \ll T$, the RWer does not accumulate significant additional activation, so the hopping rate at time $T + t$ remains close to that at time $T$. Thus, over this interval, the process behaves approximately as a standard RW with constant waiting time, leading to a diffusive scaling with $d_w^0 = 2$. Substituting this into \eqref{rel-dw-dw0} gives the explicit expression for the aging exponent:
\begin{boxedeq}
\begin{equation}
    \label{agedincr-exp-1d}
    \boxed{
    \beta = \frac{\alpha}{2 + \alpha}.
    }
\end{equation}
\end{boxedeq}

This result aligns with physical expectations: for accelerating LARWs ($\alpha > 0$), the hopping rate increases over time, so increments grow with the age $T$, leading to $\beta > 0$. Conversely, in the decelerating case ($\alpha < 0$), older trajectories experience smaller hopping rates, and increments shrink with $T$, giving $\beta < 0$.

\begin{figure}
    \centering
    \includegraphics[width=.45\textwidth]{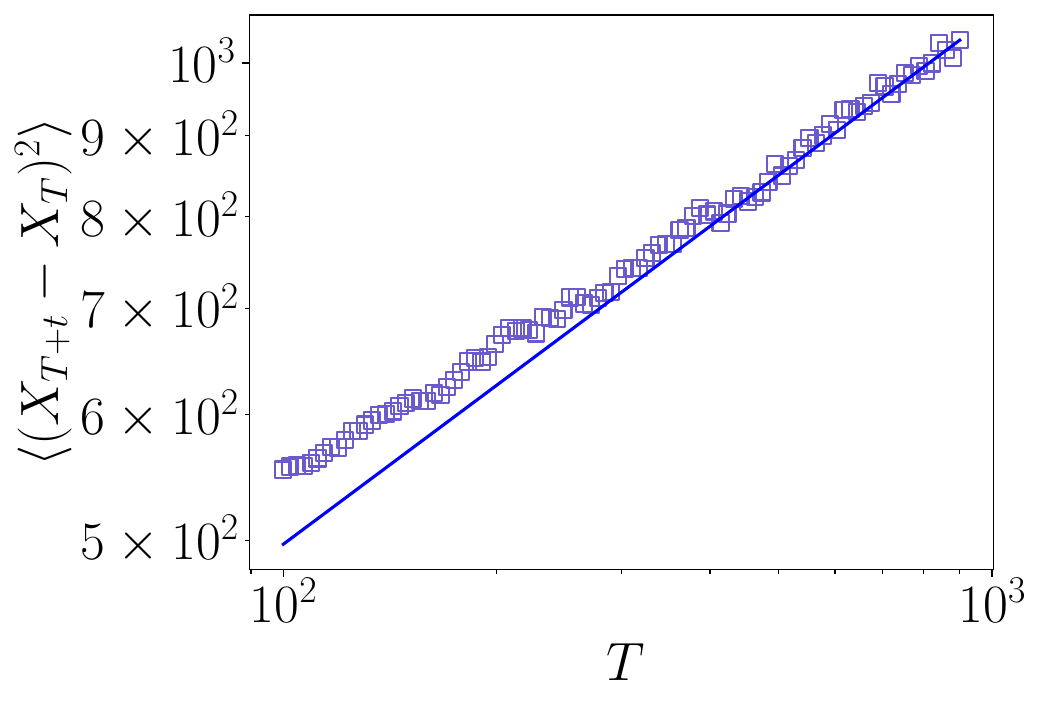}
    \includegraphics[width=.45\textwidth]{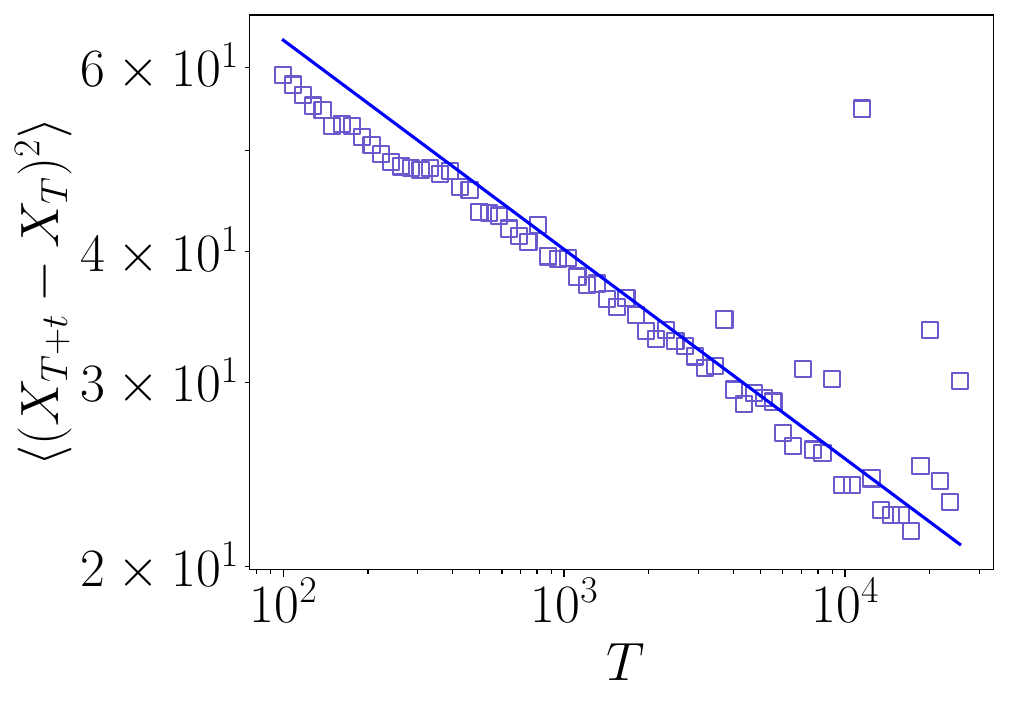}
    \caption{Aged mean squared increments $\langle (X_{T+t} - X_T)^2 \rangle$ for a $1d$ LARW with $\tau(a) = a^{-\alpha}$, shown for $\alpha = 1$ (left) and $\alpha = -1/3$ (right). Symbols denote numerical simulations; the full line corresponds to the theoretical scaling $T^{\beta}$ with $\beta$ given by \eqref{agedincr-exp-1d}. Here $t = 100$ is fixed.}
    \label{fig:aged-incr-1d}
\end{figure}

This prediction is confirmed numerically in Fig.~\ref{fig:aged-incr-1d}, which shows excellent agreement between simulation data and the theoretical scaling $T^{\beta}$.

We now turn to what is arguably the most physically relevant case of LARWs: the two-dimensional ($2d$) LARW.

\subsubsection{The $2d$ LARW}
\begin{figure}
    \centering
    \includegraphics[width=0.52\textwidth]{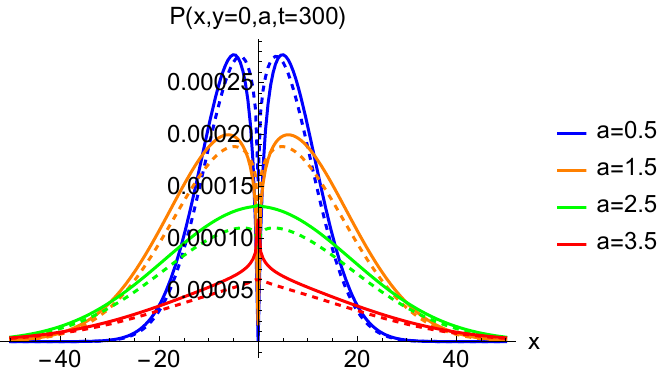}
    \includegraphics[width=0.44\textwidth]{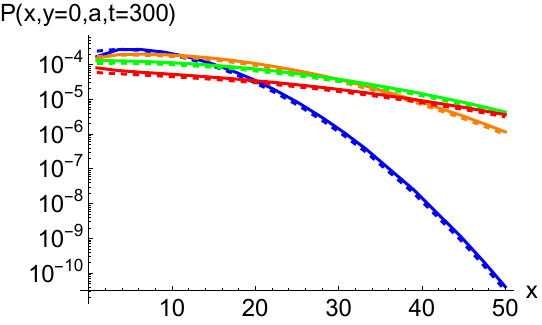}
    \caption{Joint distribution of activation $a$ (color) and position (along $x$-axis with $y=0$) for a $2d$ accelerating LARW with $\tau(a) = a^{-1}$ at time $t = 300$. Thick lines: numerical Laplace inversion of the exact joint law \eqref{jointlaw_2d_laplace_asymp}. Dashed lines: asymptotic form \eqref{jointlaw_2d_asymp}. The threshold activation is $a_*(t) \sim \sqrt{\log t} \approx 2.4$. The plot on the right shows the tails of the joint distribution.}
    \label{fig:theoretical_jointlaw}
\end{figure}

As is often the case in RW problems, the $2d$ case is the most involved and technical one. Eqs.  \eqref{jointlaw_final} together with \eqref{propagator_hypercubic} provide the exact Laplace transform
\begin{equation}
\label{fulljoint_square}
\hat{P}_{2d}(x,a,s) = 
\frac{\pi \int_0^\infty e^{-u} I_{x_1}\left(\frac{\xi_a u}{2}\right)I_{x_2}\left(\frac{\xi_a u}{2}\right) du}{2K(\xi_a^2)} \exp\left(-\frac{\pi}{2}\int_0^a \frac{db}{\tau(b)\xi_bK(\xi_b^2)}\right)
\end{equation}
where $K(k)=\int_0^{\pi/2} \frac{d\theta}{\sqrt{1 - k^2 \sin^2 \theta}}$ is the elliptic integral.
In addition, making use of  \cite{hughes1995random}, the following asymptotics can be obtained in the physically relevant regime $a s \ll 1$:
\begin{boxedeq}
\begin{equation}
\label{jointlaw_2d_laplace_asymp}
    \hat{P}_{2d}(x,a,s) \Sim{s a \ll 1} 2K_0(2\sqrt{s\tau(a)}|x|)\frac{\exp(- \frac{\pi G(a)}{\log \frac{8}{\tau(a)s}})}{\log \frac{8}{\tau(a)s}}, \quad G(a) \equiv \int_0^a \frac{db}{\tau(b)}
\end{equation} 
\end{boxedeq}
While there is a priori no simple Laplace inversion of this expression,  numerical inversion can be performed and provides the joint law for all values of parameters (see Fig.~\ref{fig:theoretical_jointlaw}, solid curves). We outline here a procedure for analytically inverting the Laplace transform \eqref{jointlaw_2d_laplace_asymp} in the regime $t \to \infty, a \ll t, x \gg \sqrt{t \tau(a)^{-1}}$, which yields the tail distribution. 
\paragraph{(i) A technical result on Laplace inversion.}  
We begin with a useful result for Laplace inversion applicable to a specific class of Laplace transforms. Let $\hat{f}(s)$ be such that $\hat{f}(s) = o(1/s)$ as $s \to 0$, and assume that $s = 0$ is its only singularity. For instance, one may consider $\hat{f}(s) = s^{-\alpha}$ with $\alpha < 1$. Additionally, assume that $\hat{f}(\overline{z}) = \overline{\hat{f}(z)}$ and that $\hat{f}$ has a branch cut along the negative real axis.

The Bromwich integral for Laplace inversion is given by:
\[
f(t) = \frac{1}{2\pi i} \int_{a - i\infty}^{a + i\infty} \hat{f}(s) e^{st} ds.
\]
To evaluate this, we deform the contour by closing it through a large arc in the left half-plane, avoiding the branch cut via a keyhole contour along the negative real axis. As $R \to \infty$ and the inner circle radius $\varepsilon \to 0$, only the integrals above and below the cut contribute, leading to the identity:
\begin{equation}
\label{theory_inverselaplace}
    f(t) = -\frac{1}{\pi} \int_0^\infty e^{-ut} \Im \hat{f}(-u + i0^+) \, du.
\end{equation}

\paragraph{(ii) Joint distribution of position $x$ and activation $a$.}  
We now apply the inversion formula \eqref{theory_inverselaplace} to the Laplace-transformed joint law \eqref{jointlaw_2d_laplace_asymp}, obtaining:
\[
P(|x, a, t) \Sim{a \ll t} -\frac{1}{\pi} \int_0^\infty e^{-ut} \Im \left[ \frac{K_0(2i|x| \sqrt{u \tau(a)})}{-i\pi + \log \frac{8}{u \tau(a)}} \exp\left( -\frac{\pi G(a)}{-i\pi + \log \frac{8}{u \tau(a)}} \right) \right] du.
\]
Changing variables to $v = \sqrt{u \tau(a)}$, this becomes:
\[
P(|x, a, t) \Sim{t \to \infty} -\frac{2}{\pi^2} \Im \int_0^\infty e^{-\frac{v^2 t}{\tau(a)}} v K_0(2i |x| v) \partial_a \exp\left( -\frac{\pi G(a)}{\log \frac{8}{v^2} - i\pi} \right) dv.
\]
The dominant contributions come from values of $v$ such that $v^2 t / \tau(a)$ is moderate, i.e., $v \sim \sqrt{\tau(a)/t}$. For $x \sqrt{\tau(a)/t} \gg 1$, the asymptotic form $K_0(2i |x| v) \sim \frac{\sqrt{\pi} e^{-2i v |x|}}{2 \sqrt{i v |x|}}$ applies, yielding:
\[
P(|x, a, t) \sim -\frac{1}{\pi^{3/2} \sqrt{|x|}} \Im \left( \frac{1}{\sqrt{i}} \int_0^\infty e^{ -\frac{v^2 t}{\tau(a)} - 2i v |x| } \sqrt{v} \, \partial_a \exp\left( -\frac{\pi G(a)}{\log \frac{8}{v^2} - i\pi} \right) dv \right).
\]
Completing the square in the exponential, we write $u = v \sqrt{\frac{t}{\tau(a)}} + i |x| \sqrt{\frac{\tau(a)}{t}}$
\begin{equation}
    \label{jointlaw_2d_contourshift}
    \begin{aligned}
    P(x, a, t) \sim 
    & -\frac{1}{\pi^{3/2} \sqrt{|x|}} \exp\left( -\frac{|x|^2 \tau(a)}{t} \right) \sqrt{ \frac{\tau(a)}{t} } \\
    & \times \Im \left[ \frac{1}{\sqrt{i}} \int_{i |x| \sqrt{\frac{\tau(a)}{t}}}^{\infty + i |x| \sqrt{\frac{\tau(a)}{t}}} 
    e^{-u^2} \sqrt{v(u)} \, \partial_a \exp\left( -\frac{\pi G(a)}{\log \frac{8}{v(u)^2} - i\pi} \right) du \right].
    \end{aligned}
\end{equation}
This integral \eqref{jointlaw_2d_contourshift} is dominated by small values of $u$ : therefore, $v^2 \sim -\frac{\tau(a)^2 |x|^2}{t^2} = \frac{\tau(a)^2 |x|^2}{t^2} e^{-i\pi}$, because we approach this value from the lower half-plane. This simplifies the $-i\pi$ in the denominator of the $\exp$ factor, and we are left with a simple Gaussian integral. We finally obtain the joint distribution in the regime $a \ll t$, $x \gg \sqrt{t / \tau(a)}$:
\begin{keyboxedeq}[Joint distribution of activation and position for the 2d LARW]
\begin{equation}
\label{jointlaw_2d_asymp}
P_{2d}(x, a, t) \sim -\frac{\tau(a)}{\pi t} \exp\left( -\frac{|x|^2 \tau(a)}{t} \right) \partial_a \exp\left( -\frac{\pi G(a)}{\log \left( \frac{8 t^2}{\tau(a)^2 |x|^2} \right)} \right).
\end{equation}
\end{keyboxedeq}

This asymptotic expression exhibits excellent agreement with numerical Laplace inversion in the regime $x \gg \sqrt{t/\tau(a)}$ (see Fig.~\ref{fig:theoretical_jointlaw}(b)). Of note, different behaviors emerge depending on whether $a$ is smaller or larger than a threshold value $a_*(t)$. For $a < a_*(t)$, trajectories have few visits to the hotspot, and $P_{2d}(x, a, t)$ exhibits a minimum at $x = 0$, corresponding to effective repulsion. For $a > a_*(t)$, the opposite occurs: frequent hotspot visits lead to a maximum at $x = 0$.

To estimate this threshold activation for $\tau(a) = a^{-\alpha}$, we analyze the sign of $\partial_x P_{2d}(x, a, t)$ at $x = 1$, using \eqref{jointlaw_2d_asymp}. Assuming $a_*(t) \propto \log^\delta t$, we write $t = e^u$ and expand:
\begin{align}
    \partial_x P_{2d}(x, a, t)_{|x = 1} < 0 &\iff \pi u^{1 + \delta(2\alpha + 1)} - 4(1 + \alpha)^2 u^{2 + \alpha \delta} < 0 \quad \text{as } u \to \infty \\
    &\iff \delta < \frac{1}{1 + \alpha},
\end{align}
which confirms the scaling:
\begin{equation}
\label{threshold-a-2d}
a_*(t) \propto \log^{\frac{1}{1 + \alpha}} t.
\end{equation}
This scaling matches the typical activation growth derived in \eqref{typical_a}. As we will see in the next paragraph, integrating out $a$ to obtain the marginal density of $x$ reveals that this sharp crossover behavior leads to effective repulsion (for $\alpha > 0$) or attraction (for $\alpha < 0$) of the RWer by the hotspot—a general feature observed in all recurrent LARWs.

\paragraph{(iii) Marginal distribution of the position $x$.}  
We now determine the marginal distribution of the position at time $t$ in the two-dimensional accelerating case. We consider $\tau(a) = a^{-\alpha}$ with $\alpha > 0$. Starting from the asymptotic joint distribution $P(x, a, t)$, valid in the regime $x \sqrt{\tau(a)/t} \gg 1$, $a \ll t$, we write:
\begin{align}
    P_{2d}(x, a, t) &\sim -\frac{\tau(a)}{\pi t} \exp\left( -\frac{x^2 \tau(a)}{t} \right) \partial_a \exp\left( - \frac{\pi G(a)}{\log \left( \frac{8 t^2}{\tau(a)^2 x^2} \right)} \right) \\
    &= \frac{\left((\alpha +1) \log \left( \frac{2 \sqrt{2} t a^\alpha}{x} \right) - \alpha \right) 
    \exp\left( -\frac{x^2 a^{-\alpha}}{t} - \frac{\pi a^{\alpha + 1}}{(\alpha + 1) \left(2 \log \left( \frac{2 \sqrt{2} t a^\alpha}{x} \right)\right)} \right)}{2(\alpha + 1) t \log^2\left( \frac{2 \sqrt{2} t a^\alpha}{x} \right)}.
\end{align}

To compute the marginal distribution, we proceed as in the $1d$ case and perform the change of variables $a = u x^{\frac{2}{1 + 2\alpha}}$, which leads to a Laplace-type integral:
\begin{equation}
\label{propag2d-sm}
    P_{2d}(x, t) \sim x^{\frac{2}{1 + 2\alpha}} \int_0^\infty du \, g(x, t, u) \exp\left( -x^{\frac{2 + 2\alpha}{1 + 2\alpha}} \phi(u) \right),
\end{equation}
where the exponent function and prefactor are given by
\begin{align}
    \phi(u) &= \frac{\pi u^{\alpha + 1}}{2(\alpha + 1) \log \left( \frac{2 \sqrt{2} t (u x^{\frac{2}{1 + 2\alpha}})^\alpha}{x} \right)} + \frac{1}{t u^\alpha}, \\
    g(x, t, u) &= \frac{(\alpha + 1) \log \left( \frac{2 \sqrt{2} t (u x^{\frac{2}{1 + 2\alpha}})^\alpha}{x} \right) - \alpha}
    {2(\alpha + 1) t \log^2 \left( \frac{2 \sqrt{2} t (u x^{\frac{2}{1 + 2\alpha}})^\alpha}{x} \right)}.
\end{align}

The function $\phi(u)$ reaches its minimum at a saddle point $u_*(t)$ satisfying
\begin{equation}
\phi'(u_*) = 0 \iff \frac{\pi u_*^\alpha}{2 \log \left( \frac{2 \sqrt{2} t (u_* x^{\frac{2}{1 + 2\alpha}})^\alpha}{x} \right)} 
\left( 1 - \frac{\alpha}{(1 + \alpha) \log \left( \frac{2 \sqrt{2} t (u_* x^{\frac{2}{1 + 2\alpha}})^\alpha}{x} \right)} \right) 
= \frac{\alpha u_*^{1 - \alpha}}{t}.
\end{equation}

While this equation cannot be solved exactly, it can be approximated for large $t$ by expanding in the logarithm. Under the assumption $t x^{-1/(2\alpha + 1)} \gg 1$, we find
\begin{equation}
    u_* \sim \left( \frac{2\alpha}{\pi t} \log \left( \frac{2 \sqrt{2} t (u_* x^{\frac{2}{1 + 2\alpha}})^\alpha}{x} \right) \right)^{\frac{1}{1 + 2\alpha}}.
\end{equation}

Substituting this into \eqref{propag2d-sm} and applying a saddle-point approximation (neglecting $\log \log$ corrections), we obtain:
\begin{equation}
\label{saddlepoint-2d-sm}
    P_{2d}(x, t) \Sim{x \gg \sqrt{t}} \frac{A}{t} \left( \frac{x}{\sqrt{t}} \right)^{\frac{1 - \alpha}{2\alpha + 1}} 
    \exp\left( -\frac{B \left(x^2 / t \right)^{\frac{1 + \alpha}{1 + 2\alpha}}}
    {\log^{\frac{\alpha}{1 + 2\alpha}} \left( C \left(t^{1 + \alpha} / x \right)^{\frac{1}{1 + 2\alpha}} \right)} \right),
\end{equation}
where
\begin{align}
    A &= \frac{ \sqrt{ \left( \frac{2}{\pi} \right)^{\frac{1 - \alpha}{2\alpha + 1}} \alpha^{\frac{\alpha + 2}{2\alpha + 1}} / (2\alpha + 1) } \left( l - \frac{\alpha}{\alpha + 1} \right)}{l^{\frac{7\alpha + 2}{4\alpha + 2}}}, \\
    B &= \frac{ \left( \frac{\pi}{2\alpha} \right)^{\frac{\alpha}{2\alpha + 1}} (2\alpha + 1) }{\alpha + 1}, \\
    C &= 2 \sqrt{2} \left( \left( \frac{2}{\pi} \right)^\alpha \alpha^\alpha \right)^{\frac{1}{2\alpha + 1}},
\end{align}
and $l = \log \left( C \left( \frac{t^{1+\alpha}}{x} \right)^{\frac{1}{1+2\alpha}} \right)$.

Unlike the $1d$ case, the propagator \eqref{saddlepoint-2d-sm} is not scale-invariant. This breakdown stems from the fact that the typical activation scales as $a(t) \sim \log^{1/(1+\alpha)} t$ rather than as a power of $t$.

In the decelerating regime ($-1 < \alpha < 0$), we choose the more physical waiting time $\tau(a)=(1+a)^{-\alpha}$ and apply a boundary Laplace method as in the $1d$ case. We finally obtain \footnote{The choice $\tau(a)=(1+a)^{-\alpha}$ does not modify the tails of the propagator in the accelerating case $\alpha>0$.}
\begin{keyboxedeq}[Propgator of the 2d LARW]
\begin{equation}
\label{marginal2d}
P_{2d}(x, t) \Sim{x \gg \sqrt{t}} 
\begin{cases}
\frac{A}{t} \left( \frac{x}{\sqrt{t}} \right)^{\frac{1 - \alpha}{2\alpha + 1}} 
\exp\left( -\frac{B (x^2 / t)^{\frac{1 + \alpha}{1 + 2\alpha}}}{\log^{\frac{\alpha}{1 + 2\alpha}} \left( C (t^{1 + \alpha} / x)^{\frac{1}{1 + 2\alpha}} \right)} \right), & \alpha > 0 \\[0.8em]
\frac{\exp( - x^2 / t )}{2|\alpha| x^2 \log(t / x)}, & -1 < \alpha < 0.
\end{cases}
\end{equation}
\end{keyboxedeq}
This result, validated numerically (see Fig.~\ref{fig:propag_2d}), shows that the distribution is strongly non-Gaussian, as in the $1d$ case. Moreover, for accelerating processes ($\alpha > 0$), the distribution reaches its maximum at a nonzero displacement $r_*(t)$, which increases with time.

\subsubsection{LARW on the $3d$ lattice.}  
We now consider the LARW on the $3d$ cubic lattice. In contrast to the $d=1,2$ cases, the standard ($\alpha = 0$) random walk in $d=3$ is transient, so that the return probability is strictly less than one. In particular, for $s \to 0$, the propagator satisfies
\begin{equation}
    P(0|0, \xi_a) \To{as \ll 1} \frac{1}{1 - R},
\end{equation}
where $R = R_3 \simeq 0.34\dots$ is the return probability to the origin on the cubic lattice \cite{hughes1995random}. 

This feature permits an explicit Laplace inversion in the limit $x \to \infty$, $a \ll t$, with $x^2 \tau(a)/t$ fixed. Using Eq.~\eqref{jointlaw_final}, and assuming $\tau(a) \ll a$ as $a \to \infty$, we find:
\begin{equation}
\label{jointlaw_3d_asymp}
P_{3d}(x,a,t) \sim (1 - R) \sqrt{\tau(a)} \left( \frac{3}{2\pi t} \right)^{3/2} 
\exp\left( -\frac{3 |x|^2 \tau(a)}{2t} - (1 - R) G(a) \right).
\end{equation}

This result contrasts sharply with the $d=1,2$ cases: the joint distribution is now Gaussian in $x$ for fixed $a$, and is always maximized at the origin, independently of the choice of $\tau(a)$. 

To obtain the marginal distribution, we consider the physical example $\tau(a) = (1 + a)^{-\alpha}$ and evaluate the integral over $a$ in the large-$x$ regime using the saddle-point method. This yields:
\begin{keyboxedeq}[Propagator of the 3d LARW]
\begin{equation}
\label{marginal3d}
P_{3d}(x, t) \Sim{x \gg \sqrt{t}}
\begin{cases}
\displaystyle \frac{A}{t^{3/2}} \left( \frac{x}{\sqrt{t}} \right)^{\frac{1 - 2\alpha}{1 + 2\alpha}} 
\exp\left( -B \left( \frac{x^2}{t} \right)^{\frac{1 + \alpha}{1 + 2\alpha}} \right), & \alpha > 0, \\[1em]
\displaystyle \left( \frac{3}{2\pi t} \right)^{3/2} \frac{2t(1 - R)}{3 |\alpha| x^2} \exp\left( -\frac{3 x^2}{2t} \right), & -1 < \alpha < 0,
\end{cases}
\end{equation}
\end{keyboxedeq}
where $A, B$ are constants. This distribution exhibits a diffusive scaling $x^2 \propto t$, despite being non-Gaussian for $\alpha \ne 0$. Remarkably, even a single localized perturbation—a hotspot at the origin—is sufficient to generate non-Gaussianity in a transient random walk.

\medskip

To elucidate the physical origin of the saddle-point behavior in $P(x,t)$, we consider the accelerated case $\tau(a) = 1/a$ as a representative example. Our goal is to estimate the typical number of returns to the hotspot made by a trajectory that ends at position $x$ at time $t$.

If the RWer returns $n$ times to the hotspot, the waiting time at the $(n+1)$-th visit satisfies $\tau_n \sim 1/a_n$, where $a_{n+1} \sim a_n + \tau_n \sim a_n + 1/a_n$, which implies $a_n \sim \sqrt{n}$. Therefore, the total time needed to complete $n$ such back-and-forth returns is:
\begin{equation}
    T_n \sim \sum_{k=1}^n \tau_k \sim \sqrt{n}.
\end{equation}

Let $p_n(x,t)$ denote the probability of reaching $x$ at time $t$ after exactly $n$ returns to the origin. The contribution from such a trajectory has the form:
\[
p_n(x,t) \asymp K^n \exp\left( -\frac{x^2 \tau_n}{2(t - T_n)} \right),
\]
where $K = 1/6$ is the probability for each return loop in $d=3$. At the exponential level, this gives:
\[
p_n(x,t) \asymp \exp\left( -\varphi_{x,t}(n) \right), \quad \text{with} \quad \varphi_{x,t}(n) \sim \frac{x^2}{2t \sqrt{n}} + n \log \left( \frac{1}{K} \right).
\]

Minimizing $\varphi_{x,t}(n)$ with respect to $n$ yields the typical number of returns:
\begin{equation}
    n_{\text{typ}}(x, t) \sim \frac{x^{4/3}}{t^{2/3}}.
\end{equation}

The corresponding contribution to the propagator is:
\[
P(x,t) \asymp \exp\left( -\frac{x^2 \tau_{n_{\text{typ}}}}{2(t - T_{n_{\text{typ}}})} \right) 
\asymp \exp\left( -A \frac{x^{4/3}}{t^{2/3}} \right),
\]
which reproduces the scaling found in Eq.~\eqref{marginal3d} from the saddle-point computation.

This interpretation highlights that the dominant contribution to the tail of $P(x,t)$ originates from rare but structured paths: the RWer first makes an optimal number of returns to the origin—thus increasing activation—then exits to large distance $x$ with small waiting times. This mechanism underlies the emergence of non-Gaussian behavior even in a transient walk.

\section{Extension to persistent LARWs}
Here, we extend our results to LARWs with persistence, defined by the following transition probabilities. The RWer performs nearest-neighbor jumps on the lattice. After a jump in direction $\sigma$, it jumps again in the same direction $\sigma$ with probability $\frac{1+\varepsilon(2d-1)}{2d}$, and in one of the remaining $2d-1$ directions with probability $\frac{1-\varepsilon}{2d}$. For $\varepsilon > 0$, the walk is persistent, favoring repeated jumps in the same direction; for $\varepsilon < 0$, it is anti-persistent. Local activation is modeled by allowing both the jump rate $1/\tau(a)$ and the persistence parameter $\varepsilon_a$ to depend on the activation $a$.

\subsection{Ergodicity breaking}

In contrast to the non-persistent LARW, the persistent LARW can exhibit three distinct dynamical regimes. Two of them—free and trapped—also occur in the non-persistent case, with the trapping criterion still given by $\tau(a) \gg a$. However, a third regime—specific to the persistent LARW—emerges: an \emph{ergodicity-breaking} phase, in which the walker undergoes an infinite sequence of back-and-forth excursions between the hotspot and one of its neighboring sites. This regime is so named to emphasize that, although the walker is not strictly trapped, it fails to explore space and thus violates ergodicity.
As we proceed to show, this regime arises when the rescaled waiting time $\tilde{\tau}(a) \equiv \frac{1-\varepsilon_a}{1+\varepsilon_a} \tau(a)$ satisfies $\tilde{\tau}(a) \gg a$.

To determine the conditions under which ergodicity breaking occurs, we consider a one-dimensional persistent LARW with $\tau(a) = \frac{1}{a^\beta}$ and $\varepsilon(a) = -1 + \frac{1}{a^\gamma}$. Then the effective waiting time becomes $\tilde{\tau}(a) = \frac{1-\varepsilon(a)}{1+\varepsilon(a)}\tau(a) \sim 2a^{\gamma - \beta}$. Let $k$ denote the number of returns to the hotspot, and let $a_k$ be the typical activation level after $k$ returns. The recurrence relation reads $a_{k+1} \sim a_k + \tau(a_k) = a_k + \frac{1}{a_k^\beta}$, which leads to $a_k \propto k^{\frac{1}{1+\beta}}$.

There exists a nonzero probability that the walker gets trapped in an anti-persistent loop between $x=0$ and $x=\pm1$ if the product $\prod_k \frac{1 - \varepsilon(a_k)}{2}$ is strictly positive. This is equivalent to the convergence of the sum $\sum_k \frac{1}{a_k^\gamma}$, which occurs for $\gamma > 1 + \beta$, since $a_k^\gamma \sim k^{\frac{\gamma}{1+\beta}}$. In this regime, $\tilde{\tau}(a) \sim a^{\gamma - \beta} \gg a$, confirming that ergodicity breaking occurs when the rescaled waiting time becomes asymptotically large compared to the activation.

\subsection{Distribution of activation and position}
Here, we demonstrate that our exact expression for the joint distribution of position $x$ and activation $a$, given in \eqref{jointlaw_final}, has applications beyond symmetric nearest-neighbor RWs. For simplicity, we focus on the one-dimensional persistent RW on $\mathbb{Z}$ \cite{ernstRandomWalks}, which serves as a discrete-space analog of the classical run-and-tumble model for active particles. Nevertheless, our analysis generalizes to $d$-dimensional persistent RWs.

\begin{figure}
    \centering
        \begin{subfigure}{.49\textwidth}
            \includegraphics[width=\textwidth]{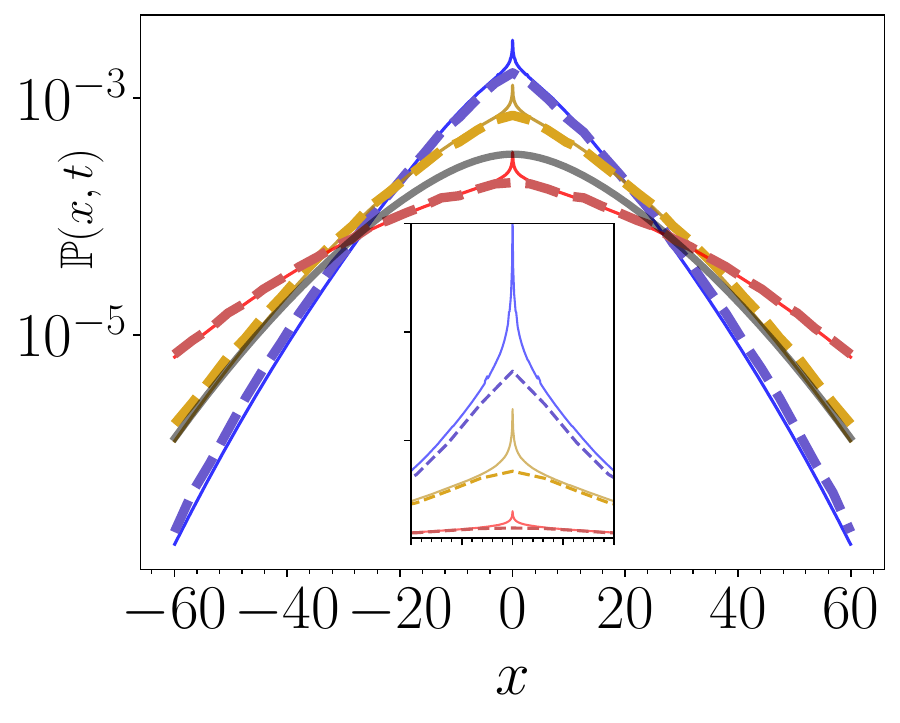}
            \begin{picture}(0,0)
            \put(-45,160){(a)}
            \end{picture}
            \label{dec_2d}
            \vspace*{-5pt}
        \end{subfigure}
        \begin{subfigure}{.49\textwidth}
            \includegraphics[width=\textwidth]{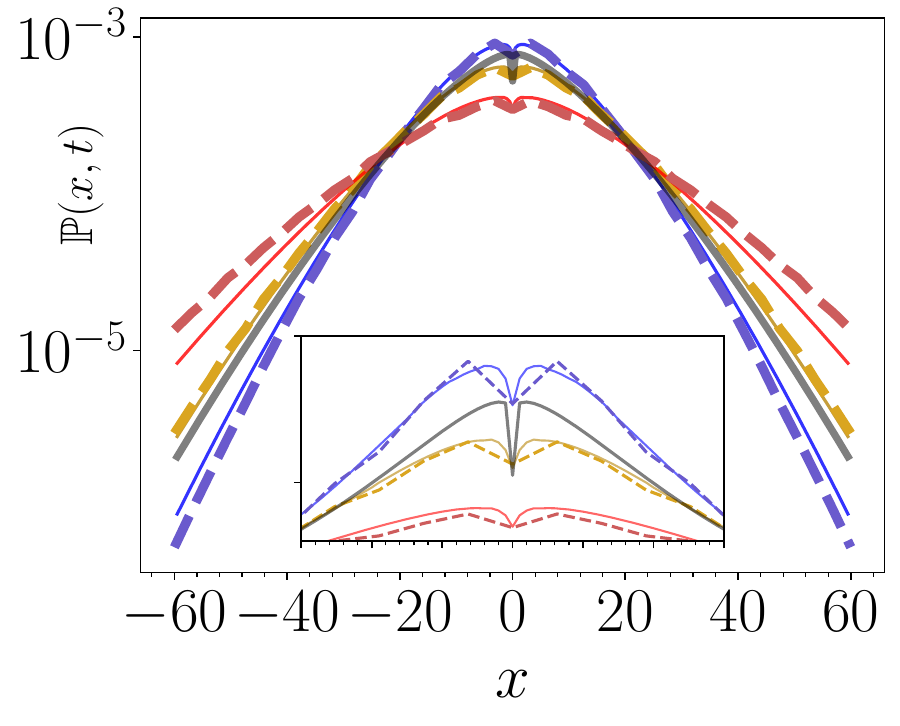}
            \begin{picture}(0,0)
            \put(-45,160){(b)}
            \end{picture}
            \vspace*{-5pt}
        \end{subfigure}
        \caption{Spatial distribution for a persistent 2d LARW with a single hotspot at the origin with persistence parameter $\varepsilon$, along a given axis $x$.  Blue, yellow and red curves correspond respectively to $\varepsilon=-0.3,0,0.4$. Grey thick lines correspond to the saddle-point approximations \eqref{marginal2d} and are drawn for $\varepsilon=0$. Thick lines are obtained from numerical integration over $a$ of the asymptotic expression \eqref{jointlaw_2d_asymp} with rescaled waiting time $\tilde{\tau}(a) = \frac{1-\varepsilon}{1+\varepsilon}\tau(a)$, while dashed lines correspond to numerical simulations. (a) Decelerating LARW, with $\tau(a) = \sqrt{1+a}$. Note the cusp at site $x=0$. (b) Accelerating LARW, with $\tau(a) = \frac{1}{a}$. Note the nonzero typical displacement scaling logarithmically with time.}
        \label{fig:propag_2d}
    \end{figure}
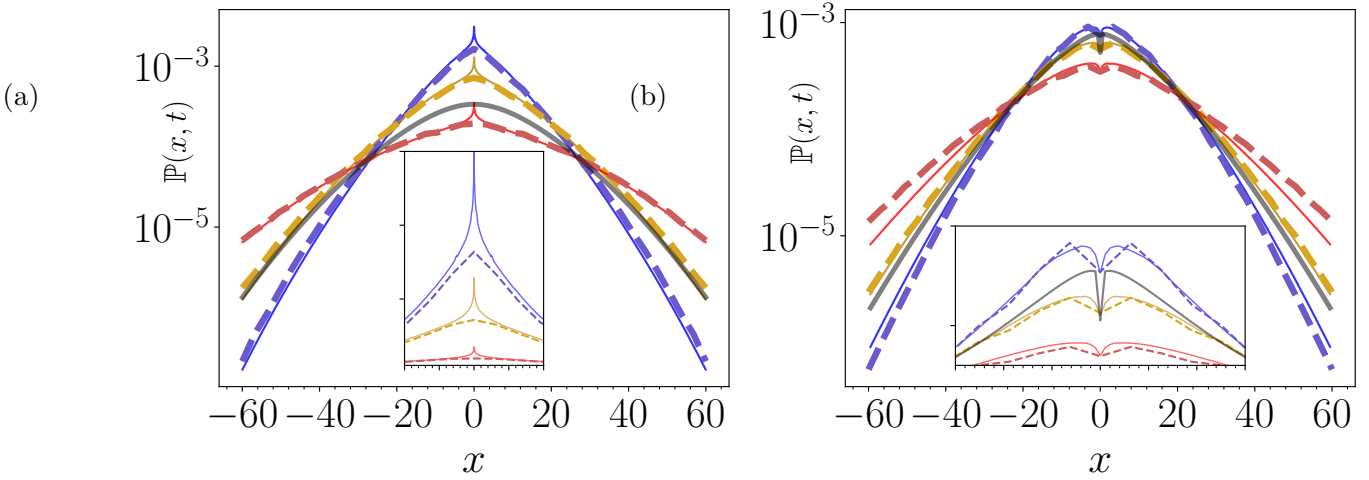

We begin by computing the propagator of the persistent, non-activated RWer on $\mathbb{Z}$. The probability to repeat the last jump direction is $\frac{1+\varepsilon}{2}$, corresponding to the 'forward jump model' in \cite{ernstRandomWalks}. The associated lattice Green's function in Fourier space reads
\begin{equation}
    \mathcal{P}(q, \xi) = \sum_{n=0}^\infty \xi^n \sum_{x \in \mathbb{Z}} e^{i q x} P(x,n) dx = \frac{1 - \varepsilon \xi \cos q}{1 + \varepsilon \xi^2 - \xi(1+\varepsilon)\cos q} = \frac{1}{1 +\varepsilon \xi^2} \frac{1-\varepsilon \xi \cos q}{1 - \alpha \cos q}, \quad \alpha = \frac{\xi(1+\varepsilon)}{1+\varepsilon \xi^2}
\end{equation}
Using the identity $\frac{1}{1-\alpha \cos q} = \int_0^\infty du\, e^{-u(1-\alpha \cos q)}$ (valid for $\alpha<1$) and the trigonometric formula $\cos(nq) \cos q = \frac{1}{2}[\cos((n+1)q) + \cos((n-1)q)]$, we derive the inverse Fourier transform in terms of modified Bessel functions:
\begin{align}
    \label{persistent-exact-sm}
    \mathcal{P}(x=n, \xi) &= \frac{1}{2\pi}\int_{-\pi}^{\pi} \mathcal{P}(q,\xi) \cos(n q)\, dq \\
    &= \frac{1}{1+\varepsilon \xi^2} \int_{0}^{+\infty} du\, e^{-u} \left[I_n(\alpha u) - \frac{\varepsilon \xi}{2} \big(I_{n+1}(\alpha u) + I_{n-1}(\alpha u)\big)\right].
\end{align}
After some algebra (performed using Mathematica), we obtain the generating function of the propagator for the persistent RW with $x \neq 0$ (noting that the $x=0$ case has a distinct form when $\varepsilon \neq 0$):
\begin{equation}
    \label{propagator-1d-persistent-sm}
    \frac{P(x, \xi)}{P(0,\xi)} = \frac{\left(\frac{\alpha }{\sqrt{1-\alpha ^2}+1}\right)^{|x|} (\xi_a  \varepsilon_a -\alpha)}{\left(1-\sqrt{1-\alpha ^2}\right) \xi_a  \varepsilon_a -\alpha }, \quad P(0,\xi) = \frac{1-\varepsilon }{\left(1-\xi ^2\right) \varepsilon -\sqrt{\left(1-\xi ^2\right) \left(1-\xi ^2 \varepsilon ^2\right)}}, \quad \alpha = \frac{\xi(1+\varepsilon)}{1+\varepsilon \xi^2}
\end{equation}
We check that this expression reduces to the propagator of the simple RW in the case $\varepsilon = 0$. Using \eqref{jointlaw_final}, we obtain the exact Laplace transform of the joint distribution of $x$ and $a$:
\begin{keyboxedeq}[Joint distribution of position and activation of the 1d persistent LARW]
\begin{equation}
    \label{p_joint_1d_persistent}
    \hat{P}(x\neq 0, a, s) = \frac{\left(\frac{1-\sqrt{1-\alpha^2}}{\alpha}\right)^{|x|} (\xi_a  \varepsilon_a -\alpha)}{\left(1-\sqrt{1-\alpha ^2}\right) \xi_a  \varepsilon_a -\alpha } \exp\left(-\int_0^a \frac{db}{\tau(b)\xi_b} \frac{\sqrt{\left(1-\xi_b ^2\right) \left(1-\varepsilon^2_b \xi_b ^2\right)}-\varepsilon_b(1- \xi_b ^2)}{1-\varepsilon_b}\right)
\end{equation}
\end{keyboxedeq}
where $\alpha = \frac{\xi_a (\varepsilon_a +1)}{\xi_a^2 \varepsilon_a +1}$.
We now use \eqref{p_joint_1d_persistent} to derive the propagator for the persistent $1d$ LARW, choosing $\tau(a) \ll a$. Expanding \eqref{propagator-1d-persistent-sm} in the limit $as \ll 1$ we find for $x\neq 0$
\begin{equation}
    \label{small-s-persistent-sm}
    \hat{P}(x,a,s) \Sim{as \ll 1} \exp \left(- \left[x(1-\varepsilon(a)) + \varepsilon(a)\right] \sqrt{\frac{2s \tau(a)}{(1+\varepsilon(a))(1-\varepsilon(a))}} -\sqrt{2s} \int_0^a da' \sqrt{\frac{1+\varepsilon(a')}{\tau(a')(1-\varepsilon(a'))}}\right)  
\end{equation}
Eq. \eqref{small-s-persistent-sm} invites us to introduce the rescaled waiting time $\tilde{\tau}(a) \equiv \frac{1-\varepsilon_a}{1+\varepsilon_a} \tau(a)$, which yields
\begin{equation}
    \label{small-s-persistent-sm-tilde}
    \hat{P}(x,a,s) \Sim{as \ll 1} \exp\left(-x \sqrt{2s \tilde{\tau}(a)} - \frac{\varepsilon_a}{1-\varepsilon_a}\sqrt{2s \tilde{\tau}(a)} -\sqrt{2s} \int_0^a da' \frac{1}{\sqrt{\tilde{\tau}(a')}}\right) 
\end{equation}
Finally, Laplace inversion of \eqref{small-s-persistent-sm-tilde} gives
\begin{equation}
\label{large-t-persistent-sm}
    P(x,a,t) \Sim{a \ll t} \frac{\sqrt{\tilde{\tau}(a)}\left(x + \frac{\varepsilon_a}{1-\varepsilon_a}\right) + \int_0^a da' \frac{1}{\sqrt{\tilde{\tau}(a')}}}{\sqrt{2\pi t^3}} \exp\left(- \frac{\left (\sqrt{\tilde{\tau}(a)}\left(x + \frac{\varepsilon_a}{1-\varepsilon_a}\right) + \int_0^a da' \frac{1}{\sqrt{\tilde{\tau}(a')}}\right)^2 }{2t}\right)
\end{equation}
We highlight the key feature of \eqref{large-t-persistent-sm}. Defining the persistence length as $L_p(a) \equiv \frac{1}{1 - \varepsilon_a}$, we observe that \eqref{large-t-persistent-sm} reduces, in the regime $x \gg \varepsilon_a L_p(a)$, to the joint distribution of a \textit{non-persistent} LARW with an effective waiting time $\tilde{\tau}(a)$. Consequently, when $L_p(a)$ remains bounded—i.e., $\varepsilon_a$ does not approach $1$ as $a$ increases—the results established for non-persistent LARWs can be directly extended to the persistent case. This observation suggests that persistence can be absorbed into a renormalization of the waiting time, and we expect this property to hold in arbitrary spatial dimensions $d$. This is supported by numerical evidence in the case $d=2$, as shown in Fig.~\ref{fig:propag_2d}. \par 
We now consider the case of a diverging persistence length, by choosing $\varepsilon_a = 1 - \frac{1}{a^\alpha}$ with $\alpha > 0$, and for simplicity take $\tau(a) = 1$, so that only the persistence depends on $a$. In this case, the effective waiting time becomes $\tilde{\tau}(a) \sim 2 a^{-\alpha}$, and \eqref{large-t-persistent-sm} reduces to
\begin{equation}\label{diverginglp}
    P(x,a,t) \Sim{1\ll a \ll t} \frac{\sqrt{2}(xa^{-\alpha/2} + a^{\alpha/2}) + \frac{a^{1+\alpha/2}}{(1+\alpha/2)\sqrt{2}}}{\sqrt{2\pi t^3}} \exp\left(- \frac{\left(\sqrt{2}(xa^{-\alpha/2} + a^{\alpha/2}) + \frac{a^{1+\alpha/2}}{(1+\alpha/2)\sqrt{2}}\right)^2}{2t} \right)
\end{equation}

For $x \gg L_p(a) = a^\alpha$, i.e., when the length scale $x$ is much larger than the persistence length, we recover the asymptotic behavior of the non-persistent LARW. However, for $x \ll a^\alpha$, this equivalence no longer holds. We now show that this discrepancy does not affect the marginal distribution of $x$. Integrating \eqref{diverginglp} over $a$, we note that the function
\[
a \mapsto xa^{-\alpha/2} + a^{\alpha/2} + \frac{a^{1+\alpha/2}}{(1+\alpha/2)\sqrt{2}}
\]
reaches its minimum, for large $x$, at $a_m \propto x^{\frac{1}{1+\alpha}}$. Define
\begin{equation}
    \phi(u,x) = \frac{\sqrt{2}(2+\alpha + u^{1+\alpha})}{u^{\alpha/2}(2+\alpha)},
\end{equation}
so that, with the change of variable $a = u x^{\frac{1}{1+\alpha}}$ and $u = O(1)$, the marginal distribution becomes
\begin{equation}
    P(x,t) \sim \frac{x^{1/(1+\alpha)}}{\sqrt{2\pi t^3}} \int_0^\infty x^{\frac{\alpha+2}{2+2\alpha}} \phi(u,x) \exp\left(-\phi(u,x)^2 \frac{x^{\frac{\alpha+2}{1+\alpha}}}{2t} \right) du.
\end{equation}
The minimum of $\phi(u,x)$ occurs at $u_*(x) \sim \alpha^{1/(1+\alpha)} + o(1)$ as $x \to \infty$, which remains $O(1)$ with respect to $x$. At leading order,
\begin{equation}
    \phi(u_*) \sim \frac{2 \sqrt{2} \alpha^{-\frac{\alpha}{2 (\alpha +1)}} (\alpha +1)}{\alpha +2}, \quad \phi''(u_*(x)) \sim \frac{\alpha^{\frac{\alpha -2}{2 (\alpha +1)}} (\alpha +1)}{\sqrt{2}}.
\end{equation}
Applying Laplace's method, we obtain the large-$x$ asymptotic behavior of the propagator:
\begin{equation}
    P(x,t) \sim \frac{A x^{\frac{1}{1+\alpha}}}{t} \exp\left(-B \frac{x^{\frac{\alpha+2}{\alpha+1}}}{2t} \right).
\end{equation}
This matches the behavior of the non-persistent LARW \eqref{eq:marginal1d} with waiting time $\tau(a)=2a^{-\alpha}$, with the same constants $A$ and $B$. The reason for this result is that the typical persistence length for a RWer reaching position $x$ is $L_p^*(x) \sim (u_*(x)\, x^{\frac{1}{1+\alpha}})^\alpha \ll x$, meaning the RWer has reached the diffusive regime by performing a large number of tumbling events. Thus, while a diverging persistence length affects the joint distribution of $x$ and $a$, it does not alter the marginal distribution of $x$ in the long-time, large-distance regime.

\chapter{LARWs with multiple hotspots}
Having analyzed the impact of a single hotspot on the dynamics and spatial exploration of LARWs, we now turn to the physically relevant case involving multiple hotspots. In this setting, activation increases at several fixed hotspots, each influencing the RWer's dynamics. As we shall see, the behavior of LARWs in the presence of multiple hotspots is rich, exhibiting new phenomena, yet remains analytically tractable via extensions of the techniques developed for the single-hotspot case.

In the following, we address three key questions:
\begin{enumerate}
    \item What is the long-time behavior of a free LARW in the presence of multiple hotspots?
    \item In the trapped regime, which hotspot captures the LARW in the long-time limit?
    \item How are the dynamics altered when activation is allowed to decay during periods when the RWer is away from hotspots?
\end{enumerate}

\section{$1d$ LARW with multiple hotspots}
We now study LARWs in the presence of multiple hotspots. The activation $a(t)$ is defined as the cumulative time spent on any hotspot up to time $t$, i.e.,
\begin{equation}
    a(t) = \sum_{x_k\ \text{hotspot}} \int_0^t \mathds{1}_{x_k}(x(t'))\,dt'.
\end{equation}

\subsection{Periodic hotspot distribution}
We first consider a $1d$ nearest-neighbor LARW with hotspots periodically arranged with spacing $L$. For analytical tractability, we take a finite number $2N-1$ of hotspots located at $x_k = (-N + k)L$, for $k = 1, \dots, 2N-1$, with the limit $N \to \infty$ in mind. Note that $x_N=0$ is the origin of the LARW.

\subsubsection{Joint distribution of position $x$ and activation $a$}
Using the formalism developed in earlier sections, we derive the joint distribution $P(x,a,t)$ of position $x$ and activation $a$. Extending the reasoning that led to \eqref{single-hotspot-transport-laplace}, we partition over the last visited hotspot. In $1d$, a RWer cannot cross a hotspot without changing its activation $a$, which effectively couples the hotspots as nearest neighbors. Note that each hotspot is also coupled to itself. This leads to the partition:
\begin{equation}
\label{partition-multiple}
\begin{aligned}
    P(x_k, a + dt, t + dt) &= \mathbb{P}(\text{at } x_k \text{ with activation } a \text{ at } t,\ \text{remain at } x_k \text{ for } dt) \\
    &\quad + \sum_{i=1}^{2N-1} \mathbb{P}(\text{from } x_i \text{ at } t' \leq t \text{ with } a+dt\ \text{to } x_k \text{ at } t+dt, \\
    &\qquad \qquad \qquad \text{without visiting any hotspot in } (t', t+dt)).
\end{aligned}
\end{equation}

Expanding \eqref{partition-multiple} for infinitesimal $dt$ and separating the contributions from $x_k$ and the remaining hotspots $x_i$ with $i \neq k$, we obtain the evolution equation:
\begin{equation}
\label{evolution-multiplehot-real}
\begin{aligned}
    \partial_t P(x_k, a, t) + \partial_a P(x_k, a, t) = \frac{1}{\tau(a)}\Bigg[ & -P(x_k, a, t) + \int_0^t \Big[ P(x_k, a, t') F_{L}^{G,a}(t - t') \\
    &+ \frac{1}{2} \sum_{i \neq k} P(x_i, a, t') F_{L}^{D,a}(t - t') \Big] dt' \Bigg],
\end{aligned}
\end{equation}
where $F_{L-1}^{G,a}(t)$ and $F_{L-1}^{D,a}(t)$ are the first-exit time distributions through the left and right boundaries, respectively, for a RWer with hopping rate $\tau(a)$ in an interval of $L-1$ sites.

We now solve \eqref{evolution-multiplehot-real} in Laplace space. Define the $(2N-1)$-dimensional vector:
\begin{equation}
    \ket{\hat{P}}(a,s) \equiv \ket{\hat{P}(x_1, a, s), \dots, \hat{P}(x_{2N-1}, a, s)},
\end{equation}
where each component is the Laplace transform of $P(x_k, a, t)$. The evolution equation becomes:
\begin{equation}
\label{periodic_local_time}
\boxed{
\partial_a \ket{\hat{P}} = \frac{1}{\tau(a)} \left[ \frac{1}{2} \mathcal{F}_{L-1}^D(\xi_a) A - \left( \frac{1}{\xi_a} - \mathcal{F}_{L-1}^G(\xi_a) \right) \mathds{1} \right] \ket{\hat{P}},
}
\end{equation}
with
\begin{equation}
    A \equiv (\delta_{|i-j|,1})_{1 \leq i,j \leq 2N-1}, \quad \xi_a(s) \equiv \left( 1 + s \tau(a) \right)^{-1},
\end{equation}
and
\begin{equation}
    \mathcal{F}_{L-1}^{G/D}(\xi) \equiv \sum_{t=0}^\infty \xi^t F_{L-1}^{G/D}(t),
\end{equation}
where $F_{L-1}^{G/D}(t)$ refers to the exit-time distributions for a simple RWer with hopping rate $\tau(a) = 1$.

As is well known \cite{hughes1995random}, these generating functions admit the closed forms:
\begin{equation}
\label{exit-times-rw}
    \mathcal{F}_{L-1}^G(\xi) = \frac{1}{\xi} - \frac{1}{\tanh(L \mu)} \frac{\sqrt{1 - \xi^2}}{\xi}, \qquad 
    \mathcal{F}_{L-1}^D(\xi) = \frac{1}{\sinh(L \mu)} \frac{\sqrt{1 - \xi^2}}{\xi},
\end{equation}
where $\mu = -\log \left( \frac{1 - \sqrt{1 - \xi^2}}{\xi} \right)$.

Since all matrices in \eqref{periodic_local_time} commute, the differential system can be diagonalized using the common eigenbasis of the Toeplitz matrix $A$. Its eigenvalues are:
\begin{equation}
    \lambda_j = 2\cos\left( \frac{\pi j}{2N} \right), \quad j = 1, \dots, 2N-1,
\end{equation}
with corresponding eigenvectors:
\begin{equation}
\label{eigenvectors}
    \ket{v_j} = \frac{1}{\sqrt{N}} \ket{ \left( \sin\left( \frac{\pi j k}{2N} \right) \right)_{k = 1, \dots, 2N-1} }.
\end{equation}

Since the matrix $A$ is symmetric, its eigenvectors form an orthonormal basis of $\mathbb{R}^{2N-1}$. Using the notation $\ket{x_k}$ for the Kronecker delta at site $x_k$, we may expand
\begin{equation}
    \label{bases-change}
    \hat{P}(x_k, a, s) = \braket{x_k}{\hat{P}} = \sum_{j=1}^{2N-1} \braket{v_j}{\hat{P}} \braket{x_k}{v_j},
\end{equation}
where the coefficients $\braket{x_k}{v_j}$ are given by the eigenvector formula \eqref{eigenvectors},
\begin{equation}
    \braket{x_k}{v_j} = \frac{1}{\sqrt{N}} \sin\left( \frac{\pi j k}{2N} \right).
\end{equation}

To compute $\braket{v_j}{\hat{P}}$, we project equation \eqref{periodic_local_time} onto the eigenvector $\ket{v_j}$:
\begin{equation}
    \partial_a \braket{v_j}{\hat{P}} = \frac{1}{\tau(a)} \left[ \frac{1}{2} \mathcal{F}_{L-1}^D(\xi_a) \lambda_j - \left( \frac{1}{\xi_a} - \mathcal{F}_{L-1}^G(\xi_a) \right) \right] \braket{v_j}{\hat{P}}.
\end{equation}
The initial condition for this ODE is given by $\braket{v_j}{\hat{P}}(a=0, s) = \braket{v_j}{x_N}$, since the walk starts at the central hotspot $x_N = 0$, implying $\ket{P} = \delta(t)\ket{x_N}$ and hence $\hat{P}(x, a=0, s) = \delta_{x,x_N}$.

Solving the ODE yields
\begin{equation}
    \braket{v_j}{\hat{P}} = \braket{v_j}{x_N} \exp\left( \int_0^a \frac{db}{\tau(b)} \left[ \frac{1}{2} \mathcal{F}_{L-1}^D(\xi_b) \lambda_j - \left( \frac{1}{\xi_b} - \mathcal{F}_{L-1}^G(\xi_b) \right) \right] \right).
\end{equation}

Returning to the original basis using \eqref{bases-change}, and letting $k = N + l$ with $l \in [-(N-1), N-1]$, we obtain
\begin{align}
    \label{calcs-periodic}
    \hat{P}(x_k, a, s) &= \sum_{j=1}^{2N-1} \braket{v_j}{\hat{P}} \braket{x_k}{v_j} \\
    &= \frac{1}{N} \sum_{j=1}^{2N-1} \sin\left( \frac{\pi j k}{2N} \right) \sin\left( \frac{\pi j}{2} \right) \exp\left( \int_0^a \frac{db}{\tau(b)} \left[ \frac{1}{2} \mathcal{F}_{L-1}^D(\xi_b) \lambda_j - \left( \frac{1}{\xi_b} - \mathcal{F}_{L-1}^G(\xi_b) \right) \right] \right). \notag
\end{align}

In the final line, we used the identity
\begin{equation}
    \sin\left( \frac{\pi j k}{2N} \right) \sin\left( \frac{\pi j}{2} \right) =
    \begin{cases}
        0, & j \text{ even}, \\
        \cos\left( \frac{\pi j l}{2N} \right), & j \text{ odd},
    \end{cases}
\end{equation}
to restrict the sum to odd indices. Changing the summation index to $j = 2m + 1$ and taking the $N \to \infty$ limit, we recognize the Riemann sum as the integral representation of the modified Bessel function of the first kind \cite{abramowitz1965handbook}:
\begin{equation}
    I_k(x) = \frac{1}{\pi} \int_0^\pi \cos(k u) e^{x \cos u} du.
\end{equation}
We thus obtain the compact expression
\begin{boxedeq}
\begin{equation}
\label{periodic_joint_law}
\hat{P}(kL, a, s) = e^{-\beta(a,s)} I_k(\alpha(a,s)),
\end{equation}
\end{boxedeq}
where, using the explicit forms \eqref{exit-times-rw} of $\mathcal{F}_{L-1}^{G,D}$,
\begin{equation}
    \label{alpha-beta-larw}
    \alpha(a,s) = \int_0^a \frac{db}{\sinh(L \mu_b)} \frac{\sqrt{1 - \xi_b^2}}{\xi_b \tau(b)}, \qquad
    \beta(a,s) = \int_0^a \frac{db}{\tanh(L \mu_b)} \frac{\sqrt{1 - \xi_b^2}}{\xi_b \tau(b)}.
\end{equation}
We emphasize that Eq.~\eqref{periodic_joint_law} provides an exact and explicit expression—up to Laplace inversion—for arbitrary choices of the waiting time function $\tau(a)$.

As in the single-hotspot case, the full Laplace-transformed joint distribution $\hat{P}(x,a,s)$ is obtained by convolving \eqref{periodic_joint_law} with the propagator $\mathcal{P}^{\dagger}$ of a simple RWer in $[1, L-1]$ with absorbing boundaries. The generating function of this propagator reads \cite{hughes1995random}, for $x \notin L\mathbb{Z}$:
\begin{equation}
    \mathcal{P}^\dagger(x | kL - 1, \xi) =
    \begin{cases}
        2\cosh(\mu) \dfrac{\sinh\left( \mu(x - (k - 1)L) \right)}{\sinh(L\mu)}, & (k - 1)L < x < kL, \\
        0, & \text{otherwise},
    \end{cases}
\end{equation}
with $\mu = \mu(\xi) = -\log\left( \dfrac{1 - \sqrt{1 - \xi^2}}{\xi} \right)$. This leads to the exact expression
\begin{keyboxedeq}[Joint distribution of position and activation for the LARW with periodic hotspots]
\begin{equation}
\label{joint-x-a-multiple}
\hat{P}(x, a, s) = \frac{1}{\sinh(L \mu_a)} \left[ \hat{P}(kL, a, s)\, \sinh\left( \mu_a[(k+1)L - x] \right) + \hat{P}((k+1)L, a, s)\, \sinh\left( \mu_a(x - kL) \right) \right]
\end{equation}
\end{keyboxedeq}
where
\begin{equation}
    \mu_a \equiv \mu(\xi_a) = -\log\left( \frac{1 - \sqrt{1 - \xi_a^2}}{\xi_a} \right).
\end{equation}

\begin{figure}
    \centering
    \includegraphics[width=.9\textwidth]{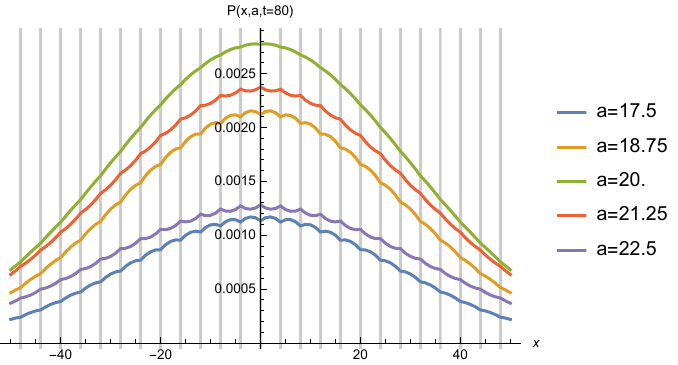}
    \caption{Joint distribution of position and activation for an accelerating LARW with hotspot sublattice $4\mathbb{Z}$ (gray vertical lines), and waiting time $\tau(a) = (1+a)^{-1}$. Values are obtained by numerically inverting \eqref{joint-x-a-multiple}. Activation $a$ is indicated by the color bar. As in the single-hotspot case, there exists a threshold $a_*(t)$ separating hotspot-dominated trajectories ($a > a_*$, convex profile between hotspots) from inter-hotspot-dominated ones ($a < a_*$, concave profile). For large times, this threshold scales as $a(t) \sim t / L$. Here, $t = 80$, yielding $a_* = t / L = 20$.}
    \label{fig:jointlaw-periodic}
\end{figure}

Crucially, the representation \eqref{joint-x-a-multiple} is explicit and well-suited for numerical Laplace inversion. Figure~\ref{fig:jointlaw-periodic} shows the result of such an inversion in the case of an accelerating LARW with periodic hotspots.

\subsubsection{Asymptotic Gaussianity of the free LARW with periodic hotspots}
We can compute the exact marginal distribution of the activation $a$ in the periodic hotspot case by extending the identity \eqref{a-from-p0a}:
\begin{equation}
    P\left(a(t)\geq a\right) = \sum_{k} \int_0^t P(kL, a, t') \, dt'.
\end{equation}
Taking the Laplace transform with respect to $t$ gives:
\begin{equation}
    \mathcal{L}_{t \to s} \, P\left(a(t)\geq a\right) = \frac{1}{s} \sum_k \hat{P}(kL, a, s).
\end{equation}
Using the exact expression \eqref{periodic_joint_law} and the identity $\sum_{k=-\infty}^{\infty} I_k(x) = e^x$ \cite{abramowitz1965handbook}, we find:
\begin{boxedeq}
\begin{equation}
    \label{cum-a-multiple}
    \mathcal{L}_{t \to s} \, P\left(a(t)\geq a\right) = \frac{e^{\alpha(a,s)-\beta(a,s)}}{s} = \frac{\exp\left(-\int_0^a \frac{\left(1 - e^{-L \mu_b}\right)\sqrt{1 - \xi_b^2}}{\xi_b \tau(b) \left(1 + e^{-L \mu_b}\right)} \, db \right)}{s}.
\end{equation}
\end{boxedeq}
We now extract the asymptotic behavior of the activation distribution in the case of \textbf{free} LARWs, i.e., when $\tau(a) \ll a$. Since the hotspots are placed with density $1/L$, we expect that the LARW spends a finite fraction of time at these sites. Therefore, we focus on the regime $a \approx t$, which corresponds to $as \approx 1$. In this limit, we have $s \tau(a) \ll 1$, so:
\begin{equation}
    \xi_a = \frac{1}{1 + s \tau(a)} \sim 1 - s \tau(a), \quad 
    \mu_a = -\log\left(\frac{1 - \sqrt{1 - \xi_a^2}}{\xi_a}\right) \sim \sqrt{2s \tau(a)}.
\end{equation}
Substituting into \eqref{cum-a-multiple}, we find:
\begin{align}
    \label{largetime-a-periodic}
    \mathcal{L}_{t \to s} \, P\left(a(t)\geq a\right) &\sim \frac{\exp\left(-\sqrt{2s} \int_0^a \frac{db}{\sqrt{\tau(b)}} \tanh\left(L \sqrt{\frac{s \tau(b)}{2}}\right)\right)}{s} 
    \sim \frac{e^{-sL}}{s}.
\end{align}
Laplace inverting \eqref{largetime-a-periodic}, we obtain:
\begin{equation}
    \label{a-delta-periodic}
    P\left(a(t) = a\right) \Sim{t \to \infty} \delta\left(a - \frac{t}{L}\right).
\end{equation}
The result \eqref{a-delta-periodic} shows that, at large times, the activation becomes sharply concentrated around the deterministic value $a = t/L$. This is natural from an ergodic perspective. Let $z(t) = x(t) \bmod L$ be the position of the LARW reduced modulo $L$. Since $a(t)$ grows unboundedly, the relative contribution of any single hotspot visit vanishes over time. As a result, the reduced process $z(t)$ becomes uniformly distributed over the $L$ sites, and the RWer spends on average a time fraction $1/L$ at the hotspot $z=0$. Thus, by the ergodic theorem \cite{mooreErgodicTheorem}, we find:
\begin{equation}
    \frac{a(t)}{t} = \sum_k \frac{\int_0^t P(kL, a, t') \, dt'}{t} \To{t \to \infty} P_s(z = 0) = \frac{1}{L},
\end{equation}
which confirms the delta-concentration result \eqref{a-delta-periodic}. \par 
At long times, the dynamics of the LARW across the full lattice effectively reduce to those of a time-changed CTRW with a deterministic, time-dependent waiting time $\tau'(t) = \tau\left( \frac{t}{L} \right)$. Consequently, on large scales, the walker behaves diffusively. More precisely, the renormalized time \eqref{renorm-time} is defined as 
\begin{equation}
    \label{renorm-t}
    \tilde{t}(t) = \int_0^t \frac{dt'}{\tau(a(t'))} \Sim{t \to \infty} \int_0^t \frac{dt'}{\tau(t'/L)},
\end{equation}
where the final asymptotic equivalence uses the fact that $\tau(a) \ll a$. The process $x(t)$ becomes Gaussian with variance $\tilde{t}(t)$, in stark contrast with the non-Gaussian behavior observed for the single-hotspot case. \par 
Finally, we stress that this conclusion extends beyond the periodic, one-dimensional setting: by ergodicity, the same asymptotic behavior holds for a LARW in any dimension with a finite hotspot density $\rho > 0$, such as Poisson-distributed hotspots on $\mathbb{Z}^d$, which we will investigate in details in Section \ref{sec:quenched}.

\subsubsection{Final position of the trapped LARW with periodic hotspots}
We now consider the case of a \emph{trapped} LARW, i.e. $\tau(a) \gg a$. In this regime, the walker is eventually trapped on a hotspot. Our goal is to determine the distribution of the final hotspot visited, denoted by $x_{\text{last}} \equiv k_l L$. This distribution is given by
\begin{equation}
    \label{trap-position-0}
    \mathbb{P}(k_l = k) = \lim_{a \to \infty} \int_0^\infty P(kL, a, t')\, dt'.
\end{equation}
This follows from the fact that the events $\{x(t) = kL,\ a(t) = a\}_{t \geq 0}$ are disjoint, since $a(t)$ increases strictly whenever the RWer is on a hotspot. For any finite $a$, the integral
\[
J_k(a) = \int_0^\infty P(kL, a, t')\, dt'
\]
isolates the unique time $t'$ at which the activation reaches $a$ while the RWer is located at site $kL$. In the trapped regime, the RWer almost surely remains indefinitely on some hotspot, so the large-$a$ limit of $J_k(a)$ yields the probability that the final trapping site is $kL$. The integral $J_k(a=\infty)$ can be computed from \eqref{periodic_joint_law} by plugging in $s=0$. Using the explicit expressions of the functions $\alpha, \beta$ in \eqref{alpha-beta-larw}, we find
\begin{equation}
    \Lim{a \to \infty}\alpha(a,s=0) = \Lim{a \to \infty} \beta(a,s=0) = \frac{1}{L} \int_0^\infty \frac{db}{\tau(b)}.
\end{equation}
Finally, we deduce that the last visited hotspot $k_l$ follows a \emph{Skellam} distribution:
\begin{keyboxedeq}[Distribution of the trapping hotspot]
\begin{equation}
    \label{trap-position}
        \mathbb{P}(k_l = k) = e^{-\frac{1}{L} \int_0^\infty \frac{db}{\tau(b)}} I_k\left(\frac{1}{L} \int_0^\infty \frac{db}{\tau(b)} \right).
\end{equation}
\end{keyboxedeq}

\begin{figure}
    \centering
    \includegraphics[width=.5\textwidth]{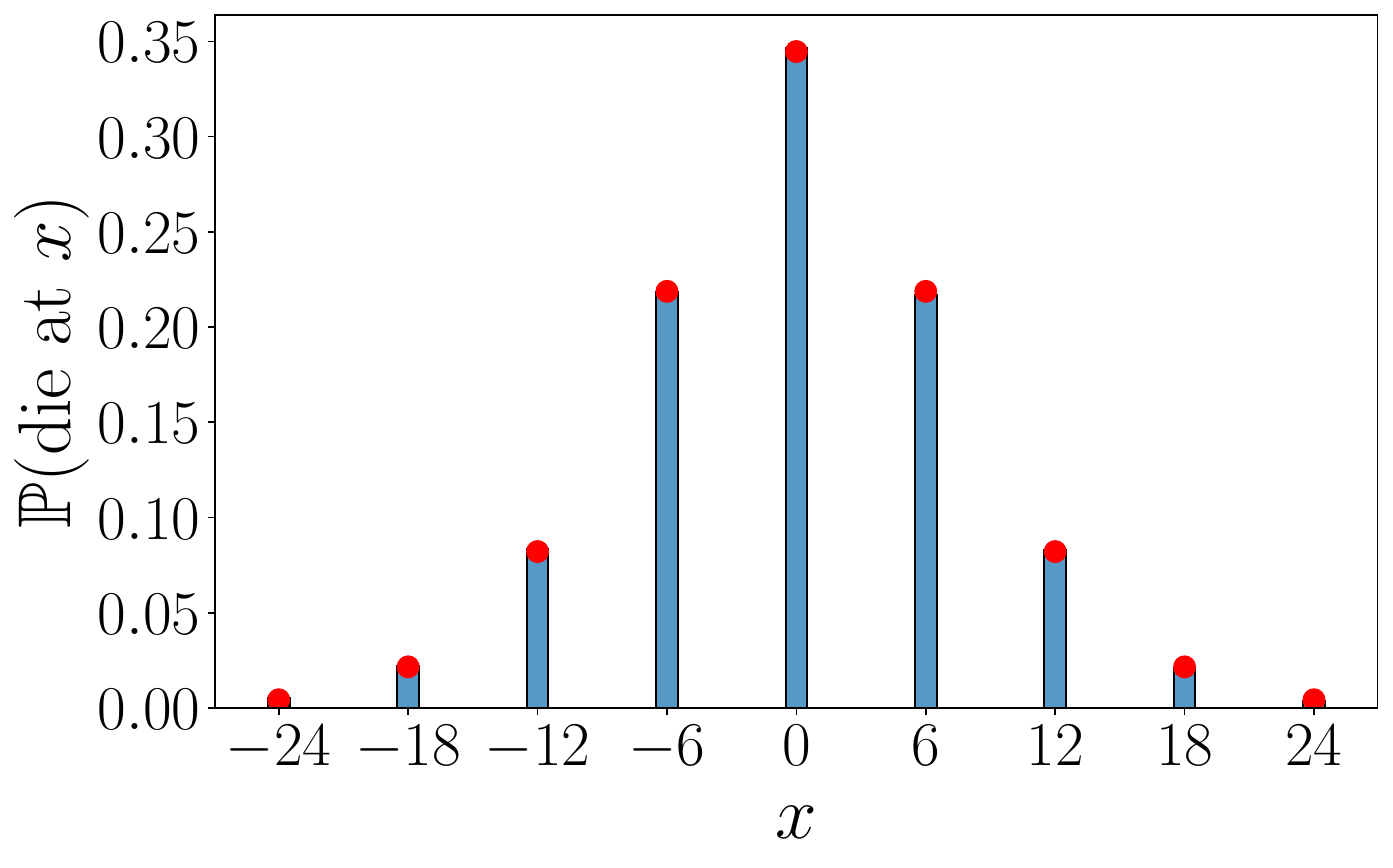}
    \caption{Probability that the walker dies at a hotspot as a function of the hotspot number. Here, there is a hotspot every $5$ sites. We observe that the particle dies over a much larger portion of the lattice in the second case, as its deceleration is much weaker. The red dots are distributed according to \eqref{trap-position} and are in excellent agreement with the numerical simulations.}
    \label{fig:p-die-1d}
\end{figure}
See Fig.~\ref{fig:p-die-1d} for a numerical confirmation of \eqref{trap-position}. \par 
Let us provide a physical argument in the form of an exact mapping to support the Skellam form of the distribution \eqref{trap-position}. Consider first a CTRW $(x_t)$ on $\mathbb{Z}$ with constant jump rate $\gamma$. The probability of finding the RWer at site $n$ at time $t$ is given by \cite{vankampenStochasticProcessesPhysics1992}
\[
P_{\text{CTRW}}(x(t) = n) = e^{-\gamma t} I_n(\gamma t),
\]
which is a Skellam distribution.
For simplicity, we first consider the case $L = 1$, where every site is a hotspot. In this setting, the activation simply grows linearly with time: $a(t) = t$. As in the time-changed CTRW analysis \eqref{renorm-time}, we reparametrize time using the renormalized time
\[
\tilde{t}(t) = G(t) \equiv \int_0^t \frac{dt'}{\tau(t')}.
\]
The trajectory of $y(\tilde{t}) \equiv x(G^{-1}(\tilde{t}))$ then describes a CTRW with jump rate $\gamma = 1$, so that
\begin{equation}
    P(y(\tilde{t}) = k) = e^{-\tilde{t}} I_k(\tilde{t}).
\end{equation}
In the trapped regime, where $\tau(a) \gg a$, the renormalized time converges to a finite limit as $t \to \infty$:
\begin{equation}
    \tilde{t}(t) \To{t \to \infty} \tilde{t}_\infty = \int_0^\infty \frac{dt'}{\tau(t')},
\end{equation}
which corresponds to the final trapping time in renormalized units. After this time, the process $y(\tilde{t})$ is trapped. We thus obtain
\begin{equation}
    \mathbb{P}(k_l = k) = \Lim{t\to \infty}\mathbb{P}(x(t) = k) = \mathbb{P}(y(\tilde{t}_\infty) = k) = e^{-\tilde{t}_\infty} I_k(\tilde{t}_\infty).
\end{equation}
We now extend this reasoning to the case $L > 1$, where only the sites of the sublattice $L\mathbb{Z}$ are hotspots. To determine the final trapping position, we may disregard the segments of the RWer’s trajectory that occur between hotspot visits. Let $t_a$ denote the first time at which the activation reaches a fixed value $a$:
\[
a(t_a) = a.
\]
We define a first coarse-grained process by sampling the RWer’s position only at times when activation increases, i.e., only during visits to hotspots:
\begin{equation}
    z(a) \equiv x(t_a),
\end{equation}
where $t_a$ denotes the first time at which the activation reaches the value $a$. The process $(z(a))_{a \geq 0}$ thus describes the RWer's trajectory restricted to hotspot locations, with activation $a$ acting as the natural time parameter. $z(a)$ thus defines a time-changed CTRW, for which we introduce a renormalized time
\begin{equation}
    \tilde{t}_a = G(a) = \int_0^a \frac{db}{\tau(b)},
\end{equation}
and define the time-rescaled trajectory as
\begin{equation}
    y(\tilde{t}) \equiv z\left( a = G^{-1}(\tilde{t}) \right).
\end{equation}
Crucially, the process $(y(\tilde{t}))_{0\leq \tilde{t} \leq \tilde{t}_\infty}$ is a CTRW on the hotspot sublattice $L\mathbb{Z}$ with \emph{constant} jump rate $\gamma = 1/L$. This rate reflects the splitting probability: upon jumping away from a hotspot, the RWer has probability $1 - \gamma = 1 - 1/L$ of returning to the same hotspot, which does not register as a jump in $y$, and only probability $1/L$ of moving to a different hotspot, which does. In this way, $y(\tilde{t})$ effectively captures the large-scale motion of the RWer through the activated sites, with activation-based renormalized time as the clock. 
Consequently, the distribution of the renormalized process is
\begin{equation}
    \mathbb{P}(y(\tilde{t}) = k) = e^{-\frac{1}{L} \tilde{t}} I_k\left( \frac{1}{L} \tilde{t} \right).
\end{equation}
Taking $t \to \infty$, that is $\tilde{t}=\tilde{t}_\infty$, we recover the distribution of the final trapping hotspot:
\begin{equation}
    \mathbb{P}(k_l = k) = \lim_{t \to \infty} \mathbb{P}(x(t) = k) = \mathbb{P}(y(\tilde{t}_\infty) = k) = e^{-\frac{1}{L} \tilde{t}_\infty} I_k\left( \frac{1}{L} \tilde{t}_\infty \right),
\end{equation}
which is exactly \eqref{trap-position}. In the following section, we use this mapping to obtain the distribution of the trapping hotspot in a disordered environment.

\subsection{$1d$ trapped LARW with quenched hotspot distribution}
\label{sec:quenched}
We now consider a $1d$ LARW in a quenched disordered hotspot landscape, where each site independently has probability $0 < p < 1$ of being a hotspot. For simplicity, we assume that the origin is always a hotspot. Denote by $x_k$ the position of the $k$-th hotspot, so that $x_0 = 0$. Our goal is to compute disorder-averaged observables, i.e., averages over all realizations of the disorder.

We denote by $\overline{O}$ the average over trajectories for a fixed disorder realization, and by $\langle O \rangle$ the average of the observable $O$ over all disorder realizations.

\begin{figure}
    \centering
    \begin{subfigure}{0.45\textwidth}
        \centering
        \includegraphics[width=\textwidth]{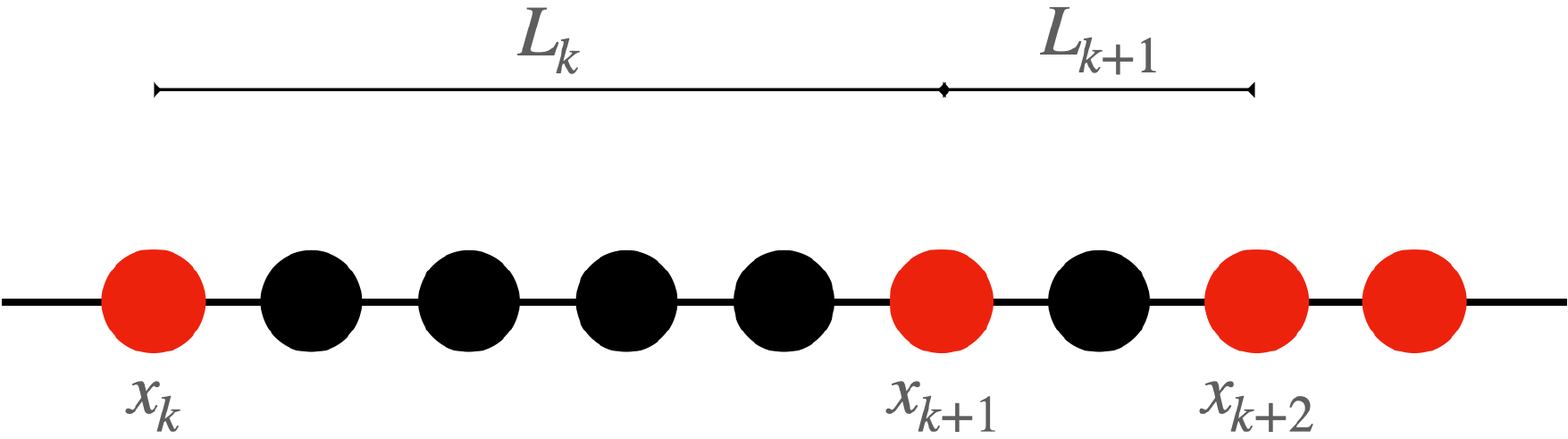}
        \caption{}
        \label{fig:poisson_og}
    \end{subfigure}
    \begin{subfigure}{0.35\textwidth}
        \centering
        \includegraphics[width=\textwidth]{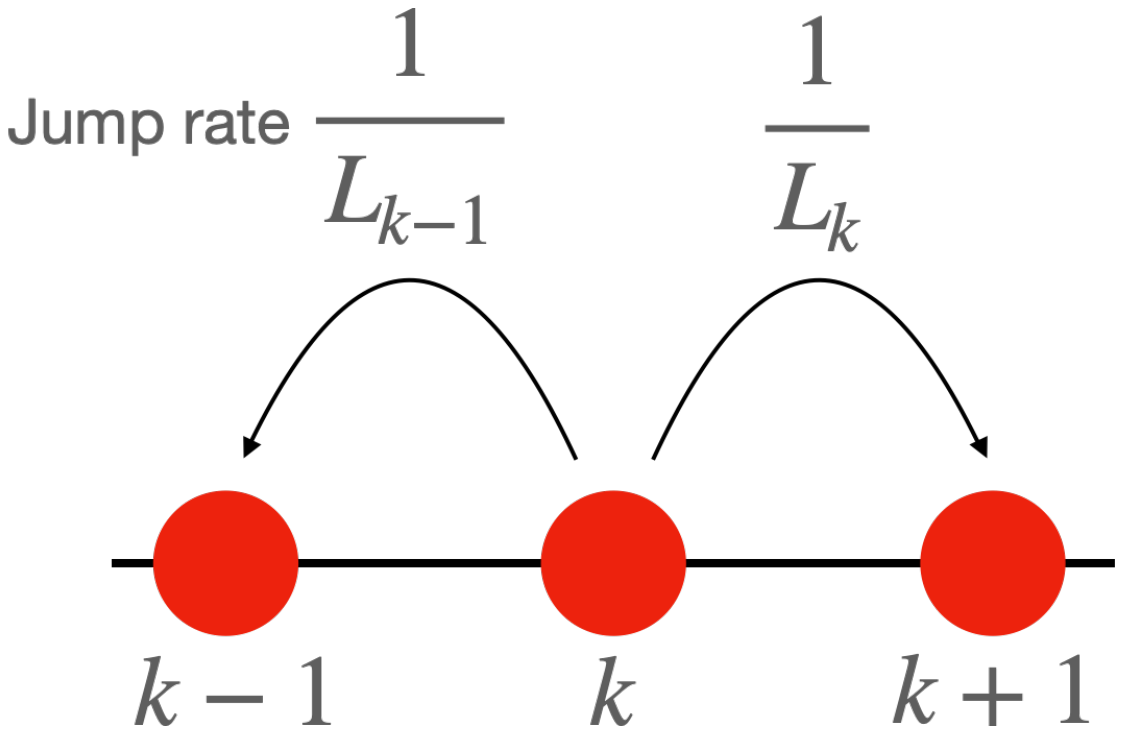}
        \caption{}
        \label{fig:poisson_mapping}
    \end{subfigure}
   \caption{Schematic mapping from the disordered LARW \ref{fig:poisson_og} to the effective CTRW on hotspot indices \ref{fig:poisson_mapping}. Hotspots are marked as filled red circles.}
   \label{fig:mapping-disorder}
\end{figure}

As in the periodic case, we map the dynamics to a renormalized nearest-neighbor CTRW $y(\tilde{t})$ defined on the hotspot indices $k \in \mathbb{Z}$, with renormalized time
\begin{equation}
    \tilde{t} = G(a) = \int_0^a \frac{db}{\tau(b)}.
\end{equation}
A schematic of this mapping is shown in Fig.~\ref{fig:mapping-disorder}. Note that this coarse-graining loses track of the real time $t$, and all observables are expressed as functions of the activation $a$. However, in the free regime $\tau(a) \ll a$, ergodicity ensures that real-time observables (such as $x(t)$) behave like those of a CTRW with constant rate and renormalized time
\begin{equation}
    s(t) = \int_0^t \frac{dt'}{\tau(p t')}.
\end{equation}

Hence, in what follows, we focus on the trapped regime $\tau(a) \gg a$, where ergodicity is broken. In this disordered model, the inter-hotspot distances $L_k \equiv x_{k+1} - x_k$ are i.i.d. and geometrically distributed with mean $1/p$:
\begin{equation}
    \label{geom-lk}
    \mathbb{P}(L_k = l) = p(1 - p)^{l - 1}.
\end{equation}
The splitting probability $\pi_k$ for a simple RWer starting from $x_k$ to reach either neighboring hotspot $x_{k\pm1}$ before returning to $x_k$ is known from standard RW theory:
\begin{equation}
    \pi_k = \frac{1}{2} \left( \frac{1}{L_k} + \frac{1}{L_{k-1}} \right),
\end{equation}
which is random and spatially-dependent. Since the distribution of the $L_k$ is known, so is the distribution of $\pi_k$.

As a result, the effective CTRW $y(\tilde{t})$ has spatially dependent waiting times $\pi_k^{-1}$, in analogy with the uniform waiting time $L$ in the periodic case. The disorder average $T$ of the waiting time is computed explicitly:
\begin{equation}
    T \equiv \langle \pi_k^{-1} \rangle 
    = 2 \sum_{l, l' \geq 1} \frac{1}{\frac{1}{l} + \frac{1}{l'}} \mathbb{P}(L_k = l) \mathbb{P}(L_{k-1} = l') 
    = \frac{2 + p}{3p}.
\end{equation}

\begin{figure}
    \centering
    \begin{subfigure}{0.4\textwidth}
        \centering
        \includegraphics[width=\textwidth]{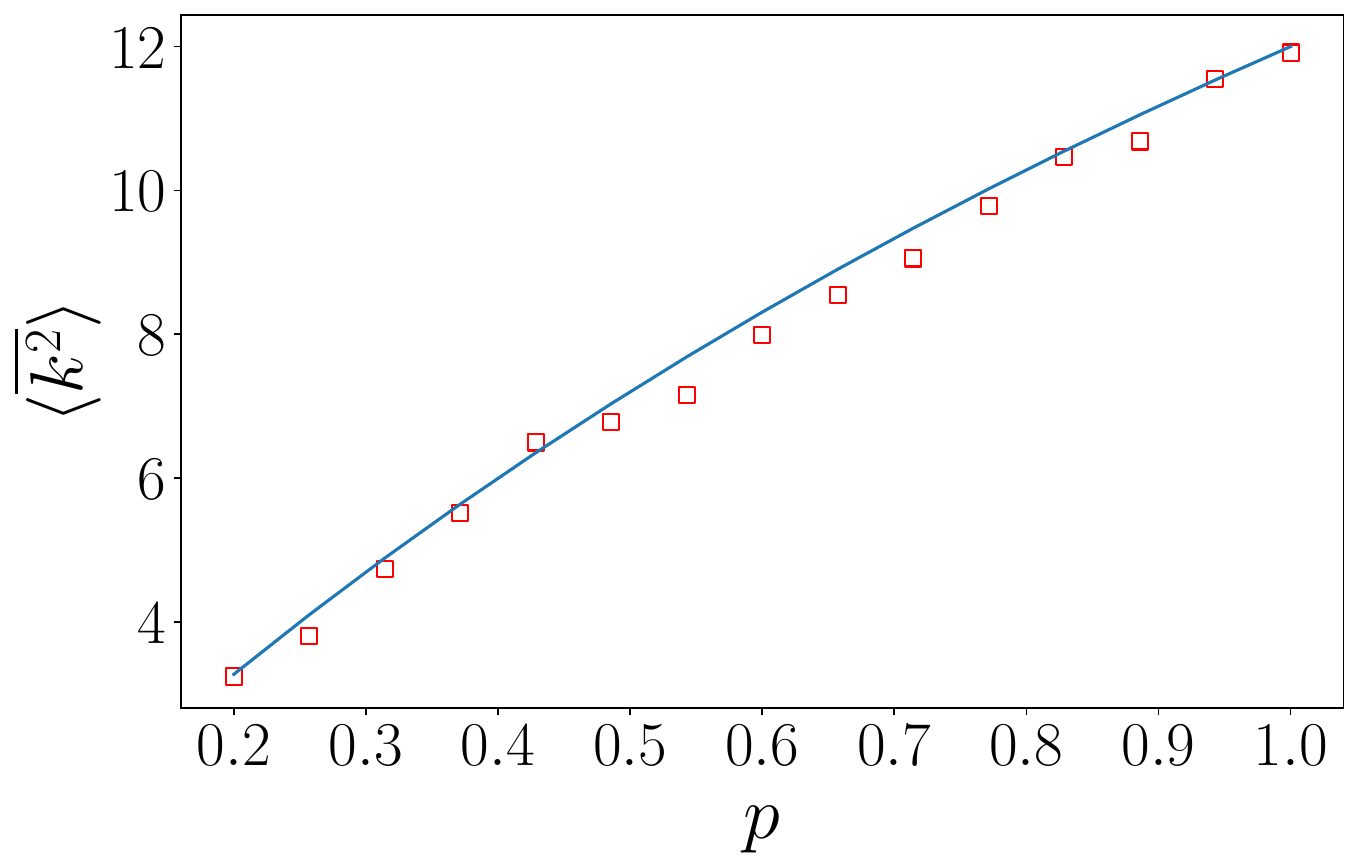}
        \caption{}
        \label{fig:msdk-accel}
    \end{subfigure}
    \begin{subfigure}{0.4\textwidth}
        \centering
        \includegraphics[width=\textwidth]{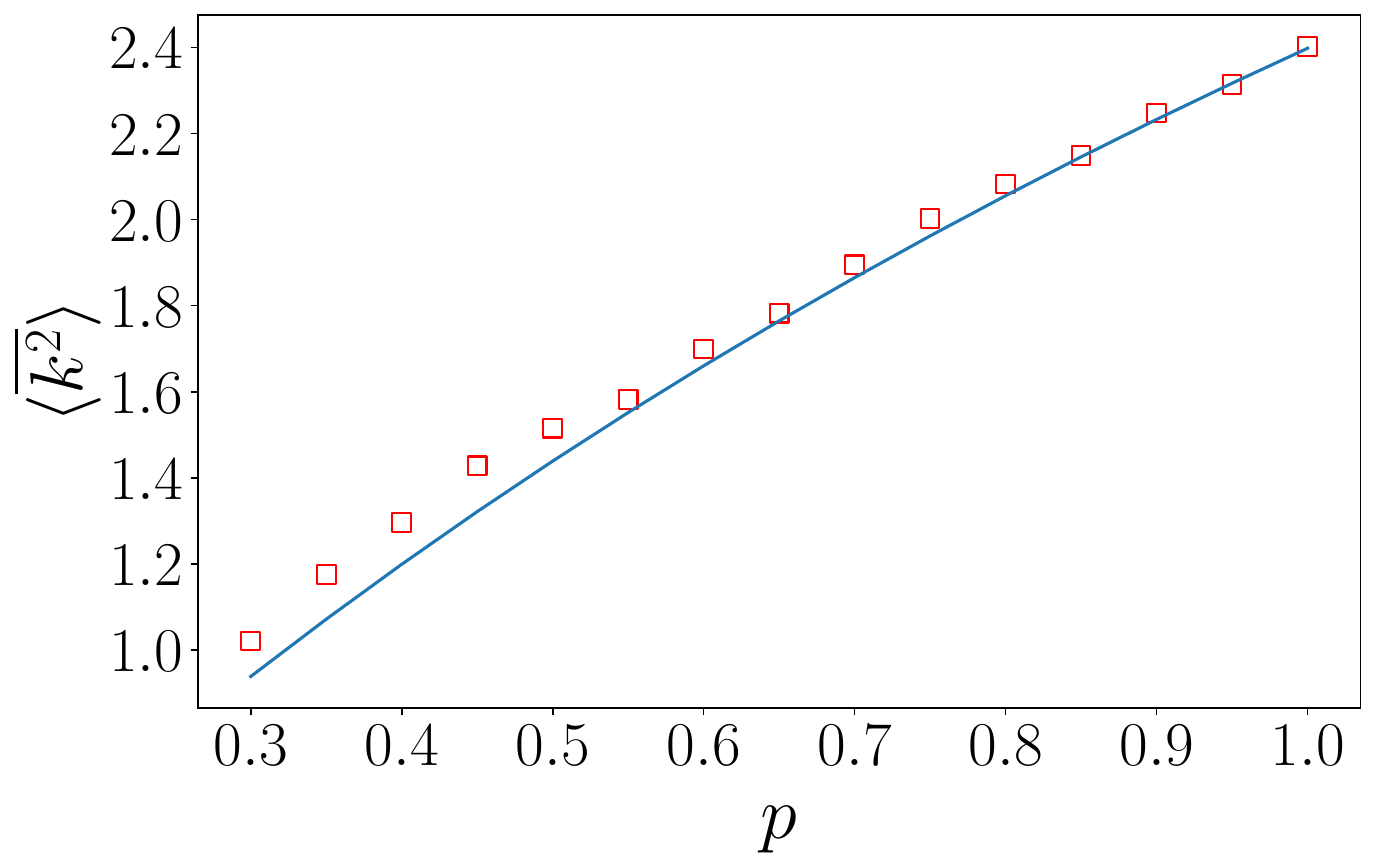}
        \caption{}
        \label{fig:msdk-decel}
    \end{subfigure}
   \caption{Quenched average of the squared trapping hotspot index $\langle \overline{k_{\text{trap}}^2} \rangle$ as a function of hotspot density $p$. Symbols are simulation data. The red curve corresponds to the prediction in \eqref{msd-p0-disorder-infty}. In \ref{fig:msdk-accel}, the LARW accelerates until it is trapped at activation $a_f$, while in \ref{fig:msdk-decel} it decelerates.}
   \label{fig:msdk}
\end{figure}

Disordered CTRWs of this type have been extensively studied \cite{alexanderExcitationDynamicsRandom1981, derridaVelocityDiffusion, bouchaudAnomalousDiffusion}. Two disorder-averaged observables of particular interest are
\begin{itemize}
    \item $\langle P_0(\tilde{t}) \rangle$: the probability to be at the origin at renormalized time $\tilde{t}$,
    \item $\langle \overline{k(\tilde{t})^2} \rangle$: the mean squared hotspot index, interpreted as a measure of spreading.
\end{itemize}
Their asymptotic behavior is known:
\begin{equation}
\label{msd-p0-disorder}
\langle P_0(\tilde{t}) \rangle \Sim{\tilde{t} \gg T} \sqrt{\frac{T}{2\pi \tilde{t}}} = \sqrt{\frac{2 + p}{6 \pi p \tilde{t}}}, \quad
\langle \overline{k(\tilde{t})^2} \rangle \Sim{\tilde{t} \gg T} \frac{\tilde{t}}{T} = \frac{3p}{2 + p} \tilde{t}.
\end{equation}
In the trapped regime, the relevant renormalized time is the finite value
\begin{equation}
    \tilde{t}_\infty = \int_0^\infty \frac{db}{\tau(b)},
\end{equation}
so that the quantities in \eqref{msd-p0-disorder} yield the disorder-averaged probability to be trapped at the origin, and the mean squared trapping hotspot index:
\begin{keyboxedeq}
\begin{equation}
\label{msd-p0-disorder-infty}
\langle \mathbb{P}(\text{trapped at } 0) \rangle \approx \sqrt{\frac{2 + p}{6 \pi p \tilde{t}_\infty}}, \quad
\langle \overline{k_{\text{trap}}^2} \rangle \approx \frac{3p}{2 + p} \tilde{t}_\infty.
\end{equation}
\end{keyboxedeq}
These expressions are valid in the weak disorder or weak trapping regime $\tilde{t}_\infty \ll T$, i.e., when $p$ is close to $1$ or trapping is slow. Fig.~\ref{fig:msdk} confirms these predictions numerically.

\section{$1d$ LARWs with relaxation and periodic hotspot distribution}
In this section, we introduce relaxation into the activation dynamics of the LARW. Specifically, we consider a LARW with a periodic hotspot distribution $L\mathbb{Z}$, where the activation $a(t)$ now decays over time with a constant relaxation rate $\gamma > 0$ when the RWer is not on a hotspot. The evolution of $a(t)$ is now governed by the differential equation
\begin{equation}
    \label{def-a-relax}
    \frac{da}{dt} = -\gamma a(t) + \sum_{k = -\infty}^{\infty} \delta(kL - x(t)).
\end{equation}
The LARW with relaxation offers a more realistic model of activation-driven dynamics in physical systems. Indeed, one does not expect a physical random walker—such as a dendritic cell—to accelerate indefinitely; rather, its speed should remain bounded due to physiological or environmental constraints.
The physics of the LARW with relaxation becomes particularly rich and nontrivial in the presence of multiple hotspots, which motivates our focus on the periodic setting. \footnote{The single-hotspot LARW with relaxation exhibits less rich physical behavior and is less amenable to systematic analytical approximations.
} As will become apparent, the qualitative phenomenology of the LARW with relaxation is the same for all choices of $\tau(a)$ which do not trap the LARW in finite time. In the following, we choose a waiting time $\tau(a)$ with $\tau(0)=1$ and assume that the LARW is free, i.e never trapped. We begin with a physical argument that clarifies the behavior of this model, and reveals the existence of two distinct dynamical regimes, determined by the parameter
\[
\nu \equiv \sqrt{\gamma} L.
\]

\subsection{Mean activation $\langle a \rangle(t)$}
Importantly, the time evolution of the mean activation $\langle a \rangle$ can be directly obtained from Eq.~\eqref{def-a-relax}, providing valuable insight into the underlying physical behavior.
Indeed, we have
\begin{align}
    \label{mean-a-relax-ode}
    \frac{d\langle a \rangle}{dt} = -\gamma \langle a \rangle(t) + P(\text{hot},t)
\end{align}
where $P(\text{hot},t)$ is the probability of being at any hotspot at time $t$. Thus,
\begin{equation}
    \label{mean-a-relax}
    \langle a \rangle(t) = \int_0^t e^{-\gamma(t-s)} P(\text{hot},s)ds.
\end{equation} 
It is already apparent from \eqref{mean-a-relax} that $\langle a(t) \rangle$ (and, a fortiori, the random variable $a(t)$ itself) is always bounded for a free LARW. Indeed, $P(\text{hot},s) \leq 1$, so 
\begin{equation}
    \langle a \rangle(t) \leq \frac{1}{\gamma},
\end{equation}
in strong contrast with non-relaxing LARWs: relaxation is always strong enough that activation stays bounded. Let us now discuss the physical picture of the model.

\subsection{Decomposition of the environment into activated intervals}

Due to relaxation, activation decreases with distance from the hotspots. As a result, we expect that $a(t)$ is significantly larger than $0$ only within 'activation intervals' $\mathcal{I}_k =  [kL - R/2, kL + R/2]$ centered around each hotspot $kL$, with some typical length $R$. If the relaxation rate increases to $\gamma' > \gamma$, the associated activation length shrinks to $R' < R$, since a faster relaxation causes the RWer to lose activation more quickly when leaving the hotspot.

We estimate the activation length $R$ as follows. The time it takes the RWer to reach the edge of the activation interval $\mathcal{I}_k$ is on the order of the relaxation time, $1/\gamma$. Since the RWer behaves diffusively within $\mathcal{I}_k$, this time also scales as $R^2$, implying:
\[
R = \frac{1}{\sqrt{\gamma}}.
\]

Two regimes then emerge, depending on how $R$ compares to the inter-hotspot distance $L$.

\begin{itemize}
    \item If $R \gg L$, i.e., $\nu \ll 1$, the activation intervals $\mathcal{I}_k$ overlap significantly, and activation decays slowly. In this regime, the RWer retains a non-negligible amount of activation between successive visits to hotspots. On top of this, $a(t)$ is bounded. Therefore, it is expected that activation saturates toward a deterministic steady-state value $a_\infty$. At long times, the system thus behaves as a simple CTRW with constant waiting time $\tau(a_\infty)$, and the spatial spreading is purely diffusive. We refer to this regime as the \emph{Gaussian phase}.
    
    \item If $R \ll L$, i.e., $\nu \gg 1$, the activation intervals $\mathcal{I}_k$ are well-separated. A RWer leaving a hotspot loses all its activation before reaching the next one, effectively resetting its memory. The activation $a(t)$ is most often close to zero and does not reach a deterministic steady-state value. We refer to this regime as the \emph{non-Gaussian phase}, for reasons that will be clarified below. Notably, single-hotspot LARWs with relaxation fall within this dynamical phase, as the length $L$ between hotspots diverges.
\end{itemize}

\subsection{Master equation for $P(x,a,t)$}
The following analysis is based on an exact evolution equation for the joint distribution of position and activation in the periodic hotspot, relaxing case. Distinguishing when the RWer is or is not on a hotspot $k L, k \in \mathbb{Z}$, we can write the following Fokker-Planck equation for $P(x,a,t)$
\begin{equation}
    \label{mastereq-relax}
    \partial_t P +\partial_a \left[ \left(\sum_{k=-\infty}^\infty \delta(x-kL)- \gamma a\right) P   \right] = \frac{1}{2\tau(a)} \Delta P 
\end{equation}
where $\Delta$ is the discrete Laplacian. Let us introduce the reduced coordinate 
\begin{equation}
    \theta(t) \equiv x(t) \bmod L
\end{equation}
which will play a key role in the following. The joint distribution $Q(\theta,a,t)$ of $\theta$ and $a$ writes, according to \eqref{mastereq-relax}
\begin{equation}
    \label{mastereq-relax-theta}
    \partial_t Q +\partial_a \left[ \left(\delta(\theta)- \gamma a\right) Q   \right] = \frac{1}{2\tau(a)} \Delta Q.
\end{equation}
We Fourier-transform $Q(\theta,a,t)$ with respect to the periodized variable $\theta$ 
\begin{equation}
    Q_n(a,t) \equiv \sum_{\theta=0}^{L-1} Q(\theta,a,t)e^{-\frac{2i \theta n\pi}{L}} \quad \iff \quad  Q(\theta, a, t) = \frac{1}{L} \sum_{n=0}^{L-1} Q_n(a,t)e^{\frac{2i \theta n\pi}{L}}
\end{equation}
We remark that $Q_0(a,t)$ is exactly the distribution $Q(a,t)$ of the activation $a$. Under Fourier transform, \eqref{mastereq-relax-theta} becomes 
\begin{equation}
    \label{mastereq-theta-ft}
    \partial_t Q_n  - \gamma(a\partial_a Q_n +Q_n) +\frac{1}{L} \partial_a \sum_{k=0}^{L-1}Q_k= -\frac{1}{\tau(a)}\left(1- \cos(\frac{2\pi n}{L})\right) Q_n.
\end{equation}
We now perform a large-$L$ analysis of \eqref{mastereq-theta-ft}. In this limit, we are only interested in the behavior of low-frequency modes $n \ll L$. Indeed, high-frequency modes $n \propto L$ are expected to vanish as the function $Q(\theta,a,t)$ does not exhibit rapid oscillations in $\theta$. In this limit, \eqref{mastereq-theta-ft} becomes, for low-frequency modes $n \ll L$
\begin{equation}
    \label{mastereq-theta-smalln}
    \partial_t Q_n  - \gamma(a\partial_a Q_n +Q_n) +\frac{1}{L} \partial_a \sum_{k=0}^{L-1}Q_k= -\frac{1}{2\tau(a)} \left(\frac{2n\pi}{L} \right)^2 Q_n.
\end{equation}
Let us analyze the consequences of \eqref{mastereq-theta-smalln}.

\subsection{Dynamical phase transition at $\nu_c = \frac{\pi}{\sqrt{2}}$}
As already hinted at, \eqref{mastereq-theta-smalln} reveals the presence of two distinct dynamical phases, governed by the magnitude of the parameter $\nu \equiv \sqrt{\gamma }L$. We now compute the critical threshold $\nu_c$ exactly in the limit $L \gg 1$ with $\nu$ fixed, starting from \eqref{mastereq-theta-smalln}. We already showed \eqref{mean-a-relax} that $a(t)$ is bounded. We thus expect the joint distribution $Q(\theta,a,t)$ of the two bounded variables $\theta$ and $a$ to converge to a stationary distribution:
\begin{equation}
    Q(\theta,a,t) \To{t \to \infty} Q^\infty(\theta,a).
\end{equation}
The small-frequency Fourier modes $Q_n^\infty(a)$ of $Q^\infty(\theta,a)$ satisfy the stationary version of \eqref{mastereq-theta-smalln}:
\begin{equation}
    \label{stationary-q-smalln}
    \gamma a \left(Q_n^{\infty}\right)' + \gamma Q_n^\infty 
    - \frac{1}{2\tau(a)} \left(\frac{2n\pi}{L} \right)^2 Q_n^\infty 
    + \frac{1}{L} \sum_{k=0}^{L-1} \left(Q_k^{\infty}\right)' = 0.
\end{equation}

To simplify this equation, we determine which of the two terms $a\left(Q_n^{\infty}\right)'$ and $Q_n^\infty$ dominates as $a \to 0$ in each phase. Note that
\begin{equation}
    a \left(Q_n^{\infty}\right)' \ll Q_n^\infty 
    \iff \frac{d}{da} \log Q_n^\infty \ll \frac{1}{a} 
    \iff Q_n(a) \ll e^{1/a}.
\end{equation}
In the Gaussian phase, that is for $\nu \ll 1$, the distribution of $a$ has almost no weight near $a=0$. Thus, we can neglect the term $a\left(Q_n^{\infty}\right)'$. Defining
\begin{equation}
    W(a) = \frac{1}{L} \sum_{k=0}^{L-1} Q_k^{\infty}(a),
\end{equation}
we find from \eqref{stationary-q-smalln}, neglecting $a(Q_n^{\infty})'$,
\begin{equation}
    Q_n^{\infty}(a) = \frac{-W'(a)}{\gamma - \frac{1}{2\tau(a)}\left(\frac{2n\pi}{L} \right)^2}.
\end{equation}
Summing over $n$, we obtain the following ODE for $W(a)$:
\begin{equation}
    \label{ode-w}
    W(a) \Sim{L \to \infty} W'(a) \frac{\sqrt{\tau(a)} \cot \left(\nu \sqrt{\frac{\tau(a)}{2}} \right)}{\sqrt{8\gamma}},
\end{equation}
where we used the identity
\begin{equation}
    \sum_{n=0}^{L-1} \frac{1}{x - n^2} \approx \sum_{n=0}^\infty \frac{1}{x - n^2} = \frac{\pi \cot\left(\pi \sqrt{x} \right)}{2\sqrt{x}} + \frac{1}{2x}.
\end{equation}

Now, the regime where $a(Q_n^{\infty})' \gg Q_n^{\infty}$ at small $a$ corresponds to $aW'(a) \gg W(a)$. From \eqref{ode-w}, we deduce
\begin{equation}
    \label{ratio-wp-w}
    \frac{a W'(a)}{W(a)} = \frac{\sqrt{8\gamma} a}{\sqrt{\tau(a)}} \tan \left(\nu \sqrt{\frac{\tau(a)}{2}} \right).
\end{equation}
The ratio in \eqref{ratio-wp-w}, evaluated near $a = 0$, probes the dynamical phase. In the non-Gaussian phase, where the typical value of $a$ is close to zero, we expect this ratio to be negative, reflecting a decay of the distribution toward $a = 0$. In contrast, in the Gaussian phase where $a$ typically remains finite, \eqref{ratio-wp-w} should be positive near the origin. Since $\tau(0) = 1$, the expression \eqref{ratio-wp-w} becomes positive as $a \to 0$ only when
\begin{equation}
    \frac{\nu}{\sqrt{2}} < \frac{\pi}{2} \quad \Longleftrightarrow \quad 
    \nu < \nu_c \equiv \frac{\pi}{\sqrt{2}}.
\end{equation}

Therefore, for $\nu < \nu_c$, the system is in the \emph{Gaussian phase} and the term $a(Q_n^{\infty})'$ is negligible. Conversely, for $\nu > \nu_c$, this term becomes dominant near $a = 0$, indicating the formation of a Dirac peak at $a = 0$ in the activation distribution — a hallmark of the \emph{non-Gaussian phase}. We conclude that in the limit $L \to \infty$ with fixed $\nu = L \sqrt{\gamma}$, the criterion separating both dynamical phases is
\begin{keyboxedeq}[Dynamical phase transition criterion]
\begin{equation}
    \label{threshold-dyn-phases}
    \nu < \nu_c \iff \text{Gaussian phase}, \quad 
    \nu > \nu_c \iff \text{non-Gaussian phase}, \quad 
    \nu_c = \frac{\pi}{\sqrt{2}}.
\end{equation}
\end{keyboxedeq}

\subsection{Detailed analysis of the non-Gaussian phase $\nu > \frac{\pi}{\sqrt{2}}$}
Having established the criterion \eqref{threshold-dyn-phases} that separates the two dynamical regimes according to the value of the parameter $\nu = \sqrt{\gamma} L$, we now focus on the more intriguing non-Gaussian regime, corresponding to $\nu > \nu_c$. In this phase, the RWer is unable to retain its activation when traveling between hotspots. Throughout this section, we concentrate on the explicit case of an accelerating, relaxing LARW with
\begin{equation}
    \tau(a) = \frac{1}{1 + a}.
\end{equation}

\subsubsection{Stationary distribution of the reduced position $\theta$}
To determine the stationary joint distribution $Q^\infty(\theta, a)$ of the activation $a$ and the reduced position $\theta$, one would in principle need to solve the stationary version of the coupled equations \eqref{stationary-q-smalln}. However, an exact analytical solution proves to be intractable. We therefore introduce a physically motivated approximation that turns out to be remarkably accurate.

Since the joint distribution is stationary, we postulate that the effective dynamics of the RWer can be described by a RW with a spatially dependent average activation $a_\theta$, where $\theta$ denotes the reduced position modulo $L$. This local activation gives rise to a position-dependent waiting time
\begin{equation}
    \tau_\theta \equiv \tau(a_\theta) = \frac{1}{1 + a_\theta}.
\end{equation}
The marginal stationary distribution $q^\infty(\theta)$ of $\theta$ satisfies local detailed balance, and the effective equilibrium condition at each site implies
\begin{equation}
    \label{ldb-q-theta}
    \frac{1}{\tau_\theta} q^\infty(\theta) = (1 + a_\theta) q^\infty(\theta) = K,
\end{equation}
where $K$ is a constant independent of $\theta$. Normalizing over the $L$ sites yields
\begin{equation}
    \label{qtheta}
    K = \frac{1 + \langle a \rangle^\infty}{L} 
    \quad \Longrightarrow \quad 
    q^\infty(\theta) = \frac{1 + a_\theta}{L(1 + \langle a \rangle^\infty)},
\end{equation}
where
\begin{equation}
    \langle a \rangle^\infty \equiv \sum_{\theta = 0}^{L - 1} a_\theta \, q^\infty(\theta)
\end{equation}
is the stationary mean activation, expressed as an average over the reduced positions.

It remains to determine the spatial profile $a_\theta$. For this, we construct a minimal toy model that captures the essential features of the LARW in the non-Gaussian regime: the \emph{pierced bag of rice} random walk.

\subsubsection{Pierced bag of rice RW}
We introduce a simple toy model designed to approximate the spatially dependent activated time $a_\theta$ discussed above. In this model, a RWer performs a discrete-time symmetric random walk on a circle with $L$ sites labeled $\theta = 0, \dots, L-1$. At each time step, the RWer jumps to the left or right with probability $\frac{1}{2}$. Crucially, the walker carries a quantity $a$ (the "rice"), which serves as an analogue of the activation in LARWs.

Every time the RWer returns to the origin $\theta = 0$, the activation is reset to a fixed value $a_0$. When the walker visits any site $\theta \neq 0$ with activation $a$, the new activation becomes $a' = e^{-\gamma} a$: a fixed fraction of the activation is lost at each step. This rule mimics the behavior of the LARW with $\tau(a) = (1+a)^{-1}$, where the activation decreases on average as $a' = a \exp(-\frac{\gamma}{1+a})$ after a jump. Expanding this expression for small $a$ (which is valid in the non-Gaussian regime), one obtains the exponential decay used in our toy model.

\vspace{0.5em}
Some remarks are in order:
\begin{itemize}
    \item In this toy model, the spatial distribution $P(\theta)$ converges to a uniform distribution at large times. This differs from the LARW, where the dynamics are governed by non-uniform waiting times. The difference arises because our toy model ignores time spent at each site (i.e., it uses discrete-time dynamics).
    
    \item Despite this simplification, we expect the model to yield a good approximation of $a_\theta$ for two main reasons:
    \begin{itemize}
        \item Assigning deterministic (but spatially varying) waiting times instead of random ones should not significantly affect long-time averages. Since the waiting times are finite, their fluctuations become negligible at large times \cite{bouchaudAnomalousDiffusion}.
        
        \item Although the activation $a$ influences the waiting time in LARWs, there is in principle a feedback: faster (i.e., more activated) walkers spend less time diffusing and thus retain more activation. However, simulations show that this feedback effect is negligible in the non-Gaussian regime, where the relaxation timescale $1/\gamma$ is short compared to the diffusion timescale $L^2$.
    \end{itemize}
\end{itemize}

Now assume that the RWer has evolved for a sufficient amount of time so that its position distribution becomes uniform, i.e., $P(\theta) = \frac{1}{L}$. We aim to compute the joint stationary distribution $Q^\infty(a, \theta)$ of the rice variable $a$ and the position $\theta$. To do so, we partition over the last return to site $\theta = 0$, and distinguish whether the RWer first moved to site $1$ or to site $L-1$ after that return. This yields the following master equation:
\begin{equation}
    \label{rice-mastereq}
    Q(a = e^{-n\gamma} a_0, \theta) = \frac{P(0)}{2} \left[ R_{n-1}(\theta|1) + R_{n-1}(\theta|L-1) \right],
\end{equation}
where $R_n(\theta|x)$ is the probability that a simple RWer starting from site $x$ reaches site $\theta$ in $n$ steps without visiting site $0$ in the meantime.

Since the position is uniformly distributed, the conditional distribution of $a$ given that the RWer is at site $\theta$ becomes
\begin{equation}
    \label{rice-mastereq-cond}
    Q(a = e^{-n\gamma} a_0 \mid \theta) = \frac{1}{2} \left[ R_{n-1}(\theta|1) + R_{n-1}(\theta|L-1) \right].
\end{equation}
Taking the expectation over $n$, we find the average value of $a$ conditioned on being at site $\theta$:
\begin{equation}
    \label{atheta-rice}
    a_\theta \equiv \mathbb{E}[a \mid \theta] = \sum_{n=1}^\infty e^{-n\gamma} a_0 \, Q(a = e^{-n\gamma} a_0 \mid \theta) = \frac{a_0 e^{-\gamma}}{2} \left[ \hat{R}(\theta|1, e^{-\gamma}) + \hat{R}(\theta|L-1, e^{-\gamma}) \right],
\end{equation}
where $\hat{R}(\theta|x, \xi) = \sum_{n=0}^\infty R_n(\theta|x) \xi^n$ is the generating function of the restricted propagator.

By definition, this generating function satisfies the discrete equation
\[
\hat{R}(\theta|x, \xi) = \delta_{\theta, x} + \frac{\xi}{2} \left[ \hat{R}(\theta+1|x, \xi) + \hat{R}(\theta-1|x, \xi) \right],
\]
together with absorbing boundary conditions $\hat{R}(0|x, \xi) = \hat{R}(x|0, \xi) = 0$.

The solution to this problem is classical (see \cite{hughes1995random}), and reads
\[
\hat{R}(\theta|x, \xi) = \frac{2}{L} \sum_{k=1}^{L-1} \frac{\sin\left( \frac{k \pi x}{L} \right) \sin\left( \frac{k \pi \theta}{L} \right)}{1 - \xi \cos\left( \frac{k \pi}{L} \right)}.
\]
Substituting into \eqref{atheta-rice}, we finally obtain an explicit expression for the mean rice $a_\theta$ conditioned on being at position $\theta$:
\begin{boxedeq}
\begin{equation}
    \label{a-theta}
    a_\theta = \frac{2 a_0 e^{-\gamma}}{L} \sum_{k=0}^{\left\lfloor \frac{L-1}{2} \right\rfloor} \frac{\sin\left( \frac{(2k+1)\pi}{L} \right) \sin\left( \frac{(2k+1)\pi \theta}{L} \right)}{1 - e^{-\gamma} \cos\left( \frac{(2k+1)\pi}{L} \right)}.
\end{equation}
\end{boxedeq}

\subsubsection{Distribution of the position $P(x,t)$}
Let us now determine the stationary distribution $q^\infty(\theta)$ using our result \eqref{a-theta}. To do this, we first need to fix the value of the reset parameter $a_0$, which controls how much rice the RWer resets to upon visiting the origin. We impose that the mean activation over all positions $\theta$ matches the stationary mean activation $\langle a \rangle^\infty$ of the original LARW:
\begin{equation}
\label{def-a0}
    \langle a \rangle^\infty = \sum_{\theta=0}^{L-1} P(\theta) a_\theta = \frac{1}{L}\sum_{\theta=0}^{L-1} a_\theta = \frac{2a_0 e^{-\gamma}}{L^2} \sum_{\theta=0}^{L-1}  \sum_{k=0}^{E\left(\frac{L-1}{2}\right)} \frac{\sin\left(\frac{(2k+1)\pi}{L} \right)\sin\left(\frac{(2k+1)\pi\theta}{L} \right)}{1-e^{-\gamma} \cos\left(\frac{(2k+1)\pi}{L} \right)}.
\end{equation}
From \eqref{mean-a-relax-ode}, we also know that
\begin{equation}
\label{stat-mean-a}
    \gamma \langle a \rangle^\infty = q^\infty(0) = \frac{1+a_0}{L(1+\langle a \rangle^\infty)},
\end{equation}
where the final equality comes from \eqref{qtheta}. The two coupled equations \eqref{def-a0} and \eqref{stat-mean-a} can be solved to determine the parameters $a_0$ and $\langle a \rangle^\infty$ in the pierced bag of rice model that approximates the LARW dynamics. Using \eqref{a-theta}, the stationary distribution of the reduced coordinate $\theta$ is then given by
\begin{equation}
\label{q-theta-f}
    q^\infty(\theta) = \frac{1+a_\theta}{L(1+\langle a \rangle^\infty)}.
\end{equation}

\begin{figure}
    \begin{subfigure}[t]{0.47\textwidth}
        \centering
        \includegraphics[width=\textwidth]{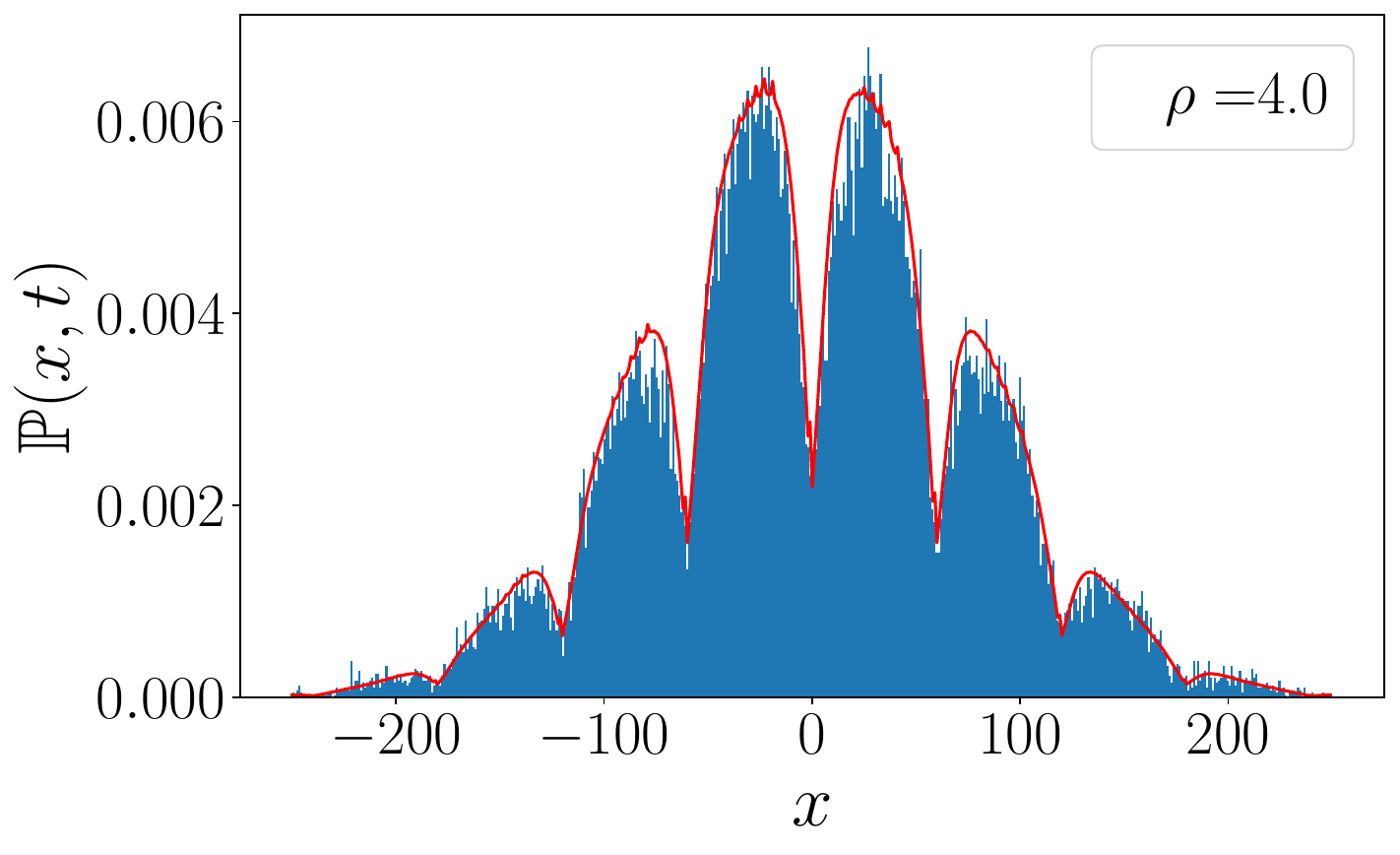}
        \caption{Distribution of the position $x$ in the non-Gaussian phase. The red curve shows our theoretical result \eqref{pxt-accel-relax}. The strongly non-Gaussian shape indicates that the RWer is most likely found between hotspots (where it moves slowly) and least likely at hotspots (where it moves quickly).}
        \label{fig:accel-relax-px}
    \end{subfigure}
    \hfill
    \begin{subfigure}[t]{0.47\textwidth}
        \centering
        \includegraphics[width=\textwidth]{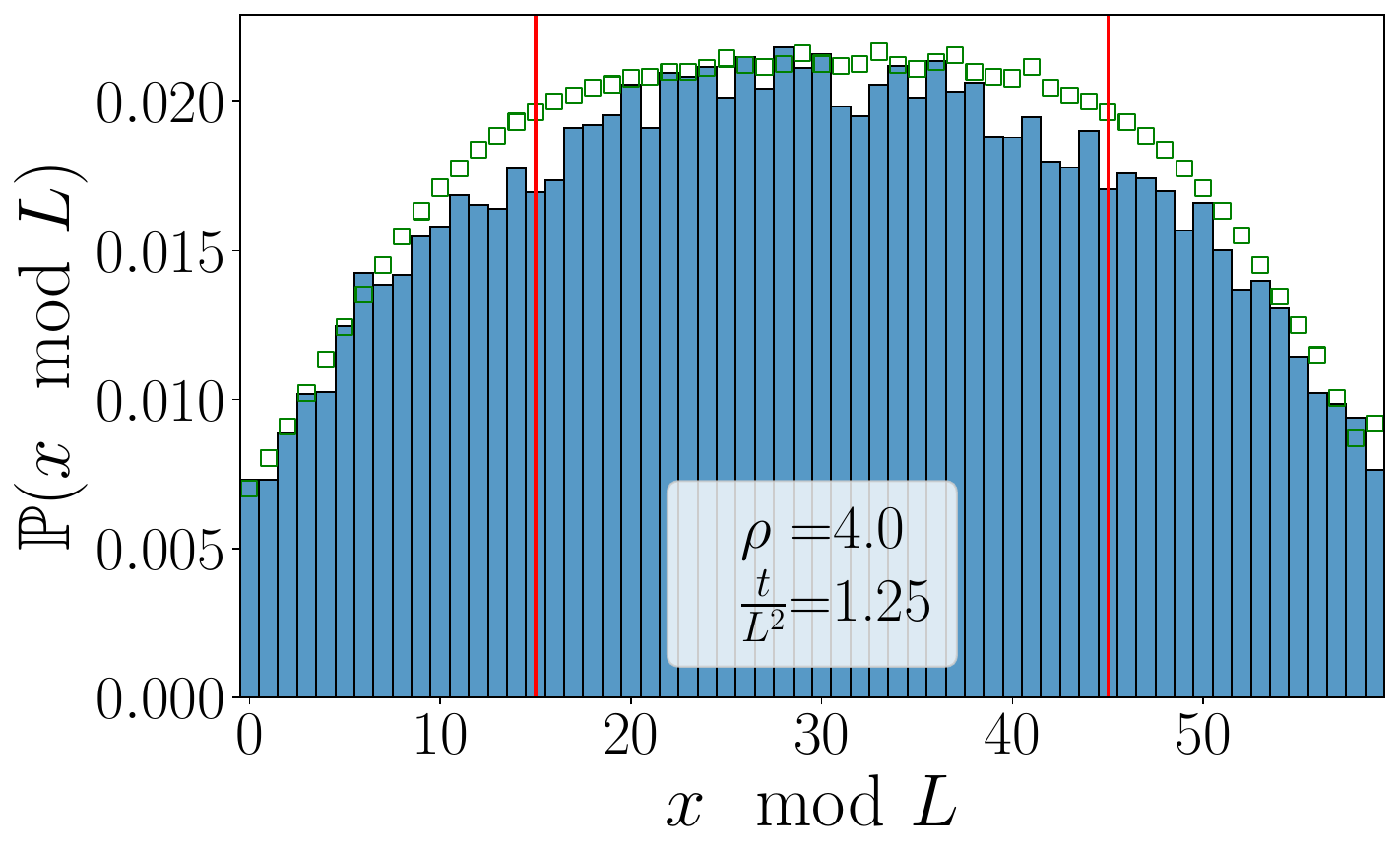}
        \caption{Stationary distribution of the reduced position $\theta = x \bmod L$. Green bars show the approximation \eqref{q-theta-f} from the pierced bag of rice model, while red lines mark the edges of the activation intervals.}
        \label{fig:accel-relax-ptheta}
    \end{subfigure}
    \caption{Numerical validation of the approximation of the LARW with relaxation by a CTRW with spatially varying waiting times $\tau(a_\theta)$. Here $L=60$, and $\rho = \nu / \nu_c = 4 > 1$, placing the system in the non-Gaussian regime.}
    \label{fig:accel-relax}
\end{figure}

See Fig.~\ref{fig:accel-relax-ptheta} for a numerical confirmation of \eqref{q-theta-f}.

To reconstruct the full position distribution $P(x,t)$, we invoke a separation of timescales argument. We decompose the position as
\begin{equation}
    x(t) = k(t) L + \theta(t),
\end{equation}
where $\theta(t) \in \{0,\dots,L-1\}$ is the reduced position. The distribution of $\theta(t)$ equilibrates over a timescale of order $L^2$ by diffusion. At large times, the large-scale coordinate $k(t)$ and local coordinate $\theta(t)$ become effectively independent, and we can factor the joint distribution as
\begin{equation}
\label{pxt-relax}
    P(x = kL + \theta, t) = P(k, t) \cdot q^\infty(\theta).
\end{equation}

We now estimate $P(k,t)$. At long times, the dynamics of $k(t)$ are described by a symmetric CTRW on $\mathbb{Z}$ with effective waiting time
\begin{equation}
\label{largescale-waiting-time}
    T = L \langle \tau(a) \rangle = L \left\langle \frac{1}{1+a} \right\rangle \approx \frac{L}{1 + \langle a \rangle},
\end{equation}
where the factor $L$ reflects the splitting probability $1/L$ at each hotspot. The last approximation follows from a Taylor expansion at small $a$, which is justified in the non-Gaussian regime. This leads to the final expression
\begin{keyboxedeq}[Propagator of the accelerating LARW with relaxation and periodic hotspots]
\begin{equation}
\label{pxt-accel-relax}
P(x = kL + \theta, t) \approx e^{-\frac{1+\langle a \rangle}{L} t} I_k\left( \frac{1+\langle a \rangle}{L} t\right)  \cdot q^\infty(\theta).
\end{equation}
\end{keyboxedeq}
Our approximation \eqref{pxt-accel-relax} compares very well with numerical simulations; see Fig.~\ref{fig:accel-relax-px}. In particular, \eqref{pxt-accel-relax} highlights a pronounced non-Gaussian character of the distribution, revealing that the LARW is highly sensitive to the spatial structure of the hotspots, and tends to be repelled from them.
This agreement with numerics justifies \emph{a posteriori} our assumptions in the non-Gaussian phase, namely:
\begin{itemize}
    \item The reduced position $\theta(t)$ performs a CTRW with spatially-dependent hopping rate $\frac{1}{1+a_\theta}$;
    \item The local activation $a_\theta$ is accurately captured by the pierced bag of rice model;
    \item Small-$a$ approximations hold in the non-Gaussian regime, where activation remains low;
    \item $\theta(t)$ and $x(t)$ are asymptotically independent for $t \gg L^2$, as $\theta(t)$ equilibrates on the diffusive timescale $L^2$.
\end{itemize}

\section{Conclusion}

The LARW emerges from this study as a minimal yet powerful framework for modeling memory-driven dynamics. Despite its simplicity, the model captures a rich variety of non-equilibrium phenomena—including aging, ergodicity breaking, anomalous diffusion, and non-Gaussian statistics—that are characteristic of complex systems far from equilibrium.

The key feature of the LARW lies in its coupling between memory and spatial dynamics: activation evolves in response to the walker’s trajectory, and in turn modulates its future motion. This feedback loop provides a versatile platform for exploring history-dependent transport in heterogeneous environments. Theoretical predictions derived from our framework exhibit strong quantitative agreement with numerical simulations, and in many cases, yield closed-form or asymptotically exact expressions.

The model also lends itself to natural extensions. In particular, we will explore in subsequent work how a spatially varying distribution of hotspots—motivated by biological processes such as chemotaxis—modifies the walker’s behavior. This direction offers a promising avenue for connecting abstract memory-based dynamics with experimentally relevant scenarios.

%% file: SIRW.tex
%!TEX root = Manuscrit.tex

\part{Self-Interacting Random Walks (SIRWs)}
\chapter{Definition and theoretical context}
\section{Context and motivations}

In many physical and biological systems, a moving agent does not explore space passively—it leaves a trace, modifies its environment, and is subsequently influenced by those modifications. A broad class of models capturing this interplay between memory and motion is known as self-interacting random walks (SIRWs). In contrast to Markovian random walks, whose future steps depend only on the present position, SIRWs evolve under the influence of their entire past trajectory. The resulting dynamics are inherently non-Markovian and history-dependent: the probability for the walker to jump from site $j$ to a neighboring site $i$ depends on how often site $i$ (or edge $\{i,j\}$, depending on the convention) has been visited in the past.

This long-range memory mechanism has proved useful for modeling a range of systems where agents actively shape and respond to their environment. Examples include ant trails and animal navigation guided by chemical cues~\cite{dussutourOptimalTrafficOrganization2004,giuggioliAnimalInteractionsEmergence2011}, Monte Carlo sampling methods~\cite{maggsNonreversibleMonte}, and, more recently, the motion of living cells that deform and chemically alter their substrate as they move~\cite{dalessandroCellMigrationGuided2021,trogerSizedependentSelfavoidanceEnables2024,flyvbjergAttractionsSet}. In such systems, memory effects induce strong localization, aging, and anomalous transport properties, even in simple geometries.

From a theoretical perspective, SIRWs occupy a rich intersection between probability theory and statistical physics. Over the past decades, they have inspired significant advances in both fields, with rigorous results on reinforced and repelled walks~\cite{pemantleSurveyRandom,tothTrueSelfAvoiding,tarresDiffusivityBounds1D2012,dumazMarginalDensities}, and physically motivated studies of non-Markovian diffusion and anomalous first-passage phenomena~\cite{amitAsymptoticBehaviorTrue1983,barbier-chebbahAnomalousPersistence,grassbergerSelfTrappingSelfRepellingRandom2017}.

This chapter develops a unified theoretical framework for describing and quantifying space exploration in one-dimensional SIRWs, for which explicit analytical results have remained remarkably limited despite growing interest. We show that a central tool in this analysis is the Ray--Knight representation, which recasts the walker's non-Markovian dynamics in terms of a Markovian evolution of its local time field—the history-dependent footprint left on the environment. Exploiting this framework, we derive exact expressions for a broad set of observables characterizing spatial exploration. To our knowledge, with very few exceptions \footnote{See \cite{dumazMarginalDensities,carmonaBetaVariables,barbier-chebbahAnomalousPersistence}.}, none of these important quantities had been previously computed, highlighting both the analytical power and versatility of the approach. These include:
\begin{itemize}
    \item The propagator of the walk;
    \item The FPT density, the persistence exponent and splitting probabilities;
    \item The distribution of the span $N(t)$—i.e., the number of distinct sites visited up to time $t$;
    \item The survival probability in finite domains.
\end{itemize}
A detailed summary of these results is provided in Section~\ref{sec:results-sirw}. We now turn to the definition of the SIRW model.

% Beyond one-time observables, we introduce a new extension of Ray--Knight theory—the \emph{aged Ray--Knight framework}—which allows us to compute two-time quantities that encode the memory structure of the walk. This enables us, for instance, to analyze how the local time field evolves between successive extremal events, or to compute aged versions of splitting probabilities.

% Finally, we show that Ray--Knight representations also extend to SIRWs on general trees. While loops introduce correlations that break the decomposition structure, trees preserve the essential independence properties of the one-dimensional case, allowing for a generalized analysis of exploration patterns in branching geometries.

% Together, these results establish a comprehensive and tractable theory of space exploration in SIRWs. They bridge rigorous probabilistic constructions and physically motivated observables, and offer analytical access to a class of non-Markovian processes that had remained largely resistant to explicit treatment.

\section{Model Definition}

We consider a class of one-dimensional self-interacting random walks (SIRWs) defined on the integers $\mathbb{Z}$. These walks are characterized by long-range memory effects: the transition probabilities at time $t$ depend on the entire history of the walk up to time $t$. More precisely, let $(X_t)_{t \in \mathbb{N}}$ be a nearest-neighbor RW on $\mathbb{Z}$, starting at $X_0 = 0$. The dynamics of the walk is governed by a weight function $w: \mathbb{N} \to \mathbb{R}_+$ and the \emph{edge-local time} $L_t(x)$, defined as the number of times the \emph{unoriented} edge $\{x, x-1\}$ has been crossed up to time $t$:
\[
L_t(x) := \#\{s < t \, | \, \{X_s, X_{s+1}\} = \{x, x-1\} \}.
\]

At each step, the RWer at site $x$ chooses to jump to $x+1$ or $x-1$ with transition probabilities depending on the current edge-local time profile:
\begin{equation}
\mathbb{P}(X_{t+1} = x\pm 1 \mid X_t = x) = \frac{w(L_t(x \pm 1))}{w(L_t(x\pm 1)) + w(L_t(x))}
\label{eq:transition}
\end{equation}

The weight function $w(n)$ encodes the form of the self-interaction:
\begin{itemize}
    \item If $w(n)$ is \textbf{increasing}, the walker is \emph{self-attracting} and tends to revisit previously used edges.
    \item If $w(n)$ is \textbf{decreasing}, the walker is \emph{self-repelling}, avoiding regions it has already explored.
\end{itemize}

To ensure the walk is not confined to a bounded region, we assume throughout that
\[
\sum_{n=0}^\infty \frac{1}{w(n)} = \infty.
\]
Under this condition, the walk is shown to be transient \cite{pemantleSurveyRandom,davisReinforcedRandom} and its mean square displacement diverges as $\langle X_t^2 \rangle \sim t^{2/d_w}$, where $d_w$ is the walk dimension, computed in a variety of works depending on the weight function $w$ \cite{amitAsymptoticBehaviorTrue1983,tothGeneralizedRayKnightTheory1996,davisReinforcedRandom}.

This general framework encompasses multiple universality classes of self-interacting random walks, each distinguished by the asymptotic ($n\to \infty$) behavior of the weight function $w(n)$. These classes are described in detail in the next section.

It is worth emphasizing that the SIRW can also be formulated in its \emph{site-reinforced} version, in which the transition probabilities depend on the number of visits to neighboring sites:
\begin{equation}
    \mathbb{P}(X_{t+1} = x \pm 1 \mid X_t = x) = \frac{w(l_t(x \pm 1))}{w(l_t(x \pm 1)) + w(l_t(x))},
    \label{eq:transition-site}
\end{equation}
where $l_t(x)$ denotes the local time at site $x$ up to time $t$. While site-reinforced and edge-reinforced variants share many qualitative features, subtle differences may arise; these will be highlighted where relevant in the following analysis.

\section{Universality Classes}

The large-scale behavior of a SIRW is governed by the asymptotic form of the weight function $w(n)$. Remarkably, all physically relevant SIRWs fall into one of three distinct universality classes, depending only on the large-$n$ behavior of $w(n)$, regardless of microscopic details \cite{tothGeneralizedRayKnightTheory1996}. Each class exhibits a specific scaling behavior and is characterized by universal values of critical exponents, such as the walk dimension $d_w$.

\vspace{1em}
\noindent The three universality classes are as follows:

\begin{description}
    \item[1. Self-Attracting Walk (SATW$_\phi$):] Defined by a \emph{saturating} weight function $w$, that tends to a non-zero constant at large $n$:
    \[
    w(n) \xrightarrow[n \to \infty]{} \frac{1}{\phi_0} > 0.
    \]
    A canonical example is the once-reinforced RW, where unvisited edges are weighted by $w(0) = 1/\phi$ and revisited edges by $w(n \geq 1) = 1$. For a general SATW however, that is, a SIRW with a saturating weight function, the reinforcement parameter $\phi$ depends on the entire weight function $w$, and is explicit in the edge-reinforced case \cite{tothGeneralizedRayKnightTheory1996}:
    \begin{equation}
        \label{phi-links}
            \phi = 1 - \sum_{j=0}^\infty \left[w(2j+1)^{-1} - w(2j)^{-1} \right].
    \end{equation} 
    We check that \eqref{phi-links} yields the correct result $\phi$ for the once-reinforced RW. In the site-reinforced version, no explicit expression of $\phi$ in terms of $w$ exists in the literature. Consequently, we estimate $\phi$ through numerical fitting when required. \par 
    The RWer can be either repelled from $(\phi<1)$ or attracted to $(\phi >1)$ its previously visited territory. Note that the usual simple RW corresponds to $\phi=1$. \par 
    The SATW$_\phi$ is diffusive \cite{davisReinforcedRandom} :
    \[
    d_w = 2.
    \] 
    
    \item[2. Polynomially Self-Repelling Walk (PSRW$_\gamma$):] Defined by a decreasing power-law weight:
    \[
    w(n) \propto n^{-\gamma}, \quad \gamma > 0.
    \]
    In this case, the RWer is repelled from previously visited edges with an algebraic strength. Strikingly, the PSRW$_\gamma$ also exhibits diffusive scaling \cite{tothGeneralizedRayKnightTheory1996} :
    \[
    d_w = 2.
    \]
    No qualitative differences exist between the site- and edge-reinforced versions.

    \item[3. Sub-Exponential Self-Repelling Walk (SESRW$_{\kappa,\beta}$):] Defined by a stretched exponential decay:
    \[
    w(n) \propto e^{-\beta n^\kappa}, \quad 0 < \kappa < 1, \ \beta > 0.
    \]
    This class includes the celebrated true self-avoiding walk (TSAW) \cite{amitAsymptoticBehaviorTrue1983,tothTrueSelfAvoiding}, corresponding to the exponential case $\kappa = 1$. The SESRW$_{\kappa,\beta}$ exhibits strong self-repulsion, leading to superdiffusive behavior. More precisely, its walk dimension is
    \[
    d_w = \frac{\kappa + 2}{\kappa + 1}<2.
    \]
    Again, there are no qualitative differences between the site- and edge-reinforced versions.
\end{description}

\section{Summary of the results}
\label{sec:results-sirw}
Our main contributions are summarized below, according to the structural organization of the chapter.
\begin{itemize}
    \item \textbf{Introduction to Ray--Knight theory for SIRWs.} We present a pedagogical and accessible presentation and derivation of the Ray--Knight theorems for SIRWs in $1D$ \cite{tothGeneralizedRayKnightTheory1996}, specifically tailored for the physics community. Although these theorems are well established in probability theory, they have remained largely underutilized in statistical physics, in part due to their abstract formulation and the absence of explicit computational frameworks. 
    \item \textbf{Explicit computations of space exploration observables for SIRWs by Ray--Knight theory.} Building on the spirit of earlier works such as \cite{dumazMarginalDensities,carmonaBetaVariables}, we take a step further by demonstrating that the Ray--Knight framework provides a versatile and powerful foundation for the analytical study of space exploration by SIRWs. Our approach naturally extends to more complex settings such as exploration on tree-like geometries, the computation of aged observables, and systems of mutually interacting SIRWs. All subsequent results, summarized below, are rooted in the Ray--Knight framework or its extensions. They illustrate both the theoretical power of this framework and its practical utility: it not only provides an exact unifying structure for describing spatial memory in SIRWs, but also enables the computation of observables that are otherwise intractable. While the framework ensures analytical accessibility, the resulting computations remain technically challenging and are a testament to the depth and flexibility of the method.
    
    \item \textbf{The joint distribution of the minimum $-m$, time $t$, and maximum $k$ as a building block for exploration observables.} We compute exactly the probability $q_+(k, -m, t)$ that a SIRW has minimum $-m$ and reaches site $k$ for the first time at time $t$. Despite its fundamental nature, this observable has received little attention in the literature — to our knowledge, it has only been explicitly discussed in the case of Markovian RWs in \cite{klingerJointStatistics}. Within the Ray--Knight framework, however, $q_+(k, -m, t)$ emerges naturally and can be derived in closed form for all classes of SIRWs. Far from being an auxiliary quantity, it serves as the fundamental building block underlying the computation of all space exploration observables computed in this chapter and presented below.

    \item \textbf{Splitting probability of all classes of SIRWs.} We compute the exact splitting probability $Q_+(z)$ that the walker, starting from $0$, visits $z$ before $-(1-z)$. We show that it can be expressed as an incomplete Beta function, which reduces to an arcsine law in the specific cases of PSRW and SESRW. This highlights the universality of this quantity across all reinforcement mechanisms.

    \item \textbf{Persistence exponents of all classes of SIRWs.} From the splitting probability, we extract the persistence exponent $\theta$ for each class of SIRW. The exponent satisfies $\theta = \phi/\dw$ \cite{zoiaAsymptoticBehaviorSelfAffine2009}, where $\phi$, the \emph{splitting exponent}, is deduced from the tail behavior of the splitting probability as 
    \[
        Q_+(z) \Propto{z \to 1} (1-z)^\phi.
    \]
    These exact values are consistent with previously known or numerically conjectured values \cite{barbier-chebbahSelfInteractingRandom,barbier-chebbahAnomalousPersistence}.

    \item \textbf{Propagators of the SATW and PSRW.} Using Ray–Knight theory and the Bessel representation, we compute the exact propagator $P(x,t)$ in the SATW and PSRW models. This includes an explicit computation of the scaling limit and reveals non-Gaussian features and bimodal shapes for self-attracting regimes.

    \item \textbf{FPT density of all classes of SIRWs.} The FPT density to a target site $x$ is derived in closed form for all classes of SIRWs, thus offering a precious analytical window on space exploration by SIRWs.

    \item \textbf{Distribution of the number of visited sites for all classes of SIRWs.} We compute the full probability distribution of the number of distinct sites visited by a SIRW up to time $t$, denoted $N(t)$. Despite its fundamental relevance, this observable had not been determined for any class of SIRW prior to this work. The resulting distribution exhibits a highly non-trivial, non-Gaussian shape, reflecting the strong memory effects inherent to the dynamics. Our analytical prediction shows excellent agreement with numerical simulations across all regimes.

    \item \textbf{Survival probability in a finite domain.} We compute the exact survival probability of a SIRW confined to a finite interval $[-a, b]$, that is, the probability that the walker remains within this domain up to time $t$. This provides a fully analytical solution to the confinement problem for SIRWs, which, despite its physical relevance, had not been addressed in the literature. Confinement problems naturally arise in diverse physical settings, as in diffusion-limited reactions where the particle must remain within a reactive region. The exact survival probability obtained here captures the interplay between memory effects and spatial boundaries, and offers a benchmark for assessing non-Markovian confinement in both theoretical models and practical applications.

    \item \textbf{Aged space exploration statistics for the SATW.} We extend the Ray--Knight framework to an \emph{aged} setting, tailored to the analysis of two-time observables, within the specific case of the SATW. This extension allows us to incorporate conditioning on the walker's past exploration when computing future probabilities. In particular, we derive the \emph{aged splitting probability} — the probability that the walker reaches one site before another, given that a certain spatial region has already been visited. This observable captures how memory and prior exploration shape the walker's future evolution and reveals subtle non-Markovian dependencies absent from classical splitting scenarios.

    \item \textbf{Space exploration statistics for a general SIRW on a tree graph.} We extend the classical one-dimensional Ray--Knight theorems for SIRWs to the broader setting of SIRWs on loopless graphs, i.e., trees. Analytical expressions are most naturally obtained in the case of the star graph — the union of $n$ half-lines joined at a common origin. In this setting, we derive exact formulas for splitting probabilities, eccentricity distributions and number of sites visited. To our knowledge, this represents the first extension of Ray--Knight-type results to SIRWs on trees, and opens a new route for computing observables that have remained inaccessible even for standard Markovian walks on such structures.

    \item \textbf{Generalized splitting probabilities for multiple interacting SATWs on the line.} We consider an ensemble of $n$ SATWs evolving on the same line and interacting through their shared edge local times. Within this setting, we define and compute exactly the generalized splitting probabilities $\mathbb{P}_z(+^k, -^{n-k})$, which quantify the probability that exactly $k$ SATW, all starting from $0$, reach site $z$ before $-(1-z)$, while the remaining $n - k$ do the opposite. This computation relies on a novel extension of the Ray--Knight framework to systems of mutually interacting SIRWs. To our knowledge, this constitutes the first exact analytical result for space exploration observables in a strongly interacting, non-Markovian many-body system of RWers.
\end{itemize}

This work is the basis of two publications \cite{bremontPersistenceExponents,bremontExactPropagators} and will include several future publications. 

\chapter{A pedagogical introduction to Ray--Knight Theory for Self-Interacting RWs}
A central theoretical tool for analyzing the space exploration properties of one-dimensional SIRWs is a generalized version of the Ray–Knight theorem. This result characterizes the statistical structure of the edge local time process. Originally developed in the context of Brownian motion and symmetric Markov processes~\cite{raySojournTimesDiffusion1963,eisenbaumRayKnightTheoremSymmetric2000}, Ray–Knight-type theorems were extended to SIRWs by B.~Tóth in a seminal work~\cite{tothGeneralizedRayKnightTheory1996}. These extensions provide a crucial link between the non-Markovian dynamics of SIRWs and the Markovian behavior of their edge-local time process, in a sense that will be made precise below.

To the best of our knowledge, the results of~\cite{tothGeneralizedRayKnightTheory1996} remain largely unknown within the physics community, despite their elegance and power in the study of SIRWs. While the original proofs in~\cite{tothGeneralizedRayKnightTheory1996,tothTrueSelfAvoiding} are technically involved, they contain fundamental insights into the structure and behavior of SIRWs. A solid understanding of these results is essential for developing the extensions of Ray–Knight theorems that we will later use to study aging in space exploration Sec.~\ref{sec:aging}, the extension to tree graphs Sec.~\ref{sec:sirw-tree}, as well as interactions between multiple SIRWs.

For these reasons, we present below a reformulation of Tóth’s generalized Ray–Knight theorems, together with a rewritten version of their original proofs, recast in the language and intuition of statistical physics. Our aim is to make these powerful results more accessible and to encourage their broader adoption within the physics community.

\section{Squared Bessel processes BESQ$_\delta$}
Before diving into the Ray--Knight description, we introduce the squared Bessel process, denoted $\mathrm{BESQ}_\delta$, which is central to the analysis. For integer $\delta = d$, the process corresponds to the squared norm of a $d$-dimensional Brownian motion:
\[
Y_d(t) = \| \mathbf{B}_t^{(d)} \|^2.
\]
For general $\delta > 0$, the process $(Y_\delta(t))_{t \geq 0}$ satisfies the stochastic differential equation \footnote{We follow the Ito convention.}:
\begin{equation}
    \mathrm{d}Y_\delta(t) = \delta \, \mathrm{d}t + 2 \sqrt{Y_\delta(t)} \, \eta_t,
\label{eq:besq_sde}
\end{equation}
where $\eta_t$ is Gaussian white noise with correlation
\[
\langle \eta_t \eta_{t'} \rangle = \delta(t - t').
\]
It is clear from \eqref{eq:besq_sde} that the BESQ$_\delta$ is a Markov process. 
We will need two key observables of BESQ$_\delta$ processes:

\paragraph{Transition probability (propagator).} The probability density $P_\delta(y, t \mid y_0)$ to go from $Y(0) = y_0$ to $Y(t) = y$ in time $t$ is given by~\cite{going-jaeschkeSurveyGeneralizationsBessel2003,lawlerNotesBessel}:
\begin{equation}
P_\delta(y, t \mid y_0) = \frac{1}{2t} \left( \frac{y}{y_0} \right)^{\delta/4 - 1/2} \exp\left( -\frac{y + y_0}{2t} \right) I_{\delta/2 - 1}\left( \frac{\sqrt{y y_0}}{t} \right),
\label{eq:besq_propagator}
\end{equation}
where $I_\alpha$ is the modified Bessel function of index $\alpha$ of the first kind.

\paragraph{First-passage time (FPT) to the origin.} For $\delta < 2$, the BESQ$_\delta$ process can hit the origin with non-zero probability. The density $f_\delta(t \mid y_0)$ of FPT to $0$ starting from $Y(0) = y_0 > 0$ is \cite{going-jaeschkeSurveyGeneralizationsBessel2003}:
\begin{equation}
f_\delta(t \mid y_0) = \frac{y_0^{1 - \delta/2}}{2^{\delta/2} \Gamma(1 - \delta/2)} t^{\delta/2 - 2} \exp\left( -\frac{y_0}{2t} \right), \qquad 0 < \delta < 2.
\label{eq:besq_fpt}
\end{equation}

\paragraph{Remarks.}
\begin{itemize}
    \item For $\delta>0$, the BESQ$_\delta$ started at $x \geq 0$ is positive, and thus reflected at $0$. 
    \item For $\delta \geq 2$, the BESQ$_\delta$ process almost surely never reaches the origin — analogous to the fact that Brownian motion in $d \geq 2$ does not hit a point.
    \item For $\delta \leq 0$, the origin is an absorbing state, and the process is killed upon hitting zero.
\end{itemize}

\section{Key Results of Ray--Knight Theory}

In this section, we summarize the central results of Tóth's work~\cite{tothGeneralizedRayKnightTheory1996}, which provide the foundation for computing several observables of interest in the study of SIRWs.

Consider a SIRW starting at $X_0 = 0$. Define the stopping time $T_k$ as the first hitting time of site $k > 0$:
\[
T_k := \inf \{ t \geq 0 \mid X_t = k \}.
\]
We are interested in the entire edge-local time profile $\{L_{T_k}(x)\}_{x \in \mathbb{Z}}$ at the moment the RWer first reaches site $k$. This profile depends on the entire trajectory of the RWer up to $T_k$. Despite the globally non-Markovian nature of the SIRW, this object exhibits a remarkably simple and universal structure in the scaling limit: it converges to a concatenation of two independent BESQ. \par 
To make this precise, consider the scaling limit where $z = k/A$ is held fixed and $A \to \infty$. In this regime, the edge-local time profile admits the scaling form
\[
L_{T_k}(x) \approx C \cdot A^\alpha \, \mathcal{L}_z(u), \qquad \text{with} \quad u = \frac{x}{A}, \quad x \in \mathbb{Z},
\]
where the scaling exponent $\alpha$, the constant $C$ and the limiting process $\mathcal{L}_z(u)$ depend on the universality class of the SIRW. On dimensional grounds, the exponent is given by $\alpha = d_w - 1$, where $d_w$ is the walk dimension. Indeed, since the local time at a fixed site, say $x = 0$, scales as $\frac{\text{(time)}}{\text{(space)}} \equiv \text{(time)}^{1 - 1/d_w}$, standard scaling arguments yield:
\[
L_{T_k}(0) \propto T_k^{1 - \frac{1}{d_w}} \sim \left(k^{d_w}\right)^{1 - \frac{1}{d_w}} \sim k^{d_w - 1}.
\]

The limiting profile $\mathcal{L}_z(u)$ is explicitly described by two backward-running squared Bessel processes:
\begin{equation}
\label{rayknight-tk}
\mathcal{L}_z(u) =
\begin{cases}
Y_\delta(z - u)^q, & \text{for } 0 \leq u \leq z, \\
\widetilde{Y}_{2 - \delta}(-u)^q, & \text{for } u \leq 0,
\end{cases}
\end{equation}
where:
\begin{itemize}
    \item $Y_\delta$ denotes a $\mathrm{BESQ}_\delta$ process of dimension $\delta$, started at zero. Since it remains non-negative for all time, it is naturally interpreted as being \emph{reflected at zero};
    
    \item $\widetilde{Y}_{2 - \delta}$ is a $\mathrm{BESQ}_{2 - \delta}$ process, started from $Y_\delta(0)$, and is \emph{absorbed at zero}. Conditioned on the interface value $Y_{\delta}(0) = \widetilde{Y}_{2 - \delta}(0)$, the process $\widetilde{Y}_{2 - \delta}$ is \emph{independent} from the process $Y_\delta$.
\end{itemize}

The parameters $\alpha$, $\delta$, $q$ and $C$ are specific to each SIRW class and are summarized in Table~\ref{tab:rayknight}. To fix ideas, consider the TSAW as an example. According to Table \eqref{tab:rayknight}, the rescaled edge-local time process in the region $u < 0$, namely $(\mathcal{L}_z(u))_{u < 0}$, is a Brownian motion absorbed at the origin. On the other hand, for $0 < u < z$, it corresponds to a Brownian motion reflected at zero.
\begin{table}[h]
    \centering
    \begin{tabular}{l|c|c|c|c}
    
    \textbf{Class} & \textbf{Exponent $\alpha$} & \textbf{BESQ dimension $\delta$} & \textbf{Power $q$} & \textbf{Constant $C$} \\
    
    SATW$_\phi$ & $1$ & $2\phi$ & $1$ & $1$ \\
    PSRW$_\gamma$ & $1$ & $1$ & $1$ & $\frac{1}{2\gamma+1}$ \\
    SESRW$_{\kappa,\beta}$ & $\frac{1}{\kappa + 1}$ & $1$ & $\frac{1}{\kappa + 1}$ & $C(\kappa,\beta)$ \cite{tothSelfinteractingRandomMotions}\\
    
    \end{tabular}
    \caption{Parameters of the Ray--Knight description for each SIRW universality class.}
    \label{tab:rayknight}
\end{table}

\begin{figure}
    \centering
    \includegraphics[width=.7\textwidth]{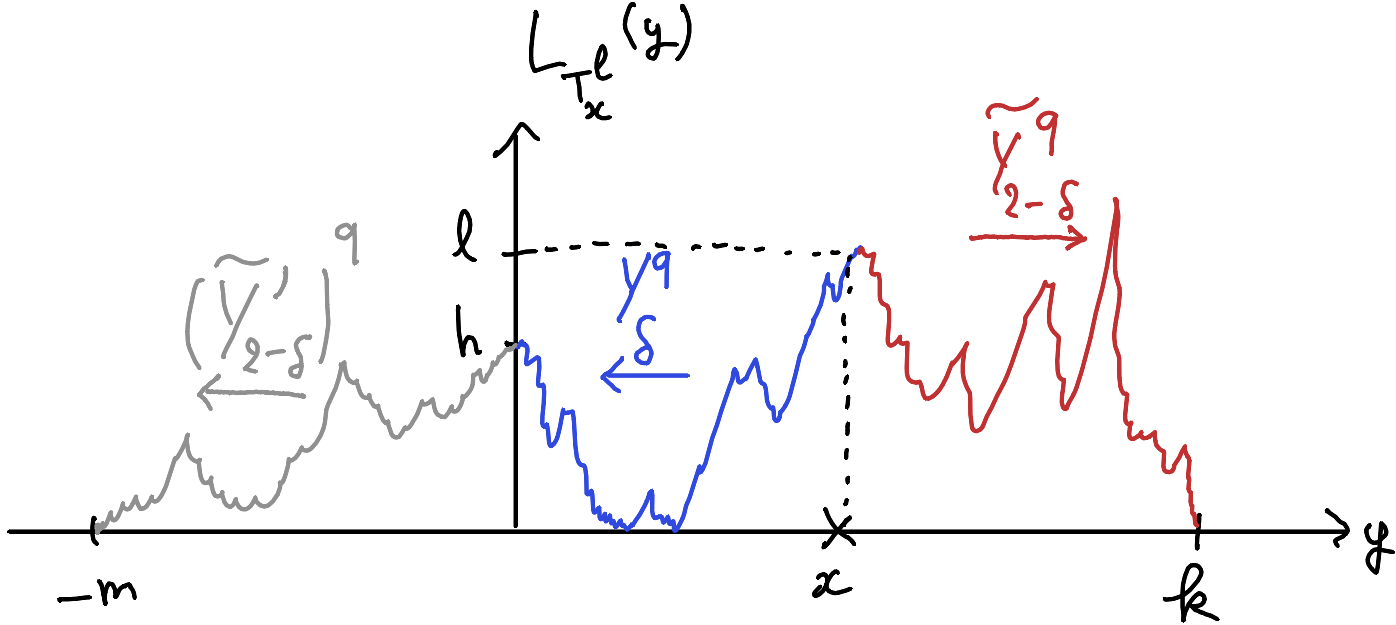}
    \caption{Ray–Knight decomposition~\eqref{rayknight-txl} of the edge-local time profile $L_{T_x^l}$ into three concatenated squared Bessel processes, all emanating from the value $l$ at site $x$. The central segment (blue), running backward from $x$ to $0$, corresponds to a $\mathrm{BESQ}_\delta^q$ process. It is reflected at the origin. The outer segments—red, running forward from $x$ to $k$, and grey, running backward from $0$ to $-m$—correspond to $\mathrm{BESQ}_{2 - \delta}^q$ processes absorbed at zero. The points $k$ and $-m$ represent the maximum and minimum of the SIRW trajectory up to time $T_x^l$.}
    \label{fig:rayknight-txl}
\end{figure}

The Ray--Knight construction reveals that although the full SIRW is non-Markovian in time, the edge-local time profile stopped at $T_k$ evolves as a Markov process in space. This spatial Markovianity, fully captured by BESQ processes in the scaling limit, also holds when stopping the process at a fixed inverse local time $T_x^l := \inf \{ t \geq 0 \mid L_t(x) = l \}$. In this case, the edge-local time profile is composed of three BESQ processes, which are mutually independent once their interface values are specified:
\begin{keyboxedeq}[Ray--Knight theorem]
\begin{equation}
    \label{rayknight-txl}
    L_{T_x^l}(y) \approx C \cdot
    \begin{cases}
    \widetilde{Y}_{2 - \delta}(y - x), & y \geq x, \\
    Y_{\delta}(x - y), & 0 \leq y \leq x, \\
    \widetilde{Y}_{2 - \delta}'(-y), & y \leq 0,
    \end{cases}
    \quad \text{with matching conditions:} \quad
    \begin{cases}
    \widetilde{Y}_{2 - \delta}(0) = Y_{\delta}(0) = l, \\
    \widetilde{Y}'_{2 - \delta}(0) = Y_{\delta}(x).
    \end{cases}
\end{equation}
\end{keyboxedeq}
See FIG.~\ref{fig:rayknight-txl} for an illustration of \eqref{rayknight-txl}. Note that we recover the special case \eqref{rayknight-tk} by taking the limit $l \to 0$ in~\eqref{rayknight-txl}.
Finally, we emphasize that the stopping time $T_x^l$ is, by definition, equal to the total area under the edge-local time profile:
\begin{equation}
\label{Tk-area}
T_x^l = \sum_{y \in \mathbb{Z}} L_{T_x^l}(y).
\end{equation}

\section{Proof of the Ray--Knight theorems for SIRWs}
\label{sec:proof}
In the following, we expose the proof of~\cite{tothGeneralizedRayKnightTheory1996} concerning Ray--Knight theorems for SIRWs in a language more suited to physicists. 

\subsection{General Results on Pòlya urns}
We begin with a fundamental identity satisfied by a generalized Pólya urn, which will be instrumental in the analysis of the local time process associated with self-interacting RWs.

Let $r(n)$ and $b(n)$ be two weight functions, interpreted respectively as the weights of red and blue balls in a generalized Pólya urn. After $i$ draws, let $\rho_i$ (resp.\ $\beta_i$) denote the number of red (resp.\ blue) balls that have been drawn. The urn evolves according to the transition rule:
\begin{align}
    &\mathbb{P}\left( (\rho_{i+1}, \beta_{i+1}) = (k+1, l) \,\middle|\, (\rho_i, \beta_i) = (k, l) \right) \equiv \frac{r(k)}{r(k) + b(l)}, \\
    &\mathbb{P}\left( (\rho_{i+1}, \beta_{i+1}) = (k, l+1) \,\middle|\, (\rho_i, \beta_i) = (k, l) \right) \equiv \frac{b(l)}{r(k) + b(l)},
\end{align}
i.e., the probability of drawing a ball depends on the current count of red and blue draws via the respective weights.

A key quantity in this setting is the time $\tau_m$ at which the $m$th red ball is drawn. Define $\mu(m) := \beta_{\tau_m}$ as the (random) number of blue balls drawn at $\tau_m$. Then, for any $\lambda < \min_{j \leq m-1} r(j)$, the following identity holds \cite{tothGeneralizedRayKnightTheory1996}:
\begin{boxedeq}
\begin{equation}
\label{lemma-polya}
    \mathbb{E} \left[ \prod_{j = 0}^{\mu(m) - 1} \left(1 + \frac{\lambda}{b(j)} \right) \right] = \prod_{j = 0}^{m - 1} \left(1 - \frac{\lambda}{r(j)} \right)^{-1}.
\end{equation}
\end{boxedeq}
This identity is crucial to the forthcoming analysis of the edge local time distribution. In particular, expanding~\eqref{lemma-polya} as a power series in $\lambda$ yields the moments of the random variable $\mu(m)$, a key step in Tóth’s proof that we are revisiting in a more accessible form.

We now present a brief proof of~\eqref{lemma-polya} by induction on $m \geq 0$. The case $m = 0$ is immediate. Since the case $m = 1$ will be needed subsequently, we establish it explicitly. First, observe that
\[
\mu(1) = \beta_{\tau_1} = \tau_1,
\]
because the first red ball is drawn at time $\tau_1$, at which point, by construction, exactly $\tau_1$ blue balls have already been drawn. The distribution of $\tau_1$ is easily computed:
\[
\mathbb{P}(\tau_1 = t) = \frac{r(0)}{b(t)+r(0)} \prod_{k=0}^{t-1} \left[\frac{b(k)}{b(k)+r(0)} \right].
\]
We therefore need to verify that
\begin{align}
\label{lemma-polya-m1}
\left(1 - \frac{\lambda}{r(0)} \right)^{-1}
&= \mathbb{E} \left[ \prod_{k = 0}^{\tau_1 - 1} \left(1 + \frac{\lambda}{b(k)} \right) \right] \notag \\
&= \sum_{t=0}^\infty \mathbb{P}(\tau_1 = t) \prod_{k = 0}^{t - 1} \left(1 + \frac{\lambda}{b(k)} \right) \notag \\
&= \sum_{t=0}^\infty \frac{r(0)}{b(t)+r(0)} \prod_{k=0}^{t-1} \left[\frac{b(k)+\lambda}{b(k)+r(0)} \right] \equiv S.
\end{align}

To evaluate $S$, define
\[
P_t = \prod_{k=0}^{t-1} \frac{b(k)+\lambda}{b(k)+r(0)}.
\]
To exploit a telescoping structure, observe that
\[
P_{t+1} - P_t = \left( \frac{b(t)+\lambda}{b(t)+r(0)} - 1 \right) P_t = \frac{\lambda - r(0)}{b(t)+r(0)} P_t.
\]
Summing both sides over $t \geq 0$, we obtain
\[
\sum_{t=0}^\infty (P_{t+1} - P_t) = \frac{\lambda - r(0)}{r(0)} S.
\]
The left-hand side telescopes to $-P_0 = -1$, since $P_0 = 1$. Hence,
\[
\frac{\lambda - r(0)}{r(0)} S = -1 \quad \Longrightarrow \quad S = \left(1 - \frac{\lambda}{r(0)} \right)^{-1},
\]
which confirms~\eqref{lemma-polya-m1}. We now proceed with the inductive step.

Suppose there is some $M \geq 1$ such that the identity holds for all $0 \leq m \leq M$ and for arbitrary weight functions $r(n)$ and $b(n)$. Consider the case $m = M+1$. By a renewal argument, we write
\[
\mu(M+1) = \mu(M) + \tilde{\mu}(1),
\]
where $\tilde{\mu}(1)$ denotes the number of blue balls drawn before the first red ball in a new urn initialized after $\mu(M)$ blue draws, with shifted weights:
\[
\tilde{r}(j) = r(M + j), \qquad \tilde{b}(j) = b(j + \mu(M)).
\]
Applying the induction hypothesis to the the cases $m=M$ and $m=1$ completes the argument, thereby establishing~\eqref{lemma-polya} for $m = M+1$ and thus for all $m \geq 0$.

\subsection{The Edge-Local Time Process as a Network of Generalized Pólya Urns}

We now describe the edge-local time process $(L_{T_{k,m}}(x))_{x \in \mathbb{Z}}$ of a self-interacting random walk (SIRW) starting at $0$, and stopped at the random time $T_{k,m}$ defined by
\begin{equation}
    T_{k,m} := \inf \left\{ t \geq 0 \mid \text{the \emph{oriented} edge } (k-1, k) \text{ has been crossed } m+1 \text{ times up to time } t \right\}.
\end{equation}
By definition, the SIRW stopped at time $T_{k,m}$ ends at site $k$, and the \emph{unoriented} edge $\{k-1, k\}$ has been crossed a total of $2m+1$ times (i.e., $m+1$ times from left to right and $m$ times from right to left). In the following, we focus on the case $k > 0$.

To describe the structure of the edge-local time profile, we associate to each integer site a generalized Pólya urn. Specifically, the index $l \in \mathbb{Z}$ labels the urn located at site $k - l$. Each urn keeps track of the steps taken by the RWer at that site, and the color of the ball drawn corresponds to the direction of the step taken from site $k - l$. Our convention, chosen to align with the spatial propagation of local times starting from the target site $k$ (i.e., from $l = 0$), is as follows:
\begin{equation}
\label{convention-colors}
    \begin{cases}
        l \in (-\infty, 0] : (\textcolor{red}{\text{jump left}}, \textcolor{blue}{\text{jump right}}), \\
        l \in (0, \infty) : (\textcolor{blue}{\text{jump left}}, \textcolor{red}{\text{jump right}}).
    \end{cases}
\end{equation}

The weight functions $r^l(\rho)$ and $b^l(\beta)$ of urn $l$ specify the weight of drawing a red or blue ball when there are $\rho$ red and $\beta$ blue balls already in the urn. These are determined by the underlying self-interaction weight function $w(n)$ of the RW. For red balls (corresponding to rightward jumps in some regions and leftward in others), we define:
\begin{equation}
\label{def-urns-red}
    r^l(\rho) =
    \begin{cases}
        w(2\rho+1), & l \in (-\infty, 0] \cup [k, \infty), \\
        w(2\rho),   & l \in [1, k-1],
    \end{cases}
\end{equation}
and for blue balls:
\begin{equation}
\label{def-urns-blue}
    b^l(\beta) =
    \begin{cases}
        w(2\beta),   & l \in (-\infty, 0] \cup [k, \infty), \\
        w(2\beta+1), & l \in [1, k-1].
    \end{cases}
\end{equation}

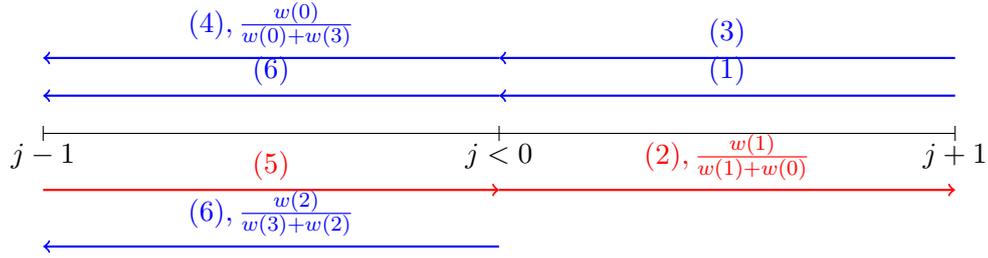
\begin{figure}[h!]
    \centering
    \begin{tikzpicture}
    % Axes
    \def\spacing{3}
    \draw (-2*\spacing,0) -- (2*\spacing,0);

    % Labels for axes
    \node[below] at (-2*\spacing,0) {$j-1$};
    \node[below] at (0,0) {$j<0$};
    \node[below] at (2*\spacing,0) {$j+1$};
    \draw (-2*\spacing,0.1) -- (-2*\spacing,-0.1); % ticks
    \draw (0*\spacing,0.1) -- (0*\spacing,-0.1); % ticks
    \draw (2*\spacing,0.1) -- (2*\spacing,-0.1); % ticks
    
    % Arrows and labels
    \draw[blue, thick, <-] (0*\spacing, 0.5) -- (2*\spacing, 0.5);
    \node[above,color=blue] at (\spacing, 1) {$(3)$};
    \draw[blue, thick, <-] (0*\spacing, 1) -- (2*\spacing, 1);
    \node[above,color=blue] at (\spacing, 0.5) {$(1)$};
    \draw[blue, thick, <-] (-2*\spacing, 0.5) -- (0*\spacing, 0.5);
    \node[above,color=blue] at (-\spacing, 1) {$(4), \frac{w(0)}{w(0)+w(3)}$};
    \draw[blue, thick, <-] (-2*\spacing, 1) -- (0*\spacing, 1);
    \node[above,color=blue] at (-\spacing, 0.5) {$(6)$};
    \draw[red, thick, ->] (0*\spacing, -0.75) -- (2*\spacing, -0.75);
    \node[above,color=red] at (-\spacing, -0.75) {$(5)$};
    \draw[red, thick, ->] (-2*\spacing, -0.75) -- (0*\spacing, -0.75);
    \node[above,color=red] at (\spacing, -0.75) {$(2), \frac{w(1)}{w(1)+w(0)}$};
    \draw[blue, thick, <-] (-2*\spacing, -1.5) -- (0*\spacing, -1.5);
    \node[above,color=blue] at (-\spacing, -1.5) {$(6), \frac{w(2)}{w(3)+w(2)}$};
    % \node[above] at (\spacing, 1) {$\frac{w(0)}{w(0) + w(3)}$};
    \end{tikzpicture}
    
    \caption{Definition of the Polya urn at a site $j<0$. Each arrow is accompanied with a labeling $(t)$ that represents time, and if the arrow departs from site $j$, the probability of jumping along this edge while on site $j$ is shown. The color of the arrow is the color of the ball drawn from the urn at the departing site and is defined by \eqref{convention-colors}. }
    \label{fig:weight}
\end{figure}

Let us briefly motivate these definitions. Consider an urn labeled by $l$ located at site $j = k - l < 0$, so that $l \in (k, \infty)$ (see Fig.~\ref{fig:weight} for an illustration). According to the convention~\eqref{convention-colors}, in this case, red balls correspond to rightward jumps across edge $(k - l, k - l + 1)$, and blue balls to leftward jumps across $(k - l, k - l - 1)$. 

Since the RWer ends at site $k > 0$ at time $T_{k,m}$, and $j < 0$, the non-oriented edge $\{k - l, k - l + 1\}$ must have been crossed an \emph{odd} number of times. This implies that, for the RWer to choose the right edge $(k - l, k - l + 1)$ for the $x$-th time, the weight is proportional to $w(2x - 1)$. On the other hand, for the $y$-th crossing of the left edge $(k - l, k - l - 1)$, the weight is proportional to $w(2y)$. This asymmetry motivates the structure of the weight functions in~\eqref{def-urns-red} and~\eqref{def-urns-blue}, which encode the memory-dependent dynamics at the level of the urns.

As an example, consider Fig.~\ref{fig:weight}. When the RWer takes a step to the right at point (2), it has not yet departed from site $j$, so $\rho = 0$ and $\beta = 0$. At this point, there has been one crossing of the edge $\{j, j+1\}$ and none of the edge $\{j-1, j\}$. Therefore, the weight to step right is $w(1)$, and the weight to step left is $w(0)$.

Now consider the RWer stepping to the left at point (6), after having previously made one rightward step at (2) and one leftward step at (4), such that $\rho = 1$ and $\beta = 1$. In this configuration, there have been a total of 3 crossings of the edge $\{j, j+1\}$ and 2 crossings of $\{j-1, j\}$. The corresponding weights are $w(3)$ for a rightward step and $w(2)$ for a leftward step.

\vspace{1em}
Let $m \geq 1$. We now define $S^>_{k,m}(l)$ to be the number of times the RWer, starting from site $0$ at time $0$, crosses the oriented edge $(k - l, k - l - 1)$ by the stopping time $T_{k,m}$—i.e., at the $(m+1)$\textsuperscript{th} arrival to site $k$ coming from the left.

As we will see, the process $S^>_{k,m}(l)$ evolves as a Markov chain in the spatial index $l$, and can be represented as a network of coupled Pólya urns. By construction,
\[
S^>_{k,m}(0) = m.
\]

Importantly, the process $S^>_{k,m}(l)$ behaves differently in the three spatial domains $(-\infty, 0)$, $(k, \infty)$, and $[0, k]$. This asymmetry arises from the fact that the RWer is guaranteed to have visited every site in $[0, k]$ at least once before time $T_{k,m}$, whereas no such condition holds for the other regions.

This leads to the following recursive representation for $l \geq 1$, using the random variable $\mu^l(n)$ defined in \eqref{lemma-polya}:
\begin{equation}
S^>_{k,m}(l) =
\begin{cases}
\mu^l\left(S^>_{k,m}(l-1) + 1 \right), & 1 \leq l \leq k, \\
\mu^l\left(S^>_{k,m}(l-1) \right), & l \geq k+1.
\end{cases}
\end{equation}

For $l \leq 0$, the process is defined in reverse as:
\begin{equation}
S^>_{k,m}(l-1) := \mu^l\left(S^>_{k,m}(l)\right).
\end{equation}

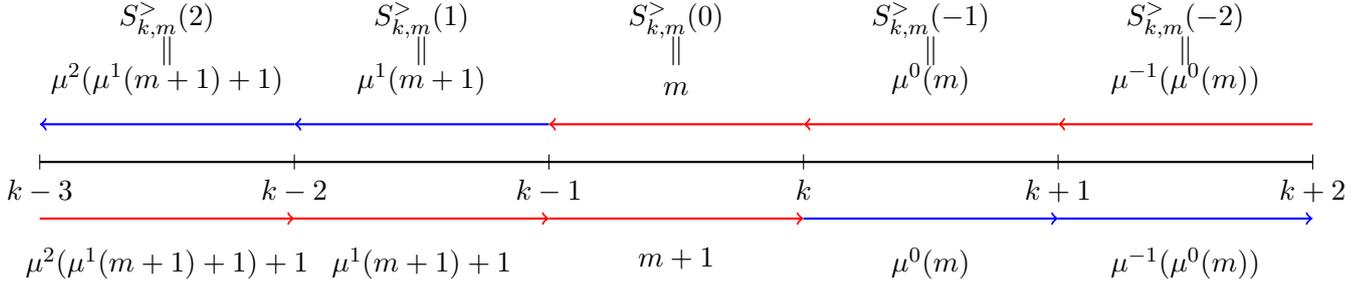
\begin{figure}
    \centering
    \begin{tikzpicture}
        % Define the spacing
        \def\spacing{3.35}
        \def\heightred{.5}
        \def\heightblue{-.75}
        
        % Draw the number line with length exactly matching the arrows
        \draw[thick] (-3*\spacing,0) -- (2*\spacing,0);
        
        % Add ticks and labels for the number line
        \foreach \x/\xtext in {-3/{$k-3$},-2/{$k-2$},-1/{$k-1$},0/$k$,1/{$k+1$},2/{$k+2$}} {
            \draw (\x*\spacing,0.1) -- (\x*\spacing,-0.1); % ticks
            \node[below] at (\x*\spacing,-0.1) {\xtext};  % labels
        }
        
        % Upper red arrows (aligned with sites on the number line)
        \draw[blue, thick, ->] (-2*\spacing, \heightred) -- (-3*\spacing, \heightred);
        \draw[blue, thick, ->] (-1*\spacing, \heightred) -- (-2*\spacing, \heightred);
        \draw[red, thick, ->] (0*\spacing, \heightred) -- (-1*\spacing, \heightred);
        \draw[red, thick, ->] (1*\spacing, \heightred) -- (0*\spacing, \heightred);
        \draw[red, thick, ->] (2*\spacing, \heightred) -- (1*\spacing, \heightred);
        
        % Lower blue arrows (aligned with sites on the number line)
        \draw[red, thick, <-] (-2*\spacing, \heightblue) -- (-3*\spacing, \heightblue);
        \draw[red, thick, <-] (-1*\spacing, \heightblue) -- (-2*\spacing, \heightblue);
        \draw[red, thick, <-] (0*\spacing, \heightblue) -- (-1*\spacing, \heightblue);
        \draw[blue, thick, <-] (1*\spacing, \heightblue) -- (0*\spacing, \heightblue);
        \draw[blue, thick, <-] (2*\spacing, \heightblue) -- (1*\spacing, \heightblue);
        
        % Upper labels (aligned with the upper arrows)
        \node[above] at (-2.5*\spacing, 1.5) {$S^>_{k,m}(2)$};
        \node[above] at (-1.5*\spacing, 1.5) {$S^>_{k,m}(1)$};
        \node[above] at (-0.5*\spacing, 1.5) {$S^>_{k,m}(0)$};
        \node[above] at (0.5*\spacing, 1.5) {$S^>_{k,m}(-1)$};
        \node[above] at (1.5*\spacing, 1.5) {$S^>_{k,m}(-2)$};
        
        % Upper equals (aligned with the upper arrows)
        \node[above] at (-2.5*\spacing, 1.15) {$\parallel$};
        \node[above] at (-1.5*\spacing, 1.15) {$\parallel$};
        \node[above] at (-0.5*\spacing, 1.15) {$\parallel$};
        \node[above] at (0.5*\spacing, 1.15) {$\parallel$};
        \node[above] at (1.5*\spacing, 1.15) {$\parallel$};
        
        % Upper \mu (aligned with the upper arrows)
        \node[above] at (-2.5*\spacing, .75) {$\mu^2(\mu^1(m+1)+1)$};
        \node[above] at (-1.5*\spacing, .75) {$\mu^1(m+1)$};
        \node[above] at (-0.5*\spacing, .75) {$m$};
        \node[above] at (0.5*\spacing, .75) {$\mu^0(m)$};
        \node[above] at (1.5*\spacing, .75) {$\mu^{-1}(\mu^0(m))$};
        
        % Lower labels (aligned with the lower arrows)
        \node[below] at (-2.5*\spacing, -1) {$\mu^2(\mu^1(m+1)+1)+1$};
        \node[below] at (-1.5*\spacing, -1) {$\mu^1(m+1)+1$};
        \node[below] at (-0.5*\spacing, -1) {$m+1$};
        \node[below] at (0.5*\spacing, -1) {$\mu^0(m)$};
        \node[below] at (1.5*\spacing, -1) {$\mu^{-1}(\mu^0(m))$};
        
        \end{tikzpicture}
    \caption{Sketch of the local time process $S_{k,m}^>(l)$ at time $T_{k,m}$, with $k \geq 3$ here. The numbers next to the left (resp right) arrows represent the number of leftwards (resp rightwards) crossings of the associated edge. The colors of the arrows represent the color of the balls associated to their direction in the urn they point out of. Notice the equality of rightwards and leftwards crossings for the edges in the halfspace $x>k$. This is due to $2$ facts : (i) we start from site $0$, (ii) we stop the process when $k$ has been visited $m+1$ times from the left. Edges $(x,x+1), x \in [0,k-1]$ have been visited once more than their counterparts $(x+1,x)$ for the same reason. There is always one red and one blue arrow pointing out of each site.}
    \label{fig:rayknight}
\end{figure}

These recursive definitions are illustrated in Fig.~\ref{fig:rayknight}.

This discussion invites us to define two Markov processes. The first one, $\mathcal{Z}(l)$, defined by 
\[
\mathcal{Z}(l+1) = \mu^{l+1}(\mathcal{Z}(l)+1),
\] 
we will identify with $S^>_{k,m}(l)$ in the interval $l \in [1,k-1]$. According to Eqs.~\eqref{def-urns-red}-\eqref{def-urns-blue} and FIG. \ref{fig:rayknight}, to make this identification, the Polya urns associated to $\mu^l$ must have weight functions 
\[r(j) = w(2j), b(j) = w(2j+1).\] The second Markov process, $\tilde{\mathcal{Z}}(l)$, defined by 
\[
\tilde{\mathcal{Z}}(l+1) = \tilde{\mu}^{l+1}(\tilde{\mathcal{Z}}(l)),
\] 
we will identify with $S^>_{k,m}(l)$ for $l \in \mathbb{Z}_-$. The Polya urns associated to $\tilde{\mu}^l$ have weight functions \[\tilde{r}(j) = w(2j+1), \tilde{b}(j) = w(2j).\] 
It now remains to compute the distribution of these processes in the continuous limit. In the following, we compute only the distribution of $\mathcal{Z}(l)$, as the computation for $\tilde{\mathcal{Z}}(l)$ is similar.
Thinking of applying \eqref{lemma-polya}, we define
\begin{equation}
    U_p(n) \equiv \sum_{j=0}^{n-1} w(2j)^{-p}, V_p(n) \equiv \sum_{j=0}^{n-1} w(2j+1)^{-p}
\end{equation}
as well as the new Markov chain $\mathcal{Y}(l) = V_1(\mathcal{Z}(l))$. We now derive the effective dynamics of the rescaled edge-local time process in the spatial variable $l$. Define $\mathcal{Y}(l) := V_1(S^>_{k,m}(l))$, where $V_1(n) = \sum_{j = 0}^{n - 1} b(j)^{-1}$, as in the previous section. The process $\mathcal{Y}(l)$ evolves according to a discrete stochastic recursion whose drift and variance can be computed explicitly using the identity~\eqref{lemma-polya}. The effective, spatially dependent drift is given by:
\[
F(x) \equiv \mathbb{E}\left[ \mathcal{Y}(l+1) \,\middle|\, \mathcal{Y}(l) = x \right] - x = U_1\left(V_1^{-1}(x) + 1\right) - x,
\]
where $U_1(n) = \sum_{j = 0}^{n - 1} r(j)^{-1}$. This expression follows directly from the properties of the generalized Pólya urn process and \eqref{lemma-polya}. Similarly, the effective variance reads:
\[
G(x) \equiv \mathbb{E}\left[ \left( \mathcal{Y}(l+1) - \mathbb{E}[\mathcal{Y}(l+1) \mid \mathcal{Y}(l)] \right)^2 \right] 
= U_2\left(V_1^{-1}(x) + 1\right) + \mathbb{E}\left[ V_2\left(V_1^{-1}(\mathcal{Y}(l+1))\right) \,\middle|\, \mathcal{Y}(l) = x \right],
\]
with $U_2(n) = \sum_{j = 0}^{n - 1} r(j)^{-2}$ and $V_2(n) = \sum_{j = 0}^{n - 1} b(j)^{-2}$. These expressions allow us to represent the stochastic increments as:
\begin{equation}
\label{increments}
    \mathcal{Y}(l+1) - \mathbb{E}\left[\mathcal{Y}(l+1) \mid \mathcal{Y}(l)\right] = \sqrt{G(\mathcal{Y}(l))} \, \epsilon_l,
\end{equation}
where the $\epsilon_l$ are zero-mean and unit-variance random variables. Crucially, since the process $\mathcal{Y}(l)$ is Markovian in $l$, the variables $\epsilon_l$ are also uncorrelated. In the scaling limit, we will interpret $\epsilon_l$ as Gaussian white noise. 
Putting everything together, we obtain the following discrete stochastic difference equation:
\begin{equation}
\label{discrete-sde}
    \mathcal{Y}(l+1) - \mathcal{Y}(l) = \sqrt{G(\mathcal{Y}(l))} \, \epsilon_l + F(\mathcal{Y}(l)).
\end{equation}
To pass to the continuum limit and identify the resulting SDE from \eqref{discrete-sde}, it remains to compute the asymptotic behavior of the functions $F(x)$ and $G(x)$. In the next sections, we will carry out this analysis for each of the universality classes of SIRWs.

\subsubsection{SATW$_\phi$}

In this simplest case, the weight function is constant at large $n$, leading to linear forms for the cumulative sums:
\[
U_p(n) = \phi^p + (n - 1), \qquad V_p(n) = n.
\]
From these expressions, we find the asymptotic behavior of the effective drift:
\[
F(x) = U_1(V_1^{-1}(x) + 1) - x \sim \phi, \qquad \text{as } x \gg 1.
\]

For the effective variance $G(x)$, since both $U_2(n)$ and $V_2(n)$ scale linearly with $n$, we write:
\begin{align}
G(x) &= U_2(V_1^{-1}(x) + 1) + \mathbb{E}\left[ V_2\left(V_1^{-1}(\mathcal{Y}(l+1)) \right) \,\middle|\, \mathcal{Y}(l) = x \right] \nonumber \\
&\sim x + \mathbb{E}[\mathcal{Y}(l+1) \mid \mathcal{Y}(l) = x] \sim x + x = 2x.
\end{align}

To take the scaling limit, we define a rescaled process:
\[
Y(t) := \lim_{A \to \infty} A^{-1} \mathcal{Y}(\lfloor At \rfloor), \qquad dW_t := \lim_{A \to \infty} A^{-1/2} \epsilon_{\lfloor At \rfloor},
\]
where $dW_t$ is interpreted as Gaussian white noise. The discrete stochastic recursion~\eqref{discrete-sde}, together with the asymptotic forms of $F$ and $G$, then leads to the stochastic differential equation
\begin{equation}
    dY(t) = \phi \, dt + \sqrt{2Y(t)} \, dW_t.
\end{equation}

Hence, $Y(t)$ evolves as (one half of) a squared Bessel process $\mathrm{BESQ}_{2\phi}$. Since in this case $\mathcal{Z}(l) = V_1^{-1}(\mathcal{Y}(l)) = \mathcal{Y}(l)$, the rescaled oriented-edge local time process
\[
Z(t) := \lim_{A \to \infty} A^{-1} \mathcal{Z}(\lfloor At \rfloor)
\]
is also a $\mathrm{BESQ}_{2\phi}$ process.

\subsubsection{PSRW$_\gamma$}

In Tóth's original work~\cite{tothGeneralizedRayKnightTheory1996}, a misprint appears in the stated asymptotics of $U_1(n)$ and $V_1(n)$ for the polynomially self-repelling walk (PSRW). We correct it here using the Euler--Maclaurin formula at first order:
\begin{equation}
\label{euler}
    \sum_{k=0}^n f(k) = \int_0^n f(x) \, dx + \frac{f(n) + f(0)}{2} + \int_0^n f'(x) \left( x - \lfloor x \rfloor - \frac{1}{2} \right) dx.
\end{equation}

The weight function can be rescaled without loss of generality. We take:
\[
w(n)^{-1} = 2^{-\gamma}(\gamma+1) n^\gamma.
\]
Applying~\eqref{euler} to the definitions of $U_1(n)$ and $V_1(n)$ yields:
\begin{equation}
\label{u-v-psrw}
    U_1(n) = n^{\gamma+1} - \frac{\gamma+1}{2} n^\gamma + o(n^\gamma), \qquad V_1(n) = n^{\gamma+1} + o(n^\gamma).
\end{equation}

From this, we find:
\[
V_1^{-1}(x) = x^{\frac{1}{\gamma + 1}} + \mathcal{O}(1).
\]
Letting $n := V_1^{-1}(x)$ and using~\eqref{u-v-psrw}, we obtain the drift:
\[
F(x) = U_1(n+1) - V_1(n) \sim \frac{\gamma + 1}{2} x^{\frac{\gamma}{\gamma + 1}}.
\]

For the variance, we use the fact that $U_2(n) \sim \frac{(\gamma+1)^2}{2\gamma+1} n^{2\gamma + 1}$ to write:
\begin{align}
G(x) &\sim \frac{(\gamma+1)^2}{2\gamma+1} \, x^{\frac{2\gamma + 1}{\gamma + 1}} + \frac{(\gamma+1)^2}{2\gamma+1} \, \mathbb{E}\left[ \mathcal{Y}(l+1)^{\frac{2\gamma+1}{\gamma+1}} \mid \mathcal{Y}(l) = x \right] \nonumber \\
&\sim 2 \frac{(\gamma+1)^2}{2\gamma+1} x^{\frac{2\gamma + 1}{\gamma + 1}},
\end{align}
assuming that $\mathbb{E}[\mathcal{Y}(l+1)] \sim x$ to leading order.

To define the scaling limit, set $\beta := \frac{1}{2\gamma + 1}$, and consider:
\[
Y(t) := (A\beta)^{-\gamma - 1} \mathcal{Y}(\lfloor At \rfloor).
\]
Then, computing the increment:
\begin{align}
Y\left(t + \frac{1}{A}\right) - Y(t) &= (A\beta)^{-\gamma - 1} \left( \mathcal{Y}_{\lfloor At \rfloor + 1} - \mathcal{Y}_{\lfloor At \rfloor} \right) \nonumber \\
&= (A\beta)^{-\gamma - 1} \left[ \sqrt{G((A\beta)^{\gamma+1} Y(t))} \, \epsilon_{\lfloor At \rfloor} + F((A\beta)^{\gamma+1} Y(t)) \right].
\end{align}

Expanding using the asymptotics of $F$ and $G$:
\begin{align}
Y\left(t + \frac{1}{A}\right) - Y(t) 
&\sim \sqrt{2}(\gamma+1) Y(t)^{\frac{2\gamma + 1}{2\gamma + 2}} \, dW_t 
+ A^{-1} \cdot \frac{(\gamma+1)(2\gamma + 1)}{2} Y(t)^{\frac{\gamma}{\gamma + 1}}.
\end{align}
Taking the limit $A \to \infty$ (i.e., $dt = 1/A$), we obtain the SDE:
\begin{equation}
\label{sde-psrw}
    dY(t) = \frac{(\gamma+1)(2\gamma+1)}{2} Y(t)^{\frac{\gamma}{\gamma+1}} dt + \sqrt{2}(\gamma+1) Y(t)^{\frac{2\gamma+1}{2\gamma+2}} dW_t.
\end{equation}
We now show that the SDE \eqref{sde-psrw} describes a certain power of a BESQ$_1$. Indeed, define $Y(t) = B_1(t)^{\gamma + 1}$, where $B_1(t)$ satisfies:
\[
dB_1(t) = \frac{1}{2} dt + \sqrt{2B_1(t)} dW_t,
\]
i.e., $2B_1(t)$ is a $\mathrm{BESQ}_1$ process. Applying Itô’s lemma to $Y(t) = B_1(t)^{\gamma + 1}$ yields:
\begin{align}
dY(t) &= (\gamma + 1) B_1(t)^{\gamma} dB_1(t) + \frac{\gamma(\gamma + 1)}{2} B_1(t)^{\gamma - 1} (2B_1(t) dt) \nonumber \\
&= (\gamma+1) B_1(t)^{\gamma} \left( \frac{1}{2} dt + \sqrt{2B_1(t)} dW_t \right) + \gamma(\gamma + 1) B_1(t)^{\gamma} dt,
\end{align}
which matches~\eqref{sde-psrw} upon substituting $B_1(t) = Y(t)^{1/(\gamma+1)}$.

Finally, recalling that $\mathcal{Z}(l) = V_1^{-1}(\mathcal{Y}(l)) \sim \mathcal{Y}(l)^{1/(\gamma+1)}$, the rescaled process
\[
Z(t) := \lim_{A \to \infty} A^{-1} \mathcal{Z}(\lfloor At \rfloor)
\]
is a $\mathrm{BESQ}_1$ process.

\subsubsection{SESRW}
The analysis of the SESRW class is technically more involved than in the SATW and PSRW cases. For clarity, we restrict our attention to the special case $\kappa = 1$, corresponding to the well-known TSAW. The original proof by Tóth~\cite{tothTrueSelfrepelling} takes a different route, relying on the \emph{Brownian web}—a central object that will be introduced and used in the third part of this thesis. For a complete treatment of the general SESRW case, we refer the reader to~\cite{tothSelfinteractingRandomMotions}.

\subsubsection{TSAW}
In the TSAW model, the weight function takes the exponential form
\[
w(n) = e^{\beta n},
\]
where the parameter $\beta > 0$ corresponds to an attractive interaction, while $\beta < 0$ indicates a repulsive one.

Using the definitions of $U_p(n)$ and $V_p(n)$, we obtain the following closed-form expressions:
\begin{equation}
\label{u-v-sesrw}
U_p(n) = \frac{e^{2p\beta n} - 1}{e^{2p\beta} - 1}, \qquad 
V_p(n) = e^{p\beta} \cdot \frac{e^{2p\beta n} - 1}{e^{2p\beta} - 1}.
\end{equation}
From this, we deduce
\[
V_1^{-1}(x) = \frac{1}{2\beta} \log \left( e^{-\beta} \left( e^{\beta} + e^{2\beta} x - x \right) \right) = \frac{\log(1 + 2\sinh(\beta) x)}{2\beta}.
\]

The drift $F(x)$ then takes the form
\begin{equation}
\label{F-tsaw}
F(x) = \mathbb{E}\left[ \mathcal{Y}(l+1) \mid \mathcal{Y}(l) = x \right] - x = U_1(V_1^{-1}(x)+1) - x = 1 + (e^\beta - 1)x.
\end{equation}

To estimate the variance of the increments $G(x)$, we employ the asymptotic forms of $U_2$ and $V_2$, leading to the expression
\begin{align}
\label{G-tsaw}
G(x) &= U_2(V_1^{-1}(x) + 1) + \mathbb{E} \left[ V_2\left( V_1^{-1}(\mathcal{Y}(l+1)) \right) \,\middle|\, \mathcal{Y}(l) = x \right] \nonumber \\
&\sim \frac{e^{2\beta} - 1}{e^{2\beta} + 1} x^2 \left( e^{2\beta} + \mathbb{E} \left[ \left( \frac{\mathcal{Y}(l+1)}{\mathcal{Y}(l)} \right)^2 \,\middle|\, \mathcal{Y}(l) = x \right] \right) \Sim{l \to \infty} g x^2,
\end{align}
where $g$ is some function of $\beta$. Unlike in the SATW and PSRW cases, here the increment $\mathcal{Y}(l+1) - \mathcal{Y}(l)$ is of the same order as $\mathcal{Y}(l)$ itself. Both the drift and the standard deviation scale linearly with $\mathcal{Y}(l)$, suggesting that $\mathcal{Y}(l)$ grows exponentially with $l$ in the large-$l$ limit.

This motivates a return to the original oriented-edge local time process $\mathcal{Z}(l) = V_1^{-1}(\mathcal{Y}(l))$, for which
\begin{equation}
\label{z-def}
\mathcal{Z}(l) = \frac{\log(1 + 2\sinh(\beta)\, \mathcal{Y}(l))}{2\beta}.
\end{equation}
The increments of this process write
\begin{equation}
\label{increments-tsaw}
\mathcal{Z}(l+1) - \mathcal{Z}(l) = \frac{1}{2\beta} \log \left( \frac{1 + 2\sinh(\beta)\, \mathcal{Y}(l+1)}{1 + 2\sinh(\beta)\, \mathcal{Y}(l)} \right).
\end{equation}

Equation~\eqref{increments-tsaw} admits two distinct regimes, depending on the magnitude of $\mathcal{Z}(l)$. In the limit $\mathcal{Z}(l) \gg 1$, we apply the discrete stochastic difference equation~\eqref{discrete-sde} in the large-$l$ regime, which gives
\begin{equation}
\label{increments-tsaw-zgg1}
\mathcal{Z}(l+1) - \mathcal{Z}(l) \sim \frac{1}{2\beta} \log \left( \frac{\mathcal{Y}(l+1)}{\mathcal{Y}(l)} \right) = \frac{1}{2\beta} \log \left( e^\beta + \sqrt{g}\, \varepsilon_l \right) = \frac{1}{2} \log \left( 1 + \sqrt{g} e^{-\beta}\, \varepsilon_l \right),
\end{equation}
where $\varepsilon_l$ are i.i.d centered random variables with unit variance. Note that the increment \eqref{increments-tsaw-zgg1} can be positive or negative. 

On the other hand, when $\mathcal{Z}(l)$ is close to $1$ (recall that $\mathcal{Z}(l) \geq 1$ since it is a positive integer), the increment~\eqref{increments-tsaw} is necessarily nonnegative:
\[
\mathcal{Z}(l+1) - \mathcal{Z}(l) \geq 0,
\]
which enforces a reflecting boundary condition at zero.

From physical considerations, we argue that the increments~\eqref{increments-tsaw} are asymptotically symmetric when $\mathcal{Z}(l) \gg 1$, i.e.,
\begin{equation}
\label{drift-z-tsaw}
r \equiv \mathbb{E}\left[ \mathcal{Z}(l+1) - \mathcal{Z}(l) \mid \mathcal{Z}(l) \gg 1 \right] = 0.
\end{equation}
We emphasize that the quantity $r$ does not depend on $l$, since the noise terms $\varepsilon_l$ are i.i.d. To establish that $r = 0$, we invoke dimensional analysis: the exponent $\alpha$ in Table~\ref{tab:rayknight} implies the scaling relation $\mathcal{Z}(l) \sim \sqrt{l}$. Therefore, the unconditional mean increment must scale as
\[
\mathbb{E}[\mathcal{Z}(l+1) - \mathcal{Z}(l)] \sim l^{-1/2}.
\]
Decomposing this expectation over the two regimes, we write
\begin{align}
\label{mean-increments-tsaw}
\mathbb{E}[ \mathcal{Z}(l+1) - \mathcal{Z}(l) ] &= \mathbb{P}(\mathcal{Z}(l) \approx 1) \cdot \mathbb{E}[ \mathcal{Z}(l+1) - \mathcal{Z}(l) \mid \mathcal{Z}(l) \approx 1] \nonumber \\
&\quad + \mathbb{P}(\mathcal{Z}(l) \gg 1) \cdot \mathbb{E}[ \mathcal{Z}(l+1) - \mathcal{Z}(l) \mid \mathcal{Z}(l) \gg 1] \nonumber \\
&= \mathbb{P}(\mathcal{Z}(l) \approx 1) + (1 - \mathbb{P}(\mathcal{Z}(l) \approx 1)) r \\ &= r + (1-r)\mathbb{P}(\mathcal{Z}(l) \approx 1)
\end{align}
Since the left-hand side is of order $l^{-1/2}$, it follows that $r = 0$.

Physically, this confirms that the positive bias in the increments appears only when the process is close to the reflecting barrier at zero—an event whose likelihood vanishes as $l^{-1/2}$. Consequently, under the diffusive rescaling
\[
z_A(t) \equiv \frac{\mathcal{Z}(\lfloor At \rfloor)}{\sqrt{A}},
\]
we conclude that $z_A(t)$ converges in distribution to reflected Brownian motion as $A \to \infty$:
\[
z_A(t) \xrightarrow{A \to \infty} C \cdot |B_t|,
\]
where $B_t$ denotes standard Brownian motion, and $C$ is a constant depending on $\beta$.
\section{Conclusion}

In this chapter, we have presented the Ray–Knight theorems for SIRWs, which provide access to the statistical structure of the edge local time process $L_{T_k^l}$ at inverse local time $T_k^l$. This process captures the hidden memory of the walker and underpins the non-Markovian nature of the dynamics of $X_t$. As such, understanding the statistics of $L_{T_k^l}$ offers a powerful route to compute key observables of SIRWs, including their space exploration properties—an analysis we pursue in the next chapter.

This framework also highlights several natural limitations and directions for generalization:

\begin{itemize}
    \item As formulated here, the theorems describe the edge local time statistics at a single inverse local time. Consequently, they are not directly applicable to the study of aging or two-time observables. In Sec.~\ref{sec:aging}, we develop an aged Ray–Knight theory tailored to the specific case of SATW$_\phi$. In Part~3 of this thesis, we will make use of the Brownian web~\cite{tothTrueSelfrepelling} to study aging in the TSAW. A general aged Ray–Knight theory applicable to all SIRWs has not yet been formulated\footnote{This question will be investigated in future work.}.
    
    \item While the results extend beyond the $1D$ setting and apply to SIRWs on trees (as discussed in Sec.~\ref{sec:sirw-tree}), they crucially rely on the absence of loops in the underlying graph. This structural assumption prevents a direct extension to the physically relevant $2D$ case.
    
    \item Finally, the Ray–Knight framework can be generalized to describe the collective edge local time profile generated by multiple interacting SIRWs, opening the door to future work on cooperative behavior in these systems.
\end{itemize}

In the following chapter, we will apply the Ray–Knight theorems developed in the present chapter to compute several key observables characterizing space exploration by SIRWs.

\chapter{Key observables of SIRWs from Ray--Knight theorems}
Despite the extensive literature on SIRWs \cite{amitAsymptoticBehaviorTrue1983,tothGeneralizedRayKnightTheory1996,barbier-chebbahSelfInteractingRandom}, very little is known about their quantitative space exploration properties. To date, explicit results are limited to the exact propagator of the TSAW~\cite{dumazMarginalDensities} and the first-passage time distribution for the SATW~\cite{carmonaBetaVariables}. Beyond these specific cases, there exists no general theoretical framework to predict how a SIRW explores space over time.

This lack of understanding stems from the intrinsic non-Markovian nature of SIRWs: their evolution depends on their entire history through reinforcement or repulsion mechanisms, making standard tools from the theory of Markov processes largely inapplicable. As a result, basic questions—such as the size of the visited territory at time $t$, how fast the RWer reaches a target site, or the asymmetry of the visited territory—remain completely open.

Yet, these questions are of central importance, both from a theoretical standpoint and for applications ranging from cellular biology~\cite{dalessandroCellMigrationGuided2021,trogerSizedependentSelfavoidanceEnables2024} to foraging~\cite{giuggioliAnimalInteractionsEmergence2011,dussutourOptimalTrafficOrganization2004} and Monte Carlo algorithms~\cite{maggsNonreversibleMonte}. Quantifying space exploration by SIRWs requires accessing observables that are inherently global and history-dependent.

In this chapter, we show that the Ray--Knight theorems presented in the previous chapter provide a powerful framework to tackle this challenge. While the mathematical literature has so far focused on proving the existence and structure of such theorems, it has yielded few explicit results of direct physical relevance. Meanwhile, the physics literature—though rich in heuristic arguments~\cite{barbier-chebbahAnomalousPersistence} and numerical investigations~\cite{barbier-chebbahSelfInteractingRandom,grassbergerSelfTrappingSelfRepellingRandom2017}—has largely lacked access to the powerful probabilistic tools needed to derive quantitative descriptions of space exploration by SIRWs. The results presented here aim to bridge this gap: using Ray--Knight theory, we derive exact expressions for several key observables that characterize the space–time dynamics of SIRWs, including:

\begin{itemize}
    \item The \emph{propagator}, which is the probability for the SIRW to be at some site $x$ at time $t$;
    \item The \emph{FPT density}, which yields the distribution of the time $T_x$ needed to hit site $x$ for the first time;
    \item the \emph{splitting probability}, which quantifies the probability for the walker to hit a point $x$ before another point $y$;
    \item the \emph{distribution of the span} $N(t)$ at fixed time $t$, which characterizes the size of the visited territory;
    \item the \emph{record age distribution}, measuring the time between successive increments of the maximum of the RWer;
    \item and the \emph{survival probability}, i.e., the probability that a walker constrained to a domain avoids touching one of its boundaries.
\end{itemize}

Together, these results offer a quantitative and predictive theory of space exploration for the broad, physically relevant class of SIRWs. We emphasize that, aside from Sec.~\ref{sec:sirw-tree}, our focus throughout this chapter is on $1D$ SIRWs. While the physically relevant extension to $2D$ remains significantly more challenging, some partial yet important results have been established in that setting~\cite{tothSuperdiffusiveBoundsSelfrepellent2010,tothDiffusiveSuperDiffusiveLimits2018}.

\section{Splitting Probability and persistence exponents of SIRWs}
\label{sec:splitting}

In this section, we compute the splitting probability of one-dimensional SIRWs, which corresponds to the probability that the walker first reaches a site $k > 0$ before hitting a site $-m < 0$, starting from the origin. This quantity provides key insights into the spatial exploration process of SIRWs and plays a crucial role in determining their persistence properties, as we show.

\subsection{Definition and Observable}

We define the splitting probability $Q_+(k, -m)$ as the probability that the SIRW hits $k$ before $-m$:
\begin{equation}
Q_+(k, -m) \equiv \mathbb{P}(T_k < T_{-m}),
\end{equation}
where $T_k$ (resp. $T_{-m}$) is the first hitting time of site $k$ (resp. $-m$), and the SIRW starts from $X_0 = 0$.

To compute $Q_+(k, -m)$, we begin by introducing the auxiliary observable $q_+(k, S)$, defined as the probability that the total number of distinct sites visited by the RWer when first hitting $k$ is exactly $S + 1$, where the additional $1$ accounts for the origin being included in the visited sites. If the walker reaches $k$ before $-m$, then it must have visited at least $k + 1$ and at most $k + m$ sites at time $T_k$. Thus, we have:
\begin{equation}
Q_+(k, -m) = \sum_{j = k}^{k + m - 1} q_+(k, j).
\label{eq:qplus-sum}
\end{equation}

\subsection{Expression of \texorpdfstring{$q_+(k, S)$}{q+(k, S)} in Terms of Edge Local Times}

To compute $q_+(k, S)$, we take advantage of the Ray–Knight representation of the edge-local time process. Let $T_k$ be the first hitting time of site $k > 0$, and denote by $L_{T_k}(x)$ the edge-local time at position $x$ at that time.

We define the support of the local time process as:
\[
I_k := \{x \in \mathbb{Z} \mid L_{T_k}(x) > 0\}.
\]
Since the walker reaches $k$ for the first time at $T_k$, it follows that $L_{T_k}(x) = 0$ for all $x > k$, and $L_{T_k}(k) = 1$. Thus, $I_k$ is necessarily of the form $[-m, k]$ for some (random) integer $m > 0$.

Then, the number of visited sites is $S + 1 = m + k + 1$, and we have:
\begin{equation}
q_+(k, S) = \mathbb{P}(I_k = [-(S - k) + 1, k]) = \mathbb{P}(\Xi_0 = S - k),
\end{equation}
where $\Xi_0$ is the first (leftmost) zero of $L_{T_k}(x)$ on the negative half-line.

To proceed, we condition on the value $a$ of the edge-local time at the origin:
\begin{equation}
q_+(k, S) = \sum_{a = 0}^\infty \mathbb{P}(\Xi_0 = S - k \mid L_{T_k}(0) = a) \, \mathbb{P}(L_{T_k}(0) = a).
\label{eq:qplus-decomposition}
\end{equation}

% ERROR BEGIN
\subsection{Scaling Limit via Ray–Knight Theory}
\begin{figure}
    \centering
    \begin{subfigure}[t]{0.48\textwidth}
        \centering
        \includegraphics[width=\textwidth]{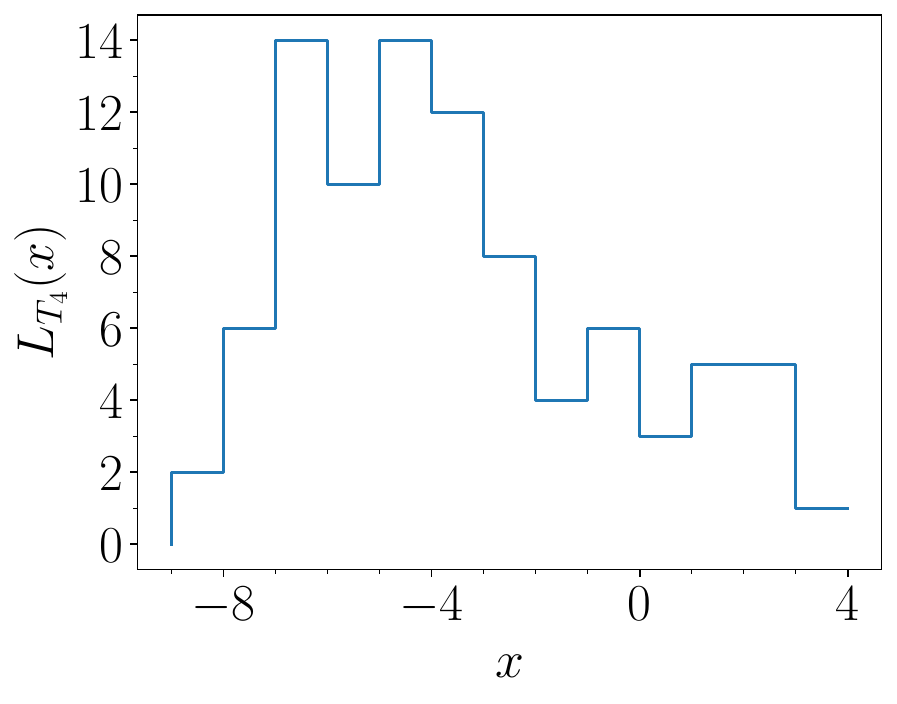}
        \caption{Edge local time process $(L_{T_4}(x))_{x\in \mathbb{Z}}$ of a SIRW. The process is stopped at the first hitting time of site $k=4$. Here, we see that the leftmost site visited by the SIRW is $-9$, as $L_{T_4}(x) = 0$ for $x \leq -9$. Thus $\Xi_0 = 9$ as defined in \eqref{eq:qplus-decomposition}. A total of $S+1 = 4 + 9 + 1 = 14$ sites have been visited: such a realization contributes to the probability $q_+(k=4, S=13)$.}
        \label{fig:loctime-disc}
    \end{subfigure}
    \hfill
    \begin{subfigure}[t]{0.48\textwidth}
        \centering
        \includegraphics[width=\textwidth]{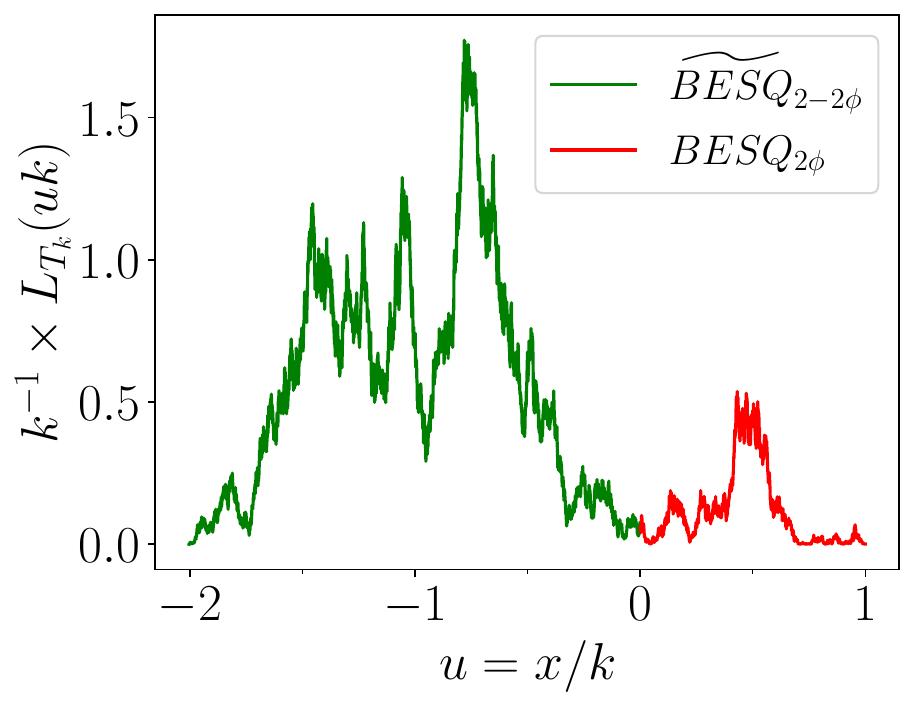}
        \caption{Scaled edge local time process $(k^{-1}L_{T_{k}}(uk))_{u\leq 1}$ of a SATW$_{\phi}$ where $\phi = \log 2$. The number of sites visited, although random, is inevitably large: $S\sim 15000$ here, and $z = k/S \sim 0.33$ is the fraction of visited sites to the right of $0$. Thus, this realization contributes to the probability $q_+(z\sim 0.33)dz$. Ray-Knight theory \eqref{rayknight-tk} states that, as $k\to \infty$, the red process converges to $(Y_{2\phi}(1-u))_{0\leq u \leq 1}$ where $Y_{2\phi}$ is a BESQ$_{2\phi}$ process conditioned on vanishing at $0$. In turn, the green process converges to $(\tilde{Y}_{2-2\phi}(-u))_{u \leq 0}$ where $\tilde{Y}_{2-2\phi}$ is a $\widetilde{\text{BESQ}}_{2-2\phi}$ process. The red and green processes are independent given their common value at $u=0$.}
        \label{fig:loctime-cont}
    \end{subfigure}
    % \caption{Visual representations of the local time process $L_{T_x}$ of a SIRW.}
    % \label{fig:loc_times}
\end{figure}

In the scaling limit $A \to \infty$ with $z = k/A$ fixed, the edge-local time profile converges to a piecewise process involving squared Bessel processes:
\[
L_{T_k}(x) \sim C \cdot A^\alpha \mathcal{L}_z(u), \qquad \text{with } u = \frac{x}{A},
\]
where
\[
\mathcal{L}_z(u) = 
\begin{cases}
Y_\delta(z - u)^q, & 0 \leq u \leq z, \\
\widetilde{Y}_{2 - \delta}(-u)^q, & u \leq 0,
\end{cases}
\]
and $Y_\delta(u)$ (respectively, $\widetilde{Y}_{2 - \delta}(u)$) denotes a $\mathrm{BESQ}_\delta$ process (respectively, a $\mathrm{BESQ}_{2 - \delta}$ process absorbed at $0$). See FIG.~\ref{fig:loctime-cont} for an illustration in the case of the SATW$_\phi$. \par  The parameters $\delta, q, C$ are specific to each SIRW class \ref{tab:rayknight}. In the following, we set $C = 1$ since we are not keeping track of time, i.e., the area under the local time profile. Indeed, the splitting probability and the quantity $q_+$ are time-integrated observables.

Let $Z_0$ be the first passage time to 0 of $\widetilde{Y}_{2 - \delta}$. Using the fact that the processes on $[0, z]$ and $]-\infty, 0]$ are independent conditioned on their value at $u = 0$, the scaling limit of $q_+(k, S)$ becomes:
\begin{equation}
q_+(z) \equiv \lim_{A \to \infty} A \cdot q_+(k = zA, S=A) 
= \int_0^\infty \mathbb{P}(Z_0 = 1 - z \mid \widetilde{Y}_{2 - \delta}(0) = a) \cdot \mathbb{P}(Y_\delta(z) = a) \, da.
\label{eq:qplus-integral}
\end{equation}

The explicit densities of the Bessel processes are known \eqref{eq:besq_propagator}, \eqref{eq:besq_fpt}:
\begin{align}
\mathbb{P}(Y_\delta(z) = a) &= \frac{1}{(2z)^{\delta/2} \Gamma(\delta/2)} a^{\delta/2 - 1} e^{-a/(2z)}, \\
\mathbb{P}(Z_0 = 1 - z \mid \widetilde{Y}_{2 - \delta}(0) = a) &= \frac{1}{(1 - z)\Gamma(\delta/2)} \left( \frac{a}{2(1 - z)} \right)^{\delta/2} e^{-a/(2(1 - z))}.
\end{align}

Substituting into \eqref{eq:qplus-integral} yields:
\begin{equation}
q_+(z) = \frac{(z(1 - z))^{-\delta/2}}{2^\delta (1 - z) \Gamma(\delta/2)^2} 
\int_0^\infty a^{\delta - 1} \exp\left( -\frac{a}{2z} - \frac{a}{2(1 - z)} \right) da.
\end{equation}

This integral evaluates to a Beta function, giving:
\begin{boxedeq}
\begin{equation}
q_+(z) = \frac{1}{B(\delta/2)} \cdot \frac{(z(1 - z))^{\delta/2}}{1 - z}.
\label{eq:qplus-final}
\end{equation}
\end{boxedeq}

\subsection{Final Expression for the Splitting Probability}

From the relation \eqref{eq:qplus-sum} and the definition of $Q_+(z)=\Lim{L \to \infty} Q_+(zL,-(1-z)L)$ in the scaling limit, we obtain
\begin{equation}
Q_+(z) = 1 - \int_0^z \frac{q_+(u)}{u} du.
\end{equation}
Finally, we obtain the explicit form of the splitting probability for each universality class:
\begin{keyboxedeq}[Splitting probability of SIRWs]
\begin{equation}
\label{splitting-sirw}
Q_+(z) =
\begin{cases}
I_{1 - z}(\phi), & \text{SATW}_\phi, \\
I_{1 - z}(1/2) = \frac{2}{\pi} \arcsin \sqrt{1 - z}, & \text{PSRW}_\gamma \text{ or } \text{SESRW}_{\kappa, \beta},
\end{cases}
\end{equation}
\end{keyboxedeq}
where $I_{x}(a)$ is the regularized incomplete Beta function:
\[
I_x(a) = \frac{1}{B(a)} \int_0^x u^{a - 1}(1 - u)^{a - 1} du.
\]
This exact result calls for several comments.
\begin{enumerate}
    \item Note that the result for SATW$_\phi$ was already computed in \cite{carmonaBetaVariables} using a completely different method. The arcsine law for the other two classes is new to the best of our knowledge and striking: the splitting probability does not depend on any of the parameters $\gamma,\kappa,\beta$.
    \item The ubiquity of $I_z$ in the expression of the splitting probability of several $1d$ scale-invariant processes was already noted in \cite{zoiaAsymptoticBehaviorSelfAffine2009}, where it was shown to appear for Lévy flights, the random acceleration process, and Sinai diffusion. However, in this article, it was also shown that expressions of splitting probabilities involving $I_z$ do not hold for all non-Markovian processes, as was checked on numerical grounds for the fBM, a paradigmatic non-Markovian process (see also \cite{wieseFirstPassageInterval2019} for analytical confirmation). In contrast, we show here that $I_z$ is involved in the splitting probability for a whole class of strongly non-Markovian processes, namely, SIRWs. This considerably extends the range of validity of the "universal" expression of the splitting probability in terms of $I_z$.
    \item We give here a new probabilistic interpretation of the integrand $(u(1-u))^{a-1}/B(a)$ involved in the splitting probability $I_z(a)$, namely, as the term $q_+(u)/u$, hinting at the fundamental nature of this observable. In other words, it provides a physical interpretation of the universal form noted in \cite{zoiaAsymptoticBehaviorSelfAffine2009}, see \eqref{splitting-from-ecc}. We will discuss this interpretation further in part 3 of this thesis.
\end{enumerate}

\begin{figure}
    \centering
    \includegraphics[width=.4\textwidth]{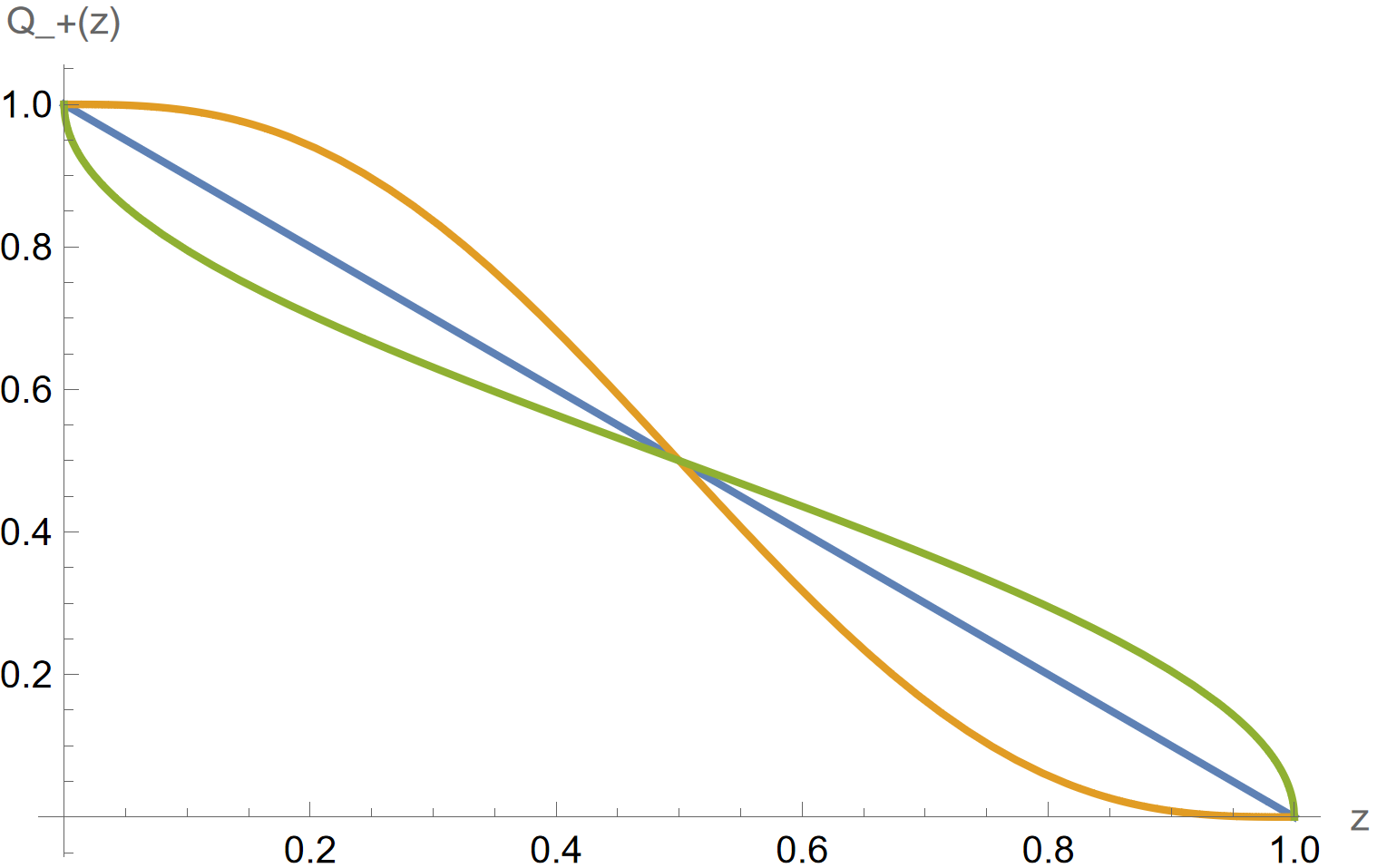}
    \caption{Plot of the splitting probability for various SIRWs. Blue : Brownian motion, orange : SATW$_{\phi=3}$, green : PSRW, SESRW$_{\kappa,\beta}$, SATW$_{\phi=1/2}$.}
    \label{fig:plot_splitting}
\end{figure}

We now exploit the exact result \eqref{splitting-sirw} to obtain the persistence exponents of SIRWs.

\subsection{Persistence exponents of SIRWs}
In this section, we compute the persistence exponent $\theta$ for each of the three classes of SIRWs. We begin by recalling a simple scaling argument that relates the behavior of the splitting probability $Q_+(z)$ as $z \to 1^-$ to the persistence exponent. We then use our exact expression for the splitting probability~\eqref{splitting-sirw} to extract the corresponding value of $\theta$ from this asymptotic relation.

\begin{figure}
    \centering
    \begin{subfigure}[t]{0.31\textwidth}
		\includegraphics[width=\textwidth]{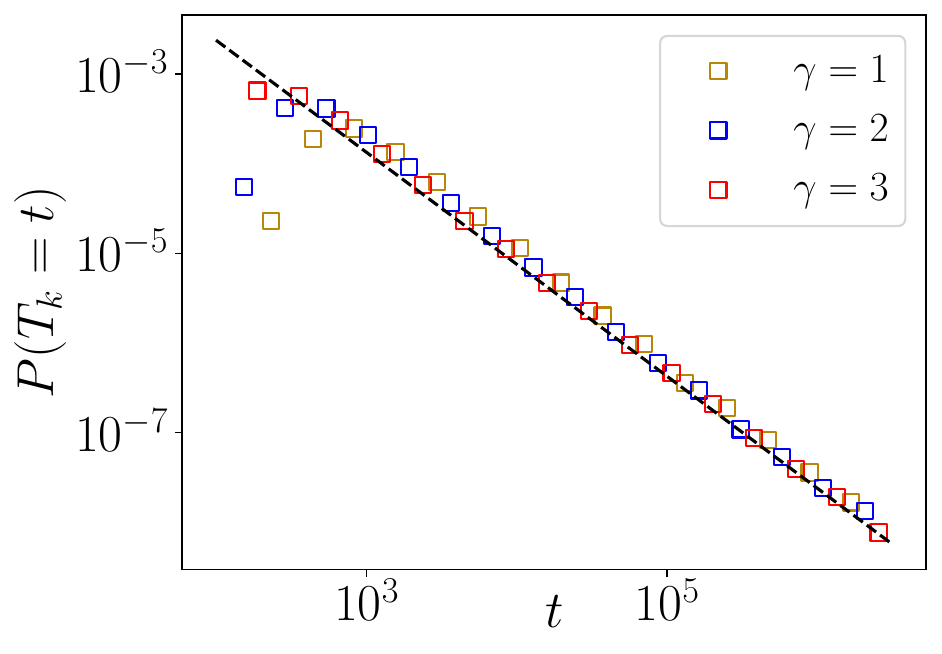}
    \end{subfigure}
    \begin{subfigure}[t]{0.31\textwidth}
        \includegraphics[width=\textwidth]{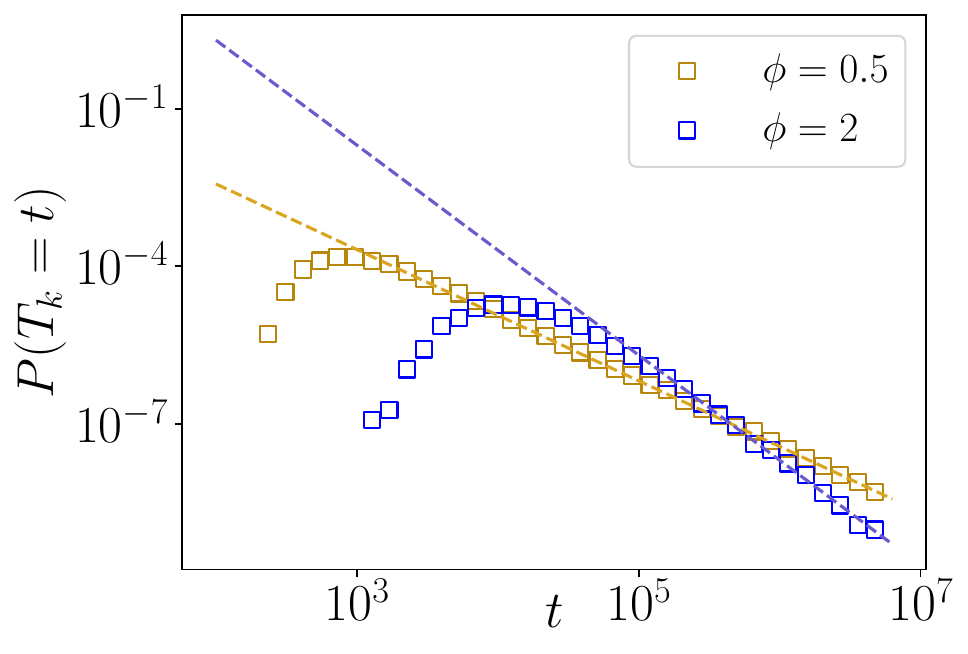}
    \end{subfigure}
    \begin{subfigure}[t]{0.31\textwidth}
        \includegraphics[width=\textwidth]{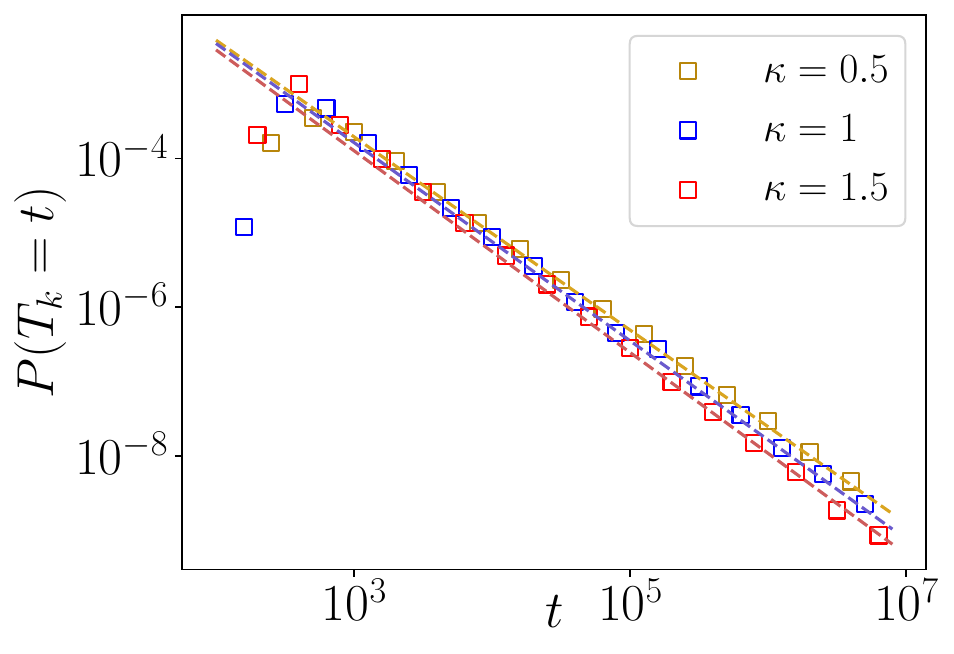}
    \end{subfigure}
    \caption{Numerical check of \eqref{persistence-exp} performed by analysing the distribution of first-hitting times $T_k$ to site $k=100$, for the three universality classes considered : (a) PSRW$_\gamma$, (b) SATW$_\phi$, (c) SESRW$_{\kappa,\beta=1}$ (Note that our results extend beyond the traditional range $0<\kappa<1$). Symbols represent numerical results while dashed lines represent the theoretical result $P(T_k = t) \Propto{t \to \infty} t^{-1-\theta}$ where $\theta$ is given by \eqref{persistence-exp}.}
    \label{fig:fpt_densities}
\end{figure}

\subsubsection{Asymptotics of the splitting probability near $z = 1$}

In~\cite{zoiaAsymptoticBehaviorSelfAffine2009}, it was shown that the splitting probability of a generic symmetric, scale-invariant one-dimensional stochastic process $X_t$ starting at $X_0 = 0$ behaves, in the limit $z \to 1$, as
\begin{equation}
\label{persistence-from-splitting}
    Q_+(z) \propto (1 - z)^{\phi}, \qquad \text{with} \quad \phi = \dw \theta,
\end{equation}
where $\dw$ denotes the walk dimension and $\theta$ is the persistence exponent of the process. In the following, we refer to $\phi$ as the splitting exponent. 

For completeness, we reproduce here a brief scaling argument leading to~\eqref{persistence-from-splitting}, in the spirit of the approach presented in~\cite{levernierUniversalFirstpassage}. Let $F(1 - z, t)$ denote the FPT density, starting from $0$, to the point $-(1 - z)$. Owing to the scale invariance of the process $X_t$, this FPT density takes the form:
\begin{equation}
    F(1 - z, t) = \frac{1}{(1 - z)^{\dw}} f\left( \frac{t}{(1 - z)^{\dw}} \right),
\end{equation}
where $f(u)$ is a scaling function that depends on the specific process.

The key observation is that, as $z \to 1$, the typical time to reach the point $-(1 - z)$ is $(1 - z)^{\dw}$, which becomes much smaller than the typical time $z^{\dw}$ required to reach $z$. Therefore, reaching $z$ before $-(1 - z)$ becomes an increasingly rare event, governed by the tails of the FPT density to the negative boundary.

The probability of reaching $z$ before $-(1 - z)$ is then given by the integrated probability that the process has not yet hit $-(1 - z)$ by the time $z$ is typically reached:
\begin{equation}
Q_+(z) \propto \int_{z^{\dw}}^{\infty} F(1 - z, t) \, dt = \int_{(z / (1 - z))^{\dw}}^{\infty} f(u) \, du.
\end{equation}

Now, by the definition of the persistence exponent $\theta$, the FPT scaling function has the asymptotic behavior:
\begin{equation}
    f(u) \propto u^{-1-\theta}, \qquad \text{as } u \to \infty.
\end{equation}
Substituting this into the integral yields:
\begin{equation}
Q_+(z) \propto \int_{(z / (1 - z))^{\dw}}^\infty u^{-\theta - 1} \, du \sim \left( \frac{z}{1 - z} \right)^{-\dw \theta} \sim (1 - z)^{\dw \theta},
\end{equation}
which confirms the scaling law~\eqref{persistence-from-splitting}.
\par 
We now apply \eqref{persistence-from-splitting} to the scale-invariant, symmetric process that is the scaling limit of SIRW considered here. We remind that $\dw$ is given in Table \ref{tab:rayknight} and $\phi$ is given by \eqref{splitting-sirw}. Namely, for SATW$_\phi$, as indicated by our notation, the reinforcement parameter $\phi$ is exactly  $\dw \theta$. For PSRW$_\gamma$ and SESRW$_{\kappa,\beta}$, we have $\phi = \tfrac{1}{2}$. 
This finally yields the persistence exponents of SIRW:
\begin{keyboxedeq}[Persistence exponents of SIRWs]
\begin{equation}
    \label{persistence-exp}
        \theta = \begin{cases}
            \frac{\phi}{2} \text{ (SATW$_\phi$)}\\
            \frac{1}{4} \text{ (PSRW$_\gamma$)} \\
            \frac{1}{2} \frac{\kappa+1}{\kappa+2}  \text{ (SESRW$_{\kappa,\beta}$).}
        \end{cases}
\end{equation}
\end{keyboxedeq}
Finally, let us comment on \eqref{persistence-exp}. 
\begin{enumerate}
    \item This result is exact and confirmed numerically in FIG. \ref{fig:fpt_densities}.
    \item We recover the persistence exponent of the SATW$_\phi$ class that was obtained using a different method in \cite{barbier-chebbahAnomalousPersistence}.
    \item We emphasize once more that obtaining the persistence exponent $\theta$ for a non-Markovian—and even non-Gaussian—process is highly uncommon. In our case, this becomes possible due to a tractable representation of the key observable $q_+(k, S)$ via Ray--Knight theory, combined with a scaling argument. To the best of our knowledge, the observable $q_+(k, S)$ has not been previously investigated in the literature, which likely explains the novelty of our result, despite its clear physical relevance.
    \item In \cite{levernierUniversalFirstpassage}, it was predicted that for processes with asymptotically stationary increments, one has $\theta = 1-\dw^{-1}$. See \cite{krugPersistenceExponents,molchanMaximumFractionalBrownian1999} in the case of the fBM. We indeed find that this holds only for the TSAW and Brownian motion, which are the only two SIRWs with asymptotically stationary increments \cite{tothTrueSelfrepelling}. 
\end{enumerate}

\subsection{Eccentricity distribution of SIRWs}

An interesting observable that can be derived from the quantities $q_{\pm}(k, m)$ is the probability $\sigma_{k,m}$ that, upon visiting $N = k + m$ distinct sites, the RW has visited $k$ sites to the right of the origin and $m$ to the left. To the best of our knowledge, this observable has not been explicitly studied in the existing RW literature, despite its clear physical interpretation as a measure of spatial asymmetry in the exploration dynamics of a general RW.

This probability naturally decomposes according to whether the walk reaches its upper boundary $k$ or its lower boundary $-(N - k)$ at the $N$-th visited site. By symmetry, we can write:
\begin{equation}
    \sigma_{k,m} = q_+(k, N) + q_+(N - k, N).
\end{equation}

In the scaling limit $N \to \infty$ with fixed eccentricity $z = k/N \in (0,1)$, the discrete probability $\sigma_{k,m}$ converges to a continuous density:
\[
N \sigma_{k,m} \To{N \to \infty, \, k/N \to z} \sigma_z,
\]
where the limiting eccentricity distribution $\sigma_z$ is given explicitly by
\begin{keyboxedeq}[eccentricity distribution of SIRWs]
\begin{equation}
    \label{eccentricity-sirw-1d}
        \sigma_z = q_+(z) + q_-(z) = \frac{q_+(z)}{z} = \frac{(z(1 - z))^{\phi - 1}}{B(\delta/2)},
\end{equation}
\end{keyboxedeq}
with $B(\delta/2)$ the Beta function and $\phi = \dw \theta$ the splitting exponent of the SIRW. The values of $\dw$ for various models are listed in Table~\ref{tab:rayknight}, while $\theta$ is given in Eq.~\eqref{persistence-exp}.

The function $\sigma_z$ represents the probability density of the eccentricity variable $z \in [0,1]$, which quantifies the asymmetry of the walk’s spatial span about the origin. Its shape encodes the effect of the underlying reinforcement mechanism:
\begin{itemize}
    \item For self-repelling SIRWs with $\phi < 1$—such as the SESRW$_{\beta,\kappa}$ and PSRW$_\gamma$ models where $\phi = \tfrac{1}{2}$—the density $\sigma_z$ diverges near $z = 0$ and $z = 1$, reflecting highly eccentric spans.
    \item For $\phi = 1$, corresponding to the simple RW, $\sigma_z$ is uniform, indicating no directional preference.
    \item For self-attracting SIRWs with $\phi > 1$, the density becomes peaked near $z = 1/2$, signaling a preference for symmetric exploration about the origin.
\end{itemize}
Finally, we note the following identity relating the eccentricity distribution $\sigma_z$ to the splitting probability $Q_+(z)$, which holds for all nearest-neighbor RWs, Markovian or not:
\begin{boxedeq}
\begin{equation}
    \label{splitting-from-ecc}
    \int_z^1 \sigma_{z'} \, dz' = Q_+(z).
\end{equation}
\end{boxedeq}
Physically, this can be understood from the fact that a RW starting at the origin and reaching $z$ before $-(1 - z)$ must necessarily have an eccentricity $z' \geq z$ by the time it attains a span of length one.

We now turn to time-dependent observables of SIRWs, beginning with the computation of the propagator—i.e., the probability distribution of the walker’s position at time $t$—for two representative and analytically tractable models: the PSRW$_\gamma$ and (a more general, asymmetric version of) the SATW$_\phi$. These models illustrate how the Ray–Knight framework can be used to extract exact dynamical information from fundamentally non-Markovian systems.

\section{Propagators of the SATW$_{\alpha,\beta}$ and PSRW$_\gamma$}
In this section, we compute analytically the propagator $\mathbb{P}(X_t = x)$ for two important classes of SIRWs: the $\mathrm{SATW}_{\alpha,\beta}$, a slightly more general and asymmetric version of $\mathrm{SATW}_\phi$, and the $\mathrm{PSRW}_\gamma$. Prior to our work, the only known analytical expression for the propagator of a SIRW was that of the TSAW, obtained by Dumaz in~\cite{dumazMarginalDensities}.

\subsection{Definition and context}
\begin{figure}
    \centering
    \includegraphics[width=.8\textwidth]{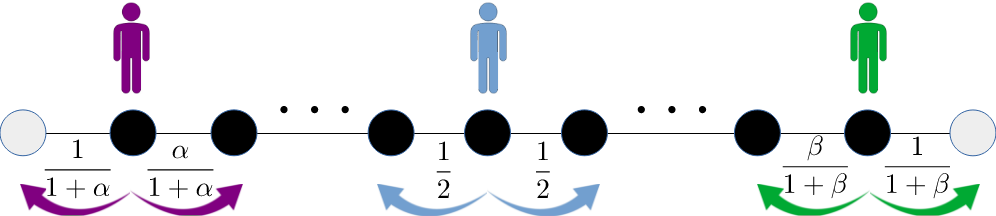}
    \caption{Sketch of a SATW$_{\alpha,\beta}$. Unvisited sites are in gray, while already visited sites are in black. Different colors represent different situations : in the purple/green case, the RW is at the left/right boundary of its span. In the blue case, it is in the bulk of its span. Jump probabilities are shown next to the arrows giving the  jump direction.}
    \label{fig:dessin-sirw}
\end{figure}
We introduce a slightly more general, asymmetric version of the SATW$_\phi$ model. As in the original SATW$_\phi$, the walker’s dynamics are governed by a weight function that saturates after an edge has been visited once. However, the present model introduces spatial asymmetry: the transition probabilities depend on whether the current site $x$ is positive or negative.

More precisely, suppose the walker is at site $x$ at time $t$. If site $x+1$ (respectively $x-1$) has not yet been visited, the walker jumps to it at time $t+1$ with probability $1/(1+\beta)$ (respectively $1/(1+\alpha)$), while it jumps to the already-visited site $x-1$ (respectively $x+1$) with probability $\beta/(1+\beta)$ (respectively $\alpha/(1+\alpha)$). If both neighboring sites have already been visited, the walker chooses either with equal probability. Thus, if $\alpha > 1$ (or $\beta > 1$), the walker is attracted to visited sites on the right (or left), and repelled if $\alpha < 1$ (or $\beta < 1$). The choice $\alpha = \beta = 1$ recovers the simple symmetric random walk (see Fig.~\ref{fig:dessin-sirw}).

The propagator $P_{\alpha,\beta}(x,t)$, defined as the probability for the walker to be at position $x$ at time $t$, plays a central role in the characterization of the dynamics. However, computing this object is notoriously difficult for non-Markovian processes. Notable exceptions where the propagator has been obtained explicitly include fractional diffusion models~\cite{metzlerAnomalousDiffusionModels2014}, the elephant random walk in its diffusive and critical regimes~\cite{bercuCenterMassElephant2021}, and its non-Gaussian behavior in the superdiffusive regime~\cite{dasilvaNonGaussianPropagator}.

Among SIRW models, only one case — the TSAW — has yielded an explicit propagator~\cite{dumazMarginalDensities}. Although the long-time behavior of the SATW class has been shown to be diffusive~\cite{davisReinforcedRandom}, and its propagator is known to admit a scaling form 
\[
P_{\alpha,\beta}(x,t) \sim \frac{1}{\sqrt{2t}} p_{\alpha,\beta}\left(\frac{x}{\sqrt{2t}}\right)
\quad \text{as } t \to \infty,\ x^2/t\ \text{fixed},
\]
even the precise value of the diffusion coefficient has remained elusive. Some progress toward determining the scaling function $p_{\alpha,\beta}$ has been made using generalized Ray–Knight theory~\cite{tothGeneralizedRayKnightTheory1996, carmonaBetaVariables}, yet no explicit expression for the propagator has been obtained through these approaches. Before proceeding to the computation of the scaling function $p_{\alpha,\beta}$, we emphasize that the universality classes established by the Ray–Knight theorems ensure the universality of the results that follow. Specifically, any SIRW with a saturating weight function will asymptotically exhibit the same behavior as the corresponding SATW$_{\alpha,\beta}$ model, where the parameters $\alpha$ and $\beta$ are determined by Eq.~\eqref{phi-links} in the edge-reinforced case.

\subsection{Computing the scaling function $p_{\alpha,\beta}$}
In the following, we derive an exact expression for the propagator of the asymmetric SATW model. Our approach differs from that of~\cite{bremontExactPropagators}, and proceeds through a more direct analysis based on Ray--Knight theory.

\subsubsection{Joint distribution of maximum, minimum and position}
Let $Q(k, m, x, t)$ denote the probability that the RWer is at position $x>0$ at time $t$, with maximum $k$ and minimum $-m$ attained over its trajectory up to time $t$. Such a configuration necessarily occurs at a time $t = T_x^l$, where $l$ is the (random) local time accumulated at site $x$.

Using the representation of $T_x^l$ as the total occupation time across all sites, we can write:
\begin{equation}
\label{Q-loctime}
Q(k, m, x, t) = \sum_{l = 0}^\infty \mathbb{P} \left( \sum_{y \in \mathbb{Z}} L_{T_x^l}(y) = t, \quad \mathrm{supp}(L_{T_x^l}) = [-m, k] \right).
\end{equation}

In the scaling limit, we apply the Ray--Knight theorems to characterize the distribution of the edge-local time profile $L_{T_x^l}$ in terms of three concatenated squared Bessel processes, as described in~\eqref{rayknight-txl}:
\begin{equation}
L_{T_x^l}(y) \approx 
\begin{cases}
\widetilde{Y}_{2 - 2\beta}(y - x), & y \geq x, \\
Y_{2\beta}(x - y), & 0 \leq y \leq x, \\
\widetilde{Y}_{2 - 2\alpha}(-y), & y \leq 0,
\end{cases}
\quad \text{with matching conditions:} \quad
\begin{cases}
\widetilde{Y}_{2 - 2\beta}(0) = Y_{2\beta}(0) = l, \\
\widetilde{Y}'_{2 - 2\alpha}(0) = Y_{2\beta}(x).
\end{cases}
\end{equation}

Here, $Y_{2\beta}$, $\widetilde{Y}_{2 - 2\alpha}$ and $\widetilde{Y}_{2 - 2\beta}$ are squared Bessel processes of respective dimensions $2\beta$, $2-2\alpha$ and $2 - 2\beta$. Together, these three processes describe the local time profile on $[x, +\infty[$, $[0, x]$, and $]-\infty, 0]$, respectively. Crucially, conditioned on their common values at $y=0$ and $y=l$, these three processes are independent of each other. Using the representation in~\eqref{Q-loctime} together with the Ray--Knight decomposition~\eqref{rayknight-txl}, we obtain the following expression for $Q(k, m, x, t)$ in the scaling limit:
\begin{align}
\label{Q-loctime-2}
Q(k, m, x, t) = \int_0^\infty \mathbb{P} \Bigg( 
& \int_0^x Y_{2\beta}(u) \, du 
+ \int_x^\infty \widetilde{Y}_{2 - 2\beta}(u) \, du 
+ \int_0^{\infty} \widetilde{Y}_{2 - 2\alpha}(u) \, du = t, \nonumber \\
& T_0^{(2 - 2\beta)} = k, \quad T_0^{(2 - 2\alpha)} = m, \quad Y_{2\beta}(0) = \widetilde{Y}_{2 - 2\beta}(0) = l 
\Bigg) \, dl,
\end{align}
where $T_0^{(\delta)}$ denotes the FPT to zero for the process $\widetilde{Y}_{2 - 2\delta}$. This integral representation expresses $Q(k, m, x, t)$ as a convolution in time, which naturally suggests taking a Laplace transform in the time variable. Applying this to~\eqref{Q-loctime-2} yields:
\begin{align}
\label{Q-loctime-laplace}
\hat{Q}(k, m, x, s) = \int_0^\infty dl \int_0^\infty dh \, 
&\mathbb{E}^l \left[ e^{-s \int_0^x Y_{2\beta}(u) \, du} \,;\, Y_{2\beta}(x) = h \right] \nonumber \\
&\times \mathbb{E}^l \left[ e^{-s \int_0^\infty \widetilde{Y}_{2 - 2\beta}(u) \, du} \,;\, T_0^{(2 - 2\beta)} = k \right] \nonumber \\
&\times \mathbb{E}^h \left[ e^{-s \int_0^\infty \widetilde{Y}_{2 - 2\alpha}(u) \, du} \,;\, T_0^{(2 - 2\alpha)} = m \right],
\end{align}
where each expectation involves a functional of a squared Bessel process (either standard or absorbed), with initial condition indicated in the superscript. The integration over $l$ and $h$ accounts for the random interface values at $y=x$ and $y=0$ that ensure continuity of the patched local time profile at these points. The functionals in \eqref{Q-loctime-laplace} are computed in the Appendix Eqs. \eqref{laplace-area-endpoint}, \eqref{joint-area-fpt}. Using $\mu = \sqrt{2s}$, we obtain 
\begin{align}
    \label{Q-loctime-laplace-2}
    \hat{Q}(k, m, x, s) 
    = \int_0^\infty \! dl \int_0^\infty \! dh \; & 
    \frac{\mu \left( \frac{h}{l} \right)^{\frac{\beta - 1}{2}} 
    e^{- \frac{1}{2} \mu (l + h) \coth(\mu x)} 
    I_{\beta - 1} \left( \frac{\mu \sqrt{hl}}{\sinh(\mu x)} \right)}{2 \sinh(\mu x)} \nonumber \\
    &\times 
    \frac{\mu \, e^{-\frac{1}{2} \mu l \coth(\mu (k - x))} 
    \left( \frac{\mu l}{2 \sinh(\mu (k - x))} \right)^\beta}{
    \Gamma(\beta) \sinh(\mu (k - x))} \nonumber \\
    &\times 
    \frac{\mu \, e^{-\frac{1}{2} \mu h \coth(\mu m)} 
    \left( \frac{\mu h}{2 \sinh(\mu m)} \right)^\alpha}{
    \Gamma(\alpha) \sinh(\mu m)}.
\end{align}
We can integrate \eqref{Q-loctime-laplace-2} by changing variables $h = \lambda l$ and using the identity 
\begin{equation}
    \int_0^\infty e^{-l B} l^{\gamma } I_{\beta -1}(l A) dl = 2^{1-\beta } A^{\beta -1} \frac{\Gamma (\beta +\gamma )}{\Gamma(\beta)} B^{-\beta -\gamma } \, _2F_1\left(\frac{\beta +\gamma }{2},\frac{1}{2} (\beta +\gamma +1);\beta ;\frac{A^2}{B^2}\right).
\end{equation}
We finally obtain 
\begin{keyboxedeq}[Joint distribution of minimum, maximum and position for SATW]
\begin{equation}
    \label{Q-laplace}
        \hat{Q}(k, m, x, s) = \frac{2\sqrt{2s}}{B(\alpha,\beta)} \frac{\sinh^\beta (m\sqrt{2s}) \sinh^\alpha(k\sqrt{2s})}{\sinh^{\alpha+\beta+1}((k+m)\sqrt{2s})} \left( \beta\frac{ \sinh(m+x)\sqrt{2s}}{\sinh(m\sqrt{2s})} + \alpha\frac{ \sinh(k-x)\sqrt{2s}}{\sinh(k\sqrt{2s})}\right).
\end{equation}
\end{keyboxedeq}
To the best of our knowledge, equation~\eqref{Q-laplace} provides the first explicit expression for the joint distribution of the maximum, minimum, and position of a SIRW. This joint distribution $Q(k,m,x,t)$ readily gives access to the propagator with absorbing boundaries $\tilde{P}_{a,b}(x,t)$, which is the probability that the RW, starting at $0$ at time $t=0$, is at $x$ at time $t$, knowing that it is absorbed if it hits the sites $-a,b$. For $x\geq 0$, this can be written as 
\[\tilde{P}_{a,b}(x,t) = \int_{x}^b dk \int_0^a Q(k,m,x,t) dm.\]
We now compute explicitly the above for $a,b \to \infty$, which yields the unconstrained propagator.

\subsubsection{The propagator from $Q(k,m,x,t)$}
Taking the marginal of equation \eqref{Q-laplace}, we obtain the Laplace transform of the propagator in the scaling limit (we first assume \( x > 0 \)):
\begin{equation}
\begin{gathered}
    \hat{P}_{\alpha,\beta}(x,s) = \int_0^\infty P_{\alpha,\beta}(x,t) e^{-st} dt = \\ 
    \frac{2\sqrt{2s}}{B(\alpha,\beta)} \int_0^\infty dm \int_{x}^\infty dk 
    \frac{\sinh^\beta (m \sqrt{2s}) \sinh^\alpha (k \sqrt{2s})}{\sinh^{\alpha+\beta+1}[(k+m) \sqrt{2s}]} \left(\alpha \frac{\sinh[(k - x)\sqrt{2s}]}{\sinh(k \sqrt{2s})} + \beta \frac{\sinh[(m + x)\sqrt{2s}]}{\sinh(m \sqrt{2s})}\right).
\end{gathered}
\end{equation}

Introducing the dimensionless scaling variable \( u = x \sqrt{2s} \) and rescaling the integration variables as \( a = m\sqrt{2s},\ b = k \sqrt{2s} \), the expression simplifies to a function we denote by \( \phi^{\alpha,\beta}(u) \), which we now aim to compute:
\begin{equation}
    \phi^{\alpha,\beta}(u) = \frac{1}{B(\alpha,\beta)} \int_0^\infty da \int_u^\infty db \frac{\sinh^\beta a \sinh^\alpha b}{\sinh^{\alpha+\beta+1}(a+b)} \left(\alpha \frac{\sinh(b - u)}{\sinh b} + \beta \frac{\sinh(a+u)}{\sinh a}\right).
\end{equation}

\paragraph{Important identities.} Several identities used throughout the computation can be derived or verified via symbolic computation tools such as \textit{Mathematica}:

\begin{equation}
\begin{gathered}
    \label{int-sinhc-sinhd}
    \int_0^\infty \frac{\sinh^c a}{\sinh^d(a+b)} da = e^{-bd} 2^{-c+d-1} B\left(c+1, \frac{d-c}{2}\right) \, _2F_1\left(d,\frac{d-c}{2};\frac{1}{2}(c+d+2);e^{-2 b}\right),
\end{gathered}
\end{equation}

\begin{equation}
\begin{gathered}
    \label{int-coth}
    \int_0^\infty \frac{\sinh^c a}{\sinh^d(a+b)} \coth(a) da = e^{-bd} 2^{-c+d-1} \left[ B\left(c,\frac{d-c}{2}\right) \, _2F_1\left(d,\frac{d-c}{2};\frac{c+d}{2};e^{-2 b}\right)\right. \\
    \left.+ B\left(c, \frac{2+d-c}{2}\right) \, _2F_1\left(d,\frac{d-c+2}{2};\frac{c+d+2}{2};e^{-2 b}\right)\right].
\end{gathered}
\end{equation}

We also recall the Euler transformation of the hypergeometric function:
\begin{equation}
    _2F_1(a,b,c;z) = (1-z)^{c-a-b} \, _2F_1(c-a,c-b,c;z).
\end{equation}

Define
\[
F = \,_2F_1\left(1 + \frac{c - d}{2},\, c;\, 1 + \frac{c + d}{2};\, e^{-2b}\right), \qquad
F_+ = \,_2F_1\left(1 + \frac{c - d}{2},\, 1 + c;\, 1 + \frac{c + d}{2};\, e^{-2b}\right).
\]
Applying the Gauss contiguous relations for the hypergeometric function, the integral \eqref{int-coth} simplifies to the more compact form
\begin{equation}
\int_0^\infty \frac{\sinh^c a}{\sinh^d(a+b)} \coth(a)\, da
= \frac{e^{-bc}}{2(c+d)\, \sinh^{d - c}(b)}\, B\left(c,\, \frac{d - c}{2} \right)
\left[ 4c\, e^{-b} \sinh(b)\, F_+ + 2(d - c)\, F \right].
\end{equation}
    
\paragraph{Telescoping series identity.} Define
\begin{equation}
    c_k = \frac{\left(\frac{1-\alpha}{2}\right)_k (\beta)_k}{k! \left(\frac{1}{2} (\alpha + 2\beta + 3)\right)_k},
\end{equation}
then
\begin{equation}
    (\alpha+k)(2k+1-\beta)c_k - (\alpha+k-1)(2k-1-\beta)c_{k-1} = (\beta-\alpha(\beta+2k)) c_k.
\end{equation}
This gives the summation identity:
\begin{equation}
    \sum_{k=0}^n (\beta - \alpha(\beta + 2k)) c_k = (-\alpha + 2n + 1)(\beta + n)c_n.
\end{equation}

\paragraph{Computing \( \phi^{\alpha,\beta}(u) \).} Define
\begin{align}
    \eta_1 &= \int_0^\infty \frac{\sinh^\beta a}{\sinh^{\alpha+\beta+1}(a+b)} da, \\
    \eta_2 &= \int_0^\infty \frac{\sinh^\beta a}{\sinh^{\alpha+\beta+1}(a+b)} \frac{\sinh(a+u)}{\sinh a} da.
\end{align}
Since \( \frac{\sinh(a+u)}{\sinh a} = \cosh(u) + \sinh(u) \coth(a) \), we get
\begin{equation}
    \eta_2 = \cosh(u) \eta_1 + \sinh(u) \psi, \quad \text{where} \quad \psi = \int_0^\infty \frac{\sinh^\beta a}{\sinh^{\alpha+\beta+1}(a+b)} \coth(a) da.
\end{equation}

Thus, for \( u>0 \),
\begin{equation}
\begin{aligned}
    \phi^{\alpha,\beta}(u) &= \frac{1}{B(\alpha,\beta)} \int_u^\infty db\, \sinh^\alpha b \left[\eta_1 \left((\alpha+\beta)\cosh u - \alpha \sinh u \coth b \right) + \beta \sinh u \psi \right].
\end{aligned}
\end{equation}

Using known identities such as \eqref{int-coth}, this becomes:
\begin{equation}
    \begin{aligned}
        \eta = \frac{\beta\, e^{-b(\beta +1)} B\left(\beta ,\frac{\alpha +1}{2}\right)}{\alpha + 2\beta + 1}
        \Bigg(&\sinh(u) \Big[\coth(b)\big((\alpha+1)F - 2\alpha F_+\big) + (\alpha+1)F + 2\beta F_+\Big] \\
        &+ 2(\alpha + \beta) \cosh(u) F_+ \Bigg)
    \end{aligned}
\end{equation}

Using the identity for the telescoping sum, we eventually find:
\begin{equation}
    \label{Phialphabeta}
    \phi^{\alpha,\beta}(u) = \frac{(\beta-1)B\left(\beta ,\frac{\alpha +1}{2}\right)}{2B(\alpha ,\beta )} \sum_{n=0}^\infty e^{-u(\beta+2n)} \frac{\left(\frac{1-\alpha}{2} \right)_n \left(\beta \right)_n}{n! \left(\frac{1+2\beta+\alpha}{2} \right)_n} \frac{n+\frac{\beta}{2}}{(n+\frac{\beta-1}{2})(n+\frac{\beta+1}{2})}.
\end{equation}

\subsubsection{Scaling limit of the propagator.} We now deduce the scaling function \( p_{\alpha,\beta}(u) \), defined in the scaling regime $x^2 \sim t$ by 
\[
P_{\alpha,\beta}(x,t) \sim \frac{1}{\sqrt{2t}} p_{\alpha,\beta}\left(u = \frac{x}{\sqrt{2t}}\right), \quad x > 0.
\]
Laplace inversion of \eqref{Phialphabeta} term by term yields finally yields the scaling function, and thus the propagator of the SATW$_{\alpha,\beta}$ in the scaling limit:
\begin{keyboxedeq}[Propagator of the SATW]
\begin{equation}
\label{propag}
    p_{\alpha,\beta}(u) = \frac{(\beta-1)B\left(\beta ,\frac{\alpha +1}{2}\right)}{B(\alpha ,\beta ) \sqrt{\pi}} \sum_{n=0}^\infty  \frac{\left(\frac{1-\alpha}{2} \right)_n \left(\beta \right)_n}{ \left(\frac{1+2\beta+\alpha}{2} \right)_n} \frac{n+\frac{\beta}{2}}{(n+\frac{\beta-1}{2})(n+\frac{\beta+1}{2})} \frac{e^{-u^2(2n+\beta)^2}}{n!}.
\end{equation}
\end{keyboxedeq}

\begin{figure}
    \centering
    \includegraphics[width=.6\textwidth]{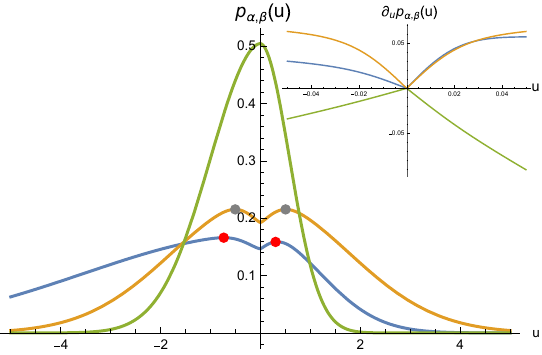}
    \caption{Scaling function of the propagator of the SATW \eqref{propag}. Different colors correspond to different values of the parameters $(\alpha,\beta)$. Blue corresponds to $(0.2,0.6)$, orange to $(0.4,0.4)$ and green to $(0.7,1.2)$. Colored dots show the local maxima in the distribution as predicted  by \eqref{approx-bump}. The inset shows the derivative $\partial_u p_{\alpha,\beta}(u)$ of the scaling function. Derivatives vanish at $u=0$.}
    \label{fig:pu}
\end{figure}

\begin{figure}
    \centering

    \begin{subfigure}[t]{0.48\textwidth}
        \includegraphics[width=\textwidth]{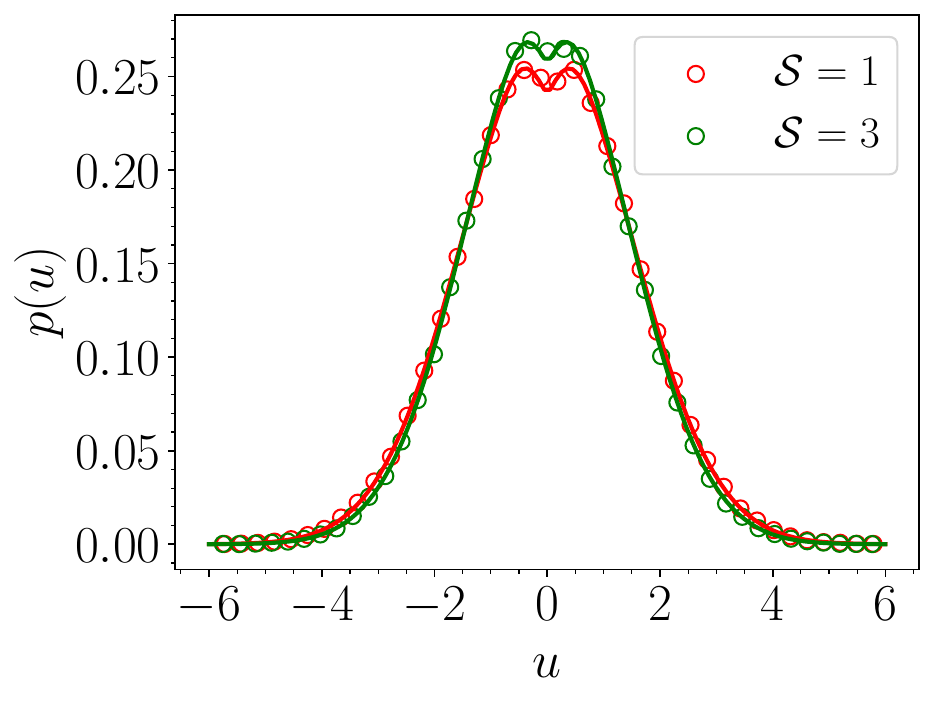}
        \caption{Distribution of the rescaled variable $u = x/\sqrt{2t}$ for a SATW with weight function $w_\mathcal{S}(n) = e$ for $n \leq \mathcal{S}$ and $w_\mathcal{S}(n) = 1$ for $n > \mathcal{S}$. The corresponding value of the reinforcement parameter $\phi$ is then given by \eqref{phi-links}. The standard once-reinforced SATW$_\phi$ is recovered in the case $\mathcal{S} = 0$, illustrating the universality of the propagator within the SATW class.}
        \label{satw_num}
    \end{subfigure}
    \hfill
    \begin{subfigure}[t]{0.48\textwidth}
        \includegraphics[width=\textwidth]{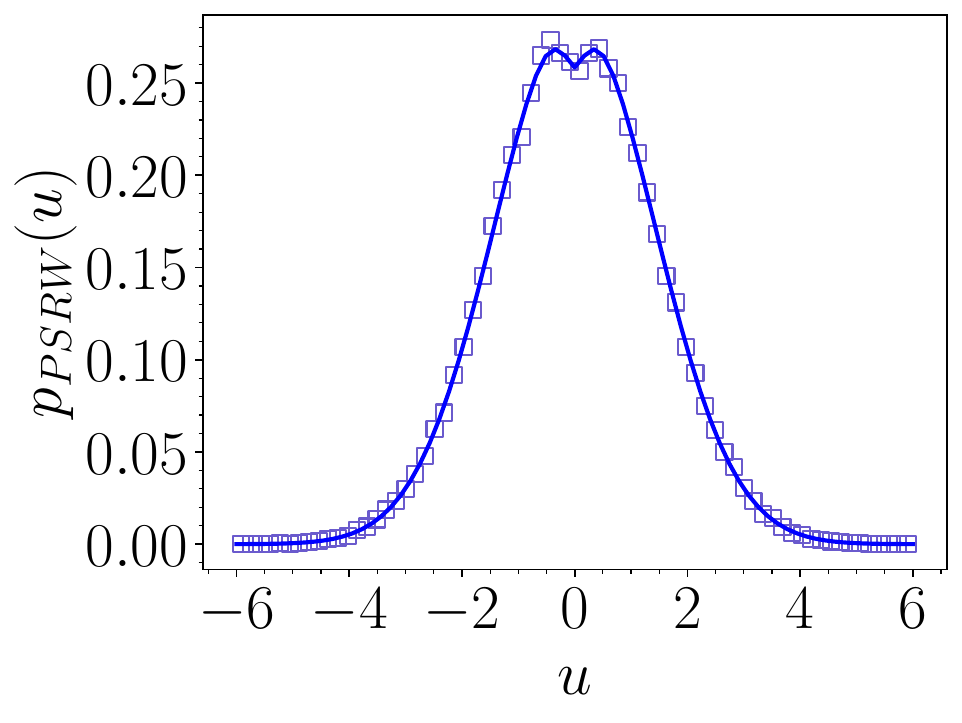}
        \caption{Distribution of the rescaled variable $u = \tfrac{x}{\sqrt{2t(2\gamma + 1)}}$ for a PSRW$_\gamma$ with exponent $\gamma = 4$, compared to the theoretical prediction $p_{1/2,1/2}(u)$ given in Eq.~\eqref{propag}.}
        \label{psrw_num}
    \end{subfigure}

    \caption{Scaling functions of the propagators for different microscopic realizations of SATW and PSRW. Solid curves correspond to the theoretical prediction~\eqref{propag}, while symbols represent results from numerical simulations.}
    \label{fig:simus-propag-sirw}
\end{figure}

By symmetry, \( P_{\alpha,\beta}(-x,t) = P_{\beta,\alpha}(x,t) \), which gives the scaling function \eqref{propag} for all \( u \in \mathbb{R} \). One can verify that the scaling function is normalized:
\[
\int_{\mathbb{R}} p_{\alpha,\beta}(u)\, du = 1.
\]
See FIG~\ref{fig:pu} for a plot of the scaling function \eqref{propag} for different values of the parameters $\alpha,\beta$, and FIG~\ref{fig:simus-propag-sirw} for a numerical confirmation. Besides its intrinsic importance, Eq.\eqref{propag} reveals several remarkable features of the SATW.

\subsubsection{Moments of the position $X_t$. } 
\begin{figure}
    \centering
    \begin{subfigure}[t]{0.48\textwidth}
        \includegraphics[width=\textwidth]{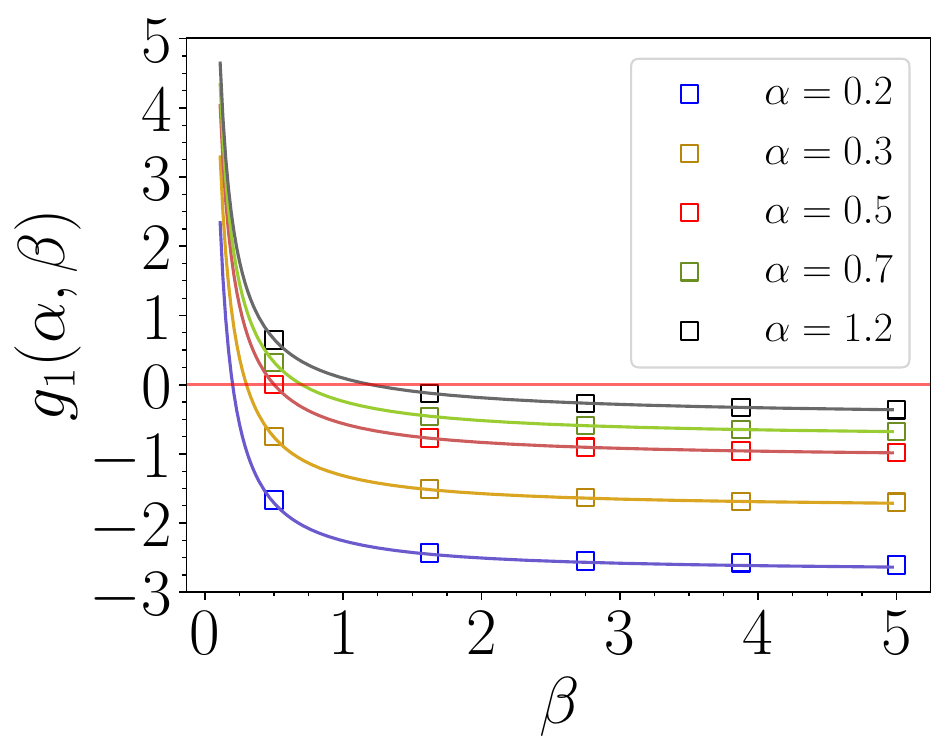}
        \caption{Scaled mean $g_1(\alpha,\beta) \sim \frac{\langle X_t \rangle}{\sqrt{2t}}$ plotted for different values of $\alpha$. As expected, it is negative for $\beta > \alpha$.}
    	\label{fig:mean_num}
    \end{subfigure}
    \hfill
    \begin{subfigure}[t]{0.48\textwidth}
        \includegraphics[width=\textwidth]{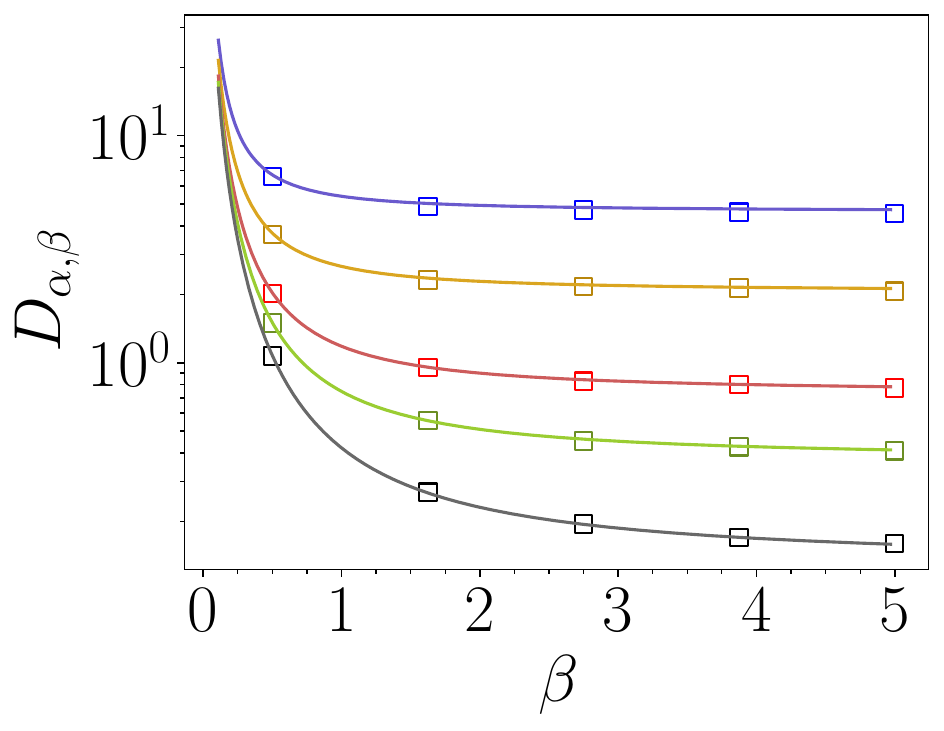}
        \caption{Diffusion coefficient $D_{\alpha,\beta} $. It is a decreasing function of both $\alpha$ and $\beta$, and diverges for either $\alpha\to 0$ or $\beta\to 0$.}
		\label{fig:diff_num}
    \end{subfigure}
    \label{fig:cumulants}
\end{figure}
First, Eq. \eqref{propag}  gives access to the moments $\langle X_t^n \rangle$ of the position of the SATW$_{\alpha,\beta}$ at time $t$.  

\paragraph{Mean of $X_t$}

In the general asymmetric case $\alpha \neq \beta$, the mean position $\langle X_t \rangle$ is nonzero and can be computed exactly from Eq.~\eqref{propag}. It takes the form
\[
\langle X_t \rangle = g_1(\alpha,\beta) \sqrt{2t},
\]
exhibiting a $\sqrt{t}$ scaling characteristic of the general form of the scaling function $p_{\alpha,\beta}$. This behavior contrasts with classical diffusion under constant drift, where the mean grows linearly in time.

The prefactor $g_1(\alpha,\beta)$ is defined via the antisymmetric combination
\[
g_1(\alpha,\beta) = h_1(\alpha,\beta) - h_1(\beta,\alpha),
\]
with
\[
h_1(\alpha,\beta) = \sqrt{2} \left( \tfrac{2^{-\alpha}(\beta-1)\Gamma(\tfrac{\beta+1}{2})\Gamma(\alpha+\beta)\, _3{F}_2(\tfrac{1-\alpha}{2},\tfrac{\beta+1}{2},\beta;\tfrac{\beta+3}{2},\tfrac{\alpha+1}{2}+\beta;1)}{\Gamma(\tfrac{\alpha}{2})\Gamma(\tfrac{\beta+3}{2})\Gamma(\tfrac{\alpha+1}{2})} + \tfrac{\alpha}{\sqrt{\pi}(\alpha+\beta)} - \tfrac{\Gamma(\tfrac{\alpha+1}{2})\Gamma(\tfrac{\beta}{2})\Gamma(\tfrac{\alpha+\beta}{2})}{\Gamma(\tfrac{\alpha}{2})\Gamma(\tfrac{\beta-1}{2})\Gamma(\tfrac{\alpha+\beta+1}{2})} \right).
\]

Numerical confirmation of this expression is provided in Fig.~\ref{fig:mean_num}. In certain limiting cases, $g_1(\alpha,\beta)$ admits simple closed forms. For instance, when $\alpha = 1$ one obtains
\[
g_1(1,\beta) = \frac{1 - \beta}{\sqrt{\pi} \beta},
\]
and in the limit $\alpha \to 0$,
\[
g_1(\alpha,\beta) \sim -\frac{1}{\alpha \sqrt{\pi}}.
\]

\paragraph{Variance of $X_t$ and diffusion coefficient $D_{\alpha,\beta}$}

The variance of the position can also be extracted from Eq.~\eqref{propag}, and grows linearly in time:
\[
V(X_t) = \langle X_t^2 \rangle - \langle X_t \rangle^2 \sim 2 D_{\alpha,\beta} \, t,
\]
thereby defining the diffusion coefficient $D_{\alpha,\beta}$. In certain special cases, $D_{\alpha,\beta}$ admits closed-form expressions. For instance:
\[
D_{1,\beta} = \frac{\pi \beta (\beta -1) - 2(\beta -1)^2 + \pi}{2\pi \beta^2},
\qquad
D_{\alpha,\beta} \sim \frac{\pi - 2}{2\pi \alpha^2} \quad \text{as } \alpha \to 0.
\]

Remarkably, even the most basic observable—the diffusion coefficient—was previously unknown in closed form for the symmetric SATW, i.e., when $\alpha = \beta$. Our approach yields an explicit expression for $D_{\alpha,\beta}$ in this case:
\begin{keyboxedeq}[Diffusiion coefficient of the SATW]
\begin{equation}
\label{diff-satw}
D_{\alpha,\alpha} = \frac{1}{2} + \frac{(1-\alpha) B\left(\tfrac{\alpha+1}{2}, \alpha \right)}{B(\alpha,\alpha)} \cdot 
\frac{
_4F_3\left( \tfrac{1-\alpha}{2}, \tfrac{\alpha}{2}, \tfrac{\alpha}{2}, \alpha \, ; \, \tfrac{\alpha}{2}+1, \tfrac{\alpha}{2}+1, \tfrac{3\alpha}{2}+\tfrac{1}{2} \, ; 1 \right)
}{\alpha^2}.
\end{equation}
\end{keyboxedeq}

This result addresses a notable gap in the literature and highlights the effectiveness of combining the structural insights provided by the Ray--Knight framework with direct, and at times technically demanding, computations. While the Ray--Knight theorems offer a powerful lens through which to view the dynamics of SIRWs, they do not in themselves yield explicit expressions for observables such as the diffusion coefficient. It is through a detailed exploitation of these theorems—along with a more brute-force approach to computing the resulting expressions—that we are able to derive a closed-form result for observables such as $D_{\alpha,\beta}$. To our knowledge, this is the first exact expression of its kind, and it demonstrates how such tools can be used not only for structural understanding but also for precise, quantitative predictions. See Fig.~\ref{fig:diff_num} for numerical confirmation of Eq.~\eqref{diff-satw} across a range of $\alpha$ and $\beta$.

\paragraph{Non-Gaussianity and Fourth Cumulant of $X_t$}

\begin{figure}
    \centering
    \includegraphics[width=0.45\textwidth]{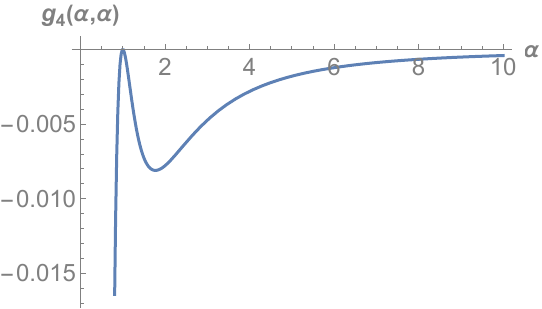}
    \caption{Scaled fourth cumulant $g_4(\alpha,\alpha)$ as a function of $\alpha$ in the symmetric case. The cumulant vanishes only at $\alpha = 1$, corresponding to the simple random walk, and is strictly negative elsewhere. This indicates that the standard RW ($\alpha = \beta = 1$) has the highest likelihood of extreme position values among all SATW$_{\alpha,\beta}$ models.}
    \label{fig:kappa4sym}
\end{figure}

Eq.~\eqref{propag} demonstrates that the propagator is non-Gaussian for general values $(\alpha, \beta) \neq (1,1)$, despite exhibiting a Gaussian tail:
\[
p_{\alpha,\beta}(u) \propto e^{-u^2 \beta^2} \quad \text{as } u \to \infty.
\]
Interestingly, when $\alpha = 2n + 1$ is an odd integer, the Pochhammer symbol $\left(\tfrac{1 - \alpha}{2}\right)_k$ vanishes for all $k > n$, implying that $p_{\alpha,\beta}(u)$ becomes a \textit{finite} sum of $n + 1$ Gaussian terms. 
To quantify deviations from Gaussianity, we consider the fourth cumulant of the position,
\[
\kappa_4(t) = \langle X_t^4 \rangle - 3 \langle X_t^2 \rangle^2 = 3g_4(\alpha,\beta)\, t^2,
\]
which captures the excess kurtosis of the distribution. In the symmetric case $\alpha = \beta$, the coefficient $g_4(\alpha,\alpha)$ admits an explicit expression
\begin{align}
    g_4(\alpha,\alpha) &=3\sqrt{\pi}\, 4^{-\alpha} (\alpha - 1) \bigg( 
    \Gamma\left(\tfrac{\alpha}{2}\right) \Gamma(2\alpha) \,
    _4\tilde{F}_3\left(
    \tfrac{1}{2} - \tfrac{\alpha}{2}, \tfrac{\alpha}{2}, \tfrac{\alpha}{2}, \alpha \, ; \,
    \tfrac{\alpha}{2} + 1, \tfrac{\alpha}{2} + 1, \tfrac{3\alpha}{2} + \tfrac{1}{2} \, ; 1
    \right) \notag \\[1ex]
    &\quad \times \left(
    2^\alpha - \sqrt{\pi}(\alpha - 1)
    \Gamma\left(\tfrac{\alpha}{2}\right) \Gamma(2\alpha) \,
    _4\tilde{F}_3\left(
    \tfrac{1}{2} - \tfrac{\alpha}{2}, \tfrac{\alpha}{2}, \tfrac{\alpha}{2}, \alpha \, ; \,
    \tfrac{\alpha}{2} + 1, \tfrac{\alpha}{2} + 1, \tfrac{3\alpha}{2} + \tfrac{1}{2} \, ; 1
    \right)
    \right) \notag \\[1ex]
    &\quad
    - \tfrac{
    2^\alpha \Gamma\left(\tfrac{\alpha}{2} + 1\right)^3 \Gamma(2\alpha + 1) \,
    _6\tilde{F}_5\left(
    \tfrac{1}{2} - \tfrac{\alpha}{2}, \tfrac{\alpha}{2}, \tfrac{\alpha}{2}, \tfrac{\alpha}{2}, \tfrac{\alpha}{2}, \alpha \, ; \,
    \tfrac{\alpha}{2} + 1, \tfrac{\alpha}{2} + 1, \tfrac{\alpha}{2} + 1, \tfrac{\alpha}{2} + 1, \tfrac{3\alpha}{2} + \tfrac{1}{2} \, ; 1
    \right)
    }{\alpha^4}
    \bigg).
\end{align}
See FIG~\ref{fig:kappa4sym} for a plot of the scaled fourth cumulant $g_4(\alpha,\alpha)$. It is clear that the propagator $P(x,t)$ is systematically \emph{platykurtic}, exhibiting a negative excess kurtosis: it is broader around the center than a Gaussian distribution—recovered in the case $\alpha = 1$—while simultaneously featuring thinner tails.

\subsubsection{Non-monotonicity of $p_{\alpha,\beta}(u)$.} A second remarkable feature of the propagator $P_{\alpha,\beta}(x,t)$—not captured by an analysis of the moments of $X_t$—is its \emph{non-monotonicity}. For certain values of $\alpha, \beta < 1$ (corresponding to repelling SATWs), the distribution develops local maxima at non-zero positions, which we denote $x_\pm(t)$. In other words, the probability density exhibits \emph{bumps} away from the origin.

Although closed-form expressions for $x_\pm$ are out of reach, a good approximation can be derived in the regime of small $\alpha$ and $\beta$ by truncating the series \eqref{propag} to its first two terms. Solving $\partial_u p_{\alpha,\beta}(u) = 0$ under this approximation yields:
\begin{boxedeq}
\begin{equation}
\label{approx-bump}
    x_+(t) \Sim{\alpha, \beta \to 0} \sqrt{ \frac{t}{2(1+\beta)} \log \left( \frac{(1-\alpha)(1-\beta)(\beta+2)^3}{\beta^2(\beta+3)(\alpha + 2\beta + 1)} \right) },
\end{equation}
\end{boxedeq}
with a symmetric expression for $x_-(t)$ obtained by interchanging $\alpha$ and $\beta$. One necessary condiition for the appearance of nonzero maxima is that the argument of the logarithm in Eq.~\eqref{approx-bump} exceeds one, so that \eqref{approx-bump} is well-defined:
\begin{equation}
    \label{criterion-bumps}
        \frac{(1-\alpha) (1-\beta) (\beta +2)^3}{\beta^2 (\beta +3) (\alpha +2 \beta +1)} > 1.
\end{equation}
Eq. \eqref{criterion-bumps} provides an explicit criterion for non-monotonicity in $P_{\alpha,\beta}(x,t)$, which we verify numerically across a broad range of parameters in Table~\ref{table-bumps}.

\begin{table}[h]
    \centering
    \begin{tabular}{|c|c|c|c|c|c|}
        \hline
        \diagbox{$\alpha$}{$\beta$} & 0.2 & 0.4 & 0.6 & 0.7 & 0.8 \\ \hline
        0.2 & $\checkmark,\checkmark$ & $\checkmark,\checkmark$ & $\checkmark,\checkmark$ & $\checkmark,\checkmark$ & $\checkmark,\checkmark$ \\ \hline
        0.4 & $\checkmark,\checkmark$ & $\checkmark,\checkmark$ & $\checkmark,\checkmark$ & $\checkmark,\checkmark$ & $\checkmark,\checkmark$ \\ \hline
        0.6 & $\checkmark,\checkmark$ & $\checkmark,\checkmark$ & $\checkmark,\times$     & $\checkmark,\times$     & $\checkmark,\times$     \\ \hline
        0.7 & $\checkmark,\checkmark$ & $\checkmark,\checkmark$ & $\checkmark,\times$     & $\checkmark,\times$     & $\times,\times$         \\ \hline
        0.8 & $\checkmark,\checkmark$ & $\checkmark,\checkmark$ & $\checkmark,\times$     & $\times,\times$         & $\times,\times$         \\ \hline
    \end{tabular}
    \caption{Validation of the bump criterion for different values of $(\alpha, \beta)$. Each cell displays two symbols $A, B$: $A = \checkmark$ if bumps are observed in the scaling function for $u > 0$, and $A = \times$ otherwise; $B = \checkmark$ if the criterion~\eqref{criterion-bumps} is satisfied, and $B = \times$ if not.}
    \label{table-bumps}
\end{table}

The accuracy of the approximation~\eqref{approx-bump} is also illustrated in Fig.~\ref{fig:pu}, confirming that strong enough self-repulsion at the boundaries of the visited region can push the walker away from the origin, resulting in a non-monotonic propagator with a local minimum at the origin. This counterintuitive effect arises despite the process remaining globally diffusive: the walker retains an infinite memory of its starting point, which subtly shapes its trajectory even at large times.

\paragraph{Regularity of $p_{\alpha,\beta}$}
We now examine the regularity properties of the scaling function $p_{\alpha,\beta}(u)$. First, we note that due to the piecewise definition of Eq.~\eqref{propag} for $\alpha \neq \beta$, requiring continuity at $u = 0$ leads to a non-trivial identity involving hypergeometric functions:
\begin{equation}
    h_2(\alpha,\beta) = \Gamma(\beta+1) \frac{\, _3F_2\left( \tfrac{3-\alpha}{2}, \tfrac{\beta + 1}{2}, \beta + 1 \, ; \, \tfrac{\beta + 5}{2}, \tfrac{\alpha + 3}{2} + \beta \, ; \, 1 \right)}{\Gamma(\tfrac{\beta + 5}{2}) \Gamma(\tfrac{\alpha + 3 + 2\beta}{2})} = h_2(\beta,\alpha).
\end{equation}
We have verified this identity numerically to very high precision.

Second, as shown in Fig.~\ref{fig:pu}, the function $p_{\alpha,\beta}(u)$ always has a continuous and vanishing derivative at $u = 0$ for all $\alpha, \beta$. Indeed, in the half-space $u > 0$, the derivative of the scaling function can be expressed as:
\begin{equation}
\label{derivative}
    \partial_u p_{\alpha,\beta}(u) \propto u \sum_{n=0}^\infty \frac{\left( \tfrac{1-\alpha}{2} \right)_n (\beta)_n}{\left( \tfrac{1+2\beta+\alpha}{2} \right)_n} \frac{(2n+\beta)^3}{\left( n + \tfrac{\beta - 1}{2} \right)\left( n + \tfrac{\beta + 1}{2} \right)} \frac{e^{-u^2(2n+\beta)^2}}{n!}.
\end{equation}
As $n \to \infty$, the summand behaves as
\begin{equation}
    \frac{\left( \tfrac{1-\alpha}{2} \right)_n (\beta)_n}{\left( \tfrac{1+2\beta+\alpha}{2} \right)_n} \frac{(2n+\beta)^3}{\left( n + \tfrac{\beta - 1}{2} \right)\left( n + \tfrac{\beta + 1}{2} \right)} \frac{e^{-u^2(2n+\beta)^2}}{n!} = O\left( n^{-\alpha} e^{-u^2(2n+\beta)^2} \right).
\end{equation}
Only the terms with $n \lesssim 1/u$ contribute significantly to the sum, since the exponential factor suppresses the rest. Therefore, Eq.~\eqref{derivative} vanishes as $u \to 0$ if and only if
\[
u \sum_{n=1}^{1/u} n^{-\alpha} \xrightarrow{u \to 0} 0.
\]
This is indeed the case, since
\[
u \sum_{n=1}^{1/u} n^{-\alpha} = u \cdot O\left( \frac{1}{u^{1-\alpha}} \right) = O(u^\alpha) \to 0.
\]
Hence, $p_{\alpha,\beta}(u)$ has a vanishing right derivative at $u = 0$. By symmetry, the left derivative also vanishes, and we conclude that $\partial_u p_{\alpha,\beta}(u)$ is continuous at $u = 0$, with
\[
\partial_u p_{\alpha,\beta}(u)\big|_{u=0} = 0.
\]

However, an extension of the same argument reveals that $p_{\alpha,\beta}(u)$ does \emph{not} possess a second derivative at $u = 0$ whenever either $\alpha < 1$ or $\beta < 1$. This lack of smoothness reflects the non-analytic nature of the scaling function at the origin in the presence of strong reinforcement or repulsion.

This smoothness property stands in contrast with other self-interacting models such as the true self-avoiding walk \cite{dumazMarginalDensities} or locally activated random walks \cite{benichouNonGaussianityDynamical,bremontAgingDynamics}, which exhibit non-differentiable cusps at the origin.

\subsubsection{Propagator of the PSRW$_\gamma$}
Finally, Eq.~\eqref{propag} also provides access to the propagator of the PSRW$_\gamma$ class. Indeed, the PSRW$_\gamma$ exhibits the same local time profile as the SATW$_{1/2, 1/2}$, up to a scaling factor $C = \frac{1}{2\gamma + 1}$. As a result, time—interpreted as the area under the local time profile—is effectively rescaled via $t \mapsto t / (2\gamma + 1)$. It follows that, if $X(t)$ denotes the position of the PSRW$_\gamma$, the rescaled variable 
\[u = \tfrac{X\left(\tfrac{t}{2\gamma+1}\right)}{\sqrt{2t}} \equiv \tfrac{X(t)}{\sqrt{2t(2\gamma + 1)}} \] 
is asymptotically distributed as $p_{1/2, 1/2}(u)$ in the scaling limit $t \to \infty$ with $u$ held fixed.
Hence, our explicit expression \eqref{propag} also yields an exact form of the PSRW$_\gamma$ propagator as a by-product; see Fig.~\ref{psrw_num} for numerical confirmation. 

\subsubsection{Conclusion}
To conclude, we have computed analytically the propagator for two important classes of SIRWs: the SATW$_{\alpha,\beta}$ and the PSRW$_\gamma$. This resolves an open question raised in~\cite{tothGeneralizedRayKnightTheory1996,carmonaBetaVariables}, where the integrals involved in the propagator expression were deemed intractable. By using our newly obtained expression for the BESQ functional \eqref{joint-area-fpt}, we managed to obtain an explicit expression for the propagator and uncovered several nontrivial analytical features. These include, notably, the emergence of non-monotonicity in the spatial profile for repelling walks, as well as access to basic transport quantities such as the diffusion coefficient—an observable that had previously remained out of reach due to the non-Markovian nature of the dynamics.

\section{Space exploration dynamics of SIRWs}
\label{sec:spacexpl-1d}
This section introduces a fundamental ingredient for understanding space exploration in SIRWs—both conceptually and analytically. We focus on the quantity \( q_+(k, -m, t) \), the joint distribution of the minimum \( -m \) and the time \( T_k \) at which the walker first reaches the maximum value \( k \). This distribution plays a central role in characterizing space-time observables and serves as a natural building block for more complex quantities, such as the FPT density and the distribution of the number \( N(t) \) of distinct sites visited up to time \( t \).

What makes \( q_+(k, -m, t) \) particularly valuable is that it is directly accessible through Ray--Knight theory, yet has received little attention in the literature. To the best of our knowledge, it has never been considered explicitly, except in the recent work~\cite{klingerJointStatistics}, where it was computed for a few Markovian RWs. Despite its apparent simplicity, this quantity encapsulates nontrivial aspects of the walk’s global memory and spatial history, making it a key object for describing the statistics of space exploration in SIRWs.

\subsection{Joint distribution of minimum $-m$ and time $T_k$ to the maximum $k$}
Define $q_+(k, -m, t)$ (resp. $q_-(k, -m, t)$) as the joint distribution of the random variables $m$ and $t$ (resp. $k$ and $t$), where $t = T_k$ (resp. $t = T_{-m}$) is the first hitting time of site $k$ (resp. $-m$), and $-m$ (resp. $k$) is the minimum (resp. maximum) of the process $(X_s)_{s \leq t}$. This distribution is of intrinsic interest, as it has been shown to play a key role in various trapping problems~\cite{klingerJointStatistics}. Also, note that we already computed a time-integrated version of $q_\pm$ in \eqref{eq:qplus-final}, which was useful in the determination of the splitting probability of SIRWs. \par 
We will take advantage of the representation of the first hitting time $T_k$ of site $k$ as the area under the local time $L_{T_k}(x)$. Namely, we have 
\begin{equation}
    \label{qplust-loctime}
    q_+(k, -m, t) = \mathbb{P}\left(\sum_{x \in \mathbb{Z}}L_{T_k}(x) = t, \mathrm{supp}(L_{T_k})=[-m,k]\right).
\end{equation}
Importantly, we note that, writing $L=k+m$, we have 
\begin{align}
    \label{span-fpt-qplus}
    &q_+(k,-m,t) + q_-(k,-m,t) \\ &= \mathbb{P}\!\left(\textnormal{among all $L\!+\!1$ possible spans of $L\!+\!1$ sites, the span $[-m,k]$ was visited exactly at time $t$}\right).
\end{align}
Thus, the probabilities $q_{\pm}$ also play a key role in the quantification of asymmetry in space exploration dynamics. We will use this fact in part 3 of this thesis.

\subsubsection{SATW$_{\alpha,\beta}$, PSRW$_\gamma$}
We first compute the observable $q_+(k, -m, t)$ in the case of the SATW$_{\alpha,\beta}$, corresponding to the constant $C=1$, and the case of the PSRW$_\gamma$, obtained by plugging in $\alpha=\beta=1/2$ and $C= \tfrac{1}{2\gamma+1}$. In the diffusive scaling limit where 
\begin{equation}
    t\to \infty, k \gg 1, \frac{k}{\sqrt{t}} = \mathrm{cte},
\end{equation}
we apply the Ray--Knight theorems to obtain the explicit distribution of $L_{T_k}$ as two patched-up BESQ processes: 
\[
L_{T_k}(u)\approx C \cdot
\begin{cases}
Y_{2\beta}(k - u), & 0 \leq u \leq k, \\
\widetilde{Y}_{2 - 2\alpha}(-u), & u \leq 0.
\end{cases}
\]
Introducing $T_0$ the first passage-time of $\widetilde{Y}_{2 - 2\alpha}$ to $0$, we have from \eqref{qplust-loctime}
\begin{equation}
    \label{qplust-loctime-2}
    q_+(k, -m, t) = \mathbb{P}\left(C\int_0^k Y_{2\beta}(u) du + C\int_{0}^\infty \widetilde{Y}_{2-2\alpha}(u) du = t, T_0 = m \right).
\end{equation}
Eq. \eqref{qplust-loctime-2} has a clear convolution structure in $t$, thanks to the conditional independence of the two processes $Y_{2\beta}, \widetilde{Y}_{2 - 2\alpha}$. Let us Laplace transform \eqref{qplust-loctime-2} with respect to time:
\begin{align}
    \label{qplus-laplace}
    \hat{q}_+(k,m,s) &= \int_0^\infty \mathbb{E}^0 \left[e^{-s C\int_0^k Y_{2\beta}(u) du}; Y_{2\beta}(k) = a\right] \mathbb{E}^a \left[e^{-sC \int_{0}^\infty \widetilde{Y}_{2-2\alpha}(u) du}; T_0 = m \right] da.
\end{align}
The two BESQ functionals introduced in \eqref{qplus-laplace} are computed explicitly in the appendix \eqref{laplace-area-endpoint}, \eqref{joint-area-fpt}. Using $\mu \equiv \sqrt{2sC}$, we obtain 
\begin{align}
    \label{qplus-laplace-2}
    \hat{q}_+(k,m,s) &= \frac{\mu}{\Gamma(\alpha) \Gamma \left(\beta \right)\sinh(\mu m)} \int_0^\infty \frac{e^{\frac{-\mu a}{2} \coth (\mu  k)} \left(\frac{\mu  a}{2 \sinh (\mu  k)}\right)^{\beta}}{a} e^{-\frac{1}{2} \mu a \coth(\mu m)} 
    \left( \frac{\mu a}{2 \sinh(\mu m)} \right)^{\alpha}da.
\end{align}
This leads to the final expression:
\begin{keyboxedeq}[Joint distribution of minimum and time to the maximum for SATW]
\begin{equation}
\label{qplus-laplace-satw}
\hat{q}_+(k,m,s) = \frac{1}{B(\alpha,\beta)} \cdot \frac{\mu \sinh^\alpha(k \mu) \sinh^{\beta -1}(\mu m)}{\sinh^{\alpha + \beta}(\mu (k + m))}.
\end{equation}
\end{keyboxedeq}

By symmetry under spatial inversion \[k \leftrightarrow m, \quad \alpha \leftrightarrow \beta,\] 
one obtains a similar expression for $\hat{q}_-(k,m,s)$, which gives rise to the following striking identity:
\begin{boxedeq}
\begin{equation}
\label{ratio-ansatz-satw}
\frac{\hat{q}_+(k,m,s)}{\hat{q}_-(k,m,s)} = \frac{\sinh(k \mu)}{\sinh(m \mu)}.
\end{equation}
\end{boxedeq}
Interestingly, this relation was derived through entirely different arguments in \cite{bremontExactPropagators}.

Note that setting $s = 0$ in Eq.~\eqref{qplus-laplace-satw} yields the time-integrated quantity:
\begin{equation}
\hat{q}_+(k,m,s=0) = \frac{1}{B(\alpha,\beta)} \cdot \frac{k^\alpha m^{\beta - 1}}{(k + m)^{\alpha + \beta}} = \frac{1}{L} q_+(k = zL), \quad \text{with } L = k + m,
\end{equation}
where $q_+(z)$ coincides with the expression in Eq.~\eqref{eq:qplus-final}, now generalized to the asymmetric SATW$_{\alpha,\beta}$ setting.

\paragraph{Connection with general scaling arguments.}
In~\cite{klingerJointStatistics}, it was argued on dimensional and scale-invariance grounds that the observable $q_+(k, m, t)$ obeys a universal scaling form in the regime where the number of sites visited before first reaching $k$ is large. More precisely, this corresponds to the asymptotic limit where $m \sim t^{1/\dw}$ and $k \ll m$, or equivalently $k \ll t^{1/\dw}$. In this regime, the joint distribution $q_+$ has been shown to take the universal form:
\begin{equation}
    \label{klinger-form}
    q_+(k,m,t) \Sim{m \sim t^{1/\dw},\; k \ll m} 
    \frac{h(k)}{m^{\dw(\theta + 1) + 1}} \, f\left( \frac{t}{m^{\dw}} \right),
\end{equation}
where $h(k)$ and $f(u)$ are scaling functions satisfying:
\begin{equation}
    \int_0^\infty f(u) \, du = 1, \qquad h(k) \propto k^{\dw \theta} \quad \text{as } k \to \infty.
\end{equation}

Taking the Laplace transform with respect to time, the scaling form~\eqref{klinger-form} becomes:
\begin{equation}
    \label{klinger-form-laplace}
    \hat{q}_+(k,m,s) \Sim{m s^{1/\dw} = \text{const},\; k s^{1/\dw} \ll 1} 
    \frac{h(k)}{m^{\dw \theta + 1}} \, \hat{f}\left( s m^{\dw} \right),
\end{equation}
where $\hat{f}(u) = \int_0^\infty e^{-u v} f(v) \, dv$ is the Laplace transform of $f$.

In the case of the SATW$_{\alpha,\beta}$ model, direct comparison with the exact expression~\eqref{qplus-laplace-satw} identifies both scaling functions explicitly:
\begin{equation}
    h(k) = \frac{k^\alpha}{B(\alpha,\beta)}, \qquad 
    \hat{f}(u) = \left( \frac{\sqrt{2u}}{\sinh(\sqrt{2u})} \right)^{\alpha + 1}.
\end{equation}
This provides a complete and exact realization of the general scaling form~\eqref{klinger-form} for SATWs, including both the prefactor $h(k)$ and the universal scaling function $\hat{f}(u)$. We note that the expression for $\hat{f}(u)$ had previously been obtained in~\cite{klingerJointStatistics} via alternative methods.

\subsubsection{TSAW$_\beta$}
We now turn to the computation of $q_+(k, -m, t)$ in the case of the TSAW. According to the Ray--Knight theorems, the local time profile $L_{T_k}(u)$ evolves as a Brownian motion run backwards from site $k$, scaled by a constant prefactor $C(\beta)$. The explicit expression for this prefactor, derived in \cite{tothTrueSelfAvoiding}, is given by
\begin{equation}
    \label{cofbeta}
    C(\beta) = \sqrt{\frac{\sum_{k=-\infty}^\infty k^2 e^{-\beta k^2}}{\sum_{k=-\infty}^\infty e^{-\beta k^2}}}.
\end{equation}
Without loss of generality, we rescale time—equivalently, the area under the edge local time process described by Brownian motion through the Ray--Knight theorems—so that the prefactor $C(\beta)$ is set to unity in the analysis that follows. To recover physical time units, one simply applies the inverse rescaling $t \mapsto C(\beta) \, t$. \par 
% \footnote{For practical purposes, one may take $\beta = 1$, since $C(\beta = 1) \approx 0.998$ is already very close to unity.}
This Brownian motion starts at the origin at $u=k$, is reflected at $0$ in the portion $0 \leq u \leq k$ and is absorbed at $0$ for $u\leq 0$. Hence, \eqref{qplus-laplace-2} becomes
\begin{align}
    \label{qplus-laplace-tsaw}
    \hat{q}_+(k,m,s) &= \int_0^\infty \mathbb{E}^0 \left[e^{-s \int_0^k |Y|(u) du}; |Y|(k) = a\right] \mathbb{E}^a_{\mathrm{abs}} \left[e^{-s \int_{0}^\infty \widetilde{Y}(u) du}; T_0 = m \right] da.
\end{align}
The two functionals appearing in Eq.~\eqref{qplus-laplace-tsaw} are evaluated explicitly in Appendix~\ref{bm-refl-fixed-t} and~\ref{bm-fpt-area-inverted}. Plugging their expressions into~\eqref{qplus-laplace-tsaw}, we obtain the Laplace transform of the distribution \( q_+(k, m, t) \) for the TSAW:
\begin{align}
    \label{qplus-laplace-tsaw-1}
    \hat{q}_+(k, m, s) &= -s \sum_{i,j=1}^\infty \exp\left[ \left( \tfrac{s^2}{2} \right)^{1/3} (k \alpha_i' + m \alpha_j) \right] 
    \int_0^\infty \frac{\mathrm{Ai}(\alpha_i' + (2s)^{1/3}a)}{\mathrm{Ai}(\alpha_i')\alpha_i'} \cdot \frac{\mathrm{Ai}(\alpha_j + (2s)^{1/3}a)}{\mathrm{Ai'}(\alpha_j)} \, da,
\end{align}
where \( \alpha_j \) and \( \alpha_i' \) respectively denote the negative zeros of the Airy function and its derivative:
\[
\mathrm{Ai}(\alpha_j) = 0, \qquad \mathrm{Ai}'(\alpha_i') = 0,
\]
with both sequences indexed increasingly over the negative real line.

Using the Airy kernel identity~\eqref{airykernel}, we can perform the integral over \( a \) explicitly, leading to the compact representation:
\begin{keyboxedeq}[Joint distribution of minimum and time to the maximum for TSAW]
\begin{equation}
    \label{qplus-laplace-tsaw-final}
    \hat{q}_+(k, m, s) = \left( \tfrac{s^2}{2} \right)^{1/3} \sum_{i,j=1}^\infty 
    \frac{
        \exp\left[ \left( \tfrac{s^2}{2} \right)^{1/3} (k \alpha_i' + m \alpha_j) \right]
    }{
        \alpha_i'(\alpha_j - \alpha_i')
    }.
\end{equation}
\end{keyboxedeq}

We verify numerically that in the limit \( s \to 0 \), this expression reproduces the known scaling form \( q_+(z) \) derived in Eq.~\eqref{eq:qplus-final} for the TSAW. 
\smallskip
\paragraph{Optional digression—Airy zero identities.}
In addition, the formula \eqref{qplus-laplace-tsaw-final} reveals a surprising mathematical structure when setting \( k = 0 \), where one expects \( q_+(0, m, t) = \delta(t)\delta(m) \). Using the identity \( \lambda \delta(\lambda m) = \delta(m) \), we deduce a nontrivial representation of the Dirac delta function in terms of the Airy zeros:
\begin{boxedeq}
\begin{equation}
\label{dirac-rep-airy}
\sum_{i,j=1}^\infty 
\frac{
    e^{m \alpha_j}
}{
    \alpha_i'(\alpha_j - \alpha_i')
} = \delta(m), \quad \text{for } m \geq 0.
\end{equation}
\end{boxedeq}
Although not directly needed for what follows, the representation~\eqref{dirac-rep-airy} is of independent mathematical interest. As far as we are aware, it is new, and we have verified it numerically to high precision. Integrating both sides over \( m \in \mathbb{R}_+ \) yields an identity resembling a Kronecker delta:
\begin{equation}
    \label{kronecker-rep-airy}
    \sum_{i,j=1}^\infty 
    \frac{
        e^{m \alpha_j}
    }{
        \alpha_j \alpha_i'(\alpha_j - \alpha_i')
    } = \mathds{1}_{m = 0}.
\end{equation}

We emphasize that the right-hand side is not a Dirac delta function but rather the indicator function of the point \( m = 0 \). In Appendix~\ref{vanishing-sum-airy}, we prove a related identity: for all \( j \geq 1 \),
\begin{equation}
\sum_{i=1}^\infty \frac{1}{\alpha_i'(\alpha_i' - \alpha_j)} = 0,
\end{equation}
which helps clarify the convergence properties of the double sums in \eqref{dirac-rep-airy} and \eqref{kronecker-rep-airy}. The well-known asymptotic behavior $\alpha_n, \alpha_n' \propto -n^{2/3}$ ensures that for $m > 0$, the exponential factor $e^{m \alpha_j}$ introduces rapid decay, rendering the double sum absolutely convergent. However, at $m = 0$, this exponential regularization is absent, and the slow decay of the Airy zeros causes the series to be only conditionally convergent. By the Riemann series theorem, the value of such conditionally convergent sums depends on the order in which terms are summed. In our case—and throughout the remainder of this work—we define such sums via a symmetric (or diagonal) cutoff:
\[
\sum_{i,j \geq 1} \equiv \lim_{N \to \infty} \sum_{1 \leq i,j \leq N}.
\]
As a result, the order of summation cannot be arbitrarily exchanged at $m = 0$ in Eq. \eqref{dirac-rep-airy}, which accounts for the singular behavior observed in the delta representation.

\subsection{First-passage time properties of SIRWs}
The FPT density $F_k(t)$ of the SIRW to a target site $k$ can be directly expressed in terms of the probability $q_+(k,m,t)$. To reach site $k$ for the first time at time $t$, the RWer must have previously attained some minimal value $-m$ by that time. Summing over all possible values of the minimum yields the following integral in the scaling limit
\begin{equation}
\label{fpt-from-q}
F_k(t) = \int_{0}^\infty q_+(k,m,t) dm.
\end{equation}

In the following, we apply this relation to derive explicit expressions for the Laplace transform $\hat{F}_k(s)$ for most classes of SIRWs, using the structure provided by Eq.~\eqref{fpt-from-q}.

\subsubsection{SATW$_{\alpha,\beta}$, PSRW$_\gamma$}
Integrating $\hat{q}_+(k,m,s)$ \eqref{qplus-laplace-satw} over $m$ yields a closed-form expression for the Laplace transform of the FPT density to site $k$, with $\mu = \sqrt{2sC}$ \footnote{Recall that $C=1$ for the SATW$_{\alpha,\beta}$ and $C=\frac{1}{2\gamma+1}$ for the PSRW$_\gamma$.}:
\begin{equation}
\hat{F}_k(s) = \frac{\mu}{B(\alpha,\beta)} \int_0^\infty \frac{\sinh^\alpha(k \mu)\, \sinh^{\beta -1}(\mu m)}{\sinh^{\alpha + \beta}(\mu (k + m))} \, \mathrm{d}m,
\end{equation}
which corresponds to a special case of the integral evaluated in Eq.~\eqref{int-sinhc-sinhd}. This leads to the expression:
\begin{equation}
\label{fpt-satw-1}
\hat{F}_k(s) = e^{-k \mu(\alpha+\beta)} (2\sinh(k \mu))^\alpha \cdot \frac{B\left(\beta, \frac{\alpha+1}{2}\right)}{B(\alpha,\beta)} \cdot {}_2F_1\left(\alpha+\beta,\tfrac{\alpha+1}{2};\tfrac{\alpha+2\beta+1}{2};e^{-2 k \mu}\right).
\end{equation}

Applying Euler's transformation of the hypergeometric function, we obtain a more convenient form:
\begin{align}
\label{fpt-satw-2}
\hat{F}_k(s) 
&= e^{-k \mu(\alpha+\beta)} (1 - e^{-2k\mu})^{-\alpha} (2\sinh(k \mu))^\alpha \cdot \frac{B\left(\beta, \frac{\alpha+1}{2}\right)}{B(\alpha,\beta)} \cdot {}_2F_1\left(\tfrac{1-\alpha}{2},\beta;\tfrac{\alpha+2\beta+1}{2};e^{-2 k \mu}\right) \\
&= e^{-k \mu \beta} \cdot \frac{B\left(\beta, \frac{\alpha+1}{2}\right)}{B(\alpha,\beta)} \cdot {}_2F_1\left(\tfrac{1-\alpha}{2},\beta;\tfrac{\alpha+2\beta+1}{2};e^{-2 k \mu}\right).
\end{align}

One can verify that the FPT density is properly normalized by setting $s = 0$ in Eq.~\eqref{fpt-satw-2}, confirming that $\hat{F}_k(0) = 1$.

Finally, performing a term-by-term inverse Laplace transform of Eq.~\eqref{fpt-satw-2}, we arrive at the explicit expression, valid for $k\geq 0$ in the scaling limit:
\begin{keyboxedeq}[FPT density of SATW and PSRW]
\begin{equation}
\label{fpt-satw-final}
F_k(t) = \frac{B\left(\beta, \tfrac{\alpha+1}{2}\right)}{B(\alpha,\beta)} \cdot \frac{k}{\sqrt{2\pi C t^3}} 
\sum_{n=0}^\infty \frac{(\beta + 2n)\left( \tfrac{1 - \alpha}{2} \right)_n (\beta)_n}{\left( \tfrac{\alpha + 2\beta + 1}{2} \right)_n n!} \cdot 
\exp\left( -\frac{k^2 (\beta + 2n)^2}{2 C t} \right).
\end{equation}
\end{keyboxedeq}
In the case SATW$_{\alpha,\beta}$, we have $C=1$, and \eqref{fpt-satw-final} matches the expression obtained in \cite{carmonaBetaVariables}, though derived there through an entirely different approach. While not immediately apparent from \eqref{fpt-satw-final}, we can check numerically that $F_k(t) \propto t^{-1-\theta}$ with the persistence exponent $\theta$ given by \eqref{persistence-exp}.

\subsubsection{TSAW}
\begin{figure}
    \centering
    \begin{subfigure}[t]{0.48\textwidth}
        \includegraphics[width=\textwidth]{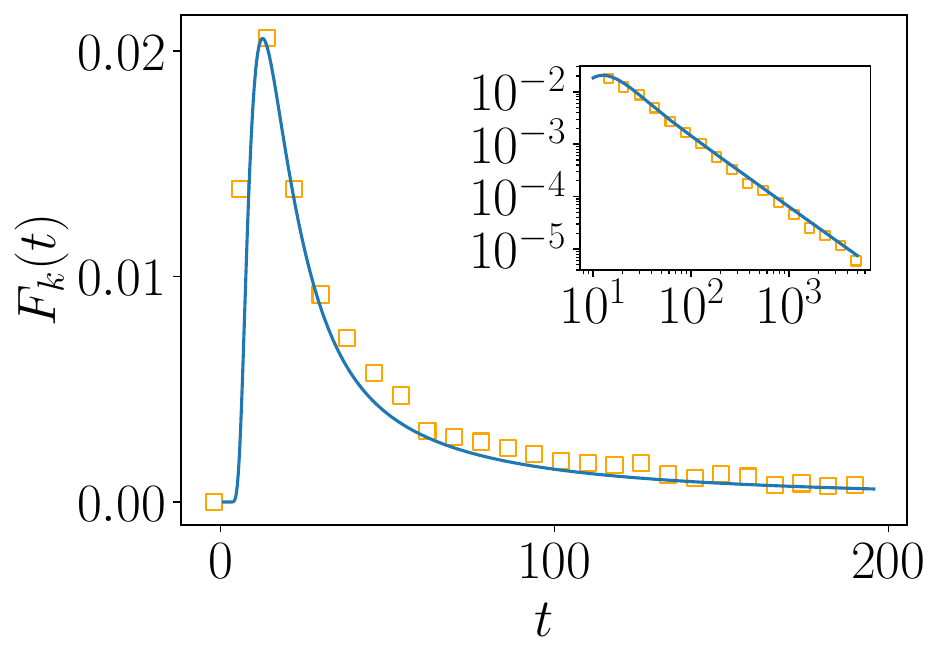}
        \caption{FPT density of the TSAW to target site $k=10$. The blue line is given by the inverse Laplace transform of the exact formula \eqref{fpt-density-tsaw}, while orange squares come from numerical simulations.}
        \label{fig:fpt-tsaw}
    \end{subfigure}
    \hfill
    \begin{subfigure}[t]{0.48\textwidth}
        \includegraphics[width=\textwidth]{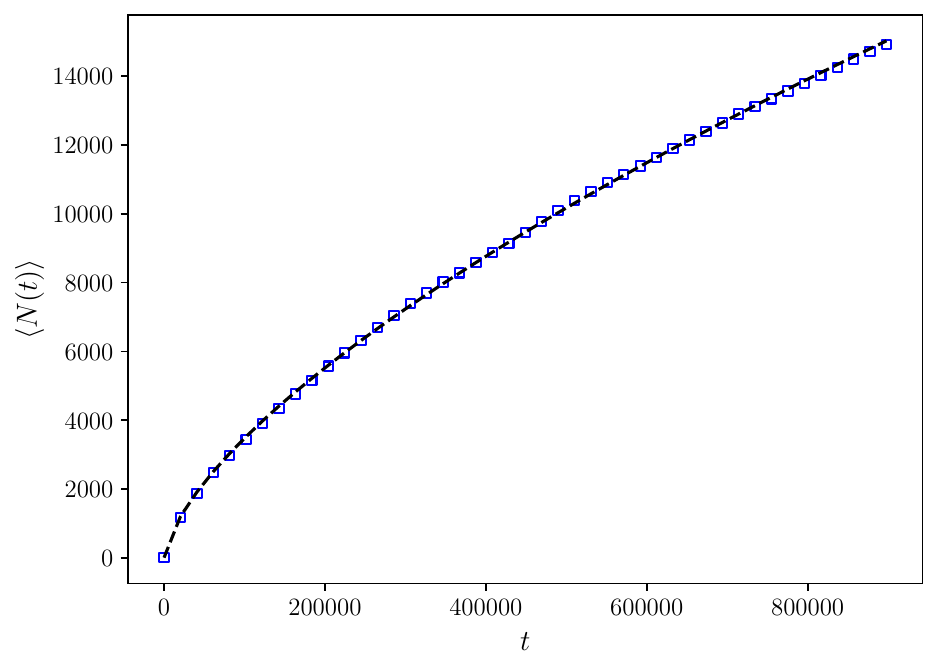}
        \caption{Mean number $\langle N(t) \rangle$ of visited sites up to time $t$ for the TSAW. The black dashed curve is the theoretical result \eqref{noft-mean-tsaw}.}
		\label{fig:mean-noft-tsaw}
    \end{subfigure}
    \begin{subfigure}[t]{0.48\textwidth}
        \includegraphics[width=\textwidth]{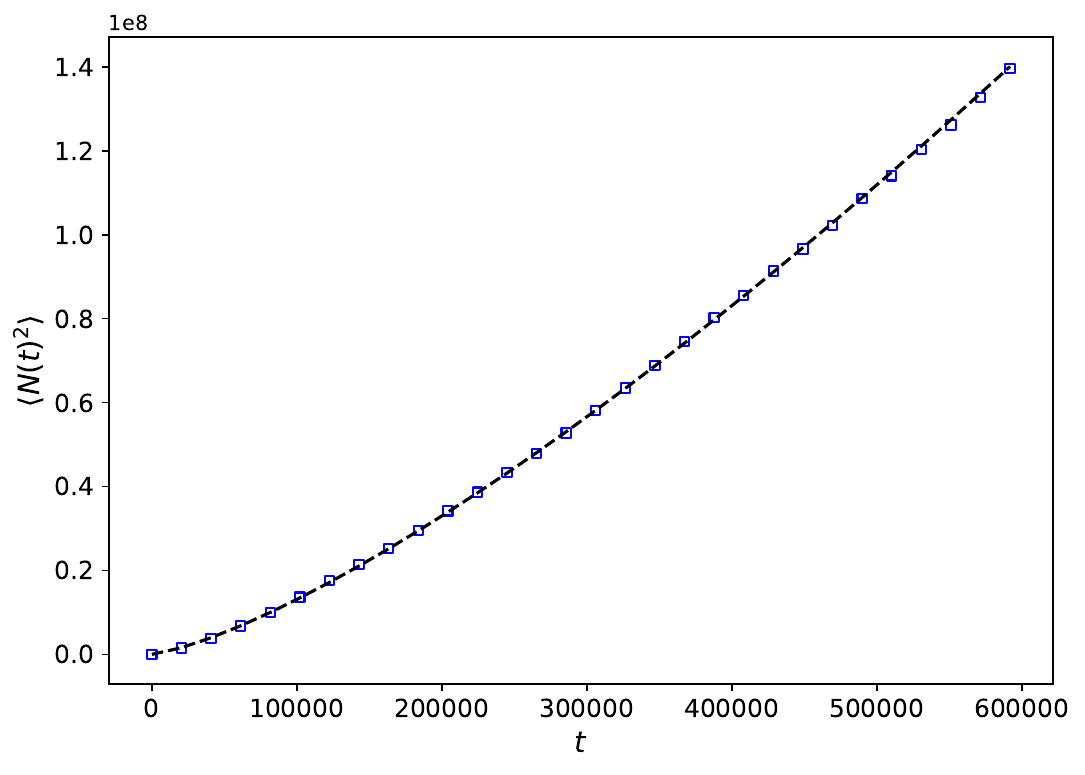}
        \caption{Second moment $\langle N(t)^2 \rangle$ of visited sites up to time $t$ for the TSAW. The black dashed curve is the theoretical result \eqref{secmom-noft-tsaw}.}
		\label{fig:secmom-noft-tsaw}
    \end{subfigure}
    \hfill
    \begin{subfigure}[t]{0.48\textwidth}
        \includegraphics[width=\textwidth]{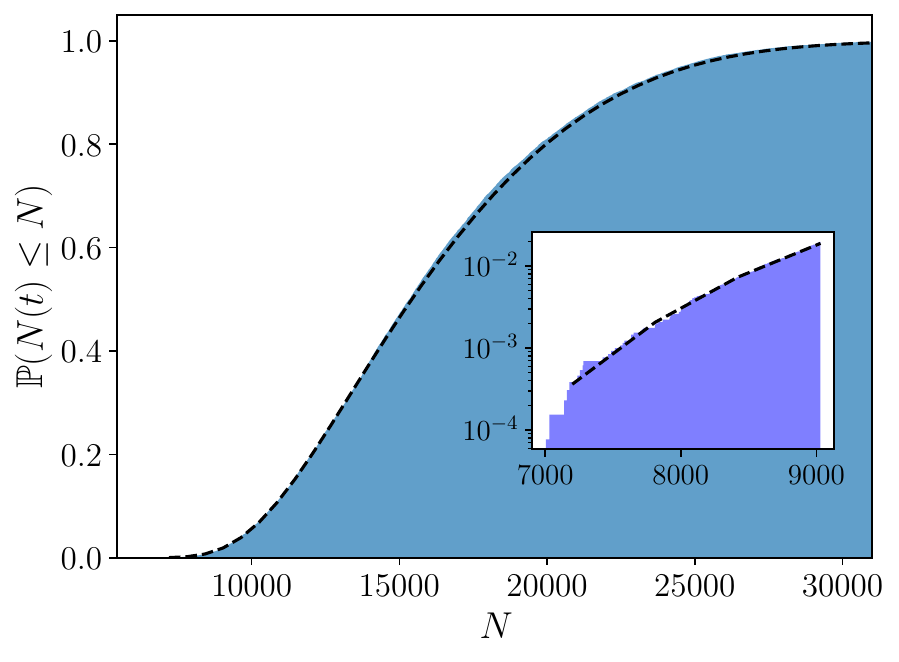}
        \caption{Cumulative distribution of the number of visited sites for the TSAW, at time $t=10^6$. The black dashed curve is the theoretical result \eqref{dist-noft-tsaw}.}
		\label{fig:cumdist-noft-tsaw}
    \end{subfigure}
    \caption{Space exploration observables for the TSAW, with parameter $\beta=1$. The scaling constant $C(\beta=1) \approx 0.706$.}
    \label{fig:figs-spaceexpl-tsaw}
\end{figure}

The integral over $m$ in the expression for $\hat{q}_+(k,m,s)$ given by \eqref{qplus-laplace-tsaw-final} can be carried out explicitly and leads to the following expression for the Laplace transform of the FPT density to site $k$:

\begin{equation}
\label{fpt-density-tsaw-0}
\hat{F}_k(s) = \sum_{i,j=1}^\infty 
\frac{
    \exp\left[ \left( \tfrac{s^2}{2} \right)^{1/3} k \alpha_i' \right]
}{
    \alpha_j \alpha_i'(\alpha_i' - \alpha_j)
}.
\end{equation}

This result reveals an intriguing and, to our knowledge, novel identity involving the zeros of the Airy function and its derivative. By setting $s = 0$ in Eq.~\eqref{fpt-density-tsaw}, one recovers the time-integrated FPT density, which must equal $1$ by normalization: this follows directly from \eqref{dirac-rep-airy}. 
% We provide an independent derivation of Eq.~\eqref{sum-alphai} in the appendix, which confirms the normalization and correctness of the Laplace-transformed expression in Eq.~\eqref{fpt-density-tsaw}. 
In the Appendix, we establish the following useful identities involving the zeros of the Airy function and its derivative:
\begin{equation}
\sum_{j=1}^\infty \frac{1}{\alpha_j(\alpha_i' - \alpha_j)} = -\frac{1}{\alpha_i'} \cdot \frac{\mathrm{Ai}'(0)}{\mathrm{Ai}(0)}, 
\qquad 
\sum_{i=1}^\infty \frac{1}{\alpha_i'^2} = -\frac{\mathrm{Ai}(0)}{\mathrm{Ai}'(0)} = \frac{\Gamma \left( \tfrac{1}{3} \right)}{3^{1/3} \Gamma \left( \tfrac{2}{3} \right)}.
\end{equation}

These results allow us to recast Eq.~\eqref{fpt-density-tsaw-0} as a single, more compact sum:
\begin{boxedeq}
\begin{equation}
\label{fpt-density-tsaw-laplace}
\hat{F}_k(s) = \frac{3^{1/3} \Gamma \left( \tfrac{2}{3} \right)}{\Gamma \left( \tfrac{1}{3} \right)} 
\sum_{i=1}^\infty 
\frac{
    \exp\left[ \left( \tfrac{s^2}{2} \right)^{1/3} k \alpha_i' \right]
}{
    \alpha_i'^2
}.
\end{equation}
\end{boxedeq}
We can write the exact FPT density $F_k(t)$ in the TSAW case by using the known result for the inverse Laplace transform of $e^{-s^{2/3}}$, which we identified as the Laplace transform of the one-sided Lévy stable law of index $\alpha=\frac{2}{3}$ \cite{montrollLevyStable}. An explicit formula is given in \cite{pensonExactExplicit}:
\begin{equation}
    \label{levy-2thirds}
    \mathcal{L}^{-1}_{s \to t}e^{-s^{2/3}} = \frac{2 e^{-\frac{2}{27 t^2}} \left(\text{Ai}\left(\frac{1}{3 \sqrt[3]{3} t^{4/3}}\right)-3^{2/3} t^{2/3} \text{Ai}'\left(\frac{1}{3 \sqrt[3]{3} t^{4/3}}\right)\right)}{3 \sqrt[3]{3} t^{7/3}}.
\end{equation}
The final result, restoring the constant $C=C(\beta)$ \eqref{cofbeta}, reads:
\begin{keyboxedeq}[FPT density of TSAW]
\begin{align}
    \label{fpt-density-tsaw}
    F_k(t) = \frac{\sqrt[3]{2} \, k \, \Gamma\left( \tfrac{2}{3} \right)}{3 \, C^{4/3} t^{7/3} \, \Gamma\left( \tfrac{1}{3} \right)} 
    \sum_{i=1}^\infty \frac{e^{\frac{k^3 \alpha_i'^3}{27 (Ct)^2}}}{\alpha_i'} \bigg[
    & k \alpha_i' \, \mathrm{Ai}\left( \frac{k^2 \alpha_i'^2}{2^{2/3} \cdot (3Ct)^{4/3}} \right) \\
    & + 2^{1/3} \cdot (3Ct)^{2/3} \, \mathrm{Ai}'\left( \frac{k^2 \alpha_i'^2}{2^{2/3} \cdot (3Ct)^{4/3}} \right)
    \bigg].
\end{align}
\end{keyboxedeq}
While the asymptotic behavior is not immediately apparent from the form of \eqref{fpt-density-tsaw}, numerical evaluation of the series reveals that $F_k(t)$ decays at large times as $F_k(t) \sim t^{-1 - \theta}$, where $\theta = 1/3$ corresponds to the persistence exponent of the TSAW, as computed in Eq.~\eqref{persistence-exp}.

Despite its intricate structure, the exact formula \eqref{fpt-density-tsaw}, confirmed numerically in Fig.~\ref{fig:fpt-tsaw}, provides the first complete and explicit description of the first-passage dynamics in the TSAW. It allows for the precise extraction of time-dependent features that are otherwise inaccessible through perturbative methods or numerical simulations alone. As such, it offers a rare analytical window into the strongly interacting regime characteristic of SIRWs.

\subsection{The distribution of the number $N(t)$ of sites visited up to time $t$}
We begin by analyzing the distribution of the first-passage time $t_L$ at which the SIRW reaches a span of length $L$ for the first time. Using the representation \eqref{span-fpt-qplus}, this distribution can be written in the scaling limit as
\begin{equation}
    \boxed{
        f_L(t) \equiv \mathbb{P}(t_L = t) = \int_0^L \left[ q_+(k, L - k, t) + q_-(k, L - k, t) \right] \, dk. 
    }
\end{equation}
This quantity is directly related to a more classical observable: the number of distinct sites $N(t)$ visited up to time $t$ \cite{regnierCompleteVisitationStatistics2022,hughes1995random,montrollRandomWalks}. Indeed, the two are linked via
\[
    \mathbb{P}(N(t) \geq N) = \mathbb{P}(t_N \leq t),
\]
which expresses the probability that at least $N$ sites have been visited by time $t$ in terms of the first time the span reaches $N$.

Taking the Laplace transform with respect to time, the (cumulative) distribution of $N(t)$ satisfies
\begin{equation}
    \boxed{
        \int_0^\infty e^{-s t} \mathbb{P}(N(t) \geq N)\, dt = \frac{1}{s} \hat{f}_N(s),
    }
\end{equation}
where $\hat{f}_N(s)$ denotes the Laplace transform of the first range time distribution for a span of length $N$. \par 
Another refined and closely related observable of space exploration, expressible explicitly in terms of the probabilities $q_\pm$, is the \emph{joint distribution of the maximum} $N_+(t)$ and \emph{minimum} $-N_-(t)$ of the SIRW. This distribution not only encodes the law of the total number of visited sites $N(t) = N_+(t) + N_-(t)$, but also determines the probability that the walker has remained confined within a given interval $[-a, b]$ up to time $t$, which occurs if and only if $N_-(t) \leq a$ and $N_+(t) \leq b$. Such survival probabilities have been extensively studied in the literature~\cite{brayPersistenceFirstPassage,benichouExitOccupationTimes2009}, and are known to be particularly challenging to compute in the case of non-Markovian processes.

To compute this observable, observe that if the walker reaches site $n_+$ (resp. $-n_-$) at some intermediate time $t'$, having previously gone beyond site $-n_-$ (resp. $n_+$), then its maximum $N_+(t)$ (resp minimum $-N_-(t)$) at any later time $t > t'$ must satisfy $N_+(t) > n_+$ (resp. $-N_-(t) < -n_-$). These considerations allow us to express the joint distribution in terms of the transition probabilities $q_\pm$ as the sum of two terms: 
\begin{equation}
    \boxed{
    \begin{aligned}
        \label{jointdist-max-min}
        \mathbb{P}(N_+(t) \geq n_+,\, N_-(t) \geq n_-) &= \mathbb{P}(N_+(t) \geq n_+,\, N_-(t) \geq n_-, \text{visit } -n_- \text{ before } n_+) + (-n_- \leftrightarrow n_+) \\ &= \int_0^t \left[ \int_{n_-}^\infty q_+(n_+, m, t')\, dm + \int_{n_+}^\infty q_-(k, n_-, t')\, dk \right] dt'.
    \end{aligned}
    }
\end{equation}

% FULLY VERIFIED USING THE 1D FORMULAE FOR JOINT DIST OF MIN AND MAX. THIS LOOKS A BIT WEIRD BUT IF WE PLUG N-=0, THE SECOND INTEGRAL IS 0 WHILE THE FIRST IS THE PROBABILITY TO HAVE HIT N+ BEFORE T.
In the following, we compute these distributions explicitly for the different universality classes of SIRWs.

\subsubsection{SATW$_{\alpha,\beta}$, PSRW$_\gamma$}
\paragraph{First-range time $t_L$}
Using the explicit formula~\eqref{qplus-laplace-satw}, we obtain the Laplace transform of the distribution of $t_L$, writing $\mu = \sqrt{2sC}$:
\begin{equation}
\hat{f}_L(s) = \frac{\mu}{B(\alpha,\beta)\, \sinh^{\alpha+\beta}(\mu L)} 
\left[ \tilde{I}_L(\alpha,\beta{-}1) + \tilde{I}_L(\alpha{-}1,\beta) \right],
\end{equation}
where we define
\begin{equation}
\tilde{I}_L(\alpha,\beta) \equiv \int_0^L \sinh^{\alpha}(k\mu)\, \sinh^{\beta}((L{-}k)\mu)\, \mathrm{d}k.
\end{equation}

The problem reduces to evaluating the integral
\begin{equation}
\label{integral-sinh}
I_L(\alpha,\beta) \equiv \int_0^L \sinh(x)^\alpha\, \sinh(L{-}x)^\beta \, \mathrm{d}x.
\end{equation}

To the best of our knowledge, no general expression for \eqref{integral-sinh} is available in the literature, and symbolic computation software (e.g., Mathematica) does not evaluate it directly. However, the convolution structure of the integral allows for a resolution via Laplace transforms. Specifically, we use
\begin{equation}
\int_0^\infty e^{-pL} \sinh^\alpha(L)\, \mathrm{d}L = 
\frac{2^{-\alpha -1} \Gamma (\alpha +1) \Gamma \left(\frac{p-\alpha }{2}\right)}{\Gamma \left(\tfrac{1}{2} (p+\alpha +2)\right)}.
\end{equation}
Applying this to both terms in the convolution gives
\begin{align}
\int_0^\infty e^{-pL} I_L(\alpha,\beta)\, \mathrm{d}L &= 
\left(\int_0^\infty e^{-pL} \sinh^\alpha(L)\, \mathrm{d}L\right)
\left(\int_0^\infty e^{-pL} \sinh^\beta(L)\, \mathrm{d}L\right) \notag \\
&= \frac{2^{-\alpha -1} \Gamma (\alpha +1) \Gamma \left(\tfrac{p-\alpha }{2}\right)}{\Gamma \left(\tfrac{1}{2} (p+\alpha +2)\right)} \cdot 
\frac{2^{-\beta -1} \Gamma (\beta +1) \Gamma \left(\tfrac{p-\beta }{2}\right)}{\Gamma \left(\tfrac{1}{2} (p+\beta +2)\right)}.
\end{align}

The inverse Laplace transform (with respect to $p$) of this product can be written in terms of the Meijer-G function \cite{gradshteynTableIntegrals}, which, in turn, can be expressed as simpler hypergeometric functions:
\begin{equation}
I_L(\alpha,\beta) = B(\alpha{+}1,\beta{+}1)\, e^{(\beta + 1) L} 
\, {}_2F_1\left(\beta{+}1,\tfrac{1}{2}(\alpha{+}\beta{+}2);\alpha{+}\beta{+}2;1 - e^{2L} \right) 
\sinh^{\alpha + \beta + 1}(L).
\end{equation}

Inserting this into the original expression for $\hat{f}_L(s)$ and using a hypergeometric transformation, we obtain ($\mu=\sqrt{2sC}$):
\begin{equation}
\boxed{
\label{firstrange-satw-diff}
\begin{aligned}
\hat{f}_L(s) = \frac{1}{\alpha + \beta} \bigg(& 
\alpha\, e^{-L \mu \beta} 
\, {}_2F_1\left(\beta,\tfrac{1}{2} (\alpha + \beta + 1);\alpha + \beta + 1;1 - e^{-2L\mu} \right) \\
&+ \beta\, e^{-L \mu \alpha} 
\, {}_2F_1\left(\alpha,\tfrac{1}{2} (\alpha + \beta + 1);\alpha + \beta + 1;1 - e^{-2L\mu} \right)
\bigg).
\end{aligned}
}
\end{equation}

In the symmetric case $\alpha = \beta$, this simplifies dramatically:
\begin{boxedeq}
\begin{equation}
    \label{firstrange-satw-equal}
    \hat{f}_L(s) = \frac{1}{\cosh^{2\alpha} \left( \tfrac{L\mu}{2} \right)},
\end{equation}
\end{boxedeq}
recovering the well-known result for the simple RW when $\alpha = \beta = 1$ \cite{borodinHandbookBrownianMotion2002}.

Setting $s = 0$ in \eqref{firstrange-satw-diff} confirms that the density $f_L(t)$ is normalized. Expanding the Laplace transform in powers of $s$ yields the moments of the first range time $t_L$:
\[
    \hat{f}_L(s) = \int_0^\infty e^{-s t} f_L(t)\, dt = 1 - s \langle t_L \rangle + \frac{s^2}{2} \langle t_L^2 \rangle + \dots
\]
From \eqref{firstrange-satw-diff}, we obtain the mean and variance of $t_L$ as
\[
    \langle t_L \rangle = C\cdot \frac{\alpha \beta}{\alpha + \beta} L^2, \qquad
    \langle t_L^2 \rangle = C^2\cdot \frac{\alpha \beta \left[\alpha (3\beta + 2) + 2\beta + 1\right]}{3(\alpha + \beta)(\alpha + \beta + 2)} L^4,
\]
\[
    \mathrm{Var}(t_L) = C^2\cdot \frac{\alpha \beta L^4 \left(2\alpha^2 - 2\alpha \beta + \alpha + 2\beta^2 + \beta\right)}{3(\alpha + \beta)^2 (\alpha + \beta + 2)}.
\]

\paragraph{Number of sites visited $N(t)$}

From \eqref{firstrange-satw-diff}, we directly obtain
\begin{boxedeq}
\begin{equation}
    \begin{aligned}
        \label{noft-satw}
        \mathcal{L}_{t \to s} \mathbb{P}(N(t)\geq N) = \frac{1}{s(\alpha + \beta)} \bigg(& 
        \alpha\, e^{-N \mu \beta} \,
        {}_2F_1\left(\beta,\tfrac{1}{2} (\alpha + \beta + 1);\alpha + \beta + 1;1 - e^{-2N\mu} \right) \notag \\
        &+ \beta\, e^{-N \mu \alpha} \,
        {}_2F_1\left(\alpha,\tfrac{1}{2} (\alpha + \beta + 1);\alpha + \beta + 1;1 - e^{-2N\mu} \right)
        \bigg).
    \end{aligned}
\end{equation}
\end{boxedeq}
In the symmetric case, \eqref{noft-satw} simplifies:
\begin{keyboxedeq}[Number of visited sites for symmetric SATW]
\begin{equation}
    \label{noft-satw-sym}
    \mathcal{L}_{t \to s} \mathbb{P}(N(t)\geq N) = \frac{1}{s \cosh^{2\alpha} \left( \tfrac{N\mu}{2} \right)}.
\end{equation}
\end{keyboxedeq}
From this expression, the mean number of sites visited at time $t$ can be obtained by integrating \eqref{noft-satw} over $N$. Using the identity
\[
    \int_0^\infty e^{-\alpha \mu N} \left(1 - e^{-2\mu N}\right)^n dN =
    \frac{\Gamma \left( \frac{\alpha}{2} \right) \, n!}{2 \mu \, \Gamma \left(n + \frac{\alpha}{2} + 1 \right)},
\]
and expanding the ${}_2F_1$ functions in \eqref{noft-satw} as power series, we find
\begin{align}
    \label{mean-noft-satw-laplace}
    \mathcal{L}_{t \to s} \langle N(t) \rangle
    = \int_0^\infty \mathcal{L}_{t \to s} \mathbb{P}(N(t)\geq N)\, dN
    = \frac{1}{\sqrt{2s^3}(\alpha \beta^2 + \alpha^2 \beta)} \bigg(
    &\alpha^2 \,
    {}_3F_2\left(
        1,\; \tfrac{\alpha + \beta + 1}{2},\; \beta;\;
        \tfrac{\beta}{2} + 1,\; \alpha + \beta + 1;\; 1
    \right) \notag \\
    + &\beta^2 \,
    {}_3F_2\left(
        1,\; \alpha,\; \tfrac{\alpha + \beta + 1}{2};\;
        \tfrac{\alpha}{2} + 1,\; \alpha + \beta + 1;\; 1
    \right)
    \bigg).
\end{align}
There is no known closed-form summation formula for the ${}_3F_2$ functions appearing in \eqref{mean-noft-satw-laplace}, but they can be efficiently evaluated numerically. Inverting the Laplace transform in \eqref{mean-noft-satw-laplace}, we obtain the mean number of sites visited at time $t$ as
\begin{equation}
    \label{mean-noft-satw-asym}
    \parbox{.75\textwidth}{%
    \begin{align*}
        \langle N(t) \rangle
        = \sqrt{\frac{2 C t}{\pi}} \cdot \frac{1}{\alpha \beta^2 + \alpha^2 \beta} \bigg(
        &\alpha^2 \,
        {}_3F_2\left(
            1,\; \tfrac{\alpha + \beta + 1}{2},\; \beta;\;
            \tfrac{\beta}{2} + 1,\; \alpha + \beta + 1;\; 1
        \right) \\
        +\; &\beta^2 \,
        {}_3F_2\left(
            1,\; \alpha,\; \tfrac{\alpha + \beta + 1}{2};\;
            \tfrac{\alpha}{2} + 1,\; \alpha + \beta + 1;\; 1
        \right)
        \bigg).
    \end{align*}%
    }%
\end{equation}
For the symmetric case $\alpha=\beta$, \eqref{mean-noft-satw-asym} simplifies:
\begin{boxedeq}
\begin{equation}
    \label{mean-noft-satw-sym}
        \langle N(t) \rangle = \frac{\Gamma (\alpha)}{\Gamma \left(\alpha +\frac{1}{2}\right) } \cdot \sqrt{2 C t}.
\end{equation}
\end{boxedeq}

Substituting $\alpha = 1 = C$ into \eqref{mean-noft-satw-sym}, we recover the well-known result for the simple RW \cite{hughes1995random} $\langle N(t) \rangle = \sqrt{\frac{8t}{\pi}}$. 
As a concrete example, for the self-repelling case $\alpha = \beta = \tfrac{1}{2}$, we find $\langle N(t) \rangle = \sqrt{2\pi t}$, which is a factor of $\tfrac{\pi}{2}$ larger than the simple RW result. This is consistent with the fact that self-repelling walks tend to explore more of the space. On the other hand, for the self-attracting case $\alpha = \beta = 2$, we obtain $\langle N(t) \rangle = \tfrac{4}{3} \sqrt{\tfrac{2t}{\pi}}$, which is only $\tfrac{2}{3}$ of the simple RW result, reflecting the walk's tendency to revisit previously visited sites more frequently. \par \vspace{1ex}
We now compute the variance of $N(t)$ in the symmetric case $\alpha = \beta$, using the expression \eqref{firstrange-satw-equal}. We begin by evaluating the second moment:
\begin{align}
    \mathcal{L}_{t \to s} \langle N^2(t) \rangle
    &= 2 \int_0^\infty \mathcal{L}_{t \to s} \mathbb{P}(N(t) \geq N) \cdot N \, dN \notag \\
    &= \frac{2}{s} \int_0^\infty \frac{N \, dN}{\cosh^{2\alpha} \left( \frac{N \mu}{2} \right)} = \frac{4^{\alpha}}{\alpha^2 s^2} \,
    {}_3F_2\left( \alpha,\; \alpha,\; 2\alpha;\; \alpha + 1,\; \alpha + 1;\; -1 \right).
\end{align}

Inverting the Laplace transform gives the time-domain expression for the second moment:
\begin{equation}
    \langle N^2(t) \rangle = \frac{4^\alpha \, {}_3F_2(\alpha,\alpha,2\alpha;\alpha + 1, \alpha + 1; -1)}{\alpha^2} \, t.
\end{equation}
Combining this result with the expression for the mean of $N(t)$, we obtain the variance in the symmetric case $\alpha = \beta$:
\begin{boxedeq}
\begin{equation}
    \label{var-noft-equal}
    \boxed{
    \mathrm{Var}(N(t)) = \left(
    \frac{4^\alpha \, {}_3F_2(\alpha, \alpha, 2\alpha; \alpha + 1, \alpha + 1; -1)}{\alpha^2}
    - \frac{2 \Gamma(\alpha)^2}{\Gamma\left(\alpha + \tfrac{1}{2} \right)^2}
    \right) Ct
    }
\end{equation}
\end{boxedeq}

Substituting $\alpha = C =1$ into \eqref{var-noft-equal} recovers the classical result for the simple RW \cite{weissAspectsApplicationsRandom1994}:
\[
\mathrm{Var}(N(t)) = \left(4 \log 2 - \frac{8}{\pi} \right) t.
\]
To quantify the spread of the RW's span, we analyze the reduced fluctuations of the number of visited sites 
\begin{equation}
    \sigma_\alpha \equiv \frac{\sqrt{\mathrm{Var}(N(t))}}{\langle N(t) \rangle},
\end{equation}
and compare it to the case of the simple RW, namely 
\[
    \sigma_1 = \sqrt{\frac{1}{2} \pi  \log (2)-1}.
\]
For the self-repelling case $\alpha = 1/2$, we obtain
\[
\sigma_{1/2} = \sqrt{\frac{4 \mathcal{C}}{\pi }-1} \approx 1.37 \sigma_1
\]
where $\mathcal{C}$ denotes Catalan's constant \cite{abramowitz1965handbook}. On the other hand, in the self-attracting case $\alpha = 2$, we find
\[
\sigma_2 = \frac{1}{4} \sqrt{\pi  (6 \log (4)-3)-16} \approx 0.71 \sigma_1.
\]
We thus conclude that, compared to simple RWs, self-repelling SIRWs visit more sites on average and exhibit larger relative fluctuations in their span, while self-attracting SIRWs explore less space with smaller relative fluctuations. Although these trends are physically intuitive, it is remarkable that the Ray--Knight framework allows us to make these comparisons fully quantitative, providing exact expressions for both the mean and variance of the span.

\subsubsection{TSAW}
In the TSAW case, we start by computing the mean number of visited sites $\langle N(t) \rangle$, as it only requires the knowledge of the FPT density $F_k(t)$. Indeed, the following identity holds:
\begin{align}
    \langle N(t) \rangle
    &= \int_0^\infty \mathbb{P}(N(t) \geq N) \, dN 
    = \int_0^t dt' \int_0^\infty \mathbb{P}(t_N \leq t') \, dN \notag \\
    &= 2 \int_0^t dt' \int_0^\infty dN \int_0^N q_+(k, N - k, t') \, dk \notag \\
    &= 2 \int_0^t dt' \iint_0^\infty q_+(k, m, t') \, dk \, dm = 2 \int_0^t \int_0^\infty F_k(t') \, dt' \, dk.
\end{align}

Using the Laplace-space expression for the first-passage density $\hat{F}_k(s)$ from Eq.~\eqref{fpt-density-tsaw-laplace}, we compute the Laplace transform of $\langle N(t) \rangle$:
\begin{align}
    \mathcal{L}_{t \to s}\langle N(t) \rangle 
    &= \frac{2}{s} \int_0^\infty \hat{F}_k(s) \, dk \notag \\
    &= -\frac{2}{s} \cdot \frac{2^{1/3}}{s^{2/3}} \cdot \frac{3^{1/3} \Gamma \left( \tfrac{2}{3} \right)}{\Gamma \left( \tfrac{1}{3} \right)} 
    \sum_{i=1}^\infty \frac{1}{(\alpha_i')^3} \notag \\
    &= \frac{2 \cdot 6^{1/3} \, \Gamma \left( \tfrac{2}{3} \right)}{s^{5/3} \Gamma \left( \tfrac{1}{3} \right)},
\end{align}
where the identity $\sum_{i=1}^\infty \frac{1}{(\alpha_i')^3} = -1$ is proven in Appendix~\eqref{sum-alphaip3}.

Inverting the Laplace transform and reinstating the scaling constant $C = C(\beta)$ yields the exact expression
\begin{boxedeq}
\begin{equation}
    \label{noft-mean-tsaw}
    \langle N(t) \rangle = \frac{6^{1/3}}{\Gamma \left( \frac{4}{3} \right)} \, (C(\beta) \cdot t)^{2/3}.
\end{equation}
\end{boxedeq}

See Fig.~\ref{fig:mean-noft-tsaw} for a numerical verification of Eq.~\eqref{noft-mean-tsaw}. \par 

Let us now compute the joint distribution of the maximum $N_+(t)$ and minimum $-N_-(t)$. According to \eqref{jointdist-max-min}, its expression in Laplace space reads, for $n_+, n_- >0$
\begin{boxedeq}
\begin{equation}
\label{jointdist-max-min-tsaw}
\begin{aligned}
    \mathcal{L}_{t \to s} \mathbb{P}(N_+(t)\geq n_+,\, N_-(t) \geq n_-) &= \frac{1}{s} \left( \int_{n_-}^\infty \hat{q}_+(n_+, m, s)\, dm + \int_{n_+}^\infty \hat{q}_-(k, n_-, s)\, dk \right) \\
    &= \frac{1}{s} \sum_{i,j \geq 1} \frac{e^{\left(\frac{s^2}{2}\right)^{1/3} (n_+\alpha_i' + n_-\alpha_j)} +e^{\left(\frac{s^2}{2}\right)^{1/3} (n_-\alpha_i' + n_+\alpha_j)}}{\alpha_i'(\alpha_i' - \alpha_j) \alpha_j}.
\end{aligned}
\end{equation}
\end{boxedeq}
Note that \eqref{jointdist-max-min-tsaw} is indeed properly normalized, in the sense that it tends to $\frac{1}{s}$ as $n_-, n_+ \to 0$. However, this is a subtle point. To see this, first set $n_+ = 0$ in \eqref{jointdist-max-min-tsaw}, yielding
\begin{align}
    \mathcal{L}_{t \to s} \mathbb{P}(N_-(t) \geq n_-) &= \frac{1}{s} \sum_{i,j \geq 1} \frac{e^{\left(\frac{s^2}{2}\right)^{1/3} (n_-\alpha_j)} +e^{\left(\frac{s^2}{2}\right)^{1/3} (n_-\alpha_i')}}{\alpha_i'(\alpha_i' - \alpha_j) \alpha_j} \\ &= \frac{1}{s} \sum_{i,j \geq 1} \frac{e^{\left(\frac{s^2}{2}\right)^{1/3} (n_-\alpha_i')}}{\alpha_i'(\alpha_i' - \alpha_j) \alpha_j} \\ &= \frac{1}{s} \hat{F}_{-n_-}(s),
\end{align}
where in the second line we used the identity \eqref{kronecker-rep-airy}, and in the third line we recognized the explicit Laplace transform of the FPT density to site $-n_-$ \eqref{fpt-density-tsaw-laplace}. \par 
From \eqref{jointdist-max-min-tsaw}, we can compute the distribution of the number of sites visited $N(t) = N_+(t) + N_-(t)$. We first derive \eqref{jointdist-max-min-tsaw} with respect to $n_+$ and $n_-$, then integrate over constant sums $n_+ + n_- = N$. Skipping simple algebra, we obtain 
\begin{equation}
    \mathcal{L}_{t \to s} \, \mathbb{P}(N(t) = N) 
    = \frac{2^{2/3}}{s^{1/3}} \sum_{i,j \geq 1} \left[
        \frac{e^{\left(\frac{s^2}{2}\right)^{1/3} N \alpha_i'} - 
        e^{\left(\frac{s^2}{2}\right)^{1/3} N \alpha_j}}{(\alpha_i'-\alpha_j)^2}
    \right].
\end{equation}

Using identities \eqref{sum-squares-airy} and \eqref{sum-squares-airy-2} from the Appendix, we obtain the following deceptively simple form for the cumulative distribution, valid for $N>0$:

\begin{equation}
    \label{dist-noft-tsaw-laplace}
    \boxed{
    \mathcal{L}_{t \to s} \, \mathbb{P}(N(t) \geq N) 
    = \frac{2}{s} \sum_{i=1}^\infty \left[
        e^{\left(\frac{s^2}{2}\right)^{1/3} N \alpha_i'} - 
        e^{\left(\frac{s^2}{2}\right)^{1/3} N \alpha_i}
    \right].
    }
\end{equation}
This expression can be inverted using term-by-term Laplace inversion, together with the following expression for the cumulative distribution function of the one-sided Lévy stable law with index $\alpha = \frac{2}{3}$, obtained by integrating \eqref{levy-2thirds}:
\begin{equation}
    \mathcal{L}^{-1}_{s \to t} \left( \frac{e^{-s^{2/3}}}{s} \right)= 1-\frac{\sqrt{3} \left(\Gamma \left(\frac{4}{3}\right) \, _2F_2\left(\frac{2}{3},\frac{7}{6};\frac{4}{3},\frac{5}{3};-\frac{4}{27 t^2}\right)+2 t^{2/3} \Gamma \left(\frac{2}{3}\right) \, _2F_2\left(\frac{1}{3},\frac{5}{6};\frac{2}{3},\frac{4}{3};-\frac{4}{27 t^2}\right)\right)}{4 \pi  t^{4/3}}.
\end{equation}
We find the following exact distribution for the number of sites visited $N(t)$ by the TSAW
\begin{keyboxedeq}[Number of visited sites for TSAW]
\begin{equation}
    \label{dist-noft-tsaw}
    \begin{aligned}
    \mathbb{P}(N(t) \geq N)
    = -\frac{1}{2\pi} \sum_{i=1}^\infty \left\{
    \frac{
    \sqrt{3}\, \alpha_i' N
    }{
    2^{2/3} (Ct)^{4/3}
    }
    \left[
    \alpha_i' N \, \Gamma\left(\tfrac{4}{3}\right)
    \, {}_2F_2\left( \tfrac{2}{3}, \tfrac{7}{6}; \tfrac{4}{3}, \tfrac{5}{3}; \tfrac{2 N^3 \alpha_i'^3}{27 (Ct)^2} \right)
    \right. \right. \\
    \left. \left.
    -2^{4/3} (Ct)^{2/3} \, \Gamma\left(\tfrac{2}{3}\right)
    \, {}_2F_2\left( \tfrac{1}{3}, \tfrac{5}{6}; \tfrac{2}{3}, \tfrac{4}{3}; \tfrac{2 N^3 \alpha_i'^3}{27 (Ct)^2} \right)
    \right]
    - (\alpha_i' \leftrightarrow \alpha_i)
    \right\}.
    \end{aligned}
    \end{equation}
\end{keyboxedeq}
    
The result \eqref{dist-noft-tsaw} is supported by numerical simulations, as shown in Fig.~\ref{fig:cumdist-noft-tsaw}. To the best of our knowledge, it constitutes the first exact and explicit expression in the literature for the range of the widely studied TSAW. While the scaling $N(t) \propto t^{2/3}$ which is apparent in \eqref{dist-noft-tsaw} was expected from scaling arguments, the exact distribution of the time-independent variable $Y = \frac{N(t)}{t^{2/3}}$ is found to have a rich, strongly non-Gaussian structure.  \par
To gain further insight into the structure of the intricate formula \eqref{dist-noft-tsaw}, we begin by demonstrating how the mean number of visited sites can be recovered from it. This is most conveniently carried out in Laplace space:
\begin{equation}
    \mathcal{L}_{t \to s} \langle N(t) \rangle = \int_0^\infty \mathcal{L}_{t \to s} \mathbb{P}(N(t) \geq N) dN = \frac{2^{4/3}}{s^{5/3}} \sum_{i=1}^\infty \left[\frac{1}{\alpha_i'} - \frac{1}{\alpha_i} \right].
\end{equation}
Note that the sum $\sum_{i=1}^\infty \left[\frac{1}{\alpha_i'} - \frac{1}{\alpha_i} \right]$ above converges (absolutely), even though the individual sums $\sum_{i=1}^\infty \frac{1}{\alpha_i'}$ and $\sum_{i=1}^\infty \frac{1}{\alpha_i}$ do not. After Laplace inversion and restoring the constant $C=C(\beta)$, we obtain 
\begin{equation}
    \label{mean-noft-2}
    \langle N(t) \rangle = \frac{2^{4/3} (Ct)^{2/3}}{\Gamma \left(\frac{5}{3}\right)}  \sum_{i=1}^\infty \left[\frac{1}{\alpha_i'} - \frac{1}{\alpha_i} \right].
\end{equation}
We verified numerically to very high precision that \eqref{mean-noft-2} is exactly the same expression as \eqref{noft-mean-tsaw}. \par 
All moments of $N(t)$ can be computed explicitly from \eqref{dist-noft-tsaw}, using the formula 
\[
    \langle N^k(t) \rangle = k \int_0^\infty N^{k-1} \mathbb{P}(N(t) \geq N) dN.
\]

Performing similar computations as above, we find  
\begin{equation}
    \boxed{
    \langle N^k(t) \rangle = \frac{2^{\frac{k}{3}+1} (Ct)^{\frac{2 k}{3}} k!}{\Gamma \left(\frac{2 k}{3}+1\right)} (-1)^{k-1} \sum_{i=1}^\infty \left[\frac{1}{\alpha_i'^k} - \frac{1}{\alpha_i^k} \right].
    }
\end{equation}
For example, the second moment can be computed explicitly using identities \eqref{sum-airyzeros}, \eqref{sum-airyprimezeros} shown in the Appendix:
\begin{boxedeq}
\begin{equation}
    \label{secmom-noft-tsaw}
    \begin{aligned}
        \langle N^2(t) \rangle &= \frac{2^{\frac{8}{3}} (Ct)^{\frac{4}{3}}}{\Gamma \left(\frac{7}{3}\right)} \sum_{i=1}^\infty \left[\frac{1}{\alpha_i^2} - \frac{1}{\alpha_i'^2} \right] \\ 
        &= \frac{2^{5/3} \sqrt[6]{3} \left(\Gamma \left(\frac{1}{3}\right)^3-3 \Gamma \left(\frac{2}{3}\right)^3\right)}{\pi  \Gamma \left(\frac{1}{3}\right) \Gamma \left(\frac{7}{3}\right)} (C t)^{4/3}.
    \end{aligned}
\end{equation}
\end{boxedeq}
See FIG~\ref{fig:secmom-noft-tsaw} for a numerical confirmation of the result \eqref{secmom-noft-tsaw}. \par \vspace{2ex} 
Having clarified several key aspects of single-time space exploration observables of SIRWs—despite their long-standing lack of understanding—we now venture into even more uncharted territory: the aging behavior of SIRWs and the structure of two-time observables. These quantities probe the memory structure of the walk and provide a deeper understanding of how past exploration influences future evolution.

% From \eqref{jointdist-max-min-tsaw}, we can compute the variance of the number of visited sites. We begin by calculating the correlator $\langle N_+(t) N_-(t) \rangle$:
% \begin{align}
%     \label{correlator-nbvisits-tsaw}
%     \mathcal{L}_{t \to s}\langle N_+(t) N_-(t) \rangle 
%     &= \iint_0^\infty \mathcal{L}_{t \to s} \, \mathbb{P}(N_+(t) \geq n_+,\, N_-(t) \geq n_-) \, dn_+ dn_- \\
%     &= \frac{2}{s} \left( \frac{s^2}{2} \right)^{-2/3} \sum_{i,j\geq 1} \frac{1}{\alpha_i'^2 \alpha_j^2 (\alpha_i' - \alpha_j)}.
% \end{align}
% Inverting the Laplace transform in \eqref{correlator-nbvisits-tsaw} and restoring the constant $C=C(\beta)$, we find
% \begin{equation}
%     \boxed{
%     \langle N_+(t) N_-(t) \rangle = \frac{2^{5/3} (Ct)^{4/3}}{\Gamma\left( \frac{7}{3} \right)} 
%     \sum_{i,j\geq 1} \frac{1}{\alpha_i'^2 \alpha_j^2 (\alpha_i' - \alpha_j)} 
%     \approx 0.411611\, (Ct)^{4/3}.
%     }
% \end{equation}

\section{Aged visitation and record statistics of the SATW$_{\alpha,\beta}$}
\label{sec:aging}
In the study of SIRWs, time-dependent observables provide crucial insight into how memory and reinforcement affect the walker’s space exploration. While one-time quantities such as the propagator or span distribution already reflect the non-Markovian nature of the process, a deeper understanding requires analyzing two-time observables—quantities that probe how spatial features evolve between time $t$ and a future time $t'>t$. Examples include the statistics of displacement increments such as $\langle (X_{t+T} - X_T)^2 \rangle$\footnote{These are not computed in this chapter but will be the focus of future work.}, the influence of previously visited territory on future spatial exploration, and various forms of record statistics.

These two-time observables are essential for quantifying the \emph{dynamics} of space exploration: they reveal how the walker balances exploration and exploitation, how newly visited territory compares to previously explored regions, and how reinforcement mechanisms affect these transitions over time. However, such observables are notoriously difficult to compute. The non-Markovian character of SIRWs means that the past influences the future in a history-dependent manner, and standard renewal techniques fail. Even in models where exact results are known for one-time quantities, little is understood about their two-time counterparts.

In this section, we introduce a new analytical framework—\emph{aged Ray--Knight theorems}—that allows us to compute such two-time observables in the class of $\mathrm{SATW}_{\alpha,\beta}$ models. Building on the classical Ray--Knight representation, which describes the local time profile at a fixed stopping time $T_k$ in terms of independent squared Bessel processes, we extend the theory to characterize how the local time evolves between two successive stopping times. To the best of our knowledge, this is the first such formulation in the context of SATW.

The construction we develop is both natural and powerful: it shows that the local time increment between $T_{k_1}$ and a later stopping time (either $T_{k_2}$ or $T_{-m_2}$) admits a layered decomposition. In regions already visited at time $T_{k_1}$, the walk behaves like a simple RW, and the increment follows classical Ray--Knight statistics for Brownian motion. In contrast, outside the original span, the process behaves as a standard SATW$_{\alpha,\beta}$, and the increment obeys the Ray--Knight description presented in the previous chapter. This separation yields a tractable and physically meaningful description of aging in the SATW.

The aged Ray--Knight theory introduced here is specific to the SATW$_{\alpha,\beta}$ class, which is defined by a weight function $w(n)$ that saturates as $n \to \infty$. This saturation ensures that, in the continuum limit, the walker behaves as Brownian motion within its span. In this setting, our framework provides a detailed analytical tool for accessing the dynamics of local time growth, and through it, a number of two-time observables relevant to space exploration.

In the remainder of this chapter, we apply the aged Ray--Knight framework to compute exact two-time statistics in SATW$_{\alpha,\beta}$. These include aged versions of the $q_\pm$ observables—which, as in the non-aged setting, serve as fundamental building blocks for more complex quantities—, aged splitting probabilities, and the distribution of time intervals between successive maxima of the walk. These results provide a rare instance of exact analytical control over two-time dynamics in a strongly non-Markovian system, bridging a key gap in the theoretical understanding of SIRWs.

\subsection{Aged Ray--Knight theorems for the SATW$_{\alpha,\beta}$}
The classical Ray--Knight representation describes the local time field of a SATW$_{\alpha,\beta}$ up to a stopping time $T_k$ as a composition of two conditionally independent BESQ processes: a forward process $(Y_{2\beta}(k - u))_{0 \leq u \leq k}$ on the positive axis, and a backward process $(\widetilde{Y}_{2 - 2\alpha}(-u))_{u \leq 0}$ on the negative axis.

\begin{figure}
    \centering
    \includegraphics[width=.7\textwidth]{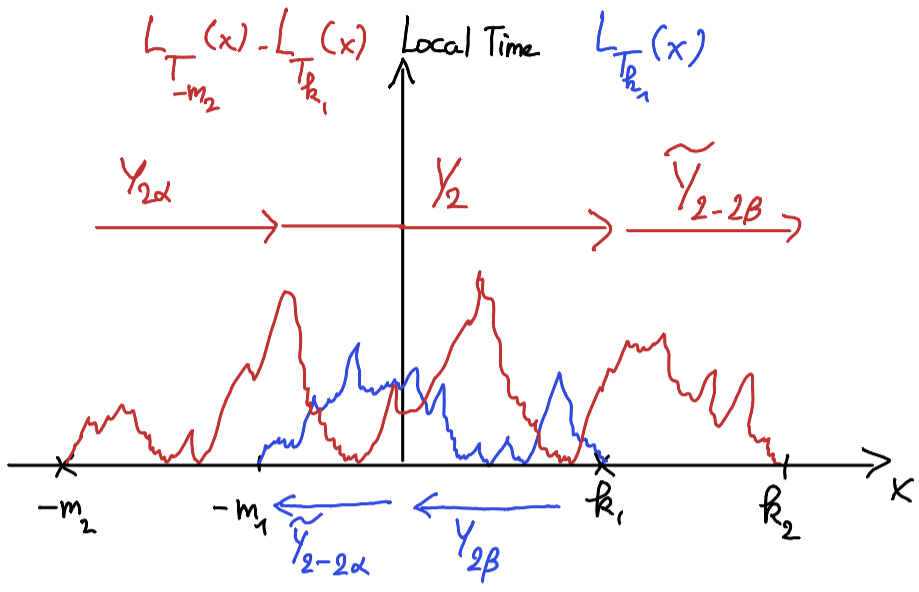}
    \caption{Ray--Knight representation of the local time process at two successive stopping times for a SATW$_{\alpha,\beta}$, illustrating the case of $q_-^{(2)}$ (leftward extension). The blue curve shows the local time profile $L_{T_{k_1}}(x)$ when the walk is stopped upon reaching site $k_1$, with $-m_1$ denoting the leftmost site visited at that time. The red curve shows the increment $\Delta(x) \equiv L_{T_{-m_2}}(x) - L_{T_{k_1}}(x)$ after the walk is continued and stopped again upon hitting site $-m_2 < -m_1$. Since the SATW$_{\alpha,\beta}$ behaves as a simple RW within its current span, the increment $(\Delta(x))_{x \in [-m_1, k_1]}$ follows the classical Ray--Knight theorem for Brownian motion and is thus distributed as a BESQ$_2$ process. Outside of this interval, the increment follows the Ray--Knight laws associated with SATW$_{\alpha,\beta}$.}
    \label{fig:rayknight-aged-qm}
\end{figure}

\begin{figure}
    \centering
    \includegraphics[width=.7\textwidth]{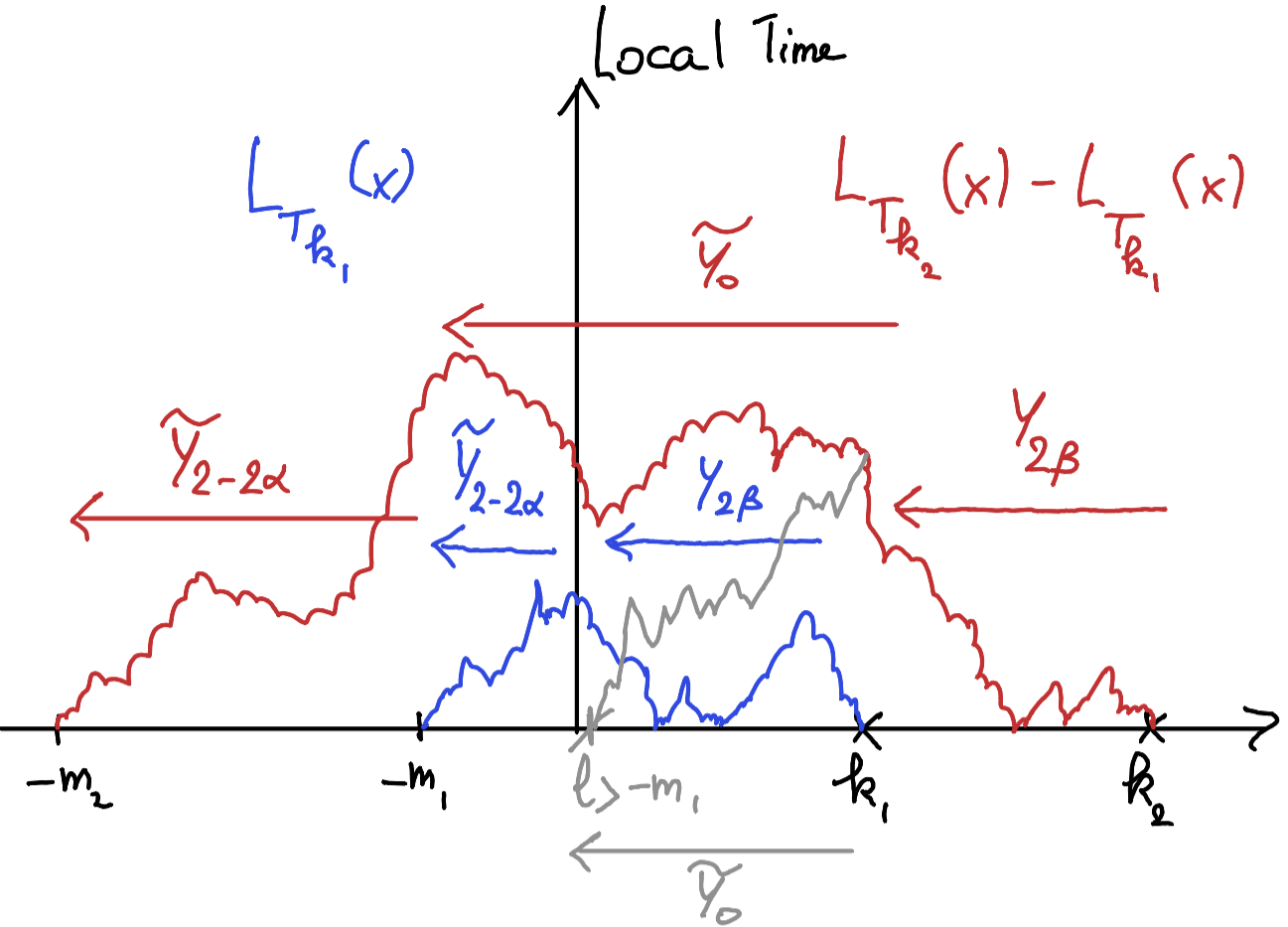}
    \caption{Ray--Knight representation of the local time process at two successive stopping times for a SATW$_{\alpha,\beta}$, illustrating the case of $q_+^{(2)}$ (rightward extension). The blue curve shows the local time profile $L_{T_{k_1}}(x)$ when the walk is stopped upon reaching site $k_1$, with $-m_1$ the minimum reached by that time. The red curve shows the increment $\Delta(x) \equiv L_{T_{k_2}}(x) - L_{T_{k_1}}(x)$ after the walk is continued and stopped again at $k_2 > k_1$. Within the interval $x \in [-m_1, k_1]$, the walk behaves as a simple RW, and the process $(\Delta(k_1 - x))_{x \in [-m_1, k_1]}$ follows a BESQ$_0$ process. In regions outside of $[-m_1, k_1]$, the increment follows the Ray--Knight description of SATW$_{\alpha,\beta}$. The gray curve illustrates an intermediate excursion that reaches a point $l > -m_1$ without extending the span further to the left.}
    \label{fig:rayknight-aged-qp}
\end{figure}

We propose here an \emph{aged} Ray--Knight theory, which extends this framework to analyze how the local time evolves between two successive stopping times. Let the walk be first stopped at $T_{k_1}$, the hitting time of site $k_1$, yielding a local time profile $L_{T_{k_1}}(x)$ with leftmost visited point $-m_1$. Suppose the walk then continues and is stopped again upon reaching either $-m_2 < -m_1$ (leftward extension, see FIG.~\ref{fig:rayknight-aged-qm}) or $k_2 > k_1$ (rightward extension, see FIG.~\ref{fig:rayknight-aged-qp}). The local time increment 
\[
\Delta(x) \equiv L_{T_{\text{new}}}(x) - L_{T_{k_1}}(x)
\]
captures the contribution of this second phase, where $T_{\text{new}} = T_{-m_2}$ or $T_{k_2}$ depending on the case. This increment admits a natural decomposition:
\begin{itemize}
    \item On the previously visited interval $x \in [-m_1, k_1]$, the SATW$_{\alpha,\beta}$ effectively behaves as a simple random walk, since the weight function $w(n)$ has saturated to its limiting value throughout this region. As a result, the local time increment $\Delta(x)$ follows the classical Ray--Knight laws associated with Brownian motion.
    \begin{itemize}
        \item In the case where the walker finishes on the left side of the origin (as is the case for $q_-^{(2)}$), the process $\Delta(x)$ is a BESQ$_2$ on $[-m_1, k_1]$.
        \item In the case where the walker finishes on the left side of the origin (as is the case for $q_+^{(2)}$), the process $\Delta(k_1 - x)$ on $[-m_1, k_1]$ is a BESQ$_0$.
    \end{itemize}
    \item Outside of the original span—i.e., in regions not yet visited at $T_{k_1}$—the increment $\Delta(x)$ follows the Ray--Knight structure associated with SATW$_{\alpha,\beta}$.
\end{itemize}

This layered decomposition provides a detailed description of aging in SATW$_{\alpha,\beta}$: as the walk extends its range, the local time grows through a sequence of increments alternating between simple RW behavior in old regions and reinforced growth governed by Ray--Knight laws in new ones. See Figs.~\ref{fig:rayknight-aged-qm} and \ref{fig:rayknight-aged-qp} for visualizations of this decomposition in both the left- and right-stopping scenarios.

\paragraph{Scope and Limitations of Aged Ray–Knight Theorems}
We emphasize that, although the Ray–Knight theorems for both $\mathrm{PSRW}_\gamma$ and $\mathrm{SATW}_{\tfrac{1}{2},\tfrac{1}{2}}$ involve $\mathrm{BESQ}_1$ processes governing the edge local times, the \emph{aged} Ray–Knight theory does not extend to $\mathrm{PSRW}_\gamma$. This is because the $\mathrm{PSRW}_\gamma$ does not reduce to a simple RW within its previously visited region. Similarly, the aged Ray–Knight theory does not extend to the $\mathrm{SESRW}_{\kappa,\beta}$. However, in the special case of the $\mathrm{TSAW}_\beta$, an analogous role to that of the aged Ray–Knight theory is played by the \emph{Brownian web}—a fundamental object in stochastic geometry introduced by Tóth and Werner \cite{tothTrueSelfrepelling}—which we will define and use extensively in the third part of this thesis.

In contrast, the aged Ray–Knight theorems \emph{do} hold for all processes in the $\mathrm{SATW}_{\alpha,\beta}$ class—that is, for all SIRWs whose weight function $w(n)$ saturates to a finite limit as $n \to \infty$, i.e., $w(n) \to l > 0$. Indeed, in the continuous-time limit, the walker typically visits each site within its span many more times than the saturation threshold, such that $w(n)$ effectively reaches its limiting value. As a consequence, the walker behaves as standard Brownian motion within its span, and aged Ray–Knight theory holds.

In the following, we apply the aged Ray--Knight framework to compute explicit two-time observables for the SATW$_{\alpha,\beta}$. This includes aged extensions of the fundamental $q_\pm$ distributions, which—as in the non-aged case—serve as essential building blocks for more complex space exploration statistics. We also derive the distribution of the walker’s eccentricity conditioned on its previously visited territory, as well as aged splitting probabilities: the likelihood that future exploration occurs on one side of the origin rather than the other, given the past trajectory. Finally, we quantify the record statistics of the SATW by computing the distribution of the time intervals between successive increments of the walker's maximum position.
These observables provide concrete, quantitative insight into how memory and reinforcement shape the future evolution of the walk. 

\subsection{Aged explorations statistics $q^{(2)}_{\pm}(k_2,m_2,t|k_1,m_1)$}
We begin by defining the \emph{aged} probabilities $q^{(2)}_{\pm}(k_2, m_2, t \,|\, k_1, m_1)$, which extend the one-time observables $q_{\pm}(k_1, m_1, t)$ to describe the statistics of a second excursion of the SIRW starting from a previously established span.

Specifically, $q^{(2)}_{+}(k_2, m_2, t \,|\, k_1, m_1)$ denotes the joint probability density that the RW, having already reached site $k_1$ at time $T_{k_1}$ with span $[-m_1, k_1]$, subsequently reaches site $k_2 > k_1$ at time $T_{k_2} = T_{k_1} + t$, and that the new span at this time is $[-m_2, k_2]$:
\begin{equation}
    q^{(2)}_{+}(k_2, m_2, t \,|\, k_1, m_1) = \mathbb{P}\left( T_{k_2} - T_{k_1} = t, \, \text{span} = [-m_2, k_2] \, \middle| \, \text{span at } T_{k_1} = [-m_1, k_1] \right).
\end{equation}
We stress that $q^{(2)}_{+}(k_2, m_2, t \,|\, k_1, m_1)$, as defined above, represents the joint distribution of the pair $(t, m_2)$. \par 
Analogously, we define the aged backward probability $q^{(2)}_{-}(k_2, m_2, t \,|\, k_1, m_1)$ for the case where the walk continues after time $T_{k_1}$ and instead reaches a more negative site $-m_2 < -m_1$:
\begin{equation}
    q^{(2)}_{-}(k_2, m_2, t \,|\, k_1, m_1) = \mathbb{P}\left( T_{-m_2} - T_{k_1} = t, \, \text{span} = [-m_2, k_2] \, \middle| \, \text{span at } T_{k_1} = [-m_1, k_1] \right).
\end{equation}
Thus, $q^{(2)}_{-}(k_2, m_2, t \,|\, k_1, m_1)$ as defined above defines the joint distribution of $(t,k_2)$. \par 
These aged observables capture the evolution of the span conditioned on the past, and reflect the aging structure of the SATW$_{\alpha,\beta}$. They serve as building blocks for two-time functionals and the analysis of memory and renewal effects in SIRWs. \par 
Using the aged Ray--Knight results, we can write $q^{(2)}_{\pm}(k_2, m_2, t \,|\, k_1, m_1)$ explicitly in Laplace space. Namely, as illustrated in Fig.~\ref{fig:rayknight-aged-qm}, we have
\begin{align}
    \hat{q}^{(2)}_{-}(k_2, m_2, s \,|\, k_1, m_1) 
    &= \iint_0^\infty \mathbb{E}^0 \left[ e^{-s \int_0^x Y_{2\alpha}(u) \, du} \,;\, Y_{2\alpha}(m_2-m_1) = a \right] \nonumber \\
    &\quad \times \mathbb{E}^a \left[ e^{-s \int_0^{k_1+m_1} Y_{2}(u) \, du} \,;\, Y_2(k_1+m_1) = b \right] \nonumber \\
    &\quad \times \mathbb{E}^b \left[ e^{-s \int_0^\infty \widetilde{Y}_{2 - 2\beta}(u) \, du} \,;\, T_0^{(2 - 2\beta)} = k_2-k_1 \right] \, da \, db.
\end{align}
Luckily, this integral can be computed exactly using \textsc{Mathematica}. Setting $\mu = \sqrt{2s}$, we obtain
\begin{keyboxedeq}[Aged distribution of maximum and time to the minimum for SATW]
\begin{equation}
\label{aged-qminus-satw}
\begin{aligned}
\hat{q}^{(2)}_{-}(k_2, m_2, s \mid k_1, m_1)
&= \beta \mu \cdot 
\frac{\sinh^{\alpha - 1}\big( \mu(k_1 + m_1) \big)}{
\sinh^{\beta + 1}\big( \mu(k_2 + m_2) \big)} \cdot
\sinh^{-\alpha + \beta + 1}\big( \mu(k_1 + m_2) \big) \\
&\quad \times
{}_2F_1\left(
1 - \alpha, \beta + 1; 1;\,
- \frac{ \sinh\big( \mu(k_2 - k_1) \big) \cdot \sinh\big( \mu(m_2 - m_1) \big) }
{ \sinh\big( \mu(k_1 + m_1) \big) \cdot \sinh\big( \mu(k_2 + m_2) \big) }
\right).
\end{aligned}
\end{equation}
\end{keyboxedeq}

As for $\hat{q}^{(2)}_{+}(k_2, m_2, s \,|\, k_1, m_1)$, as shown in Fig.~\ref{fig:rayknight-aged-qp}, we must distinguish two cases. In the first case $m_1 = m_2$, the SIRW has not gone beyond its past minimum $-m_1$ at time $T_{k_2}$, and we have:
\begin{equation}
\begin{aligned}
\hat{q}^{(2)}_{+}(k_2, m_1, s \mid k_1, m_1) 
= \int_0^{k_1 + m_1} \!\! dl \int_0^\infty \!\! da \; 
&\mathbb{E}^0 \left[ e^{-s \int_0^x Y_{2\beta}(u) \, du} \;\middle|\; Y_{2\beta}(k_2 - k_1) = a \right] \\
&\times \mathbb{E}^a \left[ e^{-s \int_0^\infty \tilde{Y}_0(u) \, du} \;\middle|\; T_0^{(0)} = l \right].
\end{aligned}
\end{equation}
This evaluates to:
\begin{equation}
\boxed{
\hat{q}^{(2)}_{+}(k_2, m_1, s \mid k_1, m_1) = 
\left( \frac{ \sinh\left( \mu(k_1+m_1) \right) }{ \sinh\left( \mu(k_2 + m_1) \right) } \right)^{\beta}.
}
\end{equation}

In the second case, when $m_2 > m_1$, the SIRW does explore new territory to the left, and we write:
\begin{align}
\hat{q}^{(2)}_{+}(k_2, m_2, s \,|\, k_1, m_1) 
&= \iint_0^\infty \mathbb{E}^0 \left[ e^{-s \int_0^x Y_{2\beta}(u) \, du} \,;\, Y_{2\beta}(k_2-k_1) = a \right] \nonumber \\
&\quad \times \mathbb{E}^a \left[ e^{-s \int_0^{k_1+m_1} \tilde{Y}_{0}(u) \, du} \,;\, \tilde{Y}_0(k_1+m_1) = b \right] \nonumber \\
&\quad \times \mathbb{E}^b \left[ e^{-s \int_0^\infty \widetilde{Y}_{2 - 2\alpha}(u) \, du} \,;\, T_0^{(2 - 2\alpha)} = m_2-m_1 \right] da \, db.
\end{align}
This yields, again using \textsc{Mathematica} and unifying both cases:
\begin{keyboxedeq}[Aged distribution of minimum and time to the maximum for SATW]
\begin{equation}
\label{aged-qplus-satw}
\begin{aligned}
\hat{q}^{(2)}_{+}(k_2, m_2, s \mid k_1, m_1) &= 
\alpha \beta \mu \cdot \frac{\sinh^{\beta - 1}\big( \mu(k_1 + m_1) \big)}{  
\sinh^{\alpha + 1}\big( \mu(k_2 + m_2) \big)} \cdot 
\sinh\big( \mu(k_2 - k_1) \big) \\
&\quad \times 
\sinh^{\alpha - \beta}\big( \mu(k_2 + m_1) \big) \\
&\quad \times 
{}_2F_1\left( 
\alpha + 1, 1 - \beta; 2;\,
- \frac{ \sinh\big( \mu(k_2 - k_1) \big) \cdot \sinh\big( \mu(m_2 - m_1) \big) }
{ \sinh\big( \mu(k_1 + m_1) \big) \cdot \sinh\big( \mu(k_2 + m_2) \big) }
\right) \\ &\quad + \delta(m_2-m_1) \left( \frac{ \sinh\left( \mu(k_1+m_1) \right) }{ \sinh\left( \mu(k_2 + m_1) \right) } \right)^{\beta}.
\end{aligned}
\end{equation}
\end{keyboxedeq}
Let us verify that the expressions \eqref{aged-qminus-satw} and \eqref{aged-qplus-satw} correctly reduce to the non-aged Laplace expressions $q_{\pm}$ given in~\eqref{qplus-laplace-satw} in the limit $k_1, m_1 \to 0$, corresponding to the case where no territory has previously been visited. This follows from the asymptotic behavior of the hypergeometric function 
~\cite{abramowitz1965handbook}:
\begin{equation}
\begin{aligned}
{}_2F_1\left( 
\alpha + 1, 1 - \beta; 2; -x\right) &\Sim{x \to \infty} \frac{x^{\beta-1}}{\alpha \beta B(\alpha,\beta)}, \\
{}_2F_1\left( 
1-\alpha, \beta+1; 1; -x\right) &\Sim{x \to \infty} \frac{x^{\alpha-1}}{\beta B(\alpha,\beta)}.
\end{aligned}
\end{equation}
By plugging in $s = 0$ in \eqref{aged-qminus-satw} and \eqref{aged-qplus-satw}, we obtain the time-integrated expressions for the aged distributions. We first give $\hat{q}^{(2)}_{+}$, the distribution of the minimum $-m_2$ upon reaching the maximum $k_2$, given that at time $T_{k_1}$, the SIRW had minimum $-m_1$:
\begin{equation}
\label{qp-aged-s0}
\begin{aligned}
\hat{q}^{(2)}_+(k_2, m_2, s = 0 \mid k_1, m_1) 
&= \alpha \beta \cdot 
\frac{(k_1 + m_1)^{\beta - 1}}{(k_2 + m_2)^{\alpha + 1}} \cdot 
(k_2 - k_1)(k_2 + m_1)^{\alpha - \beta} \\
&\quad \times 
{}_2F_1\left(
\alpha + 1,\ 1 - \beta;\ 2;\,
- \frac{(k_2 - k_1)(m_2 - m_1)}{(k_1 + m_1)(k_2 + m_2)}
\right) \\
&\quad + \delta(m_2 - m_1) \cdot 
\left( \frac{k_1 + m_1}{k_2 + m_1} \right)^{\beta}.
\end{aligned}
\end{equation}

We also give the expression for $\hat{q}^{(2)}_{-}$:
\begin{equation}
\label{qm-aged-s0}
\begin{aligned}
\hat{q}^{(2)}_-(k_2, m_2, s = 0 \mid k_1, m_1) 
&= \beta \cdot 
\frac{(k_1 + m_1)^{\alpha - 1}}{(k_2 + m_2)^{\beta + 1}} \cdot 
(k_1 + m_2)^{\beta - \alpha + 1} \\
&\quad \times 
{}_2F_1\left(
1 - \alpha,\ 1 + \beta;\ 1;\,
- \frac{(k_2 - k_1)(m_2 - m_1)}{(k_1 + m_1)(k_2 + m_2)}
\right).
\end{aligned}
\end{equation}

Although not immediately visible from the expressions \eqref{qp-aged-s0} and \eqref{qm-aged-s0}, we emphasize that both distributions are properly normalized. That is,
\begin{equation}
    \int_{m_1}^{\infty} \hat{q}^{(2)}_+(k_2, m_2, s=0 \mid k_1, m_1) \, dm_2 = 1, 
    \qquad 
    \int_{k_1}^{\infty} \hat{q}^{(2)}_-(k_2, m_2, s=0 \mid k_1, m_1) \, dk_2 = 1.
\end{equation}
This normalization strongly suggests that the expressions \eqref{qp-aged-s0} and \eqref{qm-aged-s0} can be written in a simpler form as total derivatives. This is indeed the case, as we proceed to show.

To that end, we introduce the following variables in $[0,1]$:
\begin{boxedeq}
\begin{equation}
\label{def-xp-xm-x}
x_- = \frac{m_2 - m_1}{k_1 + m_2}, \qquad
x_+ = \frac{k_2 - k_1}{k_2 + m_1}.
\end{equation}
\end{boxedeq}

We show in the Appendix \eqref{proof-deriv-qm} that we can write \eqref{qm-aged-s0} as
\begin{keyboxedeq}[Aged distribution of the maximum when hitting the minimum for SATW]
\begin{equation}
\label{qm-aged-s0-deriv}
\boxed{
\hat{q}^{(2)}_-(k_2, m_2, s = 0 \mid k_1, m_1)
= (1 - x_-)^\alpha \cdot \partial_{k_2} \left[
\sum_{n = 0}^\infty \frac{(\alpha)_n}{n!} \, I_{x_+}(n + 1,\ \beta) \, x_-^n
\right].
}
\end{equation}
\end{keyboxedeq}
where $I_z(a, b)$ denotes the regularized incomplete Beta function.

The normalization follows immediately from integrating over $k_2$, or equivalently over $x_+ \in [0,1]$:
\begin{equation}
\begin{aligned}
\int_{k_1}^{\infty} \hat{q}^{(2)}_-(k_2, m_2, s = 0 \mid k_1, m_1) \, dk_2
&= (1 - x_-)^\alpha \int_0^{1} \partial_{x_+} \left[
\sum_{n = 0}^\infty \frac{(\alpha)_n}{n!} \, I_{x_+}(n + 1,\ \beta) \, x_-^n
\right] dx_+ \\
&= (1 - x_-)^\alpha \sum_{n = 0}^\infty \frac{(\alpha)_n}{n!} \, x_-^n = 1,
\end{aligned}
\end{equation}
using the identity $I_1(a, b) = 1$ and the binomial series expansion $(1 - x)^{-\alpha} = \sum_{n = 0}^\infty \frac{(\alpha)_n}{n!} x^n$. \par 
As for $q_+^{(2)}$, we show in the Appendix \eqref{proof-deriv-qp} that it admits the following representation:
\begin{keyboxedeq}[Aged distribution of the minimum when hitting the maximum for SATW]
\begin{equation}
    \label{qp-aged-s0-deriv}
    \begin{aligned}
    \hat{q}^{(2)}_+(k_2, m_2, s = 0 \mid k_1, m_1)
    = \; (1 - x_+)^\beta \cdot \partial_{m_2} &\Bigg[
        \sum_{n = 1}^\infty 
        \frac{(\beta)_n}{n!} \,
        I_{x_-}(n,\ \alpha) \, x_+^{n} + \Theta(x_-)
    \Bigg].
    \end{aligned}
\end{equation}
\end{keyboxedeq}
    
where $\Theta(x)$ is the Heaviside step function. This representation makes the normalization of $\hat{q}^{(2)}_+$ explicit:
\begin{equation}
\begin{aligned}
\int_{m_1}^{\infty} \hat{q}^{(2)}_+(k_2, m_2, s = 0 \mid k_1, m_1) \, dm_2
&=  (1 - x_+)^\beta \int_0^{1} \partial_{x_-} \left[
\sum_{n = 1}^\infty \frac{(\beta)_n}{n!} \, I_{x_-}(n,\ \alpha) \, x_+^{n}
\right] dx_- \\
&\quad + (1 - x_+)^\beta \\
&= (1 - x_+)^\beta \left(-1+\sum_{n = 0}^\infty \frac{(\beta)_n}{n!} \, x_+^{n}\right)
+ (1 - x_+)^\beta \\
&= (1 - x_+)^\beta \left[ -1+(1 - x_+)^{-\beta}  \right] + (1 - x_+)^\beta = 1.
\end{aligned}
\end{equation}

We now consider aged space exploration observables that we can compute using the aged $q_{\pm}^{(2)}$ probabilities. 

\subsection{Aged Eccentricity Distribution $\sigma^{(2)}_l(z' \mid z)$}

We now study how the eccentricity $z$ of the previously visited territory influences the distribution of the future eccentricity $z'$. Let $L_2 > L_1$, and define $\tilde{\sigma}^{(2)}(z' L_2 \mid z L_1)$ as the probability that the walker’s span of length $L_2$ has eccentricity $z'$, meaning that its maximum is located at $k_2 = z' L_2$, given that its earlier span of length $L_1$ had eccentricity $z$, corresponding to a maximum at $k_1 = z L_1$.

We set:
\[
k_1 = z L_1, \quad m_1 = (1 - z) L_1, \quad k_2 = z' L_2, \quad m_2 = (1 - z') L_2,
\]
with the constraint that the spans satisfy $[-m_1, k_1] \subset [-m_2, k_2]$ and $k_2 + m_2 = L_2$. This requires:
\[
k_1 \leq k_2 \leq L_2 - m_1, \quad m_1 \leq m_2 \leq L_2 - k_1.
\]

The aged eccentricity distribution satisfies the identity:
\begin{equation}
    \tilde{\sigma}^{(2)}(z' L_2 \mid z L_1) = 
    \frac{ \mathbb{P}([-m_2, k_2], [-m_1, k_1]) }{ \mathbb{P}([-m_1, k_1]) },
\end{equation}
where the denominator is the total probability that the walker reaches the span $[-m_1, k_1]$:
\begin{equation}
\begin{aligned}
    \mathbb{P}([-m_1, k_1]) &= \hat{q}_+(k_1, m_1, s=0) + \hat{q}_-(k_1, m_1, s=0) \\
    &= \frac{1}{B(\alpha,\beta)} \cdot 
    \frac{k_1^{\alpha - 1} m_1^{\beta - 1}}{(k_1 + m_1)^{\alpha + \beta - 1}}.
\end{aligned}
\end{equation}

The numerator is the joint probability that the walker first reaches span $[-m_1, k_1]$, and subsequently extends it to $[-m_2, k_2]$. This can be decomposed into four disjoint events depending on which endpoint was last reached at each stage:
\begin{equation}
\begin{aligned}
    \mathbb{P}([-m_2, k_2], [-m_1, k_1]) 
    &= \mathbb{P}([-m_2, \underline{k_2}], [-m_1, \underline{k_1}]) 
     + \mathbb{P}([-m_2, \underline{k_2}], [\underline{-m_1}, k_1]) \\
    &\quad + \mathbb{P}([\underline{-m_2}, k_2], [-m_1, \underline{k_1}]) 
     + \mathbb{P}([\underline{-m_2}, k_2], [\underline{-m_1}, k_1]).
\end{aligned}
\end{equation}

Here, for instance, $\mathbb{P}([\underline{-m_2}, k_2], [-m_1, \underline{k_1}])$ denotes the probability that the span $[-m_1, k_1]$ was last extended at $k_1$, and that the larger span $[-m_2, k_2]$ was completed at $-m_2$.

By definition of the $q_\pm$ and aged $q^{(2)}_\pm$ observables, and using space-reversal symmetry — which exchanges the roles of $\alpha$ and $\beta$, as well as those of $k_2$ with $m_2$, and $k_1$ with $m_1$ — we obtain the identity:
\begin{equation}
\label{pgd-aged}
\begin{aligned}
\mathbb{P}([\underline{-m_2}, k_2], [-m_1, \underline{k_1}]) 
&= \hat{q}_+(k_1, m_1, s = 0) \cdot 
\hat{q}^{(2)}_-(k_2, m_2, s = 0 \mid k_1, m_1) \\
&= \mathbb{P}^{\alpha \leftrightarrow \beta}([-k_2, \underline{m_2}], [\underline{-k_1}, m_1]).
\end{aligned}
\end{equation}

Similarly, using the same symmetry, we have:
\begin{equation}
\label{pgg-aged}
\begin{aligned}
\mathbb{P}([\underline{-m_2}, k_2], [\underline{-m_1}, k_1]) 
&= \hat{q}_-(k_1, m_1, s = 0) \cdot 
\hat{q}^{(2),\, \alpha \leftrightarrow \beta}_+(m_2, k_2, s = 0 \mid m_1, k_1) \\
&= \mathbb{P}^{\alpha \leftrightarrow \beta}([-k_2, \underline{m_2}], [-k_1, \underline{m_1}]).
\end{aligned}
\end{equation}

By scale invariance, we introduce the ratio $l = L_1 / L_2 < 1$, which compares the size of the past span $L_1$ to that of the future span $L_2$. This ratio allows us to define the conditional distribution of the eccentricity $z'$ at the time when the total span reaches the (renormalized) unit size, given that the eccentricity was $z$ when the span was of length $l < 1$:
\begin{boxedeq}
\begin{equation}
\label{sigma2-def}
\sigma^{(2)}_l(z' \mid z) \equiv 
L_2 \cdot \tilde{\sigma}^{(2)}\big(z' \cdot L_2 \,\big|\, z \cdot l \cdot L_2\big).
\end{equation}
\end{boxedeq}

The support of $\sigma^{(2)}_l(z' \mid z)$ is constrained to:
\begin{equation}
    z' \in [z l,\; 1 - (1 - z) l],
\end{equation}
as dictated by the span inclusion condition:
\[
\frac{k_1}{L_2} \leq \frac{k_2}{L_2} = z' \leq 1 - \frac{m_1}{L_2}.
\]
In this scaling setting, using \eqref{qm-aged-s0-deriv}, we can rewrite the $x_+, x_-$ variables \eqref{def-xp-xm-x} as 
\begin{boxedeq}
\begin{equation}
    x_- = \frac{(1-z')-(1-z)l}{zl+(1-z')}, \qquad x_+ = \frac{z'-zl}{z'+(1-z)l}.
\end{equation}
\end{boxedeq}
% ALL FORMULAE BELOW ARE CHECKED.
Notice that $x_+$ and $x_-$ get exchanged under space reversal. We can now compute, for instance, using \eqref{qm-aged-s0-deriv}, \eqref{pgd-aged}: 
\begin{equation}
    \frac{\mathbb{P}([\underline{-(1-z')}, z'], [-(1-z)l, \underline{z l}]) }{\mathbb{P}([-(1-z)l,zl])} = z \cdot (1-x_-)^{\alpha} \cdot \sum_{n=0}^\infty \frac{(\alpha)_n}{n!} x_-^n \cdot \partial_{z'}\left( I_{x_+}(n+1,\beta) \right).
\end{equation}
Hence, using space-reversal symmetry and the identity $\partial_{z'} \equiv -\partial_{1-z'}$, we obtain directly % checked: we NEED the minus in front
\begin{equation}
    \frac{\mathbb{P}([-(1-z'), \underline{z'}], [\underline{-(1-z)l}, zl]) }{\mathbb{P}([-(1-z)l,zl])} = -(1-z) \cdot (1-x_+)^{\beta} \cdot \sum_{n=0}^\infty \frac{(\beta)_n}{n!} x_+^n \cdot \partial_{z'}\left(I_{x_-}(n+1,\alpha)  \right).
\end{equation}
Similarly, one has 
\begin{equation}
    \frac{\mathbb{P}([-(1-z'), \underline{z'}], [-(1-z)l, \underline{z l}]) }{\mathbb{P}([-(1-z)l,zl])} = z (1-x_+)^\beta \left[ \sum_{n=1}^\infty \frac{(\beta)_n}{n!} x_+^{n} \cdot \partial_{z'} \left(I_{x_-}(n,\alpha) \right) + \partial_{z'} \Theta(x_-) \right].
\end{equation}
Finally, using again space-reversal symmetry:
\begin{equation}
    \frac{\mathbb{P}([\underline{-(1-z')}, z'], [\underline{-(1-z)l}, z l]) }{\mathbb{P}([-(1-z)l,zl])} = -(1-z) (1-x_-)^\alpha \left[ \sum_{n=1}^\infty \frac{(\alpha)_n}{n!} x_-^{n} \cdot \partial_{z'} \left(I_{x_+}(n,\beta) \right) + \partial_{z'} \Theta(x_+) \right].
\end{equation}

From \eqref{pgd-aged} and \eqref{pgg-aged}, and after some algebra, we obtain the explicit but lengthy expression:
\begin{equation}
    \label{sigma2-aged-0}
    \begin{aligned}
    \sigma^{(2)}_l(z' \mid z) =\;
    & (1 - x_-)^{\alpha - 1} (1-x_- x_+) (1 - x_+)^{\beta - 1} \\
    & \quad \times \Big[
      \beta (1-x_+) z \cdot {}_2F_1(\alpha, \beta + 1; 1; x_- x_+) \\
    & \qquad + \alpha (1-x_-)(1-z) \cdot {}_2F_1(\alpha + 1, \beta; 1; x_- x_+) \\
    & \qquad + \alpha \beta (x_+ z + x_- (1-z) - x_- x_+) \cdot {}_2F_1(\alpha + 1, \beta + 1; 2; x_- x_+)
    \Big] \\
    & + z \delta\left(z' - [1-l(1 - z)]\right) \cdot l^{\beta} + (1 - z) \delta(z' - l z) \cdot l^{\alpha}.
    \end{aligned}
\end{equation}
As a probability density, the distribution $\sigma^{(2)}_l(z' \mid z)$ must integrate to $1$ over its support, which is constrained by the nested span condition:
\begin{equation}
    z' \in \left[z l,\; 1 - (1 - z) l\right].
\end{equation}
This condition implies that $\sigma^{(2)}_l$ should arise as the derivative (in $z'$) of a simpler primitive function. Indeed, we show in the Appendix that \eqref{sigma2-aged-0} can be expressed more elegantly as % OLIVIER A RAISON : C'EST LE MEME RESULTAT POUR TOUS LES PROCESSUS DE LA CLASSE 1 !
\begin{keyboxedeq}[Aged eccentricity distribution for SATW]
\begin{equation}
    \label{sigma2-aged}
    \begin{aligned}
    \sigma^{(2)}_l(z' \mid z) 
    &= \partial_{z'}\left[
    z \cdot f_{\alpha,\beta}(z, z') 
    - (1 - z) \cdot f_{\beta,\alpha}(1 - z, 1 - z')
    \right], \\[0.5em]
    f_{\alpha,\beta}(z, z') 
    &\coloneqq (1 - x_-)^\alpha \sum_{n=0}^\infty 
    \frac{(\alpha)_n}{n!} \, I_{x_+}(n + 1, \beta) \, x_-^n 
    + l^\beta \cdot \Theta\big(z' - [1 - l(1 - z)]\big).
    \end{aligned}
    \end{equation}
\end{keyboxedeq}
Indeed, from \eqref{sigma2-aged}, the normalization is immediate from the identities (using $I_x(1,a) = 1 - x^a$) \footnote{We adopt the convention $\Theta(0) = 1$, which is appropriate for our analysis.}:
\begin{equation}
    f_{\alpha,\beta}\big(z, z' = 1 - (1 - z)l\big) = (1 - l^\beta) + l^\beta = 1, 
    \qquad 
    f_{\beta,\alpha}\big(1-z, 1-z' = (1 - z)l\big) = 0,
\end{equation}
so that
\begin{equation}
    \int_{zl}^{1 - (1 - z)l} \sigma^{(2)}_l(z' \mid z) \, dz' = z - (-(1 - z)) = 1.
\end{equation}
    
Note that the function $f_{\alpha,\beta}$ also admits the integral representation:
\begin{equation}
    \label{f-int-rep}
    f_{\alpha,\beta}(z, z') - l^\beta \cdot \Theta(z' - (1 - l(1 - z))) = \beta (1 - x_-)^\alpha \int_0^{x_+} (1 - u)^{\beta - 1} \, 
    {}_2F_1(\alpha, \beta + 1; 1; u x_-) \, du .
\end{equation}

Several consistency checks can be performed on the expression \eqref{sigma2-aged}, and they admit natural physical interpretations:

\paragraph{Symmetry under space reversal.}
The distribution $\sigma^{(2)}_l(z' \mid z)$ is invariant under space reflection combined with the exchange of reinforcement parameters:
\begin{equation}
    \sigma^{(2)}_l(z' \mid z) = \sigma^{(2),\, \alpha \leftrightarrow \beta}_l(1 - z' \mid 1 - z).
\end{equation}
This reflects the symmetry of the model under inversion of the walker's trajectory: swapping left and right flips eccentricities $z \mapsto 1 - z$ and $z' \mapsto 1 - z'$, while simultaneously exchanging the bias parameters $\alpha \leftrightarrow \beta$.

\paragraph{Limit $l \to 0$: asymptotic irrelevance of the past.}
In the limit where the previously visited span becomes vanishingly small compared to the future one, i.e. $l = L_1 / L_2 \to 0$, the influence of the past territory disappears. Physically, this corresponds to a memoryless situation: the SIRW explores a much larger new territory than it has seen before, so the constraint of the earlier span has negligible effect on the position of the maximum and minimum of the future trajectory.

In this limit, the parameters $x_\pm$ take the form
\begin{equation}
    x_m = 1 - \frac{\ell}{1 - z'} + \mathcal{O}(\ell^2), \qquad
    x_p = 1 - \frac{\ell}{z'} + \mathcal{O}(\ell^2).
\end{equation}
We aim to show that
\begin{equation}
    \label{limit-f-beta}
    f_{\alpha,\beta}(z,z') \xrightarrow{\ell \to 0} I_{z'}(\alpha,\beta).
\end{equation}
To this end, we use the well-known asymptotic formula \cite{abramowitz1965handbook}
\begin{equation}
    \beta \,{}_2F_1(\alpha, \beta + 1; 1; x) \sim \frac{(1 - x)^{-\alpha - \beta}}{B(\alpha, \beta)} \quad \text{as } x \to 1^-,
\end{equation}
together with the integral representation \eqref{f-int-rep}. As $\ell \to 0$, the integral becomes sharply peaked near its upper boundary, and we obtain the leading-order behavior
\begin{equation}
    f_{\alpha,\beta}(z,z') \sim \frac{(\frac{\ell}{1 - z'})^\alpha}{B(\alpha, \beta)} \int_0^{1 - \frac{\ell}{z'}} (1 - u)^{\beta - 1} \left(1 - u\left[1 - \frac{\ell}{1 - z'}\right]\right)^{-\alpha - \beta} du.
\end{equation}
This integral can now be evaluated exactly in closed form using Mathematica. Taking the limit $\ell \to 0$ of the resulting expression confirms \eqref{limit-f-beta}. Consequently, we recover the unconditional eccentricity distribution \eqref{eccentricity-sirw-1d}:
\begin{equation}
    \sigma^{(2)}_\ell(z' \mid z) \xrightarrow{\ell \to 0} \sigma(z') 
    = \frac{(z')^{\alpha - 1} (1 - z')^{\beta - 1}}{B(\alpha, \beta)}.
\end{equation}

\paragraph{Case $\alpha = \beta = 1$: simple RW.}
In the case of a symmetric simple RW, corresponding to $\alpha = \beta = 1$, the unconditional eccentricity is known to be uniformly distributed on $[0,1]$ in the scaling limit. Equation \eqref{sigma2-aged} for the conditional eccentricity reproduces the expected uniform density, but also contains delta function corrections located at the boundaries of the support interval:
\begin{equation}
    \sigma^{(2)}_l(z' \mid z) 
    = 1 + z l \cdot \delta(l(z - 1) - z' + 1)
    + (1-z)l \cdot \delta(z' - l z).
\end{equation}
As explained above, these Dirac terms reflect trajectories where the eccentricity is entirely determined by the past maximum or minimum. 

\subsection{Aged Splitting Probability $Q_+(k_2, m_2 \mid k_1, m_1)$}

In this section, we examine how the previously visited territory influences the splitting probability $Q_+(k_2, m_2 \mid k_1, m_1)$. This quantity denotes the probability that a SATW$_{\alpha,\beta}$, having already explored the interval $[-m_1, k_1]$ and starting from an arbitrary position within it\footnote{This is strictly equivalent to the case where the walker has just completed visiting the interval $[-m_1, k_1]$, ending at either $k_1$ or $-m_1$.}, reaches $k_2 > k_1$ before $-m_2 < -m_1$.
, reaches $k_2 > k_1$ before $-m_2 < -m_1$. 

\begin{figure}
    \centering
    \begin{subfigure}[t]{0.48\textwidth}
        \includegraphics[width=\textwidth]{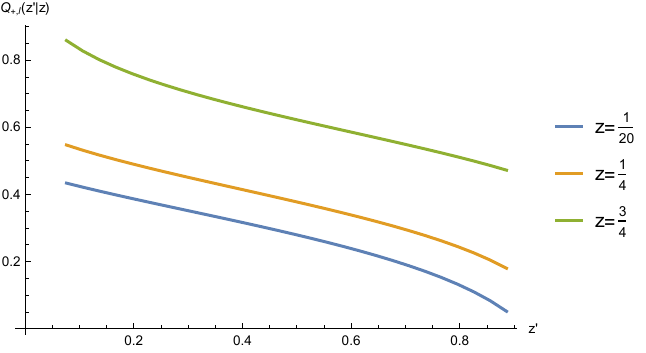}
        \caption{Aged splitting probability $Q_{+,l}(z' \mid z)$ from \eqref{aged-splitting-satw} for the self-repelling SATW$_{1/4,1/4}$ with past-to-future span ratio $l = 0.1$, shown for several values of $z$. For comparison, all curves are shown over their common support, even though their individual supports might differ. As expected, larger past eccentricity $z$ increases the likelihood of reaching the right site $z'$ before the left site $-(1 - z')$. Nevertheless, the dependence on $z$ reveals a rich structure.}
        \label{fig:splitting-diff-z}
    \end{subfigure}
    \hfill
    \begin{subfigure}[t]{0.48\textwidth}
        \includegraphics[width=\textwidth]{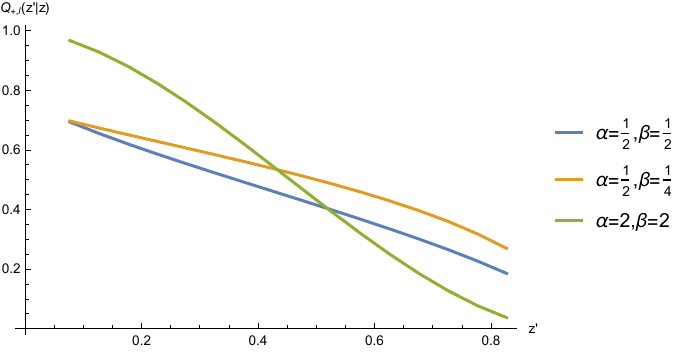}
        \caption{Aged splitting probability $Q_{+,l}(z' \mid z = 1/3)$ from \eqref{aged-splitting-satw} for SATW$_{\alpha,\beta}$ with fixed $z = 1/3$ and $l = 0.2$, plotted for various values of $\alpha$ and $\beta$. The result shows a strong dependence on the reinforcement parameters. Compared to the self-repelling SATW (blue and orange), the self-attractive SATW (green) is more likely to reach $z'$ before $-(1 - z')$ when $z'$ is small, whereas the opposite behavior emerges for larger values of $z'$.
        This intricate behavior is captured only by our exact formula.}
        \label{fig:splitting-diff-alpha}
    \end{subfigure}
    \caption{Aged splitting probability $Q_{+,l}(z' \mid z)$ from \eqref{aged-splitting-satw} for SATW$_{\alpha,\beta}$.}
    \label{fig:aged-splitting}
\end{figure}

Similarly to how the unconditional splitting probability can be obtained from the unconditional eccentricity distribution \eqref{splitting-from-ecc}, we obtain the aged splitting probability directly from the aged eccentricity distribution $\sigma^{(2)}$ in the scaling limit
\[
z' L_2 = k_2, \quad (1 - z') L_2 = m_2, \quad z l L_2 = k_1, \quad (1 - z) l L_2 = m_1
\]
via the identity:
\begin{equation}
Q_{+,l}(z' \mid z) \equiv Q_+(k_2, m_2 \mid k_1, m_1) 
= \int_{z'}^{1 - (1 - z)l} \sigma^{(2)}_l(u \mid z) \, du.
\end{equation}
From the explicit expression \eqref{sigma2-aged}, we directly obtain 
\begin{keyboxedeq}[Aged splitting probability for SATW]
\begin{equation}
    \label{aged-splitting-satw}
        Q_{+,l}(z' \mid z) = z - \left[z f_{\alpha,\beta}(z,z') - (1-z) f_{\beta,\alpha}(1-z,1-z') \right],
\end{equation}
\end{keyboxedeq}
where $f_{\alpha,\beta}$ is defined in \eqref{sigma2-aged}. An illustration of this result is shown in Fig.~\ref{fig:aged-splitting}; the integral representation \eqref{f-int-rep} of $f_{\alpha,\beta}$ is well-suited for numerical evaluation.
\par 
To better understand \eqref{aged-splitting-satw}, we first compare it to the standard splitting probability conditioned on past territory for the simple RW, corresponding to the case $\alpha = \beta = 1$, where all quantities can be computed explicitly due to the Markovian nature of the process. In this case, \eqref{aged-splitting-satw} simplifies to
\begin{equation}
    \label{aged-splitting-srw}
    Q_{+,l}(z' \mid z) = l(2z - 1) + 1 - z'.
\end{equation}

Let us now compute \eqref{aged-splitting-srw} by an independent argument. The walker has previously visited the interval $[-(1 - z)l, zl]$, having ended at $zl$ with probability $z$, and at $-(1 - z)l$ with probability $1 - z$ \eqref{eccentricity-sirw-1d}. Using the standard splitting probability formula for the simple RW, we then write:
\begin{equation}
    Q_{+,l}(z' \mid z) = z \cdot (z l + (1 - z')) + (1 - z) \cdot (1 - z' - (1 - z)l) = l(2z - 1) + 1 - z',
\end{equation}
which matches \eqref{aged-splitting-srw} exactly. \par 
As in the case of $\sigma^{(2)}$, the limit \eqref{limit-f-beta} of $f_{\alpha,\beta}$ as $l \to 0$ yields the unconditional splitting probability from \eqref{aged-splitting-satw}:
\begin{equation}
    Q_{+,l}(z' \mid z) \xrightarrow{l \to 0} z\big(1 - I_{z'}(\alpha,\beta)\big) + (1 - z) I_{1 - z'}(\beta, \alpha) = I_{1 - z'}(\beta, \alpha),
\end{equation}
which agrees precisely with \eqref{splitting-sirw}.

\subsection{Record statistics of the SATW$_{\alpha,\beta}$} 
\subsubsection{Context}
\begin{figure}
    \centering
    \begin{subfigure}[t]{0.48\textwidth}
        \includegraphics[width=\textwidth]{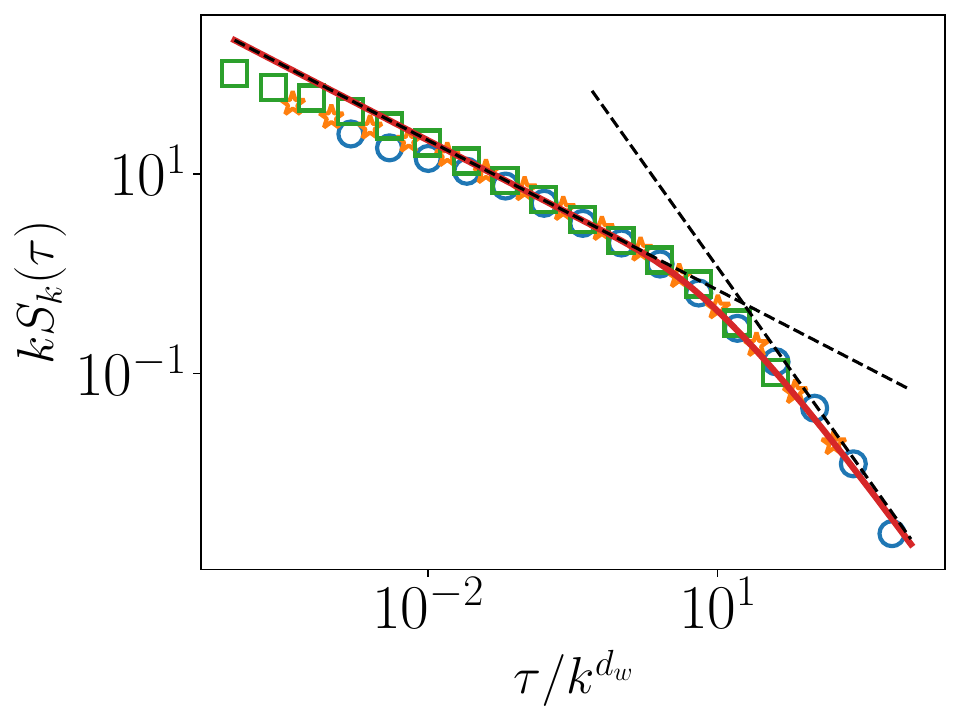}
        \caption{Self-attracting SATW$_\phi$ with $\phi=e>1$.}
        \label{fig:records-phie}
    \end{subfigure}
    \hfill
    \begin{subfigure}[t]{0.48\textwidth}
        \includegraphics[width=\textwidth]{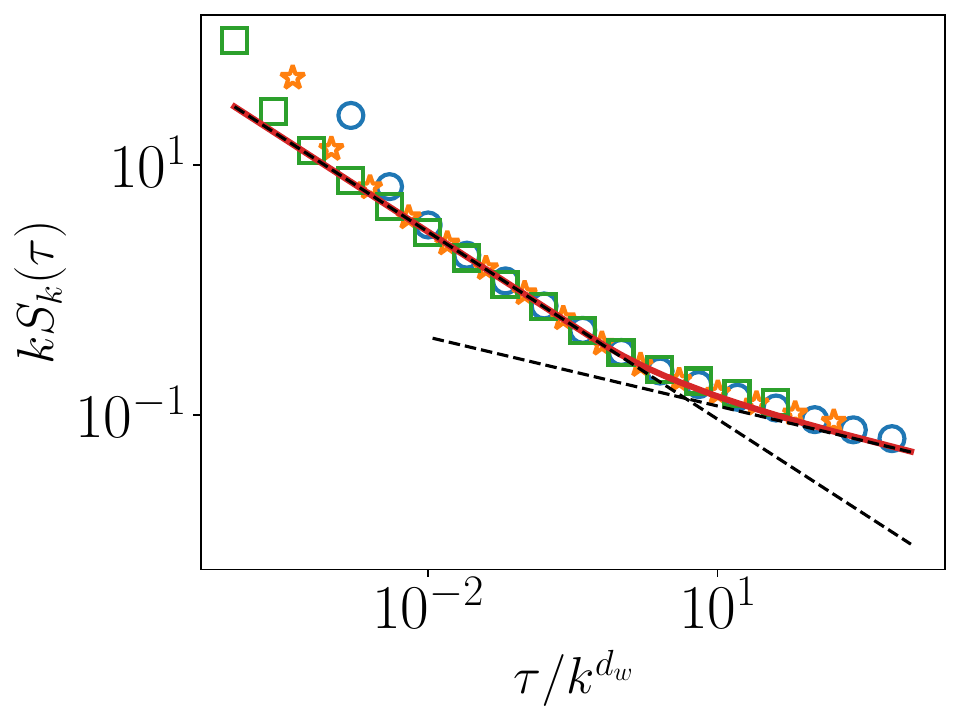}
        \caption{Self-repelling SATW$_\phi$ with $\phi=\tfrac{1}{e}<1$.}
        \label{fig:records-phiem1}
    \end{subfigure}
    \caption{Cumulative distribution $S_k(\tau)$ of the time $\tau$ to break the $k$th record for the SATW$_{\phi}$. The red curve shows the theoretical result \eqref{exact-record-time}, while symbols show numerical simulations with different symbols corresponding to different values of $k$. Dashed black lines show the exact small-$x$ \eqref{psi-smallx} and large-$x$ asymptotics \eqref{psi-largex}.}
    \label{fig:records}
\end{figure}
A key question in the study of space exploration by SIRWs is that of span dynamics, i.e how often the walker extends its visited territory. In particular, we are interested in the time interval between two successive increments of the maximum position, or \emph{records}. This aged observable captures the temporal structure of spatial growth and provides insight into the interplay between reinforcement and exploration. More precisely, if the walker reaches a new maximum $k$ at time $t$, we ask: how long does it take to reach $k+1$ for the first time?

Record age statistics are particularly sensitive to memory effects and offer a direct window into the non-Markovian character of the dynamics. In standard Markovian RWs, they are easily computed thanks to the renewal property of the process. In contrast, for non-Markovian walks such as SIRWs, the distribution of record ages reflects the influence of the entire span history on future evolution. Despite their relevance, exact results for these distributions have remained elusive. In this section, we show that the aged Ray--Knight framework developed above provides the structure required to compute them exactly in the case of SATW$_{\alpha,\beta}$.

Following \cite{regnierRecordAges}, let $t + \tau_k$ denote the time at which the walker first reaches $k + 1$, given that it reached $k$ for the first time at time $t$. We focus on the cumulative distribution function 
\begin{equation}
    S_k(\tau) \equiv \mathbb{P}(\tau_k \geq \tau)
\end{equation} 
of the time $\tau_k$ between the $k$th and $(k+1)$st records, and derive its exact form in the scaling regime $\tau \gg 1$. The final expression \eqref{exact-record-time} gives a complete, numerically tractable description of record statistics in SATW, and highlights the power of the aged Ray--Knight approach in handling nontrivial two-time quantities.

\subsubsection{Computing $S_k(\tau)$}
By integrating over all possible minima $-m_1$ encountered when first reaching $k$, and over all $-m_2 \leq -m_1$ when subsequently reaching $k+1$, the probability $S_k(\tau)$ can be expressed explicitly in terms of the aged observable $q_+^{(2)}$ and the time-integrated, unconditional observable $\hat{q}_+$:
\begin{equation}
    \label{sktau-def}
    S_k(\tau) = \int_0^\infty dm_1 \int_{m_1}^\infty dm_2 \int_{\tau}^\infty 
    \hat{q}_+(k, m_1, s = 0) \, q_+^{(2)}(k+1, m_2, \tau' \mid k, m_1) \, d\tau'.
\end{equation}
Using the identity 
\begin{equation}
    \int_0^\infty e^{-s \tau} d\tau \int_{\tau}^\infty f(\tau') d\tau' = \frac{1}{s} \left(\hat{f}(0) - \hat{f}(s) \right),
\end{equation}
the Laplace transform of \eqref{sktau-def} with respect to $\tau$ writes 
\begin{equation}
    \label{s-int-rep}
    \hat{S}_k(s) = \int_0^\infty e^{-s \tau} S_k(\tau) d\tau = \frac{1}{s} \left(1 - \int_0^\infty dm_1 \int_{m_1}^\infty dm_2
    \hat{q}_+(k, m_1, s = 0) \, \hat{q}_+^{(2)}(k+1, m_2, s \mid k, m_1)\right).
\end{equation}
For simplicity, we focus on the symmetric case $\alpha = \beta \equiv \phi$. The asymmetric case can also be treated, but leads to more cumbersome expressions. \par
We will work in the large-$k$ limit and think of $k+1$ as $k+dk$, with $dk \ll k$, and use the familiar variable $\mu = \sqrt{2s}$. Let us now obtain the full distribution $S_k(\tau)$ in the continuous limit $\tau \gg dk^2$, that is, in the Laplace regime $\mu dk \ll 1$. Plugging the explicit expression \eqref{aged-qplus-satw} for $\hat{q}_+^{(2)}$ in \eqref{s-int-rep}, we can write, to first order in $\mu dk$:
\begin{align}
    \label{full-s-ugly}
    1 - s \hat{S}_k(s) 
    &\Sim{\mu \, dk \ll 1} 
    1 - dk \cdot \frac{\mu \phi}{B(\phi, \phi)} 
    \int_0^\infty \frac{k^\phi m^{\phi - 1}}{(k + m)^{2\phi}} 
    \coth\big(\mu(k + m)\big) \, dm \notag \\
    &\quad + \frac{1}{B(\phi, \phi)} \int_0^\infty dm_1 \int_{m_1}^\infty dm_2 \,
    \frac{k^\phi m_1^{\phi - 1}}{(k + m_1)^{2\phi}} \,
    \frac{\phi^2 \mu \sinh^{\phi - 1}[\mu(k + m_1)]}
         {\sinh^{\phi + 1}[\mu(k + dk + m_2)]} \,
    \sinh(dk \mu) \notag \\
    &\quad \times 
    {}_2F_1\left(
    1 - \phi, 1 + \phi; 2; 
    - \frac{\sinh(dk \mu) \sinh[(m_2 - m_1) \mu]}
           {\sinh[(k + m_1)\mu] \sinh[(k + dk + m_2)\mu]}
    \right) \notag \\
    &\Sim{\mu \, dk \ll 1} 
    1 - dk \cdot \frac{\mu \phi}{B(\phi, \phi)} 
    \int_0^\infty \frac{k^\phi m^{\phi - 1}}{(k + m)^{2\phi}} 
    \coth\big(\mu(k + m)\big) \, dm \notag \\
    &\quad + dk \cdot \frac{\mu^2 \phi^2}{B(\phi, \phi)} 
    \int_0^\infty dm_1 \int_{m_1}^\infty dm_2 \,
    \frac{k^\phi m_1^{\phi - 1}}{(k + m_1)^{2\phi}} \,
    \frac{\sinh^{\phi - 1}[\mu(k + m_1)]}
         {\sinh^{\phi + 1}[\mu(k + m_2)]}. 
\end{align}
To proceed further, we will need the following integral, which is a direct consequence of \eqref{int-sinhc-sinhd}:
\begin{align}
    \int_{m_1}^\infty \frac{dm_2}{\sinh^{\phi+1}(k+m_2)\mu} &= \frac{1}{\mu} \int_0^\infty \frac{da}{\sinh^{\phi+1}(k\mu + \mu m_1 + a)} \\ &= \frac{1}{\mu} \frac{e^{-\mu(k+m_1)}}{\sinh^{\phi}((k+m_1)\mu)} \frac{2}{\phi+1} \,_2F_1(\frac{1-\phi}{2},1,\frac{3+\phi}{2},e^{-2(k+m_1)\mu}).
\end{align}    
This allows us to rewrite the second integral in the RHS of \eqref{full-s-ugly} as 
\begin{equation}
    \label{s-secondint}
    dk \frac{2\mu \phi^2}{B(\phi,\phi)(1+\phi)}\int_0^\infty dm  \frac{k^\phi m^{\phi-1}}{(k+m)^{2\phi}} \frac{e^{-\mu(k+m)}}{\sinh(\mu (k+m))} \,_2F_1(\frac{1-\phi}{2},1,\frac{3+\phi}{2},e^{-2(k+m)\mu}).
\end{equation}
From the identity
\[
\coth(x) = 1 + \frac{e^{-x}}{\sinh(x)},
\]
we isolate the zeroth-order term in the hypergeometric expansion from equation \eqref{s-secondint} to cancel the first integral in \eqref{full-s-ugly}. This yields the simplified expression:
\begin{multline}
\label{fulls-ugly-2}
s \hat{S}_k(s) = \mu \phi \, dk \bigg(1 - \frac{1}{B(\phi, \phi)} \int_0^\infty dm \,
\frac{k^\phi m^{\phi - 1}}{(k + m)^{2\phi}} \cdot \frac{e^{-\mu(k + m)}}{\sinh\left( \mu(k + m) \right)} \\
\times \left[ \frac{2\phi}{1 + \phi} \, {}_2F_1\left( \frac{1 - \phi}{2}, 1; \frac{3 + \phi}{2}; e^{-2\mu(k + m)} \right) - 1 \right] \bigg).
\end{multline}

The Laplace inversion $s \mapsto t$ of \eqref{fulls-ugly-2} remains analytically intractable in this form. To proceed, we simplify the hypergeometric term using a contiguous relation for ${}_2F_1$ \cite{abramowitz1965handbook}:
\begin{equation}
\label{contiguous-2f1}
(c - a - b) \, {}_2F_1(a, b; c; x)
+ a(1 - x) \, {}_2F_1(a + 1, b; c; x)
- (c - b) \, {}_2F_1(a, b - 1; c; x)
= 0.
\end{equation}

Using \eqref{contiguous-2f1} with the specific parameters
\[
a = \frac{1 - \phi}{2}, \quad b = 1, \quad c = \frac{1 + \phi}{2}, \quad x = e^{-2\mu(k + m)},
\]
we find the identity:
\begin{multline}
\label{contiguous-2f1-spec}
\frac{e^{-\mu(k + m)}}{\sinh\left( \mu(k + m) \right)} 
\left[ \frac{2\phi}{1 + \phi} \, {}_2F_1\left( \frac{1 - \phi}{2}, 1; \frac{3 + \phi}{2}; e^{-2\mu(k + m)} \right) - 1 \right] \\
= 2 \cdot \frac{\phi - 1}{\phi + 1} \cdot e^{-2\mu(k + m)} \cdot 
{}_2F_1\left( \frac{3 - \phi}{2}, 1; \frac{3 + \phi}{2}; e^{-2\mu(k + m)} \right).
\end{multline}
Finally, substituting the identity \eqref{contiguous-2f1-spec} into \eqref{fulls-ugly-2} and performing the change of variable $m = k y$, we obtain:
\begin{equation}
\label{fulls-ok-laplace}
\hat{S}_k(s) = \sqrt{\frac{2}{s}} \, \phi \, dk \left(
1 - \frac{2(\phi - 1)}{(\phi + 1) B(\phi, \phi)} 
\int_0^\infty dy \, \frac{y^{\phi - 1}}{(1 + y)^{2\phi}} 
\, e^{-2\mu k(1 + y)} \,
{}_2F_1\left( \frac{3 - \phi}{2}, 1; \frac{3 + \phi}{2}; e^{-2\mu k(1 + y)} \right)
\right).
\end{equation}

Expanding the hypergeometric function in \eqref{fulls-ok-laplace} into its power series allows us to perform the Laplace inversion term by term. This yields an exact expression for the inverse Laplace transform, valid in the asymptotic regime $\tau \gg dk = 1$:
\begin{keyboxedeq}[Distribution of the time between records for SATW]
    \begin{equation}
    \label{exact-record-time}
    \begin{aligned}
    S_k(\tau) = \phi \sqrt{\frac{2}{\pi \tau}} \bigg(
    &1 - \frac{2(\phi - 1)}{(\phi + 1) B(\phi, \phi)} 
    \sum_{n=0}^\infty \frac{\left( \frac{3 - \phi}{2} \right)_n}{\left( \frac{3 + \phi}{2} \right)_n} \\
    &\quad \times \int_0^\infty dy \, \frac{y^{\phi - 1}}{(1 + y)^{2\phi}} 
    \, \exp\left( -\frac{2 (n + 1)^2 k^2 (1 + y)^2}{\tau} \right)
    \bigg).
    \end{aligned}
    \end{equation}
\end{keyboxedeq}

This intricate result is well-suited for numerical evaluation; see Fig.~\ref{fig:records} for an illustration. The expression also invites several important remarks, which we discuss below.

\subsubsection{Connection with general results from scale invariance}

In \cite{regnierRecordAges}, it was argued that for a general scale-invariant process, the distribution $S_k(\tau)$ adopts the following universal form in the continuous-time regime $\tau \gg dk = 1$:
\begin{equation}
    S_k(\tau) \Sim{\tau \gg 1} \frac{1}{k} \, \psi\left( \frac{\tau}{k^{\dw}} \right),
\end{equation}
where $\dw$ denotes the walk dimension of the process. Based on dimensional analysis, the scaling function $\psi(x)$ is expected to exhibit two asymptotic regimes \cite{regnierRecordAges}:
\begin{equation}
    \label{psi-scaling-leo}
    \psi(x) \propto 
    \begin{cases}
        x^{-1/\dw}, & \quad x \ll 1, \\
        x^{-\theta}, & \quad x \gg 1,
    \end{cases}
\end{equation}
where $\theta$ is the persistence exponent. In the case of the self-attracting walk SATW$_\phi$ studied above, we have $\dw = 2$ and $\theta = \tfrac{\phi}{2}$, as given in \eqref{persistence-exp}.

From the exact result \eqref{exact-record-time}, we can directly identify the scaling function $\psi$ as
\begin{boxedeq}
\begin{equation}
    \label{psi-records}
    \begin{aligned}
        \psi\left(x = \frac{\tau}{k^2}\right) = \phi \sqrt{\frac{2}{\pi x}} \bigg(
            &1 - \frac{2(\phi - 1)}{(\phi + 1) B(\phi, \phi)} 
            \sum_{n=0}^\infty \frac{\left( \frac{3 - \phi}{2} \right)_n}{\left( \frac{3 + \phi}{2} \right)_n} \\
            &\quad \times \int_0^\infty dy \, \frac{y^{\phi - 1}}{(1 + y)^{2\phi}} 
            \, \exp\left( -\frac{2 (n + 1)^2 (1 + y)^2}{x} \right)
            \bigg).
    \end{aligned}
\end{equation}
\end{boxedeq}
The clear dependence of \eqref{psi-records} on the scaling variable $x = \tau / k^2$, together with the emergence of distinct asymptotic regimes for $x \ll 1$ and $x \gg 1$ shown in the following, provides a rare analytical handle on record statistics in non-Markovian processes. The fact that an exact analytical expression for the full scaling function $\psi(x)$ can be obtained in this non-Markovian setting—particularly for SIRWs—is quite remarkable, and underscores the effectiveness of the aged Ray--Knight framework. \par 
From \eqref{psi-records}, it is clear that in the small-$x$ regime, the scaling function simplifies to:
\begin{boxedeq}
\begin{equation}
    \label{psi-smallx}
        \psi(x) \Sim{x \ll 1} \phi \sqrt{\frac{2}{\pi x}}.
\end{equation}
\end{boxedeq}
Physically, this short-time regime $\tau \ll k^2$—confirmed numerically in Fig.~\ref{fig:records}—corresponds to events where the walker does not have time to return to its previous minimum before incrementing the right boundary of its span. In this regime, the behavior is fully governed by the Dirac delta contribution in $\hat{q}_+^{(2)}$, as given in \eqref{aged-qplus-satw}. Indeed, the result \eqref{psi-smallx} can be recovered directly by retaining only this delta term in the integral representation \eqref{s-int-rep} of $S_k(\tau)$.
\par 
The large-time regime $x \gg 1$ is more intricate. We focus here on the range $1 < \phi < 2$ to make computations explicit, and later extend the result to all $\phi > 0$ via analytic continuation, as supported numerically in Fig.~\ref{fig:records}.

For $1 < \phi < 2$, the following hypergeometric series converges to unity:
\begin{equation}
\label{series-Pochhammer}
\frac{2(\phi - 1)}{\phi + 1} \sum_{n=0}^\infty \frac{\left( \frac{3 - \phi}{2} \right)_n}{\left( \frac{3 + \phi}{2} \right)_n} = 1.
\end{equation}

Using \eqref{series-Pochhammer} and the integral representation of the Beta function $B(\phi, \phi)$, we obtain the following representation of $\psi(x)$ in the large-$x$ limit:
\begin{equation}
\label{largex-phiattract}
\psi(x) \Sim{x \gg 1} \phi \sqrt{\frac{2}{\pi x}} \cdot \frac{2(\phi - 1)}{(\phi + 1) B(\phi, \phi)}
\sum_{n=0}^\infty \frac{\left( \frac{3 - \phi}{2} \right)_n}{\left( \frac{3 + \phi}{2} \right)_n}
\int_0^\infty dy \, \frac{y^{\phi - 1}}{(1 + y)^{2\phi}} \left( 1 - e^{- \frac{2(n+1)^2 (1 + y)^2}{x}} \right).
\end{equation}

To extract the leading asymptotics, we convert the sum over $n$ into an integral using the substitution $n = \sqrt{x} u$ for large $x$. Using Stirling’s approximation for the ratio of Pochhammer symbols, we find:
\begin{equation}
\frac{2(\phi - 1)}{\phi + 1} \cdot \frac{\left( \frac{3 - \phi}{2} \right)_n}{\left( \frac{3 + \phi}{2} \right)_n}
\Sim{n = \sqrt{x} u,\, x \to \infty}
- x^{-{\phi}/{2}} \cdot \left( \frac{2 u^{-\phi} \Gamma\left( \frac{1 + \phi}{2} \right)}{\Gamma\left( \frac{1}{2} - \frac{\phi}{2} \right)} \right).
\end{equation}

Substituting into \eqref{largex-phiattract}, we obtain:
\begin{equation}
\label{largex-phiattract-1}
\psi(x) \Sim{x \gg 1}
- \phi \sqrt{\frac{2}{\pi x^\phi}} \cdot \frac{2}{B(\phi, \phi)} \cdot \frac{\Gamma\left( \frac{1 + \phi}{2} \right)}{\Gamma\left( \frac{1}{2} - \frac{\phi}{2} \right)}
\int_0^\infty u^{-\phi} du \int_0^\infty dy \, \frac{y^{\phi - 1}}{(1 + y)^{2\phi}} \left( 1 - e^{-2 u^2 (1 + y)^2} \right).
\end{equation}

Both integrals can be evaluated explicitly for $1 < \phi < 2$ using standard identities \cite{gradshteynTableIntegrals}:
\[
\int_0^\infty \left( 1 - e^{-2 u^2 (1 + y)^2} \right) u^{-\phi} du
= - 2^{\frac{\phi - 3}{2}} (1 + y)^{\phi - 1} \Gamma\left( \frac{1}{2} - \frac{\phi}{2} \right),
\]
\[
\int_0^\infty \frac{y^{\phi - 1} (1 + y)^{\phi - 1}}{(1 + y)^{2\phi}} \, dy
= \frac{1}{\phi}.
\]

Putting everything together, we obtain the explicit large-$x$ asymptotics:
\begin{boxedeq}
\begin{equation}
\label{psi-largex}
\psi(x) \Sim{x \gg 1} \left( \frac{2}{x} \right)^{\phi/2}
\cdot \frac{\Gamma\left( \frac{1 + \phi}{2} \right)}{\sqrt{\pi} \, B(\phi, \phi)}.
\end{equation}
\end{boxedeq}

This result, validated numerically in Fig.~\ref{fig:records} well beyond the range $1 < \phi < 2$, confirms the predicted scaling behavior $\psi(x) \propto x^{-\phi/2}$ \eqref{psi-scaling-leo} and yields an explicit expression for the prefactor. 
% A naive physical argument would imply that, as $x \gg 1$, the past visited territory is very small with respect to the future one visited at the time $t+\tau_k$ when reaching $k+1$, and as such can be neglected. This would imply that $S_k(\tau)$, in the limit $\tau \gg k^2$, can be approximated by the non-aged survival probability $S_1(\tau)$ of a SATW$_{\phi}$ starting at distance $1$ from its target. This quantity can be obtained either from \eqref{fpt-satw-final} or \cite{klingerJointStatistics}. However, the ratio between the two is not 1...
 
\section{Exploration Properties of SIRWs on a General Tree Graph}
\label{sec:sirw-tree}
\subsection{Introduction and motivations}

Thus far, we have focused on SIRWs in one dimension, where the linear structure of space enables detailed analytical control via Ray--Knight theory. In this section, we extend our analysis to a broader class of graphs—namely, general trees. These are connected graphs without loops, which retain enough structure to remain analytically tractable while introducing new geometric features absent in $\mathbb{Z}$, such as branching and anisotropic exploration.

From a theoretical perspective, tree graphs provide an important intermediary setting: more complex than $\mathbb{Z}$, yet still devoid of cycles, they allow for a natural extension of Ray--Knight representations. At the same time, the tree geometry makes the notion of spatial exploration richer and direction-dependent, raising new questions about how memory and reinforcement guide the walker along competing branches.

We define SIRWs on a general tree graph $\mathcal{T}$ as follows. At any vertex $x$ with neighbors $y_1, \dots, y_d$ (where $d$ is the degree of $x$), the transition probabilities are defined by
\begin{equation}
    \mathbb{P}(X_{t+1} = y_i \mid X_t = x) = \frac{w(L_t(\{x, y_i\}))}{\sum_{j=1}^d w(L_t(\{x, y_j\}))},
\end{equation}
where $L_t(\{a, b\})$ denotes the local time on the unoriented edge $\{a, b\}$ at time $t$, and $w$ is the weight function characterizing the universality class of the SIRW. In the following, we work directly in the continuum limit and identify the edge local time with the local time at site $a$, denoted simply by $L_t(a)$.

We show in this section that the Ray--Knight decomposition persists for SIRWs on trees. The absence of loops ensures that the independence properties used in the one-dimensional setting remain valid. As a concrete and symmetric example, exact formulae will be derived for the case of a star graph $\mathcal{S}_n$—a natural generalization of $\mathbb{Z}$ to $n$ rays joined at a common origin—and derive the corresponding Ray--Knight description of the local time field. This construction allows us to characterize how the walker allocates its exploration effort across the various branches of the star. However, we emphasize that the results derived below remain valid for arbitrary tree graphs, not just the symmetric star configuration.

\subsection{Ray--Knight theorems for SIRWs on a tree}
\begin{figure}
    \centering
    \includegraphics[width=.7\textwidth]{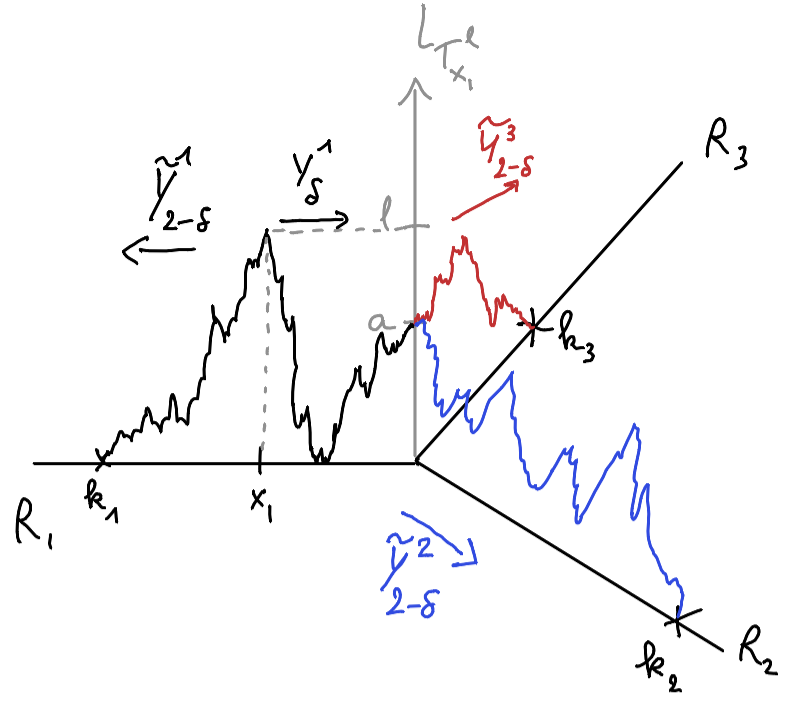}
    \caption{Ray--Knight representation of the local time process $L_{T_{x_1}^l}$ for a SIRW on a star graph $\mathcal{S}_3$ with $n = 3$ rays, given here for a SATW$_{\frac{\delta}{2},\dots,\frac{\delta}{2}}$ with $q=C=1$ (see Table \ref{tab:rayknight}). The SIRW is stopped when site $x_1$ on ray $R_1$ has been visited exactly $l$ times. The resulting distribution of the local time field $(L_{T_{x_1}^l}(x))_{x \in \mathcal{S}_3}$, both along the other branches and beyond $x_1$ on branch $R_1$, is described by the Ray--Knight decomposition given in \eqref{loctime-star}. The $k_i$ represent the maximum position of the SIRW along each ray $R_i$ at time $T_{x_1}^l$.}
    \label{fig:rayknight-star}
\end{figure}
Crucially, the Ray--Knight theorems extend naturally to the setting of a SIRW on a tree graph: the arguments given in Section~\ref{sec:proof} apply directly to tree graphs due to the absence of loops. To illustrate this concretely, we consider in what follows the case of $\mathcal{T} = \mathcal{S}_n = \cup_{i=1}^n R_i$, the $n$-branched star graph with all rays $R_i$ isomorphic to $\mathbb{R}_+$ (noting that $\mathbb{R}$ is isomorphic to $\mathcal{S}_2$). Due to the symmetry of the star, we can label the state of the SIRW as $X_t = x_i \in \mathbb{R}_+$ when it lies on ray $R_i$.

We stop the walk at time $T^l_{x_i}$, defined as the first time the SIRW has visited site $x_i$ exactly $l$ times. Then, the local time profile at this stopping time has an explicit distribution on the star, given in the continuous limit by:
\begin{keyboxedeq}[Ray--Knight theorem on a star graph]
\begin{equation}
    \label{loctime-star}
    L_{T_{x_i}^l}(y_k) \approx C \cdot
    \begin{cases}
        \tilde{Y}^{(i)}_{2-\delta}(y_i - x_i)^q, & k = i,\ y_i \geq x_i, \\
        Y^{(i)}_\delta(x_i - y_i)^q, & k = i,\ y_i < x_i, \\
        \tilde{Y}^{(k)}_{2-\delta}(y_k)^q, & k \neq i,
    \end{cases}
\end{equation}
\end{keyboxedeq}
where:
\begin{itemize}
    \item $Y^{(i)}_\delta(u)$ denotes a BESQ$_\delta$ process on $[0,x_i]$ started at $l$,
    \item $\tilde{Y}^{(i)}_{2-\delta}(u)$ is a BESQ$_{2-\delta}$ process started at $l$,
    \item For $k \neq i$, the processes $\tilde{Y}^{(k)}_{2-\delta}(u)$ are BESQ$_{2-\delta}$ processes started from the same value $a = Y_\delta(x_i)$ and absorbed at zero,
    \item The parameters $C, \delta$ and $q$ are the same as in the one-dimensional case and are listed in Table~\ref{tab:rayknight}.
\end{itemize}

The decomposition in \eqref{loctime-star}, illustrated FIG.~\ref{fig:rayknight-star}, is physically well motivated. Since the SIRW starts at the origin and is stopped at time $T_{x_i}^l$, the local time on the interval $[0, x_i]$ along ray $i$ cannot be absorbed at zero. In contrast, there is no such constraint on the remaining rays $R_k$ for $k \neq i$, nor on the half-line $[x_i, \infty)$ along ray $R_i$ beyond the stopping point. Consequently, the associated local time processes on these regions are absorbed at zero and evolve with a different BESQ dimension, just like in the $1D$ case. \footnote{Note that the presence of loops in the graph $\mathcal{T}$ obstructs any Markovian spatial description of the local time profile $L_{T_x^l}$, thereby preventing the existence of any Ray--Knight theorem.}

\subsection{Joint density of all maxima $\{k_j\}_{j=1,\dots,n}$ and first-passage time to $k_i$}

We extend the observables $q_{\pm}$, previously defined for one-dimensional SIRWs, to the case of SIRWs on the star graph $\mathcal{S}_n$.\footnote{These observables can be defined more generally for SIRWs on arbitrary tree graphs, but we restrict to the star for clarity and concreteness.}

For each $i = 1, \dots, n$, we define
\[
q_i(\{k_1, \dots, k_n\}, t)
\]
as the probability that the SIRW first hits site $k_i$ at time $t$, and that, for each $j = 1, \dots, n$, the maximum of the walk along branch $R_j$ is exactly $k_j$.

As in the one-dimensional case, this quantity admits a natural expression in terms of the support of the local time field $L_{T_{k_i}}(x)$, where $T_{k_i} = T_{k_i}^{0^+}$ is the FPT to site $k_i$. Specifically,
\begin{equation}
    \label{qi-star-loctime}
    q_i(\{k_1, \dots, k_n\}, t) = \mathbb{P}\left( \sum_{x \in \mathcal{S}_n} L_{T_{k_i}}(x) = t \ \text{and} \ \forall j = 1, \dots, n, \ \max_{\substack{s \leq t \\ X_s \in R_j}} X_s = k_j \right).
\end{equation}

As in the $1d$ case, we now partition \eqref{qi-star-loctime} over the local time $a$ at the origin, and use the explicit distribution of $(L_{T_{k_i}}(x))_{x \in \mathcal{S}_n}$ given by \eqref{loctime-star} to compute the Laplace transform of $q_i(\{k_1, \dots, k_n\}, t)$ with respect to $t$. We obtain
\begin{equation}
    \label{qi-laplace-star}
    \boxed{
    \hat{q}_i(\{k_1, \dots, k_n\}, s) = \int_0^\infty \mathbb{E}^0\!\left[e^{-s \int_0^{k_i} Y^{(i)}_\delta(u)^q du};\, Y^{(i)}_\delta(k_i) = a \right] \prod_{j \neq i} \mathbb{E}^a\!\left[e^{-s \int_0^{T_0^{(j)}} Y^{(j)}_{2-\delta}(u)^q du};\, T_0^{(j)} = k_j \right] da.
    }
\end{equation}
The time-integrated distribution is obtained by taking the limit $s \to 0$ in Eq.~\eqref{qi-satw-star}. As in the $1d$ case, the resulting expression is independent of the specific SIRW universality class, depending only on the class-specific BESQ dimensions $\alpha_1, \dots, \alpha_n$ as specified in Table~\ref{tab:rayknight}:
\begin{keyboxedeq}[Joint distribution of maxima along each ray when hitting maximum along ray i]
\begin{equation}
\label{qi-integrated-star}
\hat{q}_i(\{k_1, \dots, k_n\}, s = 0) = \frac{k_i}{B(\alpha_1, \dots, \alpha_n)} \cdot \frac{\prod_{j=1}^n k_j^{-1 - \alpha_j}}{\left( \frac{1}{k_1} + \dots + \frac{1}{k_n} \right)^{\alpha_1 + \dots + \alpha_n}}.
\end{equation}
\end{keyboxedeq}
In \eqref{qi-integrated-star}, $B(\alpha_1, \dots, \alpha_n)$ denotes the multivariate Beta function \cite{abramowitz1965handbook}. \par
We now investigate which additional observables can be derived from the fundamental quantity $q_i$, which can be obtained explicitly from \eqref{qi-laplace-star} after specifying the specific SIRW class. As in the $1d$ case, a wide range of space exploration observables—such as the splitting probability, the FPT to a site, the first range time, and the number of sites visited up to time $t$—can be expressed in terms of the 'building block' probability $q_i$. \footnote{We emphasize that the following expressions of space exploration observables in terms of the quantities $q_i$ do not rely on whether the RW is self-interacting or not. However, in the case of SIRWs, the observables $q_i$ are fully explicit, whereas for more general non-Markovian RWs, they may not be analytically accessible.
}

\subsection{Generalized splitting probability}

We define $Q^{i}_{k_1, \dots, k_n}$ as the splitting probability that the SIRW first reaches the site $k_i \in R_i$ before any of the other sites $k_j \in R_j$ for $j \neq i$. This generalizes the standard splitting probability to star graphs. As in the one-dimensional case, it admits a representation as a space- and time-integrated version of the fundamental observable $q_i$:
\begin{align}
    \label{splitting-from-qi-star}
    Q^{i}_{k_1, \dots, k_n} &= \int_0^\infty \int_0^{k_1} \dots \int_0^{k_{i-1}} \int_0^{k_{i+1}} \dots \int_0^{k_n} q_i(\{m_1, \dots, m_{i-1}, k_i, m_{i+1}, \dots, m_n\}, t) \, dt \prod_{j \neq i} dm_j \notag \\
    &= \int_0^{k_1} \dots \int_0^{k_{i-1}} \int_0^{k_{i+1}} \dots \int_0^{k_n} \hat{q}_i(\{m_1, \dots, m_{i-1}, k_i, m_{i+1}, \dots, m_n\}, s=0) \prod_{j \neq i} dm_j.
\end{align}

Using Eq.~\eqref{qi-integrated-star}, we obtain an exact expression that remains valid for all SIRWs upon specifying the BESQ dimensions $\alpha_1, \dots, \alpha_n$ Table \ref{tab:rayknight}:
\begin{keyboxedeq}[Splitting probability for a SIRW on a star]
\begin{equation}
    \label{splitting-exact-star}
        Q^{i}_{k_1, \dots, k_n} = \int_0^\infty \left[ -\frac{d}{du} \Gamma\left(\alpha_i, \frac{u}{k_i} \right) \right] \cdot \frac{\prod_{j \neq i} \Gamma\left(\alpha_j, \frac{u}{k_j} \right)}{\Gamma(\alpha_1) \dots \Gamma(\alpha_n)} \, du.
\end{equation}
\end{keyboxedeq}
In \eqref{splitting-exact-star}, we used the incomplete Gamma function 
\[\Gamma(a,z) = \int_z^\infty t^{a-1} e^{-t} dt. \]
While Eq.~\eqref{splitting-exact-star} does not generally admit a closed-form solution, it is well-suited for numerical evaluation due to the smooth and exponentially decaying integrand. \par 
Let us perform three consistency checks of Eq.~\eqref{splitting-exact-star}.

\medskip
\noindent
First, consider the case $n = 2$. In this setting, the integral in Eq.~\eqref{splitting-exact-star} can be computed exactly and reproduces the $1d$ result Eq.~\eqref{splitting-sirw}, namely:
\begin{equation}
    Q^{1}_{k_1,k_2} = I_{1 - \frac{k_1}{k_1 + k_2}}(\alpha_1, \alpha_2),
\end{equation}
where $I_{x}(a,b)$ denotes the regularized incomplete Beta function.

\medskip
\noindent
Second, for the simple RW with $\alpha_1 = \dots = \alpha_n = 1$, we can use the identity $\Gamma(1, z) = e^{-z}$ to compute the integral exactly. Substituting into Eq.~\eqref{splitting-exact-star} yields
\[
Q^{i}_{k_1, \dots, k_n} = \frac{\frac{1}{k_i}}{\sum_{j=1}^n \frac{1}{k_j}},
\]
which corresponds to a well-known result for the splitting probabilities of simple RWs on star graphs \cite{benichouExitOccupationTimes2009}.

\medskip
\noindent
Third, consider the symmetric case where $\alpha_1 = \dots = \alpha_n \equiv \alpha$ and $k_1 = \dots = k_n \equiv k$. In this case, Eq.~\eqref{splitting-exact-star} simplifies to
\[
Q^{i}_{k, \dots, k} = \int_0^\infty \frac{1}{n} \frac{\partial_u \left[-\Gamma(\alpha, u)^n\right]}{\Gamma(\alpha)^n} \, du = \frac{1}{n},
\]
as expected by symmetry. A similar argument confirms the normalization of the splitting probabilities in the general case:
\[
\sum_{i=1}^n Q^i_{k_1, \dots, k_n} = \int_0^\infty \left[ -\frac{d}{du} \left( \frac{\prod_{j=1}^n \Gamma\left(\alpha_j, \frac{u}{k_j} \right)}{\Gamma(\alpha_1) \dots \Gamma(\alpha_n)} \right) \right] du = 1.
\]

\subsubsection{A surprising consequence: splitting via independent variables}
Note that the splitting probability \eqref{splitting-exact-star} admits the following interpretation. Let $X_i \sim \text{Gamma}(\alpha_i)$ be $n$ independent Gamma random variables, and define
\[
T_i = \frac{X_i}{k_i}, \quad T = \min_i T_i.
\]
Then, the probability that $T_i = T$ is exactly $Q^i_{k_1, \dots, k_n}$. 
% Because exactly one of the $T_i$ attains the minimum, the sum over $i$ of these probabilities satisfies
% \[
% \sum_{i=1}^n \mathbb{P}(T_i = \min_j T_j) = \sum_{i=1}^n Q^i_{k_1, \dots, k_n} = 1.
% \]
At first glance, it may seem that the auxiliary variables $T_i = X_i / k_i$, with $X_i \sim \text{Gamma}(\alpha_i)$ independent, represent the actual hitting times $\tau_i$ for the SIRW to reach site $k_i$. However, this is not the case: the true hitting times $\tau_i$ are neither independent nor Gamma distributed--as the $1d$ case clearly showed \eqref{fpt-density-tsaw}, \eqref{fpt-satw-final}. Nevertheless, and quite strikingly, the splitting probabilities
\[
Q^i_{k_1, \dots, k_n} = \mathbb{P}(\tau_i = \min_{j=1, \dots, n} \tau_j)
\]
coincide exactly with the simpler quantity
\[
\mathbb{P}(T_i = \min_{j=1, \dots, n} T_j),
\]
where the $T_j$ are independent and analytically tractable.

This equivalence is highly nontrivial. It reveals that, although the internal dynamics of the SIRW are history-dependent and globally coupled, splitting events can be described as if governed by a system of independent timers. In essence, the complex competition between reinforced branches of the star graph reduces to a race between independent clocks — a feature that both simplifies analysis and sheds light on the hidden structure underlying SIRWs.

\subsection{First-passage time density}

Analogously to the one-dimensional case, the first-passage time (FPT) density to site $x_i$ is given by
\begin{equation}
    F_{x_i}(t) = \int_0^{\infty} q_i(\{m_1, \dots, m_{i-1}, x_i, m_{i+1}, \dots, m_n\}, t) \prod_{j \neq i} dm_j.
\end{equation}

\subsection{First-range time, eccentricity distribution and number of sites visited}

\subsubsection{First-range time and eccentricity distribution}

We define the \emph{first-range time} $T_{k_1, \dots, k_n}$ as the first time at which the SIRW has reached maximal displacement $k_i$ along each ray $R_i$. Its probability density is given by summing the observables $q_i$ over all branches:
\begin{equation}
    \label{firstrange-star}
    f_{k_1, \dots, k_n}(t) = \mathbb{P}(T_{k_1, \dots, k_n} = t) = \sum_{i=1}^n q_i(\{k_1, \dots, k_n\}, t).
\end{equation}

Integrating this density over time yields the \emph{eccentricity distribution}, defined as the probability that the SIRW attains the span $\{k_1, \dots, k_n\}$:
\begin{align}
    \label{span-prob-star-0}
    \sigma_{k_1, \dots, k_n} &\equiv \int_0^\infty f_{k_1, \dots, k_n}(t) \, dt \notag \\
    &= \mathbb{P}\left(\text{among all spans of total size } L = \sum_{i=1}^n k_i,\ \text{the walk visited } \{k_1, \dots, k_n\}\right).
\end{align}

This defines a probability distribution on the simplex
\[
\mathcal{K}_n(L) \equiv \left\{(k_1, \dots, k_n) \in \mathbb{R}_+ \ \big| \ \sum_{i=1}^n k_i = L \right\}.
\]
Using Eq.~\eqref{qi-satw-star}, and denoting $L = \sum_{i=1}^n k_i$, we obtain an explicit expression:
\begin{keyboxedeq}[Probability of a given span of length L for a SIRW on a star]
\begin{equation}
    \label{span-prob-star}
        \sigma_{k_1, \dots, k_n} = \frac{L}{B(\alpha_1, \dots, \alpha_n)} \cdot \frac{\prod_{j=1}^n k_j^{-1 - \alpha_j}}{\left( \frac{1}{k_1} + \dots + \frac{1}{k_n} \right)^{\alpha_1 + \dots + \alpha_n}}.
\end{equation}
\end{keyboxedeq}
To our knowledge, even in the simple RW case, the eccentricity distribution on a star has not been studied in the literature. In the $1d$ case, recall that this distribution \eqref{eccentricity-sirw-1d} was uniform for the simple RW: this is not the case on a star as shown by \eqref{span-prob-star}. More generally, for a star with $n\geq 3$ rays, regardless of the values of the $\alpha_i$, the least probable eccentricity is for the symmetric span $k_1 = \dots k_n = \frac{1}{n}$, and the density at this point is
\[\sigma_{\frac{1}{n}, \dots, \frac{1}{n}} = \frac{n^{\sum_{i=1}^n (1-\alpha_i)}}{B(\alpha_1, \dots, \alpha_n)}. \]

\subsubsection{Eccentricity distribution via independent variables}
By scale invariance, we restrict attention to the case where the span is fixed to $L = 1$. The normalization of the eccentricity distribution $\sigma_{k_1, \dots, k_n}$ then leads to the following nontrivial integral over the standard $(n-1)$-simplex:
\begin{equation}
    \int_{k_1 + \dots + k_n = 1, \, k_i \geq 0} \frac{\prod_{j=1}^n k_j^{-1 - \alpha_j}}{\left( \frac{1}{k_1} + \dots + \frac{1}{k_n} \right)^{\alpha_1 + \dots + \alpha_n}} \, dk_1 \dots dk_n = B(\alpha_1, \dots, \alpha_n),
\end{equation}
where $B(\alpha_1, \dots, \alpha_n)$ is the multivariate Beta function. This integral has appeared in the literature under the name of the \emph{Schlömilch integral}~\cite{schlom}.

Remarkably, this expression admits a probabilistic interpretation in terms of independent random variables. Let $Z_i \sim \Gamma(\alpha_i, 1)$ be independent Gamma-distributed random variables. Define
\begin{equation}
    \label{gammarep}
    k_i \equiv \frac{\frac{1}{Z_i}}{\frac{1}{Z_1} + \dots + \frac{1}{Z_n}}.
\end{equation}
Then, \cite{schlom} shows that the random vector $(k_1, \dots, k_n)$ follows the same law as the eccentricity of a SIRW with span normalized to $1$. That is,
\begin{equation}
    \label{schlomilch}
    \sigma_{k_1, \dots, k_n} = \mathbb{P}(k_1, \dots, k_n).
\end{equation}

This provides yet another instance where the statistics of a SIRW can be computed via a transformation of independent random variables. One may interpret Eq.~\eqref{gammarep} through an electrical network analogy: imagine each ray of the star as a resistive wire with random resistance $Z_i \sim \Gamma(\alpha_i, 1)$. All wires are held at potential $V = 1$ at infinity, while the central node (site $0$) is grounded at potential $V = 0$. In this setting, the quantity $k_i$ represents the fraction of the total current that flows through wire $i$.

\subsubsection{Number of sites visited $N(t)$}

In the continuum limit, the total number of sites visited at time $t$ is given by the total span $L = \sum_{i=1}^n k_i$. The probability that the walk has visited at least $N$ sites by time $t$ can be written as
\begin{equation}
    \label{nbsites-star-eq}
    \mathbb{P}(N(t) \geq N) = \int_{k_1 + \dots + k_n = N} \mathbb{P}(T_{k_1, \dots, k_n} \leq t) \, dk_1 \dots dk_n = \int_0^t \int_{k_1 + \dots + k_n = N} f_{k_1, \dots, k_n}(t') \, dt' \, dk_1 \dots dk_n.
\end{equation}
In what follows, we quantify the number of visited sites on the star graph for both the SATW${\alpha_1,\dots, \alpha_n}$ and PSRW$\gamma$ models. The TSAW case, by contrast, presents significantly greater technical challenges, owing to the intrinsic complexity of the explicit formulae involved.

\subsection{SATW$_{\alpha_1,\dots, \alpha_n}$, PSRW$_\gamma$}
The natural extension of the asymmetric SATW$_{\alpha,\beta}$ to the star graph $\mathcal{S}_n$ is the SATW$_{\alpha_1, \dots, \alpha_n}$, defined by a ray-dependent weight function
\[
w_i(\ell) = 
\begin{cases}
    \frac{1}{\alpha_i}, & \ell = 0, \\
    1, & \ell \geq 1.
\end{cases}
\]
We can compute explicitly the Laplace transform of the distribution $q_i(\{k_1, \dots, k_n\}, t)$ for this class of SIRWs, as well as for the PSRW$_\gamma$ class. Letting $\mu = \sqrt{2sC}$,\footnote{Recall that $C = 1$ for SATW$_{\alpha_1, \dots, \alpha_n}$, and $C = \frac{1}{2\gamma + 1}$ with $\alpha_1 = \dots = \alpha_n = \frac{1}{2}$ for PSRW$_\gamma$.} and using Eqs.~\eqref{qi-laplace-star}, \eqref{joint-area-fpt}, and \eqref{laplace-area-endpoint}, we obtain, following similar calculations to those in the one-dimensional case,
\begin{keyboxedeq}[Joint distribution of maxima along rays j and time to the maximum along ray i]
\begin{equation}
\label{qi-satw-star}
\hat{q}_i(\{k_1, \dots, k_n\}, s) = 
\frac{\mu^{n-1}}{B(\alpha_1, \dots, \alpha_n)} \,
\frac{ \sinh(k_i \mu) }{ \prod_{j=1}^n \sinh(k_j \mu)^{1 + \alpha_j} } \,
\left( \sum_{j=1}^n \coth(\mu k_j) \right)^{-\sum_{j=1}^n \alpha_j}.
\end{equation}
\end{keyboxedeq}
We check that \eqref{qi-satw-star} reduces to \eqref{qplus-laplace-satw} for $n=2$. Indeed, one has the hyperbolic identity 
\begin{equation}
    \frac{
    \sinh(a \mu) \left( \coth(a \mu) + \coth(b \mu) \right)^{-\alpha - \beta}
    }{
    \sinh^{\alpha + 1}(a \mu) \, \sinh^{\beta + 1}(b \mu)
    }
    = 
    \frac{
    \sinh^{\beta}(a \mu) \, \sinh^{\alpha - 1}(b \mu)
    }{
    \sinh^{\alpha + \beta}(\mu (a + b))
    }.
\end{equation}
Due to the complicated structure of Eq.~\eqref{qi-satw-star}, it is generally difficult to obtain explicit expressions for time-dependent quantities in real time, except in the form of moments of the number of visited sites. 

\subsubsection{Number of visited sites $N(t)$}
To obtain the moments of $N(t)$, we employ an identity previously used in the one-dimensional case, which remains valid for $l \geq 1$:
\begin{equation}
    \mathcal{L}_{t \to s} \langle N(t)^l \rangle = l \int_0^\infty \mathcal{L}_{t \to s} \mathbb{P}(N(t) \geq N) \cdot N^{l - 1} \, dN.
\end{equation}
According to \eqref{nbsites-star-eq}, this expression can be evaluated explicitly in terms of the Laplace-transformed probability $\hat{q}_1$ as
\begin{equation}
    \label{moments-noft-star-satw}
    \mathcal{L}_{t \to s} \langle N(t)^l \rangle 
    = \frac{l}{s} \sum_{i=1}^n \int_0^\infty \dots \int_0^\infty \hat{q}_i(\{k_1, \dots, k_n\}, s) \cdot (k_1 + \dots + k_n)^{l - 1} \, dk_1 \dots dk_n.
\end{equation}
This formulation serves as a starting point for computing the mean and fluctuations of the number of sites visited at time $t$. For clarity, we restrict to the symmetric case $\alpha_1 = \dots = \alpha_n = \alpha$; however, the calculations can be extended to the asymmetric case without conceptual difficulty.
\paragraph{Mean $\langle N(t) \rangle$}
\begin{figure}
    \centering
    \begin{subfigure}[t]{0.48\textwidth}
        \includegraphics[width=\textwidth]{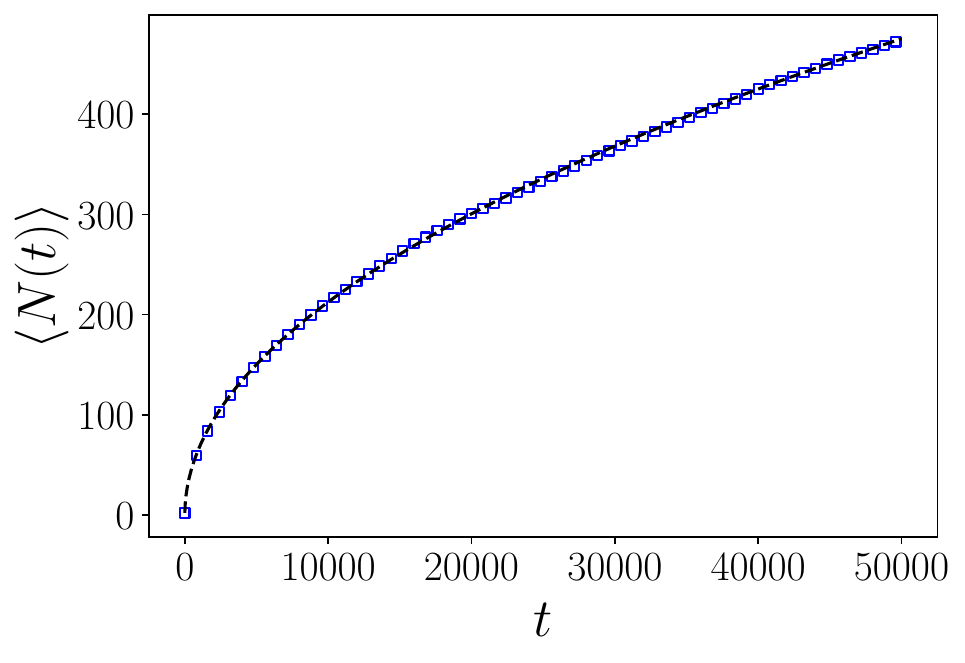}
        \caption{Simple RW ($\alpha = 1$). The black dashed line shows the exact theoretical result from Eq.~\eqref{meannoft-simple-star}. In this case, $\langle N(t) \rangle \sim 2.13 \sqrt{t}$.}
        \label{fig:mean-noft-star-simplerw}
    \end{subfigure}
    \hfill
    \begin{subfigure}[t]{0.48\textwidth}
        \includegraphics[width=\textwidth]{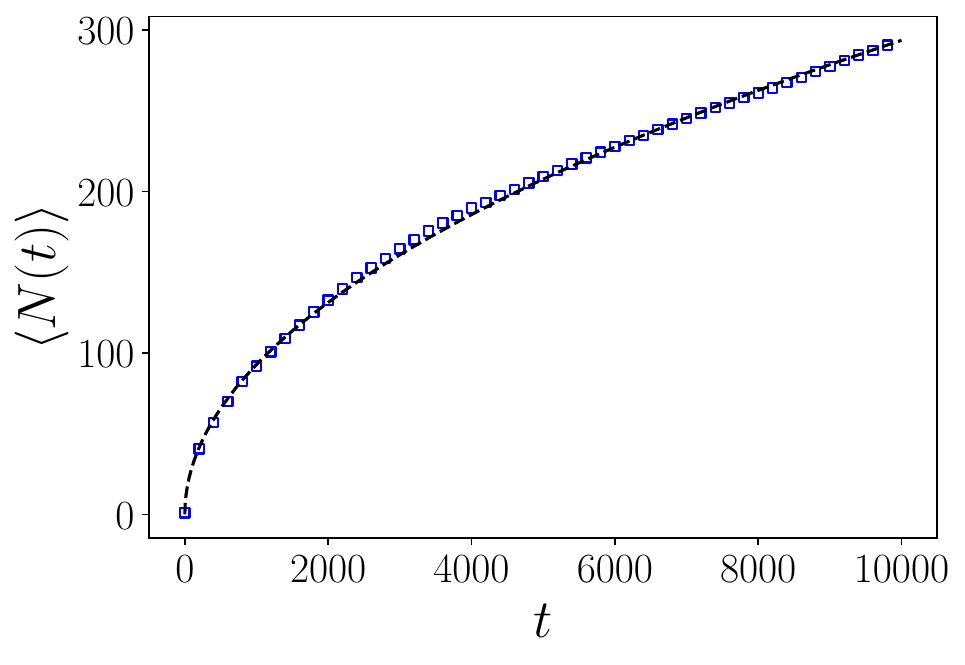}
        \caption{Self-repelling SATW with $\alpha = \tfrac{1}{2}$. The black dashed line shows the theoretical prediction from Eq.~\eqref{mean-noft-star}, with the integral $\mathcal{J}_n(\alpha)$ evaluated numerically. Here, $\langle N(t) \rangle \sim 2.93 \sqrt{t}$.}
        \label{fig:mean-noft-star-satw}
    \end{subfigure}
    \caption{Numerical verification of the mean number of visited sites $\langle N(t) \rangle$ for the SATW$_\alpha$ on a star graph with $n = 5$ rays.}
    \label{fig:mean-noft-star}
\end{figure}

For $l = 1$, Eq.~\eqref{moments-noft-star-satw} simplifies significantly. Performing the change of variables $\mu k_i \to x_i$ and evaluating the integral in Laplace space, we obtain:
\begin{equation}
\begin{aligned}
\mathcal{L}_{t \to s} \langle N(t) \rangle 
&= \frac{n}{s} \int_0^\infty \dots \int_0^\infty \hat{q}_1(\{k_1, \dots, k_n\}, s) \, dk_1 \dots dk_n \\
&= \frac{n}{\sqrt{2 s^3} \, B(\alpha, \dots, \alpha)} 
\int_0^\infty \dots \int_0^\infty 
\frac{ \sinh(x_1) }
{ \prod_{j=1}^n \sinh(x_j)^{1 + \alpha} } \,
\left( \sum_{j=1}^n \coth(x_j) \right)^{-n \alpha} 
\, dx_1 \dots dx_n.
\end{aligned}
\end{equation}

Laplace-inverting the above, and restoring the scaling constant $C = 1$ for SATW and $C = \frac{1}{2\gamma+1}$ for PSRW$_\gamma$, we obtain:
\begin{boxedeq}
\begin{equation}
\label{mean-noft-star}
\langle N(t) \rangle = \sqrt{\frac{2t}{\pi}} \cdot \frac{n}{B(\alpha, \dots, \alpha)} \cdot \mathcal{J}_n(\alpha)
\end{equation}
\end{boxedeq}

where the dimensionless integral $\mathcal{J}_n(\alpha)$ is defined as
\begin{align}
\label{jintegral}
\mathcal{J}_n(\alpha) 
&= \int_0^\infty \dots \int_0^\infty 
\frac{ \sinh(x_1) }
{ \prod_{j=1}^n \sinh(x_j)^{1 + \alpha} } 
\left( \sum_{j=1}^n \coth(x_j) \right)^{-n \alpha} 
\, dx_1 \dots dx_n \notag \\
&= \int_0^\infty \dots \int_0^\infty 
\frac{1}{\sqrt{v_1(2 + v_1)}} \cdot 
\frac{\prod_{i=1}^n (v_i(2 + v_i))^{\frac{\alpha - 1}{2}}}
{(n + \sum_{i=1}^n v_i)^{n\alpha}} \, dv_1 \dots dv_n,
\end{align}
after the change of variables $\coth(x_i) = 1 + v_i$, so that $\sinh(x_i) = \frac{1}{\sqrt{v_i(2 + v_i)}}$ and $dx_i = -\frac{dv_i}{v_i(2 + v_i)}$. See FIG.~\ref{fig:mean-noft-star-satw} for a numerical confirmation of \eqref{mean-noft-star}.

The integral $\mathcal{J}_n(\alpha)$ can be evaluated numerically for moderate values of $n$, but appears analytically intractable for general $\alpha$. The special case $\alpha = 1$, corresponding to the simple RW, is particularly noteworthy, as the number of visited sites on the star does not appear to have been computed in prior literature. An exception is the work of Csáki et al.~\cite{csakiLimitTheorems}, where certain observables for the RW on the star are analyzed. In particular, they determine the distribution of $N_i(t)$, the number of sites visited along a single ray $R_i$. This quantity is directly related to our object of interest, since $N(t) = \sum_{i=1}^n N_i(t)$. At the level of the mean, both observables are equivalent: $\langle N(t) \rangle = n \langle N_i(t) \rangle$, and their result is compatible with our expression \eqref{meannoft-simple-star}, although it is not stated in that form. \footnote{For higher moments, however, the two observables differ due to nontrivial correlations between the $N_i(t)$, and to our knowledge, our approach is the only one that captures these dependencies, see below. For completeness, we rederive the result in \cite{csakiLimitTheorems} using our Ray-Knight approach Eq. \eqref{dist-oneray-star}.}

We now compute $\mathcal{J}_n(1)$ explicitly:
\begin{equation}
\begin{aligned}
\mathcal{J}_n(1) &= \int_0^\infty \dots \int_0^\infty 
\frac{1}{\sqrt{v_1 2 + v_1}} \cdot 
\frac{1}{\left(n + \sum_{i=1}^n v_i \right)^n} 
\, dv_1 \dots dv_n \\
&= \int_0^\infty \frac{dv_1}{\sqrt{v_1 (2 + v_1)}} 
\int_0^\infty \frac{dS}{(n + v_1 + S)^n} 
\int_{\substack{v_2 + \dots + v_n = S \\ v_i \geq 0}} dv_2 \dots dv_n \\
&= \int_0^\infty \frac{dv_1}{\sqrt{v_1 (2 + v_1)}} 
\int_0^\infty \frac{S^{n-2}}{(n-2)! (n + v_1 + S)^n} \, dS \\
&= \int_0^\infty \frac{dv_1}{\sqrt{v_1 (2 + v_1)} (n - 1)! (n + v_1)} \\
&= \frac{2 \cosh^{-1}\left( \sqrt{\frac{n}{2}} \right)}
{\sqrt{(n - 2)n} \cdot (n - 1)!}.
\end{aligned}
\end{equation}

Thus, for the simple RW on the star graph, we obtain the following exact (and apparently new) formula:
\begin{boxedeq}
\begin{equation}
\label{meannoft-simple-star}
\langle N(t) \rangle = \sqrt{\frac{8nt}{(n - 2)\pi}} \cdot \cosh^{-1}\left( \sqrt{\frac{n}{2}} \right).
\end{equation}
\end{boxedeq}
See FIG.~\ref{fig:mean-noft-star-simplerw} for a numerical confirmation of \eqref{meannoft-simple-star}. 

This reproduces the well-known 1D result $\langle N(t) \rangle = \sqrt{\frac{8t}{\pi}}$ in the limit $n \to 2$. In the opposite limit of many rays, $n \gg 1$, we find:
\[
\langle N(t) \rangle \sim \sqrt{\frac{2t}{\pi}} \cdot \log(2n).
\]
\paragraph{Variance $\mathrm{Var}(N(t))$}
We start by computing the second moment from \eqref{moments-noft-star-satw}. Performing the same steps as above and using the symmetry between the indices $k_2, \dots, k_n$, we obtain:
\begin{align}
    \mathcal{L}_{t \to s} \langle N^2(t) \rangle
    &= \frac{2n}{s} \left(
        \int_0^\infty \dots \int_0^\infty
        \left(k_1 + (n - 1)k_2\right)
        \hat{q}_1(\{k_1, \dots, k_n\}, s)
        \, dk_1 \dots dk_n
    \right).
\end{align}

Writing $k_i = \sqrt{2s} x_i$ and inverting the Laplace transform, we find:
\begin{equation}
\label{secmom-noft-star}
\boxed{
\langle N^2(t) \rangle =
\frac{n t}{B(\alpha,\dots, \alpha)}
\left(
    \mathcal{B}_n^{(1)}(\alpha)
    + (n - 1)\mathcal{B}_n^{(2)}(\alpha)
\right)
}
\end{equation}
where we define the integrals:
\begin{align}
\mathcal{B}_n^{(1)}(\alpha) &=
\int_0^\infty \dots \int_0^\infty
\frac{x_1 \sinh(x_1)}
     {\prod_{j=1}^n \sinh(x_j)^{1 + \alpha}}
\left( \sum_{j=1}^n \coth(x_j) \right)^{-n\alpha}
\, dx_1 \dots dx_n \notag \\
&=
\int_0^\infty \dots \int_0^\infty
\frac{\sinh^{-1}\left( \frac{1}{\sqrt{v_1(2 + v_1)}} \right)}
     {\sqrt{v_1(2 + v_1)}}
\cdot
\frac{
    \prod_{i=1}^n (v_i(2 + v_i))^{\frac{\alpha - 1}{2}}
}{
    (n + \sum_{i=1}^n v_i)^{n\alpha}
}
\, dv_1 \dots dv_n,
\end{align}

\begin{align}
\mathcal{B}_n^{(2)}(\alpha) &=
\int_0^\infty \dots \int_0^\infty
\frac{x_2 \sinh(x_1)}
     {\prod_{j=1}^n \sinh(x_j)^{1 + \alpha}}
\left( \sum_{j=1}^n \coth(x_j) \right)^{-n\alpha}
\, dx_1 \dots dx_n \notag \\
&=
\int_0^\infty \dots \int_0^\infty
\frac{\sinh^{-1}\left( \frac{1}{\sqrt{v_2(2 + v_2)}} \right)}
     {\sqrt{v_1(2 + v_1)}}
\cdot
\frac{
    \prod_{i=1}^n (v_i(2 + v_i))^{\frac{\alpha - 1}{2}}
}{
    (n + \sum_{i=1}^n v_i)^{n\alpha}
}
\, dv_1 \dots dv_n.
\end{align}

We evaluate these integrals explicitly in the Appendix for the simple RW case $\alpha = 1$, which turns out to be an interesting analytical challenge. See \eqref{secmom-noft-star} for the exact expression. We focus on the simple RW for the end of this section. We give the large-$n$ asymptotics, accurate up to order $1/n$:
\begin{equation}
\label{largen-secmom-noft}
\langle N^2(t) \rangle = t \cdot \left( \log^2 n - \log n + \frac{5\pi^2}{12} - 1 - \log^2 2 + \log 2 \right) + o\left(\frac{1}{n}\right)
\end{equation}

This large-$n$ approximation is remarkably accurate: even for $n = 3$, the error is only about $3\%$ off.

Combining \eqref{largen-secmom-noft} with \eqref{meannoft-simple-star}, we obtain the large-$n$ asymptotics of the variance, again up to order $1/n$:
\begin{boxedeq}
\begin{equation}
\begin{aligned}
\mathrm{Var}(N(t)) =\; t \cdot \frac{1}{12\pi} \bigg[&
    (12\pi - 24)\log^2 n
    - (48 \log 2 + 12\pi)\log n
    + 5\pi^3 - 12\pi - 12\pi \log^2 2 \\
    &+ 12\pi \log 2 - 24 \log^2 2
    \bigg] + o\left(\frac{1}{n}\right)
\end{aligned}
\end{equation}
\end{boxedeq}

While higher moments appear challenging to compute due to rapidly growing expression sizes, numerical integration remains effective for moderate values of $n$, and generalizes to genuine SIRWs with arbitrary $\alpha \neq 1$.

%% file: UniversalExploration.tex
\part{Universal Exploration Properties of non-Markovian Random Walks}
\chapter{Introduction}
Random walks with memory offer a versatile framework for modeling non-Markovian effects in stochastic exploration. In this thesis, we have so far focused on specific classes of such processes, notably LARWs and SIRWs. Unlike classical memoryless walks, these models feature history-dependent mechanisms. We have shown that, depending on the memory structure, the walker can be biased either toward or away from previously visited regions. This results in a diverse range of non-Markovian behaviors and gives rise to a nontrivial exploration process, which we have characterized in detail for the classes considered. \par 
\vspace{2ex}
In this part of the thesis, we shift focus to the statistical properties and evolution of the explored interval $[x_{\rm min}, x_{\rm max}]$ for \emph{completely general} one-dimensional, scale-invariant, symmetric, non-Markovian RWs. Unlike the specific classes previously considered, our analysis does not rely on any particular rule or mechanism. We will identify features that are universal across models, and when certain behaviors depend on model-specific details, we will quantify them explicitly for representative classes.

We introduce the notion of \emph{flips}—events where the walker, after reaching one extremity of its visited domain, subsequently expands the opposite boundary. Central to our analysis is the \emph{flip probability} $\pi$, defined as the probability that the next extension of the domain occurs at $x_{\rm min}-1$, given that the last extension was at $x_{\rm max}$. This quantity, which coincides with the classical splitting probability $Q_+$ in the Markovian case, becomes a powerful probe of memory-induced dynamics in non-Markovian regimes. \par 
\vspace{2ex}
A first key result derived in this chapter is the emergence of a universal asymptotic scaling law for the flip probability, $\pi \propto 1/n$, where $n = x_{\rm max} - x_{\rm min}$ denotes the current size of the explored interval. Remarkably, this decay law emerges independently of the specific stochastic details of the underlying dynamics, signaling a universal aging behavior rooted in the process’s non-Markovian character. In stark contrast, the classical splitting probability $Q_+$—which differs from $\pi$ in that it does not condition the process on having visited the domain $[x_{\rm min}, x_{\rm max}]$—does \emph{not} exhibit such universality. Instead, as already discussed in earlier chapters, it decays as $1/n^{\phi}$, where the exponent $\phi = \dw \theta$ depends explicitly on the process, reflecting both its walk dimension $\dw$ and persistence exponent $\theta$~\cite{zoiaAsymptoticBehaviorSelfAffine2009}. \par 
\vspace{2ex}
The universal scaling of the flip probability $\pi$ thus illustrates a broader and largely unexplored principle: that conditioning a non-Markovian process on the size of its visited domain can give rise to universal statistical behavior, largely insensitive to the microscopic details of the dynamics. In spirit, this phenomenon is reminiscent of the central limit theorem, yet it emerges here in a setting dominated by strong temporal correlations, revealing an unexpected route to universality in non-Markovian systems. The influence of model-specific features persists only through a process-dependent prefactor $A$ in the expression for $\pi$. In the following, we compute this prefactor explicitly for a wide diversity of non-Markovian processes, thereby quantifying how microscopic dynamics modulate—but do not break—the underlying universal behavior. \par 
\vspace{2ex}
Beyond this, we explore how the flip statistics can be used to reconstruct full trajectories and access further observables characterizing the persistence and anisotropy in space exploration. These include:
\begin{itemize}
    \item The \emph{persistence of visitation}, defined as the probability of no flip occurring after a given number of sites visited;
    \item The eccentricity distribution $\sigma_z$, already discussed in earlier chapters, shown here to be very closely related to the concept of flips;
    \item The number of flips throughout the trajectory, quantifying the dynamics of exploration, parametrized not by time but by number of sites visited.
\end{itemize} \par 
Central to this analysis is the function $\Phi(z)$, which encodes the conditional probability of a flip for a walker whose visited domain has eccentricity $z$. This function captures the interplay between aging and memory, effectively linking local flipping events to the global geometry of the trajectory. Through $\Phi$, we gain insight into how geometric imbalance shapes future exploration. Crucially, computing $\Phi$ also yields the model-dependent prefactor $A$ appearing in the expression for $\pi$, thereby linking microscopic dynamics to the amplitude of the universal scaling law. \par 
\vspace{2ex}
To quantify model-dependent features, we categorize the most physically relevant scale-invariant stochastic processes into three distinct classes suited to our analysis.

Class (I) consists of random walks in which flips are independent of one another. As shown in the Appendix, this class includes not only Markov processes but also several key non-Markovian examples, such as Lévy Walks\footnote{We recall that we adopt the crossing prescription for visitation.}, the Random Acceleration Process (RAP), and the saturating SIRW that is the SATW. In these models, the absence of flip memory reflects a simplifying structural independence.

Outside of Class (I), flip events exhibit memory, reflecting the underlying non-Markovian character of the dynamics. Class (II) comprises the fractional Brownian Motion (fBM), the only scale-invariant Gaussian process with stationary increments. These increments satisfy
\begin{equation}
    \langle (X_{T+t} - X_T)^2 \rangle = t^{2H}.
\end{equation}
For this class, we derive explicit results perturbatively to first order in $\varepsilon = H - 1/2$. The walk dimension of this process is $\dw = \tfrac{1}{H}$, while its persistence exponent is $\theta = 1-H$ \cite{molchanMaximumFractionalBrownian1999}. Importantly, its splitting exponent $\phi = \dw \theta = \dw - 1$ can be expressed in terms of $\varepsilon$ as 
\begin{equation}
    \label{phi-fbm}
    \phi = 1-4\varepsilon+O(\varepsilon^2).
\end{equation}

Finally, Class (III) includes the broader family of non-saturating SIRWs. Examples include the TSAW and the PSRW. From our earlier results \eqref{persistence-exp} and \eqref{splitting-sirw}, we know that this class possesses the splitting exponent $\phi = \frac{1}{2}$.

\par 
\vspace{2ex}
Finally, the theoretical framework developed here proves robust beyond one-dimensional settings, extending naturally to higher-dimensional and more intricate geometries. This highlights the broad applicability of flip-based observables as fundamental tools for characterizing memory-driven exploration dynamics.

\chapter{Flip-based theory of space exploration}
\begin{figure}
    \centering
    \begin{subfigure}[t]{\textwidth}
        \centering
        \includegraphics{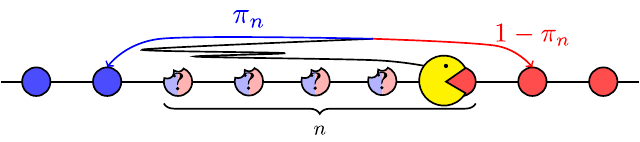}
        \subcaption{In this example, as denoted by the question marks, no information is available regarding the eccentricity of the visited domain, i.e., how many of the first $n$ units consumed were red. A flip occurs when the next unit eaten is blue, which happens with probability $\pi_n$.}
        \label{fig:illus-pin}
    \end{subfigure}
    	
    \begin{subfigure}[t]{\textwidth}
        \centering
        \includegraphics{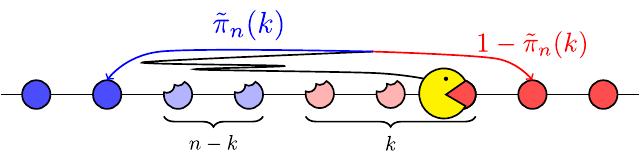}
        \subcaption{This scenario is similar to (1a), with the key difference being that we now know how many of the first $n$ units consumed were red ($k=3$ out of $n=5$ in this case). In this situation, a flip occurs with probability $\tilde{\pi}_n(k)$.}
        \label{fig:illus-pintilde}
    \end{subfigure}
    	
    \begin{subfigure}[t]{\textwidth}
        \centering
        \includegraphics{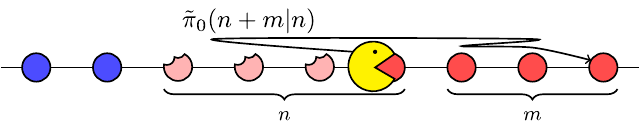}
        \subcaption{Here, the situation mirrors that of (1b), but we have explicit knowledge of the eccentricity of the visited territory. Specifically, all the first $n=4$ units consumed are red. The probability that the next $m=3$ units eaten will also be red is given by $\tilde{\pi}_0(n+m|n)$.}
        \label{fig:illus-pers-expl}
    \end{subfigure}
    	
    \begin{subfigure}[t]{\textwidth}
        \centering
        \includegraphics{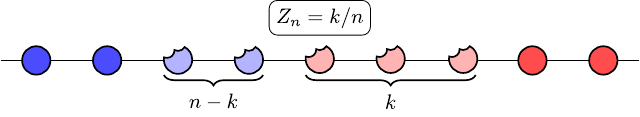}
        \subcaption{In this case, no information is provided about the color of the last unit consumed; only the visited territory is displayed. The example corresponds to a fraction $Z_n = 3/5$ of the consumed food being red.}
        \label{fig:illus-eccentricity}
    \end{subfigure}
    	
    \caption{Illustration of the various concepts related to space exploration considered in this part of the thesis. All depictions correspond to the moment when the walker (in yellow) consumes its $n^\text{th}$ unit of food: previously eaten food presents bite marks.}
    \label{fig:illus-univ-explo}
\end{figure}

In this chapter, we consider a general symmetric one-dimensional non-Markovian process $(X_t)$. To quantify exploration properties, we consider the integer lattice $\mathbb{Z}$ as a collection of \emph{sites}, which are deemed \emph{visited} by the RW when crossed. The crossing prescription ensures that the visited sites form an interval.  

A relevant and pedagogically valuable example of an exploration process is foraging, where sites are interpreted as units of food, represented by filled balls, eaten by the walker upon crossing (see Fig.~\ref{fig:illus-univ-explo}). Crucially, even if the RW is symmetric, the visited territory typically exhibits asymmetry relative to the starting point. To account for this asymmetry, which fully determines the geometry of visited territories of the same size, we color food on positive (resp. negative) sites red (resp. blue). 
\par 
\vspace{2ex}
We introduce several observables related to the notion of a \emph{flip}, which we define as a reversal in the direction of exploration of the process $X_t$—for instance, when an increment of the minimum of $X_t$ follows an increment of its maximum, or vice versa. In the context of the foraging analogy, this corresponds to the consumption of a blue ball immediately after a red one, or the reverse. The first observable we consider is $\pi_n$, the probability that a flip occurs as the visited domain expands from size $n$ to $n+1$. See Fig.~\ref{fig:illus-pin} for an illustration. \par 
\vspace{2ex}
We propose $\pi_n$ as a novel metric for quantifying random explorations. To refine this description, we also define two conditional probabilities: 
\begin{itemize}
    \item The \emph{eccentricity-informed} flip probability $\tilde{\pi}_n(k)$, which conditions on the asymmetry of the visited territory—quantified by the number $k$ of red balls eaten—and is illustrated in Fig.~\ref{fig:illus-pintilde};
    \item The \emph{flip-informed} flip probability $\overline{\pi}_n(f)$, which conditions on the total number of flips $f$ that have occurred throughout the walker's trajectory.
\end{itemize}
\par 
\vspace{2ex}
For Markovian RWs, computing $\pi_n$ and its refined counterparts $\tilde{\pi}_n(k)$ and $\overline{\pi}_n(f)$ is straightforward. Due to the memoryless nature of the dynamics, all these quantities reduce to the classical splitting probability $Q_+$--which is the probability that the process reaches $-(n-k)$ before $k+1$, independently of the past trajectory. In contrast, non-Markovian dynamics introduce memory effects that manifest both in the asymmetry of the explored region and in the number of flips, making these observables essential for a complete description of the process.

Despite the broad relevance of non-Markovian random walks across scientific domains, the influence of memory remains so pronounced that a general, quantitative understanding of space exploration in such systems has long remained out of reach. The framework developed here aims to bridge that gap.

\section{\texorpdfstring{Universal behavior of the flip probabilities $\pi$, $\tilde{\pi}$, $\overline{\pi}$}{Universal behavior of the flip probabilities pi, pi-tilde, pi-bar}}
\begin{figure}[h!]
    \centering
    % Left panel: 3 rows, 2 columns of small figures
    \begin{minipage}[t]{\textwidth} % Adjust width for smaller figures
        \centering
        \begin{subfigure}[t]{0.495\textwidth}
            \centering
            \includegraphics[width=\textwidth]{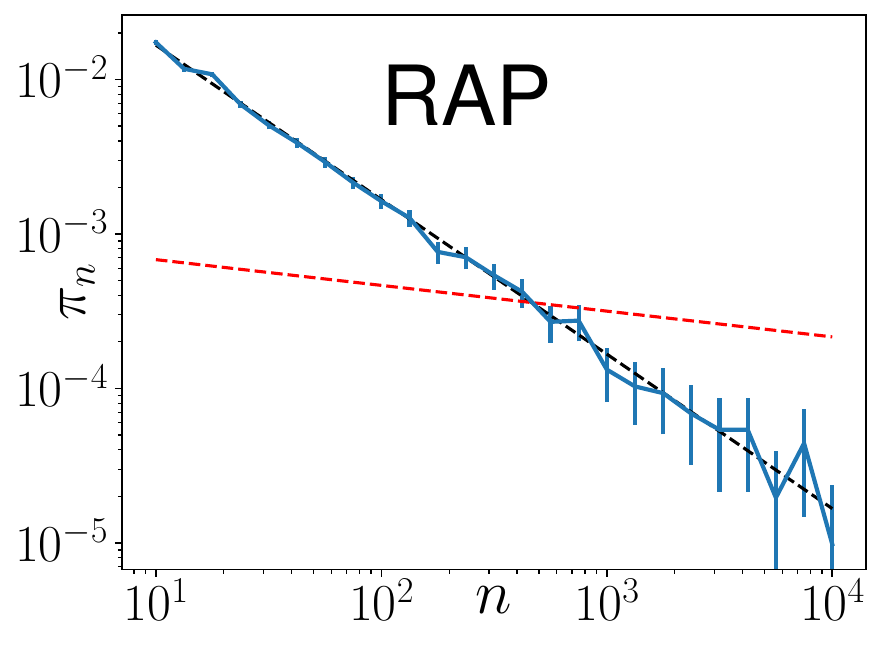}            \label{fig:small1}
        \end{subfigure}%
        \hfill
        \begin{subfigure}[t]{0.495\textwidth}
            \centering
            \includegraphics[width=\textwidth]{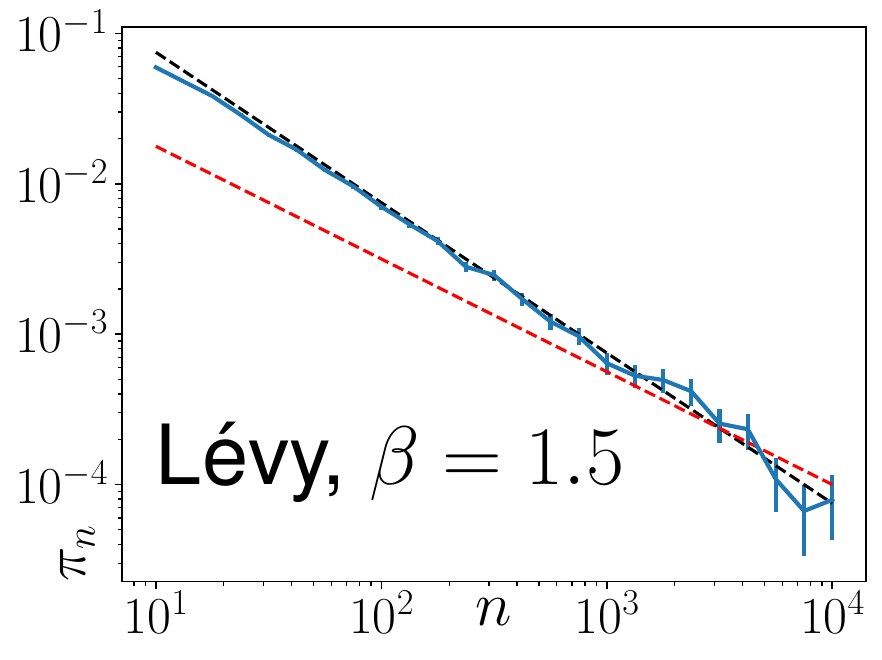}
            \label{fig:small2}
        \end{subfigure}\\[-1em]
        \begin{subfigure}[t]{0.495\textwidth}
            \centering
            \includegraphics[width=\textwidth]{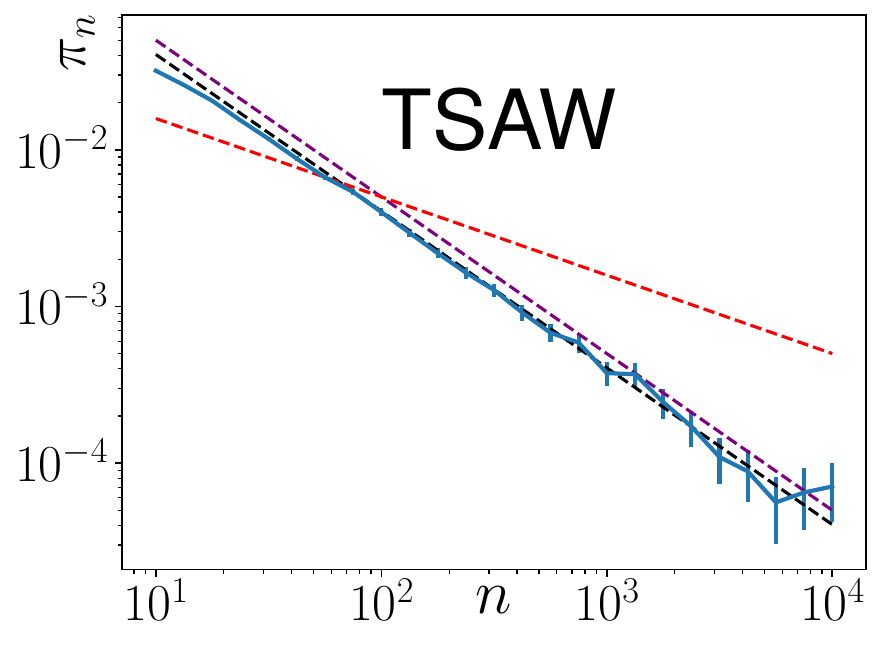}
            \label{fig:small3}
        \end{subfigure}%
        \hfill
        \begin{subfigure}[t]{0.495\textwidth}
            \centering
            \includegraphics[width=\textwidth]{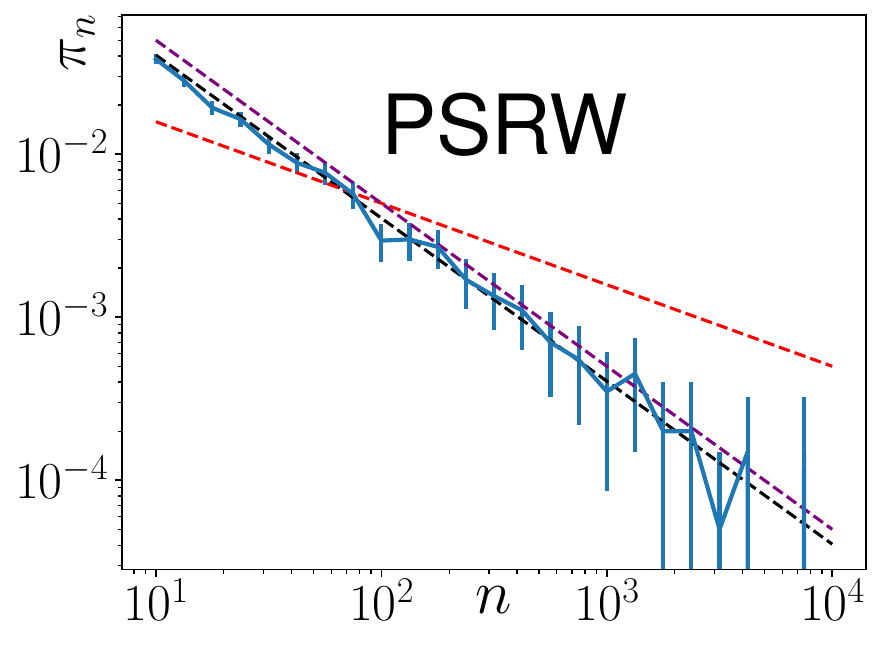}
            \label{fig:small4}
        \end{subfigure}\\[-1em]
        \begin{subfigure}[t]{0.495\textwidth}
            \centering
            \includegraphics[width=\textwidth]{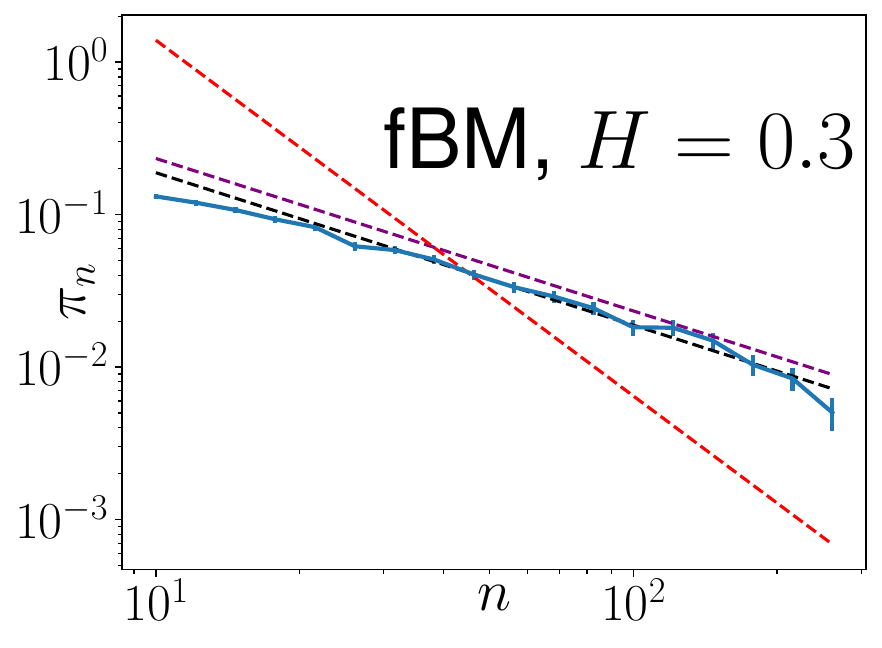}
            \label{fig:small5}
        \end{subfigure}%
        \hfill
        \begin{subfigure}[t]{0.495\textwidth}
            \centering
            \includegraphics[width=\textwidth]{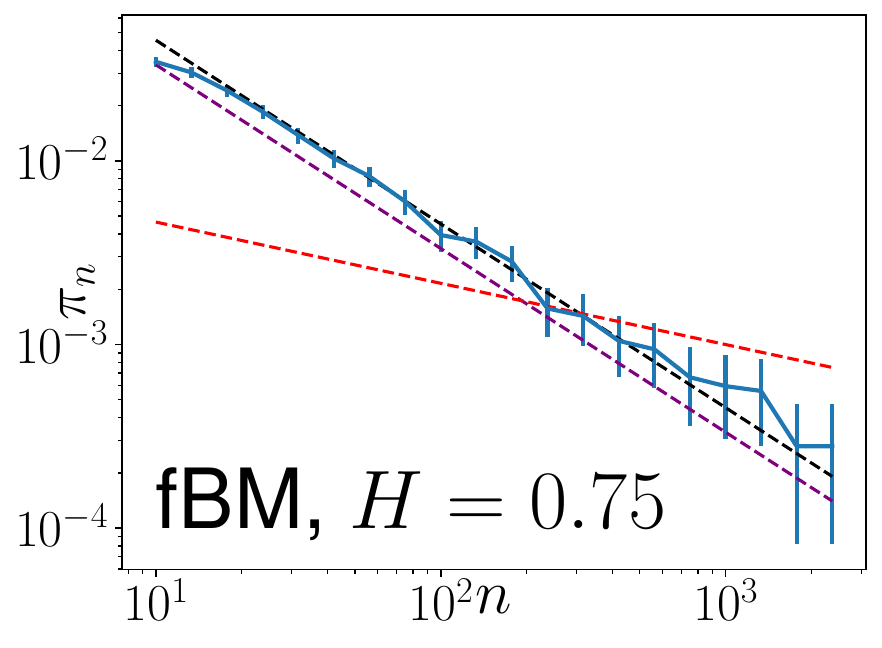}
            \label{fig:small6}
        \end{subfigure}
    \end{minipage}
    
    % Global caption
    \caption{Flip probability $\pi_n$ for paradigmatic examples of non-Markovian RWs. Rows correspond to distinct classes: the first to class (I), the second to class (II), and the third to class (III). The black line denotes the theoretical decay $A/n$ with $A$ determined by \eqref{pin}, while the red line represents the non-aged splitting probability, decaying as $1/n^\phi$. In turn, the purple dashed line shows the naive guess $\pi_n \sim \frac{\phi}{n} = \frac{\dw \theta}{n}$, which is wrong outside of class (I) as it assumes a constant function $\Phi(z)$.}
    \label{fig:splitting}
\end{figure}

We first determine, relying on a simple scaling analysis, the scaling with $n$ of the probability $\pi_n$ of a flip occurring after $n$ sites have been visited, that is, the probability of visiting the $(n+1)^{\text{th}}$ site on the other side of the $n^\text{th}$ visited site. After $n$ sites have been visited, an average of $\langle f_n \rangle \equiv \sum_{i=1}^n \pi_i$ flips have occurred. Since the RW is recurrent, we must have $\langle f_n \rangle \to \infty$: otherwise, the RW would eventually stop visiting sites on one side. So if $\pi_n\propto n^{-\alpha}$, then necessarily $\alpha \leq 1$. Let us assume that $\alpha<1$. Then an average of $\langle f_n \rangle \propto n^{1-\alpha}$ flips have occurred after $n$ sites have been visited, and, conversely, at the $k^\text{th}$ flip, $k^{1/(1-\alpha)}$ sites have been visited. \par 
Note that the only relevant length of the problem is the size of the visited domain. Hence, due to scale invariance, the number of visited sites between the $(k+1)^{\text{th}}$ and $k^{\text{th}}$ flips, of the order of $(k+1)^{1/(1-\alpha)} - k^{1/(1-\alpha)} \propto k^{\alpha/(1-\alpha)}$, must be proportional to the length $k^{1/(1-\alpha)}$ of the entire visited interval at the $k^\text{th}$ flip. This implies $\frac{\alpha}{1-\alpha}=\frac{1}{1-\alpha}$, which contradicts $\alpha<1$. This finally yields that $\alpha=1$, yielding the universal behavior. More precisely, our analysis shows that 
\begin{boxedeq}
    \begin{equation}
        \langle f_n \rangle = A \log n,
        \label{mean-flips}
    \end{equation}
\end{boxedeq}
meaning the number of sites visited at the $k^\text{th}$ flip is exponential with $k$. This is consistent with the proportionality between the number of newly and previously visited sites. \par The determination of the corresponding class-dependent prefactor $A$ is performed in the following sections. We obtain \footnote{In comparisons with numerical results for class (II), we use exponential approximations of the form $e^{B \varepsilon} + O(\varepsilon^2)$, which reproduce the correct first-order expansion $1 + B \varepsilon + O(\varepsilon^2)$, but have the added benefit of remaining strictly positive for all $\varepsilon$.
}:
\begin{keyboxedeq}[Flip probability $\pi_n$]
    \begin{align}
        \label{pin}
            \pi_n \sim \frac{A}{n}, \; A = \begin{cases} \phi = \dw \theta  &\text{(I)} \\ 1 -3.16198 \varepsilon + \mathcal{O}(\varepsilon^2) & \text{(II)} \\ 4/\pi^2 &\text{(III)} . \end{cases}
        \end{align}
\end{keyboxedeq}
This central, exact result, illustrated Fig.~\ref{fig:splitting}, calls for important comments.
\begin{itemize}
    \item The $1/n$ scaling of $\pi_n$ is both strikingly simple and completely universal, holding for all asymptotically self-similar RWs considered above. 
    \item As outlined in the introduction, the quantity $\pi_n$ generalizes the classical splitting probability $q_n$, which is defined for a RW starting at the boundary of a \emph{fixed} interval of length $n$. In contrast, $\pi_n$ is defined for a RW at the edge of its dynamically evolving visited territory. Strikingly, the process-dependent scaling $q_n \propto 1/n^\phi$ contrasts sharply with the universal scaling $\pi_n \propto 1/n$, highlighting the substantial and universal impact of memory on the exploration process. Note that $\phi$ can be either bigger or smaller than $1$, so that $\pi_n$ can decrease slower or faster than the non-aged splitting probability: see Fig.~\ref{fig:splitting}.
    \item The case of Lévy walks $J_t$, which are part of class (I), is striking and warrants a deeper examination. Suppose $J_t$ has just visited—or, in our convention, crossed—its $n^{\text{th}}$ site, $x$. Two outcomes are possible: either $J_t$ stops its jump within the interval $[x, x+1[$, which occurs with probability $s_n$, or it continues its jump, crossing sites beyond $x+1$. The probability of a flip is thus given by $\pi_n = s_n q_n + (1-s_n) \times 0$, as $J_t$ cannot flip if it continues its jump. As established in \cite{regnierUniversalExploration}, $s_n$ has the scaling behavior $s_n \propto n^{\phi-1}$. Substituting this into $\pi_n = s_n q_n$ recovers our general scaling relation $\pi_n \propto 1/n$. This inertial-like memory, characteristic to Lévy walks, is fundamentally distinct from the memory present in other non-Markovian RWs: it is striking that the $1/n$ scaling holds for any type of memory.
\end{itemize} \par \vspace{2ex}
Importantly, the more refined flip probabilities—conditioned either on the eccentricity of the visited domain, $\tilde{\pi}_n(k)$, or on the number of flips, $\overline{\pi}_n(f)$—exhibit the same universal $1/n$ scaling as their averaged counterpart $\pi_n$. This further extends the applicability of the scaling relation in Eq.~\eqref{pin}. More precisely, we have:
\begin{keyboxedeq}[Universal scaling of refined flip probabilities]
\begin{align}
\label{Phi-def}
    \tilde{\pi}_n(k) \sim \frac{\Phi\left(\tfrac{k}{n}\right)}{n}, \quad \overline{\pi}_n(f) \sim \frac{\Psi\left(\frac{f}{\log n} \right)}{n},
\end{align}
\end{keyboxedeq}
where $\Phi$ and $\Psi$ are class-dependent scaling functions that capture the influence of eccentricity and flip history, respectively.

Before showing the scaling relations \eqref{Phi-def}, let us focus on its implications for $\tilde{\pi}_n(k)$. Remarkably, this quantity coincides with the splitting probability for a walker initially located at one of the boundaries of the fixed interval $[-(n-k), k]$. In this setting, $\tilde{\pi}_n(k)$ corresponds exactly to the classical splitting probability $Q_+$ that the RW starting from $k$ visits $-(n-k)-1$ before $k+1$, with one crucial difference: the process is now conditioned on having already explored the entire interval $[-(n-k), k]$. This distinction proves essential. The fact that $\tilde{\pi}_n(k)$ exhibits the universal $1/n$ scaling underscores a key insight—conditioning the walker on having visited $[-(n-k),k]$ produces universal scaling laws. In contrast, when no such conditioning is imposed, universality is lost: the asymptotic behavior of $Q_+$ becomes model-dependent, with $Q_+ \propto n^{-\phi}$ and $\phi = \dw \theta$ \cite{zoiaAsymptoticBehaviorSelfAffine2009}. 

\subsection{\texorpdfstring{Scaling of $\tilde{\pi}_n(k)$ and the $\Phi$ function}{Scaling of pitilde and the Phi function}}
\subsubsection{Scaling arguments and expression of the $\Phi$ function}
\begin{figure}[htbp]
    \centering
    \begin{subfigure}[t]{.485\textwidth}
        \centering
        \includegraphics[width=\textwidth]{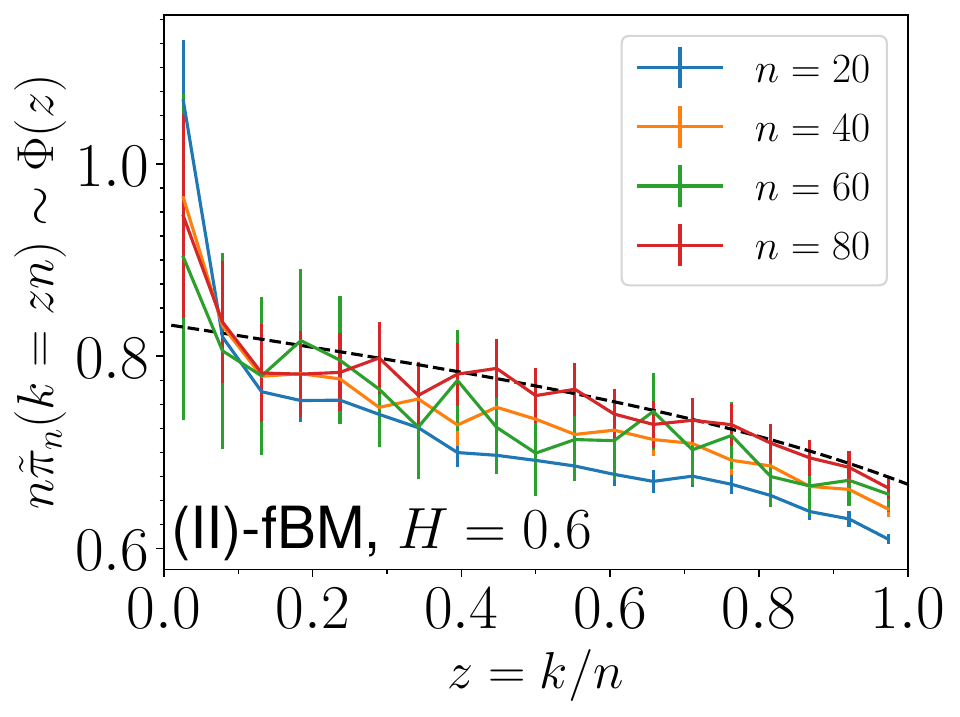}
        \subcaption{Class (II): fBM with Hurst index $H=0.6$ (superdiffusive). The $\Phi$ function decreases, reflecting fewer flips after visiting many sites on the same side—consistent with positive correlations.}
        \label{fig:phi-fbm06}
    \end{subfigure}
    \hfill
    \begin{subfigure}[t]{.485\textwidth}
        \centering
        \includegraphics[width=\textwidth]{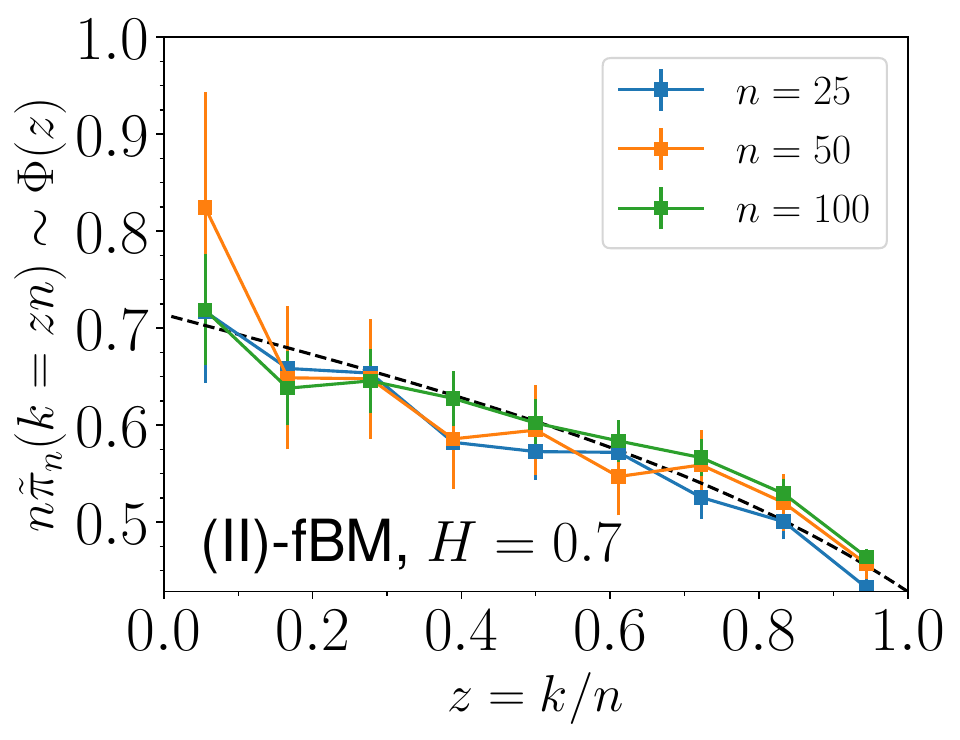}
        \subcaption{Class (II): fBM with Hurst index $H=0.7$ (superdiffusive). The $\Phi$ function is also decreasing.}
        \label{fig:phi-fbm07}
    \end{subfigure}
    \begin{subfigure}[t]{.485\textwidth}
        \centering
        \includegraphics[width=\textwidth]{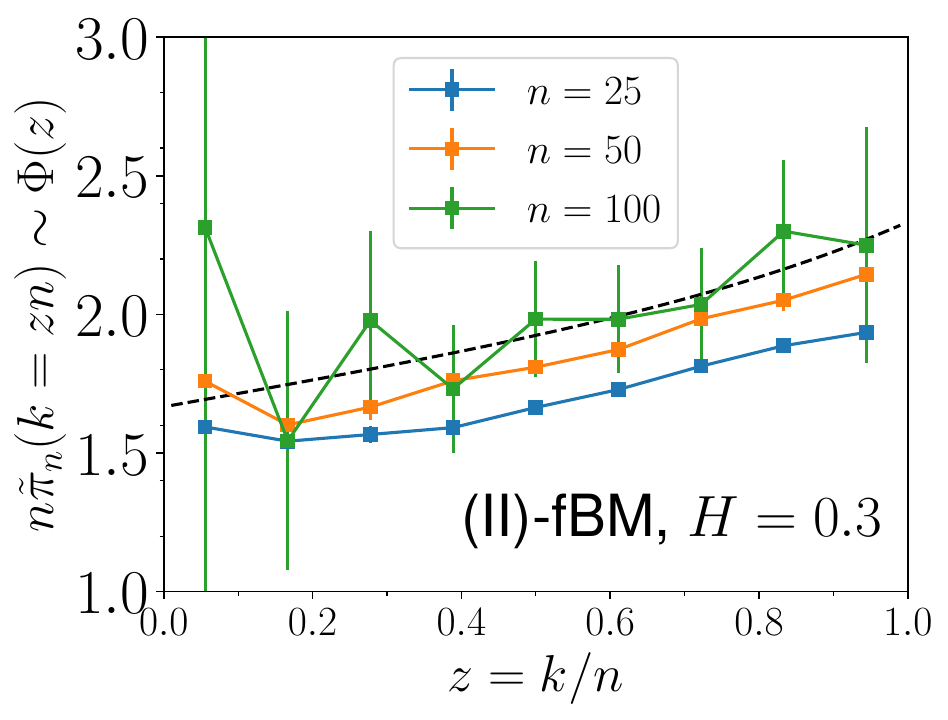}
        \subcaption{Class (II): fBM with Hurst index $H=0.3$ (subdiffusive). The $\Phi$ function increases, consistent with stronger flip tendencies due to negative correlations.}
        \label{fig:phi-fbm03}
    \end{subfigure}
    \hfill
    \begin{subfigure}[t]{.485\textwidth}
        \centering
        \includegraphics[width=\textwidth]{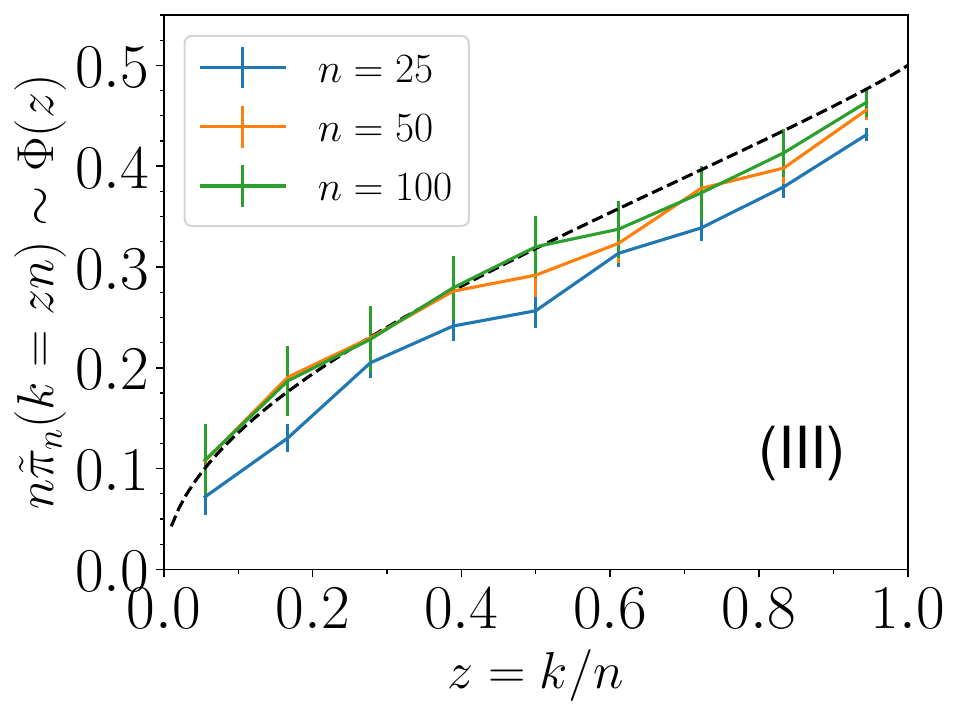}
        \subcaption{Class (III): example of the TSAW. We remind that all Class (III) processes have the same $\Phi$ function. As showed, $\Phi(z)$ increases with $z$, which is surprising given that the TSAW is superdiffusive—opposite to the behavior observed in superdiffusive fBM. This reveals subtle memory effects in space exploration that go beyond simple sub- or super- diffusive scaling.}
        \label{fig:phi-tsaw}
    \end{subfigure}
    \caption{Numerical convergence of $n \tilde{\pi}_n(k = zn)$ to the function $\Phi(z)$ for class (II) and class (III) processes. Dashed black lines indicate the theoretical expression from Eq.~\eqref{phi-expr}.}
    \label{fig:phi-func}
\end{figure}

Recall that $\pi_n$ denotes the probability of a flip occurring immediately after the $n^\text{th}$ site $x_n$ is visited. Two cases must be distinguished: either $x_n$ is the current maximum of the trajectory, i.e., $x_n = k$, or it is the current minimum, $x_n = -(n-k)$. As introduced in earlier sections, we define 
\[q_+(k,n-k) \equiv \mathbb{P}(x_n = k = \max X_t).\] By symmetry of the process $X_t$, we then have \[\mathbb{P}(x_n = -(n-k) = \min X_t) = q_+(n-k,k).\] This leads to the decomposition:
\begin{equation}
\label{partition-pi}
    \pi_n = \sum_{k=0}^n \left[ \tilde{\pi}_n(k) q_+(k,n-k) + \tilde{\pi}_n(n-k) q_+(n-k,k) \right] = 2\sum_{k=0}^n \tilde{\pi}_n(k) q_+(k,n-k),
\end{equation}
where the second equality follows from the symmetry of $X_t$ and a relabeling of the summation index.

Assuming scale invariance and applying dimensional analysis, the general scaling forms hold: 
\[\tilde{\pi}_n(k) \sim n^{-\gamma} \Phi(k/n), \quad q_+(k,n-k) \sim n^{-\delta} q_+(z = k/n),\]
where $\Phi$ and $q_+$ are process-specific functions.

Due to the symmetry of the process, conditioned on being at its boundary, the probability $\int_0^1 q^+(z) \, dz$ that the walker is at its maximum is equal to that of being at the minimum. This yields
\[
\int_0^1 q^+(z) \, dz = \frac{1}{2},
\]
from which we deduce that $\delta = 1$. Substituting the scaling forms into Eq.~\eqref{partition-pi} and using $\pi_n \propto 1/n$, we find:
\begin{align}
\label{avg-pi-continuous}
    \frac{A}{n} &\sim \pi_n = 2\sum_{k=0}^n \tilde{\pi}_n(k) q_+(k,n-k) \sim \frac{2}{n^{\gamma + 1}} \sum_{k=0}^n \Phi(k/n) q_+(k/n) \notag \\
    &\sim \frac{2}{n^{\gamma}} \int_0^1 \Phi(z) q_+(z) \, dz.
\end{align}
Matching the scaling on both sides, we conclude that $\gamma = 1$, thus showing the first scaling relation of \eqref{Phi-def}. See Fig.~\ref{fig:illus-pintilde} for a numerical confirmation. In the next chapter, we will show the following exact expressions for the $\Phi$ function, illustrated Fig.~\ref{fig:phi-func}: 
\begin{keyboxedeq}[Explicit expression of the $\Phi$ function]
    \begin{equation}
        \label{phi-expr}
        \Phi(z) =
        \left\{
        \begin{array}{ll}
        \phi = \dw \theta & \text{(I)} \\[1.2em]
        
        \displaystyle
        1 + \varepsilon \left( 
        \frac{ \psi\left( \tfrac{z+3}{2} \right) + \psi\left( \tfrac{3-z}{2}\right) + 2\gamma - 4 + 2 \log 4 }{z}
        + 2z \big[ 
        -\psi\left( \tfrac{z+3}{2} \right) + \psi(z + 2) \right. \\[0.4em]
        \left. \quad -\psi\left( \tfrac{3-z}{2} \right) + \psi(2 - z) - \log 4 \big] - 2 \right)
        + O(\varepsilon^2)
        & \text{(II)} \\[1.2em]
        
        \displaystyle
        \frac{ \sqrt{z(1 - z)} + (2z - 1)\, \arcsin\left( \sqrt{z} \right) }{\pi z}
        & \text{(III)}
        \end{array}
        \right.
        \end{equation}
\end{keyboxedeq}
In \eqref{phi-expr}, $\psi = \frac{\Gamma'}{\Gamma}$ is the Digamma function. \par
\vspace{2ex}
Finally, we demonstrate that the value of the function $\Phi$ at $z = 1$ is \emph{universal}, taking the value $\phi = \dw\theta$ independently of the specific process considered:
\begin{keyboxedeq}[Universal value of $\Phi$ at $z=1$]
    \begin{equation}
        \label{phi1}
        \Phi(1) = \phi = \dw\theta.
    \end{equation}
\end{keyboxedeq}

To establish Eq.~\eqref{phi1}, it suffices to show that the probability of a flip occurring at the $(n+1)^{\mathrm{st}}$ site visited—under the condition that no flip has occurred previously—scales as
\begin{equation}
    \label{pintildek0}
    \tilde{\pi}_n(n) \sim \frac{\phi}{n}.
\end{equation}

This asymptotic behavior can be deduced from the classical splitting probability $Q_+(n)$: the probability that the walker, starting at the origin, reaches site $n$ before site $-1$. For the walker to reach $n+1$ before $-1$, it must first reach $n$ before $-1$, and then proceed from $n$ to $n+1$ without flipping, conditioned on the absence of flips throughout its entire trajectory. This yields the recursive relation:
\begin{equation}
    \label{pintildek0-0}
    Q_+(n+1) = Q_+(n)\, (1-\tilde{\pi}_n(k)).
\end{equation}

Using the known asymptotic form $Q_+(n) \sim B n^{-\phi}$ \cite{zoiaAsymptoticBehaviorSelfAffine2009} for large $n$, we immediately deduce from Eq.~\eqref{pintildek0-0} that
\[
\tilde{\pi}_n(k) = 1-\frac{Q_+(n+1)}{Q_+(n)} \sim \frac{\phi}{n},
\]
thus recovering the claimed result \eqref{pintildek0} and completing the proof of the universality relation \eqref{phi1}. We check easily that the explicit forms \eqref{Phi-def} satisfy \eqref{phi1}. \par \vspace{2ex}
It is important to emphasize that the proof of Eq.~\eqref{phi1} fundamentally depends on the existence of a unique sequence of visited sites corresponding to the scenario in which no flip occurs during the $n$ visits, or, equivalently, that the RW has eccentricity $z=1$. In contrast, for values $z \neq 1$, there are infinitely many distinct visitation sequences that can result in an asymmetry $z$. This multiplicity of contributing paths is precisely why only the value $\Phi(z = 1)$ is universal, while the full function $\Phi(z)$ depends a priori on the detailed properties of the underlying process.

\subsubsection{The prefactor $A$}
We stress that Eq.~\eqref{avg-pi-continuous} also yields an exact expression for the prefactor $A$ in the relation $\pi_n = A/n$, namely:
\begin{keyboxedeq}[Prefactor $A$ of the flip probability]
\begin{equation}
    \label{expr-A-phi}
    A = 2 \int_0^1 \Phi(z) q_+(z) \, dz.
\end{equation}
\end{keyboxedeq}
Along with \eqref{phi-expr}, the identity \eqref{expr-A-phi} is instrumental in the explicit computation of the prefactor $A$ given in \eqref{pin}.

\subsection{Scaling of $\overline{\pi}_n(f)$}
\begin{figure}[t]
    \centering
    \begin{subfigure}[t]{.48\textwidth}
        \centering
        \includegraphics[width=\textwidth]{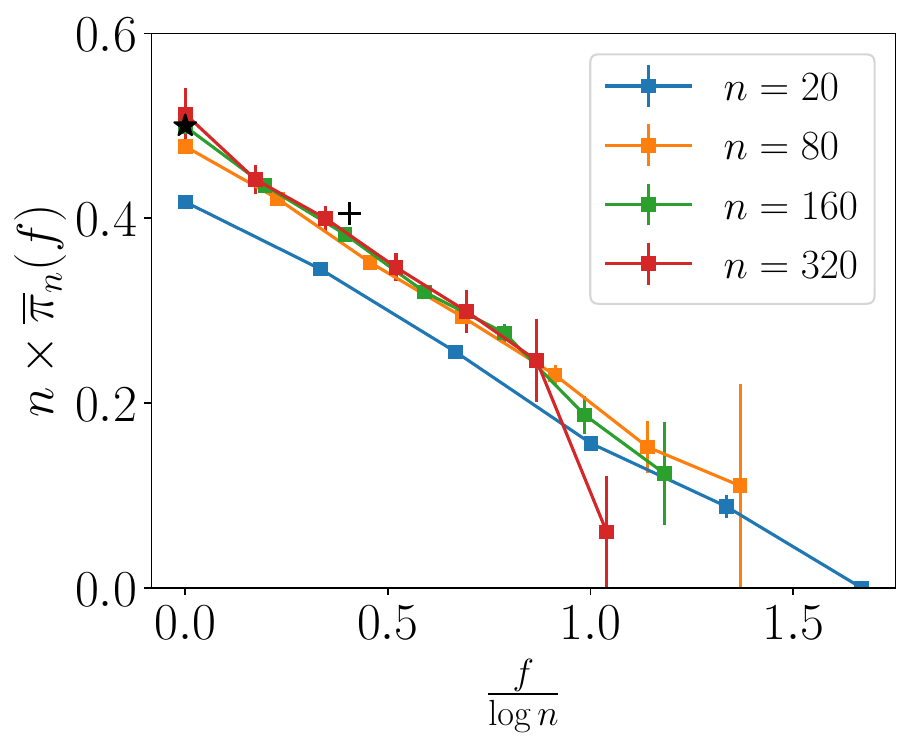}
        \subcaption{Numerical convergence of $n \overline{\pi}_n(f)$ to $\Psi\left(\tfrac{f}{\log n} \right)$ for the TSAW. The black star marks the expected value $\Psi(0) = \phi$, and the black cross indicates $\Psi(A) = A$. Convergence with $n$ is slow due to the logarithmic growth of the number of flips, $f_n \propto \log n$.}
        \label{fig:psi-tsaw}
    \end{subfigure}
    \hfill
    \begin{subfigure}[t]{.48\textwidth}
        \centering
        \includegraphics[width=\textwidth]{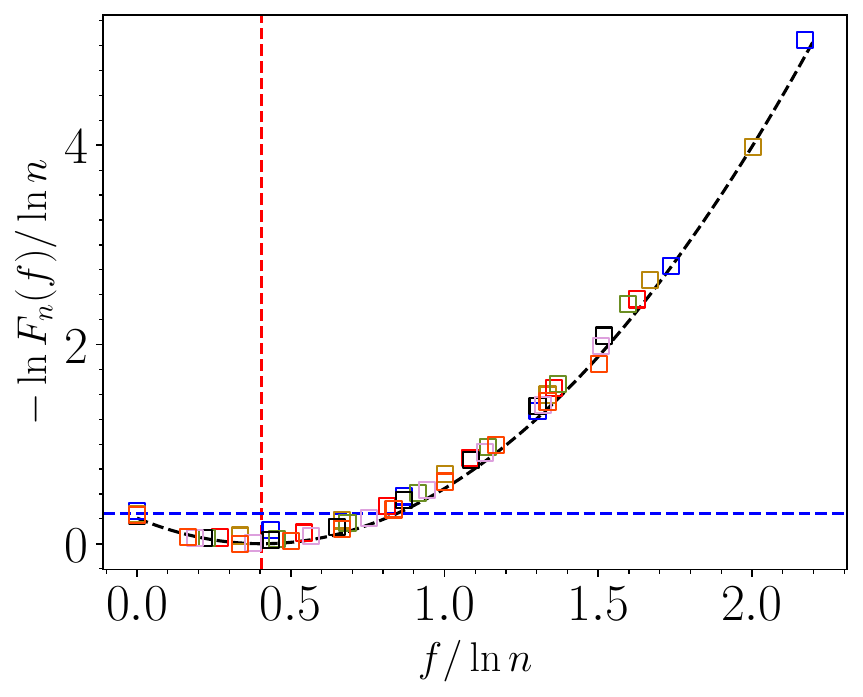}
        \subcaption{Numerical confirmation of the large deviation function \eqref{ldp-flips} for the TSAW. The black dashed curve shows the Gaussian result \eqref{gaussian-flip} and has mean $A = \frac{4}{\pi^2}$ from Eq.~\eqref{pin}. This Gaussian distribution is valid in the regime of typical fluctuations. Its standard deviation $\nu = \sqrt{\frac{A}{1 - 2\Psi'(A)}}$ is computed using the numerically estimated value of $\Psi'(A)$ from panel~\ref{fig:psi-tsaw}. The red dashed line marks the mean value $A$, while the blue dashed line indicates $I(0) = \phi = \frac{1}{2}$.}
        \label{fig:ldp-tsaw}
    \end{subfigure}
    \caption{Numerical validation of the scaling behavior of the flip-conditioned flip probability \eqref{Phi-def}, and the large deviation form of the number of flips \eqref{ldp-flips}, for a class (III) process: the TSAW.}
\end{figure}

From equation~\eqref{mean-flips}, we observe that the number of flips after $n$ visits scales as $f_n \propto \log n$. Combined with scale invariance, this suggests that the scaled quantity $\frac{f_n}{\log n}$ becomes asymptotically independent of $n$. Therefore, in the limit of large $n$, we expect
\begin{equation}
    \tilde{\pi}_n(f) \sim_{n \to \infty} \frac{\Psi\left(\frac{f}{\log n} \right)}{n^\varepsilon},
\end{equation}
for some exponent $\varepsilon > 0$. \par In turn, as a consequence of normalization, the probability that exactly $f$ flips occurred upon the $n$th visit should scale as
\begin{equation}
    \mathbb{P}(f_n = f) \sim \frac{p\left(\frac{f}{\log n} \right)}{\log n}.
\end{equation}

Now, expressing the flip probability $\pi_n$ as
\begin{equation}
    \frac{A}{n} \sim \pi_n = \sum_{f=0}^n \mathbb{P}(f_n = f) \cdot \overline{\pi}_n(f) \sim \frac{1}{n^\varepsilon} \int_0^{\infty} p(u) \Psi(u)\, du,
\end{equation}
we deduce that $\varepsilon = 1$, which leads directly to the scaling relation~\eqref{Phi-def} for $\overline{\pi}_n(f)$. We also find that
\[
A = \int_0^{\infty} p(u) \Psi(u)\, du,
\]
although this expression is of limited practical use compared to~\eqref{expr-A-phi}, since $\Psi$ is generally more difficult to compute than $\Phi$. \par 
Analogously to Eq.~\eqref{phi1} and for the same physical reasons, the function $\Psi$ also satisfies a universal identity at the origin:
\begin{keyboxedeq}[Universal value of $\Psi(0)$]
    \begin{equation}
        \Psi(0) = \phi = \dw \theta.
    \end{equation}
\end{keyboxedeq}

\subsection{Beyond One-Dimensional Non-Markovian Processes}
\begin{figure}[htbp]
    \centering
    \begin{subfigure}[t]{.32\textwidth}
        \centering
        \includegraphics[width=\textwidth]{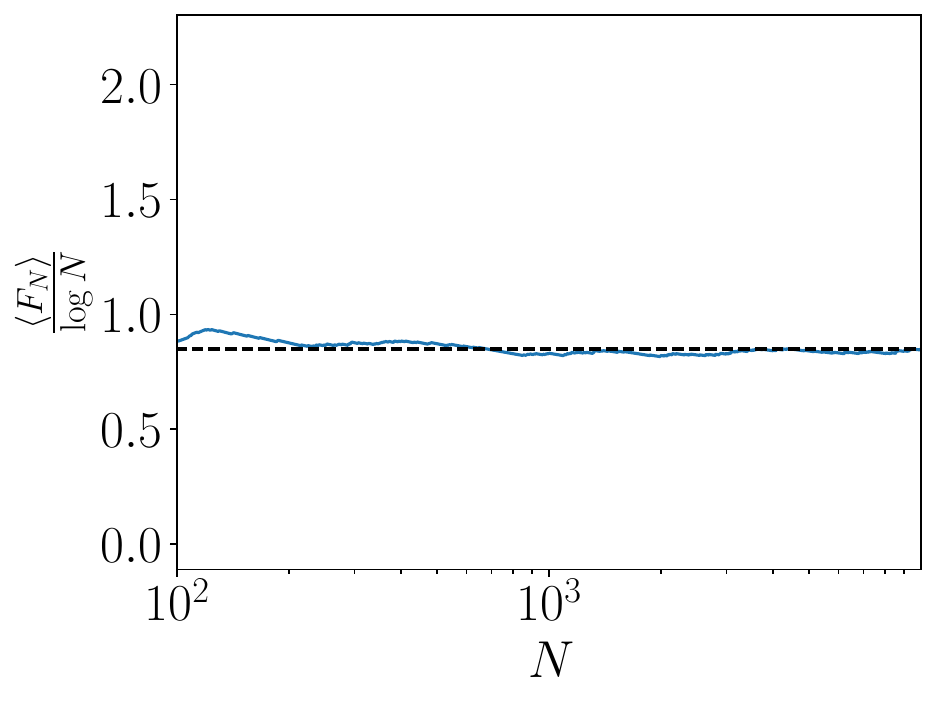}
        \subcaption{$\langle f_n \rangle \propto \log n$ for a compact RW on a Sierpinsky triangle.}
        \label{fig:flips-sierp}
    \end{subfigure}
    \hfill
    \begin{subfigure}[t]{.32\textwidth}
        \centering
        \includegraphics[width=\textwidth]{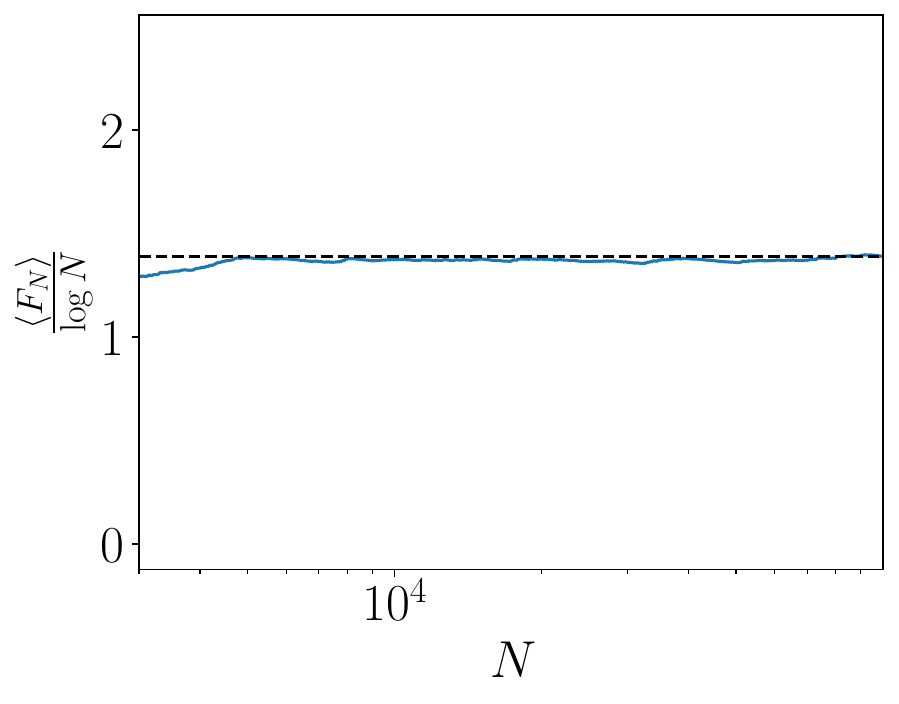}
        \subcaption{$\langle f_n \rangle \propto \log n$ for a compact RW on a percolation cluster.}
        \label{fig:flips-perc}
    \end{subfigure}
    \begin{subfigure}[t]{.32\textwidth}
        \centering
        \includegraphics[width=\textwidth]{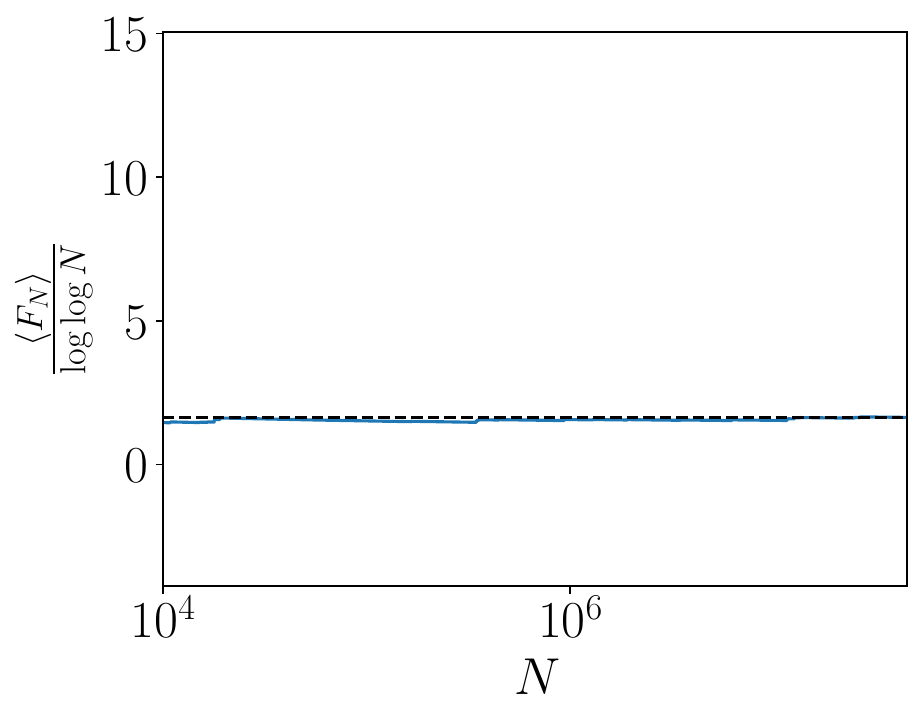}
        \subcaption{$\langle f_n \rangle \propto \log \log n$ for a marginal RW on a 2d square lattice.}
        \label{fig:flips-2d}
    \end{subfigure}
    \caption{Numerical confirmation of the extension of our results on flips to compact RWs on fractals and marginal RWs, through the scaling of the mean number of flips $f_n$.}
    \label{fig:flips-higherd}
\end{figure}
Crucially, our results extend beyond the one-dimensional case. They apply to all recurrent, asymptotically scale-invariant non-Markovian RWs, including physically relevant systems such as generic non-Markovian RWs on deterministic or random fractals—e.g., critical percolation clusters, which model transport in crowded or disordered environments.

In these settings, we define a flip as the event where the walker returns to the origin after visiting the \( n \)th new site but before reaching the \( (n+1) \)st. This preserves the structure of the one-dimensional analysis: new sites correspond to blue balls, while the origin acts as an infinite reservoir of red balls. In foraging models, where each site carries one unit of food and the origin represents a nest, this translates into the key observable that is the distribution of the total number \( f_n \) of returns to the nest after collecting \( n \) units of food. \par \vspace{2ex} 

Importantly, we stress that our key results—such as the scaling of \( \pi_n \) \eqref{pin} and the large deviation principle for the number of flips \eqref{ldp-flips}—continue to hold in this generalized setting. See Figs.~\ref{fig:flips-sierp} and ~\ref{fig:flips-perc} for a numerical confirmation. Let us show that the scaling of $\pi_n$ is unchanged for compact RWs.

\subsubsection{Case of Compact Random Walks}

For compact random walks, the distribution of the time \( \tau_n \) between the \( n \)th and \( (n+1) \)st site discoveries is known to follow the asymptotic form \cite{regnierCompleteVisitationStatistics2022}:
\begin{equation}
    \label{leo}
    \mathbb{P}(\tau_n = \tau) = P_n(\tau) \sim \frac{1}{n^{1 + 1/\mu}} \, \psi\left( \frac{\tau}{n^{1/\mu}} \right),
\end{equation}
where \( \mu = \frac{d}{d_w} < 1\), with \( d \) the fractal dimension of the medium and \( d_w \) the walk dimension.

Let \( V_n \) denote the set of sites visited by the walker after \( n \) distinct sites. Since the walk is compact, this region spans a linear size \( R \sim n^{1/d} \) \cite{benichouFirstpassageTimes}. Now, consider the condition for a flip, i.e., for the walker to return to the origin before visiting the \( (n+1) \)st new site.

This occurs only if the walker, starting near the boundary \( \partial V_n \), avoids reaching \( \partial V_n \) before returning to the origin. This requires that the discovery time \( \tau_n \) be larger than the time to wander across the domain, i.e., \( \tau_n \gtrsim R^{d_w} = n^{1/\mu} \). After this timescale, the walker is effectively well-mixed within \( V_n \), with high probability of being closer to the origin than the boundary.

Because the walk is compact, the walker is more likely to hit the origin than to escape. Therefore, the flip probability scales as:
\begin{keyboxedeq}[Universal scaling of the flip probability for compact RWs]
    \begin{equation}
        \pi_n \propto \int_{n^{1/\mu}}^\infty P_n(\tau) \, d\tau \propto \frac{1}{n},
    \end{equation}
\end{keyboxedeq}
with the last $\propto$ sign being a consequence of \eqref{leo}.

\subsubsection{Case of marginal RWs}
In the case of marginal RWs such as the $2D$ simple RW, the scalings are modified due to the presence of holes in the visited territory. We find (see Fig.~\ref{fig:flips-2d}):
\begin{keyboxedeq}[Universal scaling of the flip probability for marginal RWs]
    \begin{equation}
        \label{flip-2d}
        \pi_n^{\text{marginal}} \propto \frac{1}{n \log n}, \qquad \langle f_n^{\text{marginal}} \rangle \propto \log \log n.
    \end{equation}
\end{keyboxedeq}

\section{Distribution of the eccentricity $z$}

In this section, we show how the full functional form of $\Phi$ governs the distribution of the eccentricity, as illustrated in Fig.~\ref{fig:illus-eccentricity}. Within the foraging analogy, we consider the fraction $Z_n = k_n / n$ of red food units consumed, where $k_n$ denotes the number of red units eaten after $n$ total units have been consumed. This distribution was computed explicitly for class (I) and class (III) processes in Eq.~\eqref{eccentricity-sirw-1d}. Our aim here is to show that it is directly related to $\Phi$, reinforcing the role of $\Phi$ as a fundamental building block for space-exploration observables.

Using a decomposition based on whether a flip occurred at the $n$-th visit, we obtain the following exact recurrence relations for the quantities $q_\pm(k, n-k)$, which denote the probability of having consumed $k$ red units and $n-k$ blue units, ending in the $+$ or $-$ state, respectively:
\begin{align}
\begin{cases}
q_+(k+1,n-k) = \left(1 - \tilde{\pi}_n(k) \right) q_+(k,n-k) + \tilde{\pi}_n(n-k) q_-(k,n-k), \\[0.5em]
q_-(k,n+1-k) = \left(1 - \tilde{\pi}_n(n-k)\right) q_-(k,n-k) + \tilde{\pi}_n(k) q_+(k,n-k).
\end{cases}
\label{eq:DiscreteEq}
\end{align}

Remarkably, the explicit knowledge of the flip probabilities $\tilde{\pi}_n(k)$, as expressed in Eqs.~\eqref{Phi-def} and~\eqref{phi-expr}, enables a Markovian-style analysis of these fundamentally non-Markovian dynamics.

Assuming the scaling forms
\[
q_\pm(k, n-k) \sim \frac{1}{n} q_\pm\left(z = \frac{k}{n}\right), \qquad \tilde{\pi}_n(k) \sim \frac{1}{n} \Phi\left(z = \frac{k}{n}\right),
\]
the discrete evolution equations \eqref{eq:DiscreteEq} yield the following coupled differential equations in the scaling limit:
\begin{equation}
    \begin{cases}
    \partial_z\left( (1-z) q_+(z)\right) = \Phi(1-z)\, q_-(z) - \Phi(z)\, q_+(z), \\[0.5em]
    -\partial_z\left( z q_-(z)\right) = \Phi(z)\, q_+(z) - \Phi(1-z)\, q_-(z).
    \end{cases}
    \label{ode-qp}
\end{equation}

Adding the two equations in \eqref{ode-qp} leads to a remarkable identity:
\begin{keyboxedeq}[Duality relation]
    \begin{equation}
        \frac{q_+(z)}{q_-(z)} = \frac{z}{1-z}.
        \label{ratio-ansatz-general}
    \end{equation}
\end{keyboxedeq}

This duality holds for \emph{any} symmetric, scale-invariant, non-Markovian process. Though surprising at first glance, its implications become clear when considering the distribution of the eccentricity:
\begin{boxedeq}
\begin{equation}
    \label{ecc-def}
    \mathbb{P}(Z_n = z = \tfrac{k}{n}) = q_+(k,n-k) + q_-(k,n-k) \sim \frac{1}{n} \sigma_z \sim \frac{1}{n} (q_+(z) + q_-(z)).
\end{equation}
\end{boxedeq}
Substituting the duality relation \eqref{ratio-ansatz-general} into \eqref{ecc-def}, we arrive at:
\begin{keyboxedeq}[Eccentricity distribution]
    \begin{equation}
        \sigma_z = \frac{q_+(z)}{z}.
        \label{ecc-expr-gen}
    \end{equation}
\end{keyboxedeq}

This means that, when conditioning the process to have explored the interval $[-(1-z), z]$, the probability that it reached $z$ after $-(1-z)$ is simply
\begin{boxedeq}
    \begin{equation}
\mathbb{P}(\text{reach $z$ before $-(1-z)$|visit $[-(1-z),z]$}) = \frac{q_+(z)}{\sigma_z} = z.
\end{equation}
\end{boxedeq}
In other words, the process $X_t$, though non-Markovian, behaves exactly like Brownian motion for this specific observable. This is yet another manifestation of our broader principle: conditioning on the set of visited sites often unveils universal, Markovian-like behavior, even in non-Markovian systems. \par \vspace{2ex}
We now solve Eq.~\eqref{ode-qp}. Using the duality relation in Eq.~\eqref{ratio-ansatz-general}, we obtain a single differential equation:
\begin{equation}
    \partial_z\left( (1 - z) q_+(z) \right) = \left( \Phi(1 - z)\, \frac{1 - z}{z} - \Phi(z) \right) q_+(z).
\end{equation}

This equation can be solved directly and yields:
\begin{boxedeq}
    \begin{equation}
        \label{gen-expr-qp}
        q_+(z) = \mathcal{N} \frac{\mathcal{G}(z) \mathcal{G}(1 - z)}{1 - z}, \quad \mathcal{G}(z) \equiv \exp\left(-\int_0^z \frac{\Phi(u)}{1 - u} \, du\right),
    \end{equation}
\end{boxedeq}
where $\mathcal{N}$ is a normalization constant chosen such that $\int_0^1 q_+(z) \, dz = \frac{1}{2}$. Combined with Eq.~\eqref{expr-A-phi}, Eq.~\eqref{gen-expr-qp} shows that knowledge of $\Phi(z)$ is sufficient to determine the prefactor $A$ of the flip probability $\pi_n$, and thus fully characterizes its exact large-$n$ behavior.
\par 
From Eq.~\eqref{ecc-expr-gen}, we directly obtain the eccentricity distribution $\sigma_z$ in terms of $\Phi(z)$:
\begin{keyboxedeq}[Universal expression for the distribution of the eccentricity]
\begin{align}
\label{pofz}
    \sigma_z = \mathcal{N} \frac{\mathcal{G}(z) \mathcal{G}(1 - z)}{z(1 - z)}, \quad \mathcal{G}(z) = \exp\left(-\int_0^z \frac{\Phi(u)}{1 - u} \, du\right).
\end{align}
\end{keyboxedeq}
Here, $\mathcal{N}$ ensures normalization over the interval $z \in (0,1)$.  \par  

Using Eqs.~\eqref{pofz} and~\eqref{phi-expr}, we find the explicit forms of $\sigma_z$ for classes (I)–(III):
\begin{keyboxedeq}[Eccentricity distribution for classes (I)–(III)]
\begin{equation}
\label{pofz-explicit}
\sigma_z = \mathcal{N}
\begin{cases}
(z(1 - z))^{\phi - 1} & \text{(I), (III)} \\
1 - 4\varepsilon \log\left( \dfrac{\Gamma \left(1 - \frac{z}{2}\right) \Gamma \left(\frac{z + 1}{2}\right)}{\Gamma \left(\frac{1 - z}{2}\right) \Gamma \left(\frac{z}{2}\right)}\right) + \mathcal{O}(\varepsilon^2) & \text{(II)}
\end{cases}
\end{equation}
\end{keyboxedeq}

This key result calls for several comments.

\begin{itemize}
    \item Surprisingly, despite the $\Phi$ functions of classes (I) and (III) being entirely distinct, both yield the same $\sigma_z$, as established earlier in Eq.~\eqref{eccentricity-sirw-1d}. From the general result \eqref{pofz}, it is straightforward to see that any $\Phi$ function leading to a Beta distribution for $z$, i.e., $\sigma_z \propto (z(1-z))^{\phi - 1}$, must satisfy the following symmetry relation:
    \begin{boxedeq}
    \begin{equation}
        \label{phi-sym}
        z \Phi(z) = (1 - z)\Phi(1 - z) + (2z - 1)\phi.
    \end{equation}   
    \end{boxedeq} 
    We verify that \eqref{phi-sym} holds for class (III) processes using \eqref{phi-expr}.
    \item As shown previously in Eq.~\eqref{splitting-from-ecc}, the derivative of the splitting probability $Q_+(z)$ is directly related to the eccentricity distribution $\sigma_z$. Our result in Eq.~\eqref{pofz} thus helps resolve a previously open question about the physical origin of the seemingly "super-universal" form of splitting probabilities~\cite{zoiaAsymptoticBehaviorSelfAffine2009} as incomplete Beta functions. For processes with independent flips—i.e., class (I) processes—our formula Eq.~\eqref{pofz} confirms that the splitting probability takes the form of an incomplete Beta function. More generally, any process whose $\Phi$ function satisfies the symmetry relation \eqref{phi-sym} (such as class (III) processes) also exhibits this property.

    \item Importantly, Eq. \eqref{pofz-explicit} shows that qualitative features of the eccentricity distribution are ruled by the single exponent $\phi$. For $\phi > 1$, the most probable values of $z$ are located at the edges. In other words, sites are typically discovered on the same side. Conversely, for $\phi < 1$, trajectories typically discover an equivalent number of sites on both sides. Note that for $\phi = 1$, which is realized only for Brownian Motion, all eccentricities are statistically equivalent.
\end{itemize}

\section{Distribution of the number of flips $f_n$}
It is now clear that the number of flips $f_n$ after visiting $n$ sites is a fundamental observable for characterizing how a random walker explores space. Each flip corresponds to a directional reorientation, and their accumulation reflects the evolving geometry of the trajectory. As such, $f_n$ captures the degree of directional persistence and serves as a sensitive probe of memory effects and non-Markovian dynamics. Accessing the full distribution of $f_n$ provides insights far beyond those available from traditional, Markovian descriptions. Remarkably, we are able to determine the full distribution of $f_n$ in complete generality for all scale-invariant, symmetric non-Markovian processes. Even more striking is the fact that, despite strong correlations between flips for processes outside class (I), the typical fluctuations of $f_n$ around its mean are Gaussian. This counter-intuitive result reveals an unexpected universality in the statistical structure of trajectory geometry. It underscores both the analytical power and the broad applicability of the flip-based framework developed in this part of the thesis.

\subsection{The case of class (I) processes}
We recall that class (I) processes are those in which flips occur independently. This class includes several important examples such as Lévy walks, the RAP, the SATW, and Markovian processes. For such systems, the distribution of the number of flips after $n$ sites have been visited, denoted by $F_n(f) \equiv \mathbb{P}(f_n = f)$, can be computed exactly thanks to the independence of flips.

We begin by expressing the total number of flips as a sum of indicator variables:
\begin{equation}
    f_n = \sum_{i=1}^n \delta_i,
\end{equation}
where $\delta_i = 1$ if a flip occurs upon discovering the $i$-th site, and $\delta_i = 0$ otherwise.

For class (I) processes, the variables $\delta_i$ are uncorrelated. While they are not identically distributed, their distribution has a simple form:
\begin{equation}
    \mathbb{P}(\delta_i = 1) = 1 - \mathbb{P}(\delta_i = 0) = \pi_i \sim \frac{\phi}{i}.
\end{equation}

From this, we immediately deduce that:
\begin{keyboxedeq}[Distribution of $f_n$ for class (I) processes]
    \begin{equation}
        \text{$f_n$ is Poisson distributed with mean } \lambda_n = \phi \log n.
    \end{equation}
\end{keyboxedeq}

Explicitly, this gives:
\begin{boxedeq}
    \begin{equation}
        \label{poisson-dist}
        \mathbb{P}(f_n = f) = F_n(f) \sim n^{-\phi} \cdot \frac{(\phi \log n)^f}{f!}.
    \end{equation}
\end{boxedeq}

From Eq.~\eqref{poisson-dist}, we see that $f_n$ satisfies a large deviation principle with speed $\log n$, and an explicit rate function:
\begin{boxedeq}
\begin{equation}
    \label{ldp-flips-classI}
    F_n(f) \asymp n^{-I\left(\frac{f}{\log n} \right)}, \quad I(x) = x \log \left(\frac{x}{\phi} \right) + \phi - x.
\end{equation}
\end{boxedeq}

This rate function has the following key properties\footnote{Recall that for class (I) processes only, $A = \phi$. However, we distinguish between $A$ and $\phi$ in Eq.~\eqref{general-features-ldp} to allow for a broader generalization beyond class (I).}:

\begin{boxedeq}
    \begin{equation}
        \label{general-features-ldp}
        I(0) = \phi, \quad I(A) = 0, \quad I'(A) = 0.
    \end{equation}
\end{boxedeq}

As we will see, both the existence of the large deviation principle in Eq.~\eqref{ldp-flips-classI} and the key features of the rate function in Eq.~\eqref{general-features-ldp} are universal and extend beyond class (I) processes.

\subsection{A general large deviation principle for $f_n$}
We recall that the mean number of flips $\langle f_n \rangle$ after visiting $n$ sites is given by
\begin{equation}
    \langle f_n \rangle = A \log n,
\end{equation}
with $A$ defined in Eq.~\eqref{pin}. In this section, we go beyond the mean and analyze the full distribution of $f_n$. We show that it satisfies a large deviation principle with speed $\log n$, namely:
\begin{keyboxedeq}[Large deviation principle for the number of flips $f_n$]
    \begin{equation}
        \label{ldp-flips}
        F_n(f) \equiv \mathbb{P}(f_n = f) \asymp n^{-I\left(\frac{f}{\log n} \right)}.
    \end{equation}
\end{keyboxedeq}

To establish Eq.~\eqref{ldp-flips}, we start from a simple recurrence relation. It partitions over whether a flip occurs at the $(n+1)^{\text{st}}$ visited site:
\begin{equation}
    \label{rec-flips}
    F_{n+1}(f+1) = \left(1 - \overline{\pi}_n(f+1)\right) F_n(f+1) + \overline{\pi}_n(f)\, F_n(f).
\end{equation}

We now insert the large deviation ansatz 
\[
F_n(f) \asymp n^{-I\left(\frac{f}{\log n} \right)}.
\]
Since $f_n$ grows as $\log n$, we expand $F_{n+1}(f+1)$ to leading order in $1/n$:
\begin{equation}
    \label{ldp-1morestep}
    F_{n+1}(f+1) \asymp (n+1)^{-I\left(\frac{f+1}{\log(n+1)} \right)} 
    \approx n^{-I\left(\frac{f+1}{\log n} \right)} \left[ 1 - \frac{1}{n} I\left(\frac{f+1}{\log n} \right) + \frac{f+1}{n \log n} I'\left(\frac{f+1}{\log n} \right) \right].
\end{equation}

Next, we use the scaling form of the flip-informed flip probability:
\[
    \overline{\pi}_n(f) \sim \frac{\Psi\left(\frac{f}{\log n} \right)}{n}.
\]

Substituting this into the recurrence relation \eqref{rec-flips} and using the asymptotics \eqref{ldp-1morestep} confirms that the large deviation form \eqref{ldp-flips} holds, provided that the rate function $I(x)$ satisfies:
\begin{keyboxedeq}[Universal identity satisfied by the rate function $I$]
    \begin{equation}
        \label{eq-rate-flips}
        I(x) - x I'(x) = \Psi(x) \left(1 - e^{I'(x)}\right).
    \end{equation}
\end{keyboxedeq}
For class (I) processes, $\Psi(x) \equiv \phi$ is constant. Substituting into Eq.~\eqref{eq-rate-flips} and solving the resulting differential equation, we recover the Poissonian rate function given in Eq.~\eqref{ldp-flips-classI}. \par 
The rate function $I(x)$ is non-negative and vanishes only at $x = A$, consistent with the mean behavior $\langle f_n \rangle = A \log n$:
\begin{equation}
    I(A) = I'(A) = 0.
\end{equation}

Evaluating Eq.~\eqref{eq-rate-flips} in the limit $x \to A$ gives the curvature of the rate function in terms of $\Psi$:
\begin{boxedeq}
    \begin{equation}
    \label{curvature-I}
    I''(A) = \frac{1 - 2\Psi'(A)}{A}.
    \end{equation}
\end{boxedeq}

This curvature is particularly relevant, as it is directly related to the variance $\nu^2$ of the variable $f_n$. Indeed, defining the rescaled deviation
\[
    y \equiv \frac{f_n}{\log n} - A,
\]
we find that typical fluctuations $y \propto \frac{1}{\sqrt{\log n}}$ are distributed according to
\begin{equation}
    F_n(y \log n) \asymp n^{-I(y)} \asymp n^{-\frac{y^2}{2} I''(A)}.
\end{equation}

Thus, in the regime of typical fluctuations, the distribution of $f_n$ takes the Gaussian form:
\begin{keyboxedeq}[Gaussian typical fluctuations of the number of flips $f_n$]
    \begin{equation}
        \label{gaussian-flip}
        \mathbb{P}(f_n = f) = F_n(f) \sim \frac{1}{\nu \sqrt{2\pi \log n}} \exp\left( -\frac{\left( \frac{f}{\log n} - A \right)^2}{2\nu^2 \log n} \right),
        \quad \nu = \frac{1}{\sqrt{I''(A)}} = \sqrt{ \frac{A}{1 - 2\Psi'(A)} }.
    \end{equation}
\end{keyboxedeq}
This result, Eq.~\eqref{gaussian-flip}, is illustrated in Fig.~\ref{fig:ldp-tsaw}. Crucially, since the value of $A$ is known for classes (I)–(III), Eq.~\eqref{gaussian-flip} shows that the full distribution of the typical fluctuations of the number of flips $f_n$—a highly nonlocal and physically meaningful observable in non-Markovian processes—is now determined up to a single quantity: the derivative $\Psi'(A)$. \par \vspace{2ex}

Although a general method for computing $\Psi'(A)$ remains elusive, this chapter has demonstrated that our flip-based theory of space exploration is not only analytically tractable but also quantitatively predictive and broadly applicable. To our knowledge, no other existing framework provides such predictive power across all scale-invariant, symmetric non-Markovian processes. The development of this theory marks a significant step forward, and its refinement and extension will be a central focus of our continued work. \par \vspace{2ex}
In the next chapter, we carry out the analytical computations leading to the explicit expressions \eqref{phi-expr} for the central object of our framework: the $\Phi$ function. While these calculations are technically and conceptually involved, they are crucial—$\Phi$ encodes the core statistical structure of the flip process and enables the analytical tractability of our approach. The predictive power of our framework stems directly from the explicit computation of $\Phi$.

\chapter{\texorpdfstring{Computing the $\Phi$ function for classes (II) and (III)}{Computing the Phi function for classes (II) and (III)}}
Computing the $\Phi$ function is a central step in validating the analytical power of the flip-based framework developed in the previous chapter. While its computation proves technically challenging, this function plays a pivotal role: it encodes how the probability of a flip depends on the asymmetry of the visited territory, thereby capturing how memory influences the walker’s future exploration. Once $\Phi$ is known for a given process, a wide array of physical observables—such as the distribution of eccentricity \eqref{pofz}, or the number of flips $f_n$ \eqref{gaussian-flip}—follow from universal relations derived in the previous chapter. In this sense, $\Phi$ serves as a key object of the theory: its explicit form provides direct access to the non-trivial statistical structure of non-Markovian trajectories. The computations that follow, although intricate and technical, are what allow the flip-based framework to go beyond qualitative insights and deliver exact, testable predictions for a wide variety of non-Markovian processes.

\section{\texorpdfstring{Class (II) processes: the $\Phi$ function for the fBM}{Class (II) processes: the Phi function for the fBM}}
\begin{figure}[h!]
    \centering
    \includegraphics[width=.7\textwidth]{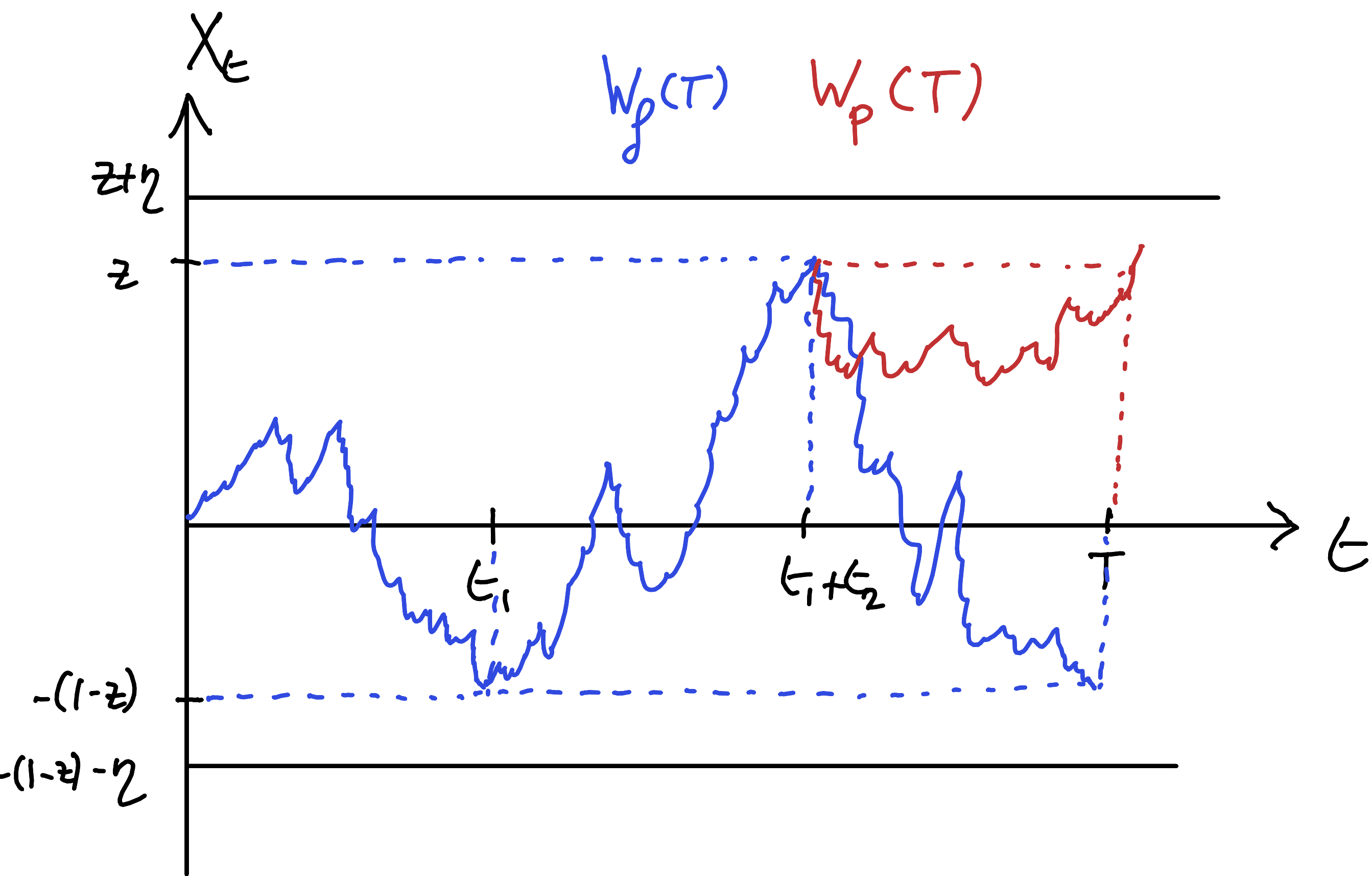}
    \caption{Schematic representation of two fBM paths, blue and red, contributing respectively to the observables $W_f(T)$ and $W_p(T)$.}
    \label{fig:scheme-fbm}
\end{figure}

In this section, we compute analytically to order $\mathcal{O}(\varepsilon^2)$ the $\Phi$ function for the fBM, given in \eqref{phi-expr}. For this computation, we will use the formalism developed by K.Wiese in \cite{wieseFirstPassageInterval2019}. This is a diagrammatic, field-theoretical approach based on the explicit expression of the Gaussian action for the fBM at first order in $\varepsilon$. \par \vspace{2ex}
To compute $\Phi(z)$, we introduce two observables: $W_f(T)$ and $W_p(T)$. The first, $W_f(T)$, is defined as the joint probability that the span of the process is $[-(1 - z), \underline{z}]$ and that a flip occurs at time $T$, i.e., the fBM crosses its minimum $-(1 - z)$ at time $T$. The second observable, $W_p(T)$, corresponds to the joint probability that the span is also $[-(1 - z), \underline{z}]$, but the fBM instead crosses its maximum $z$ at time $T$.

To calculate these probabilities, we regularize the problem by introducing absorbing boundaries at $-(1 - z) - \eta$ and $z + \eta$, where the regularization parameter $\eta \to 0$ at the end of the computation. See Fig.~\ref{fig:scheme-fbm} for an illustration. Consider, for example, a path $X_t$ contributing to $W_f(T)$. A crucial observation is that whenever $X_t$ approaches within a distance $\eta$ of either absorbing boundary, its contribution acquires a factor of order $O(\eta)$. To contribute non-trivially to Eq.~\eqref{phi-fbm-compute}, such a path must approach the boundaries only three times at $O(\eta)$ distance. For instance, a path contributing to $W_f(T)$ may first cross $-(1 - z)$ at some random time $t_1$, then reach $z$ at $t_1 + t_2$, and finally cross below $-(1 - z) - \eta/2$ at $T = t_1 + t_2 + t_3$. \par

The probability that a flip occurs, given that the span is $[-(1 - z), \underline{z}]$, is then expressed as:
\begin{equation}
    \label{phi-fbm-compute}
    \Phi(z) = \lim_{\eta \to 0} \frac{1}{\eta^3} \cdot \frac{W_f}{q_+(z)}, \quad
    W_f = \int_0^\infty W_f(T)\, dT, \quad
    W_p = \int_0^\infty W_p(T)\, dT.
\end{equation}

In Eq.~\eqref{phi-fbm-compute}, we identify the familiar quantity:
\begin{equation}
    q_+(z) = \frac{W_f + W_p}{\eta^2} = z + \varepsilon\, \alpha(z) + O(\varepsilon^2),
\end{equation}
where $\alpha(z)$ denotes the leading-order correction in $\varepsilon$. This correction can be computed from the known relation between $q_+(z)$ and the splitting probability:
\begin{equation}
    Q_+'(z) = -\frac{q_+(z)}{z},
\end{equation}
where $Q_+(z)$ has been obtained to first order in $\varepsilon$ in~\cite{wieseFirstPassageInterval2019}. From this, we find:
\begin{equation}
    \alpha(z) = -4z \log\left( \frac{ \Gamma \left(1 - \frac{z}{2} \right) \Gamma \left( \frac{z+1}{2} \right) }{ \Gamma \left( \frac{1 - z}{2} \right) \Gamma \left( \frac{z}{2} \right) } \right) + zC,
\end{equation}
where $C$ is a normalization constant ensuring that $\int_0^1 q_+(z)\, dz = \frac{1}{2}$ at all orders in $\varepsilon$. Importantly, we will not need to compute $C$ explicitly: it can be fixed by using the physical constraint from Eq.~\eqref{phi1}, which fixes $\Phi(1) = \phi = 1 - 4\varepsilon + O(\varepsilon^2)$.

Similarly, we define the leading-order correction to $W_f$ as:
\begin{equation}
    W_f = \eta^3 \left(z + \varepsilon\, \gamma(z) + O(\varepsilon^2)\right).
\end{equation}

Substituting into Eq.~\eqref{phi-fbm-compute}, we find that the first-order correction to $\Phi(z)$ takes the form:
\begin{boxedeq}
\begin{equation}
    \label{phi-fbm-diagram}
    \Phi(z) = \frac{z+\varepsilon \gamma(z) + O(\varepsilon^2)}{z + \varepsilon\alpha(z) + O(\varepsilon^2)} = 1 + \varepsilon\, \beta(z) + O(\varepsilon^2), \quad \text{with} \quad \beta(z) = \frac{\gamma(z) - \alpha(z)}{z}.
\end{equation}
\end{boxedeq}
Thus, the problem reduces to computing $\gamma(z)$, which corresponds to the first-order correction to the observable $W_f$.

\subsection{The observable $W_f(T)$}
By construction, the observable $W_f(T)$ admits the following path integral representation:
\begin{equation}
    W_f(T) = \int_{X_0 = 0}^{X_T = -(1 - z)} \mathcal{D}X \, \prod_{t = 0}^{T} \left[ \Theta(X_t + 1 - z + \eta)\, \Theta(z + \eta - X_t) \right] \int_{t_1 + t_2 \leq T} \delta(X_{t_1} + 1 - z)\, \delta(X_{t_1 + t_2} - z)\, e^{-S[X]},
\end{equation}
where $S[X]$ is the effective action governing the fBM.

At leading order in $\varepsilon$, this action has been derived in~\cite{wieseFirstPassageInterval2019} and takes the form:
\begin{equation}
\label{action-exp}
    S = \frac{1}{D} \left(S_0 - \frac{\varepsilon}{2} S_1 + \dots \right), \quad
    D \equiv e^{2\varepsilon(1 + \log \omega)}, \quad
    S_1 \equiv \int_0^{\Lambda} dy \int_0^T dr_1 \int_{r_1}^T dr_2\, \dot{x}(r_1)\, \dot{x}(r_2)\, e^{-y(r_2 - r_1)}.
\end{equation}
Here $\Lambda$ is the large-momentum cutoff, equivalent to the small-time cutoff $\omega$ by $\Lambda = e^{-\gamma}/\omega$. Eventually $\Lambda \to \infty$. An important point is that, from the small-time cutoff, all integrals involving some time $r_1$ as a lower bound actually involve the regularized time $r_1 + \omega$, and hence \eqref{action-exp} involves only the \emph{normal-ordered} velocities $ \dot{x}(r_1) \dot{x}(r_2)$ with $r_2 > r_1$: there is no self-contraction $r_1 = r_2$. \par \vspace{2ex}
From the expressions above, it is clear that we will need to evaluate Brownian velocity correlators of the form $\langle \dot{x}(r_1) \dot{x}(r_2) \rangle$ in the presence of absorbing boundary conditions. However, care must be taken: the relative positions of the times $t_1$, $t_1 + t_2$, $r_1$, and $r_2$ significantly affect the structure of these correlators. For instance, if $r_1 < t_1 < r_2$, then the velocities at $r_1$ and $r_2$ are uncorrelated, because the position $X_{t_1} = z$ is fixed. In this case, the correlator factorizes as
\[
\langle \dot{x}(r_1) \dot{x}(r_2) \rangle = \langle \dot{x}(r_1) \rangle \langle \dot{x}(r_2) \rangle.
\]
Importantly, neither of these averages vanishes: since the endpoints of the path are fixed, the system no longer possesses space-reversal symmetry, and nonzero average velocities can emerge.

To proceed, we introduce the following key functions. First, we define the propagator $Z_t(x_1, x_2)$ as the probability amplitude for a Brownian path to go from $x_1$ at time 0 to $x_2$ at time $t$, in the presence of absorbing boundaries at $-(1 - z) - \eta$ and $z + \eta$. The Laplace transform of this propagator with respect to $t$ is well known \cite{borodinHandbookBrownianMotion2002} and will be used extensively.

Second, we define the class of functionals $J_t(u, v; y_1, \dots, y_n)$, which capture the contributions from paths with velocity insertions:
\begin{equation}
    J_t(u,v; y_1,\dots, y_n) = \int_{0 < r_1 < \dots < r_n < t} \prod_{i=1}^n e^{-y_i r_i} \int_{X_0 = u}^{X_t = v} \mathds{1}_{\text{$X$ stays in } [-(1 - z) - \eta,\, z + \eta]} \mathcal{D}X\, \dot{x}(r_1) \dots \dot{x}(r_n)\, e^{-S_0/D}.
\end{equation}

A central tool in the calculations that follow is the Markov property of Brownian paths, which allows us to decompose the path integral over relevant time intervals. This decomposition will be crucial for organizing the contributions to the velocity correlators and computing the first-order corrections to $W_f$ and thus to $\Phi(z)$.

\subsection{Computing the \texorpdfstring{$J$}{J} functions}

\subsubsection{\texorpdfstring{$J_t(u,v;y)$}{J\_t(u,v;y)}}

By definition, the single-velocity insertion functional reads:
\begin{equation}
    J_t(u,v;y) = \int_0^t e^{-y r} \, dr \int_{X_0 = u}^{X_t = v} \mathds{1}_{\text{$X$ stays in } [-(1 - z) - \eta, z + \eta]} \mathcal{D}X \, \dot{x}(r) \, e^{-S_0/D}.
\end{equation}

We use the known expression for the average velocity of a Brownian path on $[0,t]$, conditioned to start at $u$ and end at $v$:
\begin{equation}
    \langle \dot{x}(r) \rangle = 2D \int_{-\infty}^\infty Z_r(u,x) \, \partial_x Z_{t - r}(x,v) \, dx,
\end{equation}
where $Z_t(x_1,x_2)$ is the propagator in the presence of absorbing boundaries at $-(1 - z) - \eta$ and $z + \eta$.

Taking the Laplace transform in time $t \to s$ yields:
\begin{equation}
    \tilde{J}_s(u,v;y) = 2D \int_{-(1 - z)}^z \tilde{Z}_{s + y}(u,x) \, \partial_x \tilde{Z}_s(x,v) \, dx.
\end{equation}

This function satisfies the symmetry relation:
\begin{equation}
    \tilde{J}_s(u,v;y) = -\tilde{J}_{s + y}(v,u; -y).
\end{equation}

For the case $u < v$, the expression becomes explicitly:
\begin{multline}
    \frac{\tilde{J}_s(u,v;y)}{2D} =
    \frac{\text{csch}(\sqrt{s + y}) \, \sinh\left(\sqrt{s + y}(u - z + 1)\right)}{y}
    \left[\text{csch}(\sqrt{s}) \, \sinh\left(\sqrt{s}(v - z + 1)\right) - \cosh\left(\sqrt{s + y}(v - z)\right)\right] \\
    - \frac{\text{csch}(\sqrt{s}) \, \sinh\left(\sqrt{s}(v - z)\right)}{y}
    \left[\text{csch}(\sqrt{s + y}) \, \sinh\left(\sqrt{s + y}(u - z)\right) + \cosh\left(\sqrt{s}(u - z + 1)\right)\right].
\end{multline}

\subsubsection{\texorpdfstring{$J_t(u,v;y_1, y_2)$}{J\_t(u,v;y1,y2)}}

The two-velocity insertion functional is defined as:
\begin{equation}
    J_t(u,v;y_1,y_2) = \int_0^t dr_1 \int_{r_1}^t dr_2 \, e^{-y_1 r_1 - y_2 r_2} \int_{X_0 = u}^{X_t = v} \mathds{1}_{\text{$X$ stays in } [-(1 - z) - \eta, z + \eta]} \mathcal{D}X \, \dot{x}(r_1)\dot{x}(r_2) \, e^{-S_0/D}.
\end{equation}

We apply the Markov property of Brownian paths to evaluate the velocity correlator for $r_1 < r_2$:
\begin{equation}
    \langle \dot{x}(r_1) \dot{x}(r_2) \rangle = (2D)^2 \int_{-(1 - z) - \eta}^{z + \eta} dx_1 \, dx_2 \, Z_{r_1}(u,x_1) \, \partial_{x_1} Z_{r_2 - r_1}(x_1,x_2) \, \partial_{x_2} Z_{t - r_2}(x_2,v).
\end{equation}

After Laplace transforming (and using $e^{-y_1 r_1 - y_2 r_2} = e^{-(y_1 + y_2) r_1 - y_2 (r_2 - r_1)}$), we obtain:
\begin{equation}
    \tilde{J}_s(u,v;y_1,y_2) = (2D)^2 \int_{-(1 - z) - \eta}^{z + \eta} dx_1 \, dx_2 \, \tilde{Z}_{s + y_1 + y_2}(u,x_1) \, \partial_{x_1} \tilde{Z}_{s + y_2}(x_1,x_2) \, \partial_{x_2} \tilde{Z}_s(x_2,v).
\end{equation}

In practice, we often only require the special case $\tilde{J}_s(u,v;-y,y)$, which simplifies to:
\begin{equation}
    \tilde{J}_s(u,v;-y,y) = (2D)^2 \int_{-(1 - z) - \eta}^{z + \eta} dx_1 \, dx_2 \, \tilde{Z}_s(u,x_1) \, \partial_{x_1} \tilde{Z}_{s + y}(x_1,x_2) \, \partial_{x_2} \tilde{Z}_s(x_2,v).
\end{equation}

When the endpoint $v$ lies near a boundary, for instance at $v = z - \eta$, we use:
\begin{equation}
    \tilde{Z}_s(x_2, z - \eta) = \tilde{Z}_s(x_2, z) - \eta \left. \partial_a \tilde{Z}_s(x_2, a) \right|_{a = z} = -\eta \left. \partial_a \tilde{Z}_s(x_2, a) \right|_{a = z} \equiv \eta \tilde{\mathcal{J}}_s(x_2, z),
\end{equation}
where $\tilde{\mathcal{J}}_s(x_2, z)$ is the Laplace-transformed probability current exiting at the boundary $z$.

Inserting this into the expression above yields:
\begin{equation}
\label{j_y_my}
    \tilde{J}_s(x,z;-y,y) = \eta (2D)^2 \int_{-(1 - z)}^{z} dx_1 \int_{-(1 - z)}^{z} dx_2 \, \tilde{Z}_s(x,x_1) \, \partial_{x_1} \tilde{Z}_{s + y}(x_1,x_2) \, \partial_{x_2} \tilde{\mathcal{J}}_s(x_2, z).
\end{equation}

Equation~\eqref{j_y_my} matches Eq.~(60) of~\cite{wieseFirstPassageInterval2019}, and is given explicitly in Eq.~(61) of the same reference.

\subsection{Diagrammatic Expansion of \texorpdfstring{$W_f(T)$}{W\_f(T)}}
We identify six distinct diagrammatic contributions to $W_f(T)$, categorized according to the relative positions of the velocity insertion times $r_1 < r_2$ with respect to the key times $t_1$, $t_1 + t_2$, and $T$.

The first three diagrams, denoted $A_1$, $A_2$, and $A_3$, involve correlated velocity insertions:
\begin{itemize}
    \item $A_1$: $0 < r_1 < r_2 < t_1$,
    \item $A_2$: $t_1 < r_1 < r_2 < t_1 + t_2$,
    \item $A_3$: $t_1 + t_2 < r_1 < r_2 < T$.
\end{itemize}
These contributions involve the connected correlator $\langle \dot{x}(r_1) \dot{x}(r_2) \rangle$.

The remaining three diagrams, denoted $B_1$, $B_2$, and $B_3$, involve uncorrelated velocity insertions due to the separation of $r_1$ and $r_2$ into distinct time intervals:
\begin{itemize}
    \item $B_1$: $0 < r_1 < t_1 < r_2 < t_1 + t_2$,
    \item $B_2$: $0 < r_1 < t_1 < t_1 + t_2 < r_2 < T$,
    \item $B_3$: $t_1 < r_1 < t_1 + t_2 < r_2 < T$.
\end{itemize}
In these cases, the velocity correlators factorize: $\langle \dot{x}(r_1) \dot{x}(r_2) \rangle = \langle \dot{x}(r_1) \rangle \langle \dot{x}(r_2) \rangle$.

Fixing the diffusion coefficient to $D = 1$ \footnote{This is the convention chosen by Wiese in his formalism \cite{wieseFirstPassageInterval2019}, and we keep it here for consistency.}, so that $(2D)^2 = 4$ (since we are not tracking absolute time units and will perform integration over $T$ in the end), the first-order expansion of $W_f(T)$ becomes:
\begin{equation}
    W_f(T) = \eta^3 z - \frac{4\varepsilon}{2} \left(A_1 + A_2 + A_3 + B_1 + B_2 + B_3\right).
\end{equation}

This implies the following expression for the first-order correction $\gamma(z)$:
\begin{equation}
    \gamma(z) = -2\int_0^\infty \frac{A_1 + A_2 + A_3 + B_1 + B_2 + B_3}{\eta^3} \quad dT.
\end{equation}

We now proceed to evaluate each of the six diagrams and extract their leading-order contributions in $\eta$. Since lower-order terms vanish, we focus on terms of order $\eta^3$, which provide the dominant non-zero corrections.

\subsubsection{The \texorpdfstring{$A$}{A} Diagrams}

We now compute the three diagrams $A_1$, $A_2$, and $A_3$, which correspond to contributions from velocity correlators $\langle \dot{x}(r_1)\dot{x}(r_2) \rangle$ when both insertion times lie within the same time interval.

\paragraph{Diagram $A_1$} 
This diagram corresponds to $0 < r_1 < r_2 < t_1$, and reads:
\begin{equation}
    A_1 = \int_{t_1 + t_2 + t_3 = T} \int_0^{\Lambda} dy \, J_{t_1}(0, -(1 - z); -y, y) \, Z_{t_2}(-(1 - z), z) \, Z_{t_3}(z, -(1 - z)) \, dt_1 dt_2 dt_3.
\end{equation}
Taking the Laplace transform in $T \to s$ gives:
\begin{equation}
    \tilde{A}_1(s) = \tilde{Z}_s(-(1 - z), z)^2 \int_0^{\Lambda} \tilde{J}_s(0, -(1 - z); -y, y) \, dy.
\end{equation}
For the time-integrated case ($s = 0$), we use $\tilde{Z}_0(-(1 - z), z) = \eta$ (splitting probability of a Brownian motion starting at distance $\eta$ from the boundary), yielding:
\begin{equation}
    \tilde{A}_1(0) = \eta^2 \int_0^{\Lambda} \tilde{J}_0(0, -(1 - z); -y, y) \, dy.
\end{equation}
This integral is given explicitly in Eq.~(61) of~\cite{wieseFirstPassageInterval2019}.

\paragraph{Diagram $A_2$} 
For the case $t_1 < r_1 < r_2 < t_1 + t_2$, we write:
\begin{equation}
    A_2 = \int_{t_1 + t_2 + t_3 = T} \int_0^{\Lambda} dy \, Z_{t_1}(0, -(1 - z)) \, J_{t_2}(-(1 - z), z; -y, y) \, Z_{t_3}(z, -(1 - z)) \, dt_1 dt_2 dt_3,
\end{equation}
and in Laplace space:
\begin{equation}
    \tilde{A}_2(s) = \tilde{Z}_s(0, -(1 - z)) \, \tilde{Z}_s(-(1 - z), z) \int_0^{\Lambda} \tilde{J}_s(-(1 - z), z; -y, y) \, dy.
\end{equation}
For $s = 0$, we use $\tilde{Z}_0(0, -(1 - z)) = z$, so:
\begin{equation}
    \tilde{A}_2(0) = z \eta \int_0^{\Lambda} \tilde{J}_0(-(1 - z), z; -y, y) \, dy.
\end{equation}

\paragraph{Diagram $A_3$}
Similarly, for $t_1 + t_2 < r_1 < r_2 < T$, we have:
\begin{equation}
    A_3 = \int_{t_1 + t_2 + t_3 = T} \int_0^{\Lambda} dy \, Z_{t_1}(0, -(1 - z)) \, Z_{t_2}(-(1 - z), z) \, J_{t_3}(z, -(1 - z); -y, y) \, dt_1 dt_2 dt_3,
\end{equation}
leading to:
\begin{equation}
    \tilde{A}_3(s) = \tilde{Z}_s(0, -(1 - z)) \, \tilde{Z}_s(-(1 - z), z) \int_0^{\Lambda} \tilde{J}_s(z, -(1 - z); -y, y) \, dy,
\end{equation}
and at $s = 0$:
\begin{equation}
    \tilde{A}_3(0) = z \eta \int_0^{\Lambda} \tilde{J}_0(z, -(1 - z); -y, y) \, dy.
\end{equation}

\paragraph{Connection to Wiese's Results}
These integrals were evaluated in~\cite{wieseFirstPassageInterval2019}, where the first-order correction to the outgoing current at the boundary $z$, for a path starting at $x - (1 - z)$, was defined as:
\begin{equation}
    \tilde{\mathcal{A}}(x) = \int_0^{\infty} \tilde{J}_0(x - (1 - z), z; -y, y) \, dy, \quad \text{with } 0 \leq x \leq 1.
\end{equation}
Notably, $\tilde{\mathcal{A}}(0) = \tilde{\mathcal{A}}(1) = 0$. Using this, the $A$ diagrams can be expressed as:
\begin{equation}
    A_1 = -\eta^3 \tilde{\mathcal{A}}(1 - z), \quad A_2 = A_3.
\end{equation}
Here we used the spatial symmetry of the current: $\tilde{\mathcal{A}}(x) = -\tilde{\mathcal{A}}(1 - x)$. In fact, both $A_2$ and $A_3$ diverge in the limit $\Lambda \to \infty$:
\begin{equation}
    \label{a2a3}
    A_2 = A_3 = \eta^3 z \int_0^\Lambda \frac{\left(\coth\left(\sqrt{y}\right) - \text{csch}\left(\sqrt{y}\right)\right)^2}{y} \, dy.
\end{equation}
Indeed, the integrand in \eqref{a2a3} decays as $y^{-1}$ for large $y$, leading to a logarithmic infrared divergence as $\Lambda \to \infty$. However, this divergence poses no problem: as we will show, the contributions from the $B$ diagrams precisely cancel those from the $A$ diagrams, ensuring a finite result.

\subsubsection{The \texorpdfstring{$B$}{B} Diagrams}

We now compute the diagrams $B_1$, $B_2$, and $B_3$, which correspond to contributions where the velocity insertions $r_1$ and $r_2$ lie in different time intervals. In such cases, the velocity correlator factorizes: $\langle \dot{x}(r_1) \dot{x}(r_2) \rangle = \langle \dot{x}(r_1) \rangle \langle \dot{x}(r_2) \rangle$.

\paragraph{Diagram $B_1$} 
Here, $r_1 \in [0, t_1]$ and $r_2 \in [t_1, t_1 + t_2]$, so we write $r_2 = t_1 + r_2'$, with $r_2' \in [0, t_2]$. The weight in the action contributes an additional exponential factor $e^{-y t_1}$ due to this time separation:
\begin{equation}
    B_1 = \int_{t_1 + t_2 + t_3 = T} \int_0^\Lambda dy \, e^{-y t_1} \, J_{t_1}(0, -(1 - z); -y) \, J_{t_2}(-(1 - z), z; y) \, Z_{t_3}(z, -(1 - z)) \, dt_1 dt_2 dt_3.
\end{equation}
In Laplace space $T \to s$, this becomes:
\begin{equation}
    \tilde{B}_1(s) = \tilde{Z}_s(-(1 - z), z) \int_0^\Lambda \tilde{J}_{s + y}(0, -(1 - z); -y) \, \tilde{J}_s(-(1 - z), z; y) \, dy.
\end{equation}
For $s = 0$, using $\tilde{Z}_0(-(1 - z), z) = \eta$, we obtain:
\begin{equation}
    \tilde{B}_1(0) = \eta \int_0^\Lambda \tilde{J}_y(0, -(1 - z); -y) \, \tilde{J}_0(-(1 - z), z; y) \, dy.
\end{equation}
Applying the symmetry relation $\tilde{J}_y(a,b;-y) = -\tilde{J}_0(b,a;y)$, we rewrite:
\begin{equation}
    \tilde{B}_1(0) = -\eta \int_0^\Lambda \tilde{J}_0(-(1 - z), 0; y) \, \tilde{J}_0(-(1 - z), z; y) \, dy.
\end{equation}
Explicitly, the expression for $\tilde{B}_1(0)$ reads:
\begin{align}
    \label{b1-exp}
    \tilde{B}_1(0) = -\eta^3 \int_0^\Lambda \bigg[ &
    \frac{z \left( \coth(\sqrt{y}) - \text{csch}(\sqrt{y}) \right)^2}{y} \notag \\
    & + \frac{ \left( \coth(\sqrt{y}) - \text{csch}(\sqrt{y}) \right)
    \left( \text{csch}(\sqrt{y}) - \text{csch}(\sqrt{y}) \cosh(\sqrt{y} z) \right)}{y}
    \bigg] dy.
\end{align}

The term proportional to $z$ in \eqref{b1-exp} leads to a divergent integral as $\Lambda \to \infty$. However, this divergence exactly cancels against the contribution from either $A_2$ or $A_3$. 

\paragraph{Diagram $B_2$}
Here, $r_1 \in [0, t_1]$ and $r_2 \in [t_1 + t_2, T]$. Writing $r_2 = t_1 + t_2 + r_2'$, we find two exponential weights: $e^{-y t_1}$ from $r_1$ to $r_2$, and an additional $e^{-y t_2}$ separating the time intervals:
\begin{equation}
    B_2 = \int_{t_1 + t_2 + t_3 = T} \int_0^\Lambda dy \, e^{-y (t_1 + t_2)} \, J_{t_1}(0, -(1 - z); -y) \, Z_{t_2}(-(1 - z), z) \, J_{t_3}(z, -(1 - z); y) \, dt_1 dt_2 dt_3.
\end{equation}
Transforming to Laplace space:
\begin{equation}
    \tilde{B}_2(s) = \int_0^\Lambda \tilde{J}_{s + y}(0, -(1 - z); -y) \, \tilde{Z}_{s + y}(-(1 - z), z) \, \tilde{J}_s(z, -(1 - z); y) \, dy.
\end{equation}
At $s = 0$:
\begin{equation}
    \tilde{B}_2(0) = \int_0^\Lambda \tilde{J}_y(0, -(1 - z); -y) \, \tilde{Z}_y(-(1 - z), z) \, \tilde{J}_0(z, -(1 - z); y) \, dy.
\end{equation}
Applying symmetry to both $\tilde{J}_0$ terms, we obtain:
\begin{equation}
    \tilde{B}_2(0) = \int_0^\Lambda \tilde{J}_0(-(1 - z), 0; y) \, \tilde{Z}_y(-(1 - z), z) \, \tilde{J}_y(-(1 - z), z; -y) \, dy.
\end{equation}
Explicitly, this writes 
\begin{equation}
    \label{b2-exp}
    \tilde{B}_2(0) =\eta ^3  \int_0^\Lambda  \frac{\text{csch}\left(\sqrt{y}\right) \left(z \cosh \left(\sqrt{y}\right)-\cosh \left(\sqrt{y} z\right)-z+1\right)}{\sqrt{y} \left(\cosh \left(\sqrt{y}\right)+1\right)} dy.
\end{equation}
This integral has a finite $\Lambda \to \infty$ limit. 

\paragraph{Diagram $B_3$}
This case involves $r_1 \in [t_1, t_1 + t_2]$ and $r_2 \in [t_1 + t_2, T]$. Writing $r_1 = t_1 + r_1'$, $r_2 = t_1 + t_2 + r_2'$, the weights from the action are $e^{-y r_2 + y r_1} = e^{-y(t_2 + r_2' - r_1')}$, and we find:
\begin{equation}
    B_3 = \int_{t_1 + t_2 + t_3 = T} \int_0^\Lambda dy \, Z_{t_1}(0, -(1 - z)) \, e^{-y t_2} \, J_{t_2}(-(1 - z), z; -y) \, J_{t_3}(z, -(1 - z); y) \, dt_1 dt_2 dt_3.
\end{equation}
In Laplace space:
\begin{equation}
    \tilde{B}_3(s) = \tilde{Z}_s(0, -(1 - z)) \int_0^\Lambda \tilde{J}_{s + y}(-(1 - z), z; -y) \, \tilde{J}_s(z, -(1 - z); y) \, dy.
\end{equation}
Evaluated at $s = 0$, we use $\tilde{Z}_0(0, -(1 - z)) = z$:
\begin{equation}
    \tilde{B}_3(0) = z \eta \int_0^\Lambda \tilde{J}_y(-(1 - z), z; -y) \, \tilde{J}_0(z, -(1 - z); y) \, dy.
\end{equation}
Applying symmetry:
\begin{equation}
    \tilde{B}_3(0) = -z \eta \int_0^\Lambda \tilde{J}_0(z, -(1 - z); y)^2 \, dy.
\end{equation}
Or, in preferred form:
\begin{equation}
    \tilde{B}_3(0) = -z \eta \int_0^\Lambda \tilde{J}_y(-(1 - z), z; -y)^2 \, dy.
\end{equation}
Explicitly, we find:
\begin{equation}
    -z \eta^3 \int_0^\Lambda \frac{\left( \coth(\sqrt{y}) - \text{csch}(\sqrt{y}) \right)^2}{y} \, dy.
\end{equation}
As before, the integral diverges in the infrared limit $\Lambda \to \infty$. However, this divergence is exactly canceled by the corresponding divergence in one of the $A_2$ or $A_3$ diagrams.

\subsection{Summing All Diagram Contributions}

As previously discussed, only the diagrams $A_1, B_1,B_2$ contribute in the $\Lambda \to \infty$ limit. We therefore write:
\begin{align}
    \label{gamma-final}
    \gamma(z) &= -\frac{2}{\eta^3} \left(\tilde{A}_1(0) + \tilde{B}_2(0) + \tilde{B}_1(0)\big|_{\text{finite}} \right) \notag \\
    &= 2\bigg(\tilde{\mathcal{A}}(1-z) -\int_0^\infty \frac{\text{csch}\left(\sqrt{y}\right) 
    \left( z \cosh\left(\sqrt{y}\right) - \cosh\left(\sqrt{y} z\right) - z + 1 \right)}{
    \sqrt{y} \left( \cosh\left(\sqrt{y}\right) + 1 \right)} \, dy \notag \\
    &\hspace{3cm}
    - \int_0^\infty \frac{\left( \coth\left(\sqrt{y}\right) - \text{csch}\left(\sqrt{y}\right) \right) 
    \left( \text{csch}\left(\sqrt{y}\right) - \text{csch}\left(\sqrt{y}\right) \cosh\left(\sqrt{y} z\right) \right)}{y} \, dy.\bigg)
\end{align}

Changing variables via \( y = -\log^2 r \), the first integral in Eq.~\eqref{gamma-final} can be evaluated explicitly. It yields:
\begin{align*}
    \frac{1}{2} \bigg( 
        -2 z^2 H_{1-z}
        - 2 z^2 H_{z+1}
        + (2 z^2 - 1) H_{\frac{z+1}{2}} 
        + (2 z^2 - 1) H_{\frac{1}{2} - \frac{z}{2}} 
        + z^2 \log 16 
        - 2z + 4 
        - 2 \log 4 
    \bigg),
\end{align*}
where \( H_z \) denotes the analytic continuation of the harmonic numbers~\cite{abramowitz1965handbook}.

The second integral in Eq.~\eqref{gamma-final} can also be computed using the same substitution. It evaluates to:
\begin{align}
\label{int-b2}
    &12 \log A 
    - 4 \psi^{(-2)}\left( \tfrac{z+1}{2} \right) 
    + 4 \psi^{(-2)}\left( \tfrac{z}{2} \right) 
    + 4 \psi^{(-2)}\left( 1 - \tfrac{z}{2} \right)
    - 4 \psi^{(-2)}\left( \tfrac{1}{2} - \tfrac{z}{2} \right) \notag \\
    &\quad + 2z \log \left[ \Gamma \left( 1 - \tfrac{z}{2} \right) \Gamma \left( \tfrac{z+1}{2} \right) \right]
    - 2z \log \left[ \Gamma \left( \tfrac{1}{2} - \tfrac{z}{2} \right) \Gamma \left( \tfrac{z}{2} \right) \right] 
    - \tfrac{1}{3} \log 2 - 2z.
\end{align}

Now, recalling Eq.~\eqref{phi-fbm-diagram}, the first-order correction to the \(\Phi\) function is given by:
\begin{equation}
    \beta(z) = \frac{\gamma(z) - \alpha(z)}{z},
\end{equation}
where
\begin{equation}
\label{alpha-fbm}
    \alpha(z) = -4z \log\left( 
    \frac{ \Gamma \left( 1 - \tfrac{z}{2} \right) \Gamma \left( \tfrac{z+1}{2} \right) }{
           \Gamma \left( \tfrac{1 - z}{2} \right) \Gamma \left( \tfrac{z}{2} \right) } \right) + zC.
\end{equation}

Here, \( C \) is a constant determined by enforcing the universal condition:
\begin{equation}
    \Phi(1) = \phi = 1 - 4\varepsilon + \mathcal{O}(\varepsilon^2),
\end{equation}
as required by Eq.~\eqref{phi1}.

Crucially, the term $2z \log \left[ \Gamma \left( 1 - \tfrac{z}{2} \right) \Gamma \left( \tfrac{z+1}{2} \right) \right]
- 2z \log \left[ \Gamma \left( \tfrac{1}{2} - \tfrac{z}{2} \right) \Gamma \left( \tfrac{z}{2} \right) \right]$ in Eq.~\eqref{int-b2} matches exactly the structure of \eqref{alpha-fbm}, and thus cancels in the expression for \(\beta(z)\). Similarly, the PolyGamma sum in Eq.~\eqref{int-b2} arises from the splitting probability \(\tilde{\mathcal{A}}(1-z)\), as established in~\cite{wieseFirstPassageInterval2019}:
\begin{align}
    \tilde{\mathcal{A}}(1-z) =\; & 4 \psi^{(-2)}\left( \tfrac{z+1}{2} \right) 
    - 4 \psi^{(-2)}\left( \tfrac{z}{2} \right) 
    - 4 \psi^{(-2)}\left( 1 - \tfrac{z}{2} \right) 
    + 4 \psi^{(-2)}\left( \tfrac{1}{2} - \tfrac{z}{2} \right) 
    + C'(2z - 1),
\end{align}
where the constant \( C' \) is exactly the same as in \eqref{int-b2}, as it is given by:
\begin{equation}
    C' = -\frac{1}{3} \left[ \log(2) - 3 + 36 \zeta'(-1) \right] = 12 \log A - \tfrac{1}{3} \log 2.
\end{equation}

Putting all pieces together, we finally see that only $B_2$ really contributes :
\begin{align}
    \beta(z) =\; & C'' 
    - \frac{1}{z} \bigg( 
        -2 z^2 H_{1-z} 
        - 2 z^2 H_{z+1} 
        + (2 z^2 - 1) H_{\frac{z+1}{2}} 
        + (2 z^2 - 1) H_{\frac{1}{2} - \frac{z}{2}} \notag \\
        &\quad + z^2 \log 16 
        - 2z + 4 - 2 \log 4 
    \bigg).
\end{align}

Finally, fixing the constant \( C'' \) to match the condition \(\Phi(1) = 1 - 4\varepsilon + \mathcal{O}(\varepsilon^2)\), we arrive at the main result announced in Eq.~\eqref{phi-expr}:
\begin{keyboxedeq}[$\Phi$ function for the fBM]
\begin{align}
    \label{phi-fbm}
    \Phi(z) = 1 + \varepsilon \Bigg( 
        &\frac{ \psi\left( \tfrac{z+3}{2} \right) + \psi\left( \tfrac{3-z}{2} \right) + 2\gamma - 4 + 2 \log 4 }{z} \notag \\
        &+ 2z \left[ 
            -\psi\left( \tfrac{z+3}{2} \right) 
            + \psi(z + 2) 
            - \psi\left( \tfrac{3 - z}{2} \right) 
            + \psi(2 - z) 
            - \log 4 
        \right] - 2 \Bigg) + \mathcal{O}(\varepsilon^2).
\end{align}
\end{keyboxedeq}
Let us now verify Eq.~\eqref{phi-fbm} using the general framework developed in this part of the thesis. According to our general expression for the eccentricity distribution \(\sigma_z\) given in Eq.~\eqref{pofz}, and using the identity \(\sigma_z = -Q_+'(z)\), the $\Phi$ function must satisfy the relation:
\begin{boxedeq}
\begin{equation}
\label{sym-phi-fbm}
    \frac{\Phi(z)}{1 - z} - \frac{\Phi(1 - z)}{z} 
    = \frac{-1}{z} + \frac{1}{1 - z}
    - 4\varepsilon \, \partial_z \log \left( 
        \frac{ \Gamma\left(1 - \tfrac{z}{2}\right) \Gamma\left( \tfrac{z+1}{2} \right) }{
               \Gamma\left( \tfrac{1 - z}{2} \right) \Gamma\left( \tfrac{z}{2} \right) } \right) 
    + \mathcal{O}(\varepsilon^2).
\end{equation}
\end{boxedeq}

Using the explicit expression for \(\Phi(z)\) in Eq.~\eqref{phi-fbm}, we confirm that Eq.~\eqref{sym-phi-fbm} is indeed satisfied. This serves as a nontrivial consistency check of our result and further validates the predictive power of the flip-based analytical framework.

\section{The $\Phi$ function for Class (III) processes}
In this section, we present the explicit derivation of the exact $\Phi$ function for class (III) processes~\eqref{phi-expr}, corresponding to non-saturating SIRWs. This result, along with its derivation, stands out as analytically precise, physically meaningful, and mathematically elegant. It provides direct insight into the geometric structure of non-Markovian trajectories. It also plays a pivotal role in our flip-based theory of space exploration. Notably, this was our first exact computation of the $\Phi$ function for a process beyond the simplest—though important—class (I). As such, it marked a key step in extending the reach of our framework and demonstrating its broader analytical power. As we will show, it is sufficient to perform the computation for the TSAW, since all class (III) processes share the same $\Phi$ function.

\subsection{The Brownian Web}
To construct the scaling limit of the TSAW in a mathematically rigorous way, Tóth and Werner introduced a powerful probabilistic object known as the \emph{Brownian web}~\cite{tothTrueSelfrepelling}. Intuitively, the Brownian web is a collection of reflecting and coalescing Brownian paths starting from every point in the upper half-space \(\mathbb{R} \times \mathbb{R}_+\), where \(\mathbb{R}\) represents space and \(\mathbb{R}_+\) represents time. It can be seen as the scaling limit of systems of reflecting and coalescing lattice random walks.

Importantly, the Brownian web gives a complete description of the space-time local time field \( (L_t(x))_{x \in \mathbb{R},\, t > 0} \) associated with a TSAW trajectory. Specifically, the time \( L_x(t) \) that the TSAW has spent at site \( x \) up to time \( t \) is given by the height at position \( x \) of a particular Brownian path in the web. This path the unique one for which the area between it and the real line \(\mathbb{R}\) equals \( t \). Uniqueness is ensured by the coalescing-reflecting structure of the web. For more details on the Brownian Web, we refer the reader to the original and elegant presentation in~\cite{tothTrueSelfrepelling}, as well as the insightful articles~\cite{dumazCleverSelfrepellingBurglar2012,dumazLargeDeviationsPath2012}, which greatly aided our understanding of this object.
\par \vspace{2ex} 
Crucially, the Ray--Knight theorems for the TSAW~\cite{tothTrueSelfAvoiding}, such as Eq.~\eqref{rayknight-txl}, describe only the \emph{single-time marginals} of the Brownian web. In contrast, the complete Brownian web encodes the \emph{entire} space-time evolution of the local time field. As a result, it captures dynamical features such as aging in the TSAW, which lie beyond the reach of the Ray--Knight framework. Despite its power, the Brownian web has, to our knowledge, never been used to study aging properties in the TSAW. \par
\vspace{2ex}
For our purposes, the Brownian web provides a concrete and tractable framework to study aging in the TSAW. Its structure—an ensemble of coalescing and reflecting Brownian trajectories—naturally encodes the full history of the walker's past visits. This makes it especially well-suited to capturing the long-range temporal correlations and memory effects that underlie aging. Although initially introduced for mathematical rigor, the Brownian web offers a physically transparent picture of how the TSAW explores space while preserving detailed information about its entire trajectory.

In the following, we take a physicist’s perspective on the Brownian web and show how it enables the computation of a key observable: the $\Phi$ function.

\subsection{The $\Phi$ function from the Brownian web}
\begin{figure}[h!]
    \centering
    \includegraphics[width=.85\textwidth]{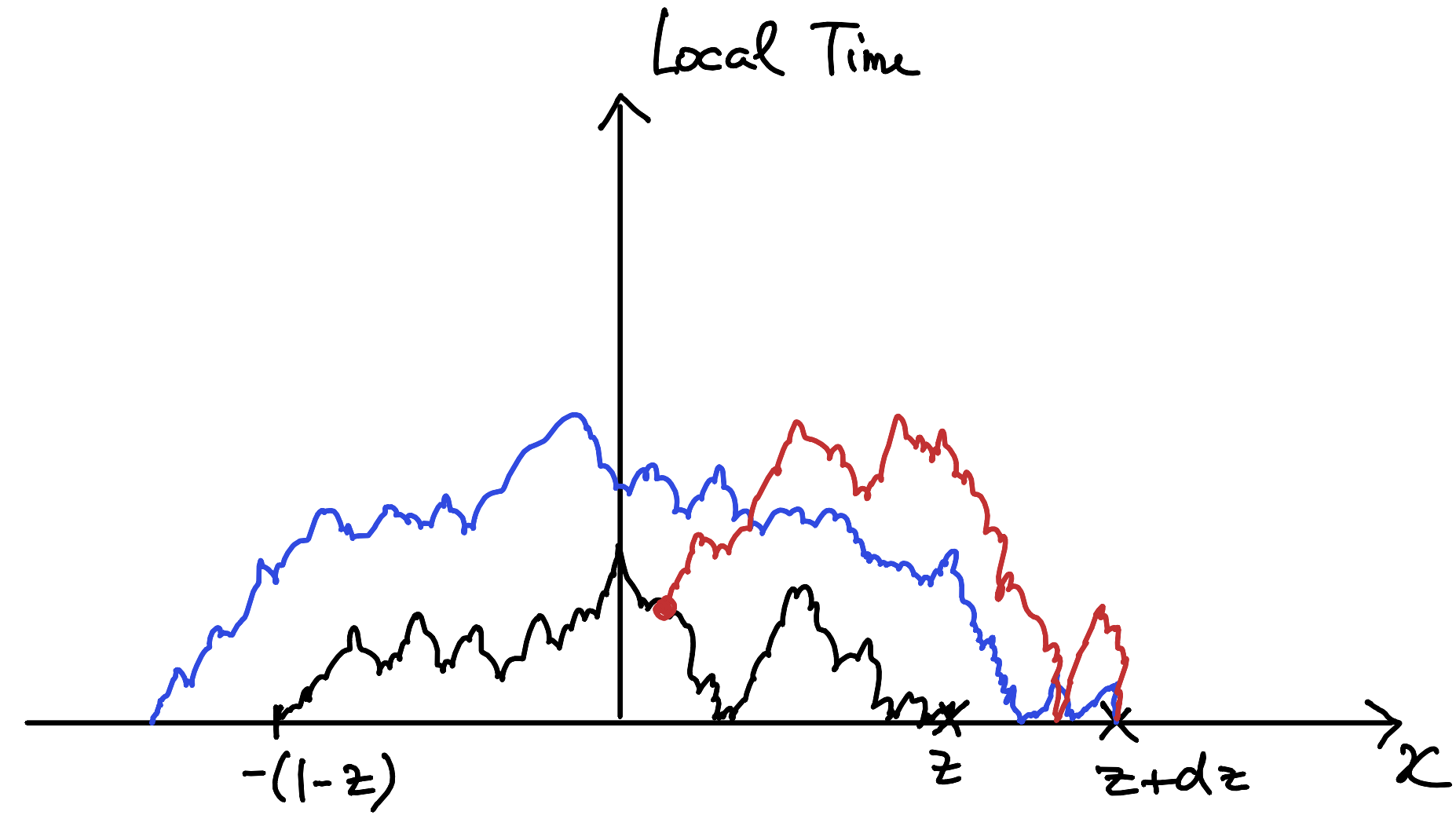}
    \caption{Schematic representation of the computation of the $\Phi$ function for the TSAW using the Brownian web. The black curve shows the initial local time profile when the walker first reaches site \( z \), extending left to \( -(1-z) \), with the initially visited interval rescaled to length 1. Its distribution is given by the Ray--Knight theorem: a reflecting Brownian motion from \( z \) to 0 (right) and a Brownian motion conditioned to hit 0 at \( -(1-z) \), propagating backward from 0 (left).
    The red and blue curves represent possible local time profiles when the walker reaches \( z + dz \). The red path coalesces with the black one before time \( -(1-z) \), indicating that no flip occured when extending the range. On the contrary, the blue path does not coalesce with the initial black profile, meaning the walker flipped—i.e., it reached \( -(1-z)-dz \) before \( z + dz \).
    Crucially, the Brownian web ensures that the red and blue future profiles are Brownian motions starting at \( z + dz \), reflected at 0 within the narrow interval \( [z, z + dz] \), and propagating backward in time until they either coalesce with the black initial profile or are absorbed at the real line. The value of $\Phi(z)$ thus corresponds to the probability that the future local time profile behaves like the blue one—avoiding coalescence with the initial profile—which signals a flip when extending the range.
    }
    \label{fig:scheme-tsaw}
\end{figure}
Our strategy for computing the $\Phi$ function is summarized in Fig.~\ref{fig:scheme-tsaw}. As illustrated, the problem reduces to evaluating the coalescence probability between two Brownian paths with different statistical laws. Specifically, we express $\Phi(z)$ as a path integral:

\begin{boxedeq}
\begin{equation}
    \label{strat-phi-tsaw}
    \Phi(z) = \frac{\mathbb{P}([\underline{-(1-z)-dz}, z], [-(1-z), \underline{z}])}{\mathbb{P}([-(1-z), \underline{z}])} 
    = \Lim{\varepsilon \to 0} \frac{\int \mathcal{D}L_1^\varepsilon \, \mathcal{D}L_2 \, \mathds{1}_{\{L_2 \text{ does not coalesce with } L_1^\varepsilon\}}}{\mathbb{P}([-(1-z), \underline{z}])}.
\end{equation}
\end{boxedeq}

In this expression:
\begin{itemize}
    \item $L_1^\varepsilon(t)$ denotes the initial local time profile at the moment the TSAW reaches \( z \). According to the Ray--Knight theorem, this profile is a Brownian motion starting at 0 at time \( t = z \), reflected on the interval \( [0, z] \). When conditioning on the eccentricity being exactly \( z \), it is further conditioned to hit 0 at time \( t = -(1 - z) \). In the numerator of Eq.~\eqref{strat-phi-tsaw}, however, there is no such conditioning. Instead, we integrate over all Brownian paths \( L_1^\varepsilon \) that remain positive on the interval \( [0, -(1 - z)] \) and end at a value \( 0<L_1^\varepsilon(-(1 - z)) \leq \varepsilon \). As in the fBM computation, we remove the regulator at the end by taking the limit \( \varepsilon \to 0 \).
    
    \item $L_2(t)$ denotes the future local time profile at the moment the TSAW reaches \( z + dz \). According to the Brownian web, it is a Brownian motion starting from 0 at time \( t = z + dz \), reflected on the short interval \( [z, z + dz] \), and propagating backward in time until it coalesces with either the initial profile \( L_1(t) \) or the real line—whichever it encounters first. A flip event occurs if it coalesces with the real line, or, equivalently, if it does not coalesce with $L_1^\varepsilon$.
\end{itemize}
\par \vspace{2ex}
At this point, we make a key observation: raising the local time profiles \( L_1^\varepsilon \) and \( L_2 \) in Eq.~\eqref{strat-phi-tsaw} to any power \( q > 0 \) leaves the value of \( \Phi(z) \) unchanged. This holds because:
\begin{enumerate}
    \item All paths start and end at zero, and this remains unchanged when raised to a power \( q \);
    \item Coalescence between \( L_1^\varepsilon \) and \( L_2 \) occurs if and only if the same is true for \( (L_1^\varepsilon)^q \) and \( (L_2)^q \).
\end{enumerate}
As suggested by the Ray--Knight theorems \eqref{rayknight-txl}, the Brownian web constructions for SESRW and PSRW involve powers of Brownian motions, rather than standard Brownian paths, as shown in~\cite{tothTrueSelfrepelling}. This confirms that the \( \Phi \) function is fully universal across all class (III) processes. This argument justifies our focus on the TSAW case.

\subsection{Computation of the Coalescence Probability}
\begin{figure}[h!]
    \centering
    \includegraphics[width=.85\textwidth]{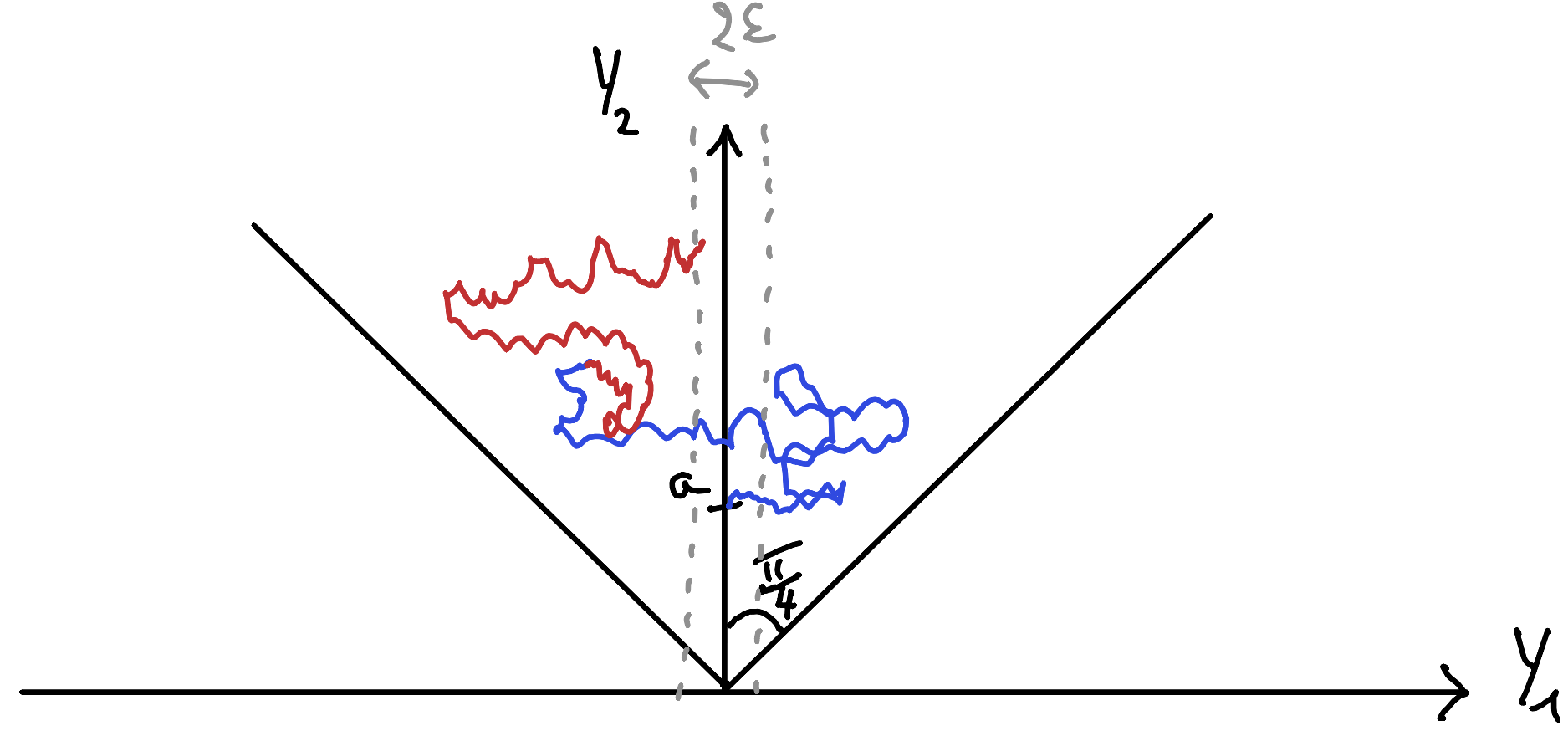}
    \caption{Schematic representation of the mapping to $2d$ Brownian motion. In this figure, the blue path represents the $2d$ Brownian motion $Y(t)$ run from time $t=0$ to time $t=z$. It is started from some point $a$, distributed as $p(a|0,dz)$ where $p(x|y,t)$ is the propagator of the $1d$ reflected Brownian motion. The red path represents $Y(t)$ run from time $t=z$ to time $t=1$. Our aim is now to compute the probability that $Y(t)$, starting from $a$, never hits the wedges of respective angles $\frac{\pi}{2}$ (for $0<t<z$) and $\frac{\pi}{4}$ (for $z<t<1$), and ends in the vertical strip $|Y_1| \leq \varepsilon$ at time $t=1$.
    }
    \label{fig:2dbm-tsaw}
\end{figure}
We now proceed to compute the path integral in Eq.~\eqref{strat-phi-tsaw}. To do so, we consider a two-dimensional Brownian motion constructed from the local time profiles:
\begin{equation}
    Y_0(t) = \big( L_1^\varepsilon(t),\,L_2(t)\big).
\end{equation}
Because \( L_2 \) may coalesce with \( L_1^\varepsilon \), the process \( Y_0(t) \) is absorbed when it hits the diagonal \( x = y \).

The time variable \( t \) lies in the interval \( [-dz, 1] \), and the behavior of the Brownian motion \( Y_0(t) \) depends on the specific subinterval:

\begin{itemize}
    \item For \( -dz < t < 0 \), the first coordinate of \( Y_0(t) \) is identically zero, while the second coordinate evolves as a reflected $1d$ Brownian motion started from $0$, since $L_2(t)$ is reflected at $0$ over the interval $[z,z+dz]$.
    
    \item For \( 0 < t < z \), the boundary \( x = 0 \) acts as a reflecting wall, since \( L_1^\varepsilon \) is reflected at zero over this interval.
    
    \item For \( z < t < 1 \), the boundary \( x = 0 \) becomes absorbing, as \( L_1^\varepsilon \) remains strictly positive and is no longer reflected at zero.
\end{itemize}
\par \vspace{2ex}
To simplify the analysis, we remove the reflecting boundary in the first time window by identifying $Y_0$ with its mirror image across the \( x = 0 \) axis. This leads to an equivalent two-dimensional Brownian motion \( Y(t) \) with the following properties:
\begin{itemize}
    \item It is absorbed when it hits either diagonal \( x = y \) or \( x = -y \). Equivalently, $Y(t)$ is a Brownian motion in an absorbing wedge of angle $\frac{\pi}{2}$ \cite{redner2001guide}.
    \item In the second time window \( z < t < 1 \), it is additionally absorbed at \( x = 0 \). Equivalently, $Y(t)$ is a Brownian motion in an absorbing wedge of angle $\frac{\pi}{4}$ \cite{redner2001guide}.
\end{itemize}
This mapping to $2d$ Brownian motion in wedges is illustrated Fig.~\ref{fig:2dbm-tsaw}. Summing up, we showed that we can write 
\begin{equation}
    \label{phi-tsaw-propag}
    \begin{aligned}
    \Phi(z) 
    &= \lim_{\varepsilon \to 0} \int_0^\infty p(a\,|\,0, dz) \cdot 
    \frac{
        \mathbb{P}^a\left(
            \text{$Y(t)$ ends in the vertical strip $|x| \leq \varepsilon$ at $t=1$ and avoids the wedges}
        \right)
    }{
        \mathbb{P}^a\left(
            \text{$Y(t)$ ends in the vertical strip $|x| \leq \varepsilon$ at $t=1$}
        \right)
    } \, da \\
    &= \lim_{\varepsilon \to 0} 
    \int_0^\infty p(a\,|\,0, dz) 
    \cdot \frac{
        \int_0^\infty db \int_0^\infty dx \int_x^\infty dy \, 
        \Tilde{p}_{\frac{\pi}{2}}\left((x, y) \,|\, (0, a), z\right)
        \Tilde{p}_{\frac{\pi}{4}}\left((\varepsilon, b) \,|\, (x, y), 1 - z\right)
    }{
        \dfrac{\varepsilon}{\pi} \sqrt{\dfrac{z}{1 - z}}
    } \, da.
    \end{aligned}
    \end{equation}
In Eq.~\eqref{phi-tsaw-propag}, we denoted by $p(x\,|\,y, t)$ the propagator of one-dimensional reflected Brownian motion, and by $\tilde{p}_{\theta}((x_2, y_2)\,|\,(x_1, y_1), t)$ the propagator of two-dimensional Brownian motion in an absorbing wedge of angle $\theta$. We also recognized that the denominator of the integrand in Eq.~\eqref{phi-tsaw-propag} corresponds to $\varepsilon q_+(z)$, as computed in Eq.~\eqref{eq:qplus-final}.

We will now derive the explicit formulae for $2d$ Brownian motion in an absorbing wedge using the image method \cite{redner2001guide}.

\subsection{The image method}

Let us define the free propagator of two-dimensional Brownian motion as 
\[
G(x, y, t) = \frac{e^{-\frac{x^2 + y^2}{2t}}}{2\pi t}.
\]

Using the image method, the propagator in a $\frac{\pi}{2}$ absorbing wedge is given by
\[
\tilde{p}_{\frac{\pi}{2}}((a,b)\,|\,(x,y),t) = G(a - x, b - y, t) + G(a + x, b + y, t) - \left[ G(a - y, b - x, t) + G(a + y, b + x, t) \right].
\]

For the $\frac{\pi}{4}$ absorbing wedge, we have:
\begin{align*}
\tilde{p}_{\frac{\pi}{4}}((a,b)\,|\,(x,y),t) &= G(a - x, b - y, t) + G(a + x, b + y, t) + G(a - y, b + x, t) + G(a + y, b - x, t) \\
&\quad - \left[ G(a - x, b + y, t) + G(a + x, b - y, t) + G(a - y, b - x, t) + G(a + y, b + x, t) \right].
\end{align*}

From this, we obtain the small-$\varepsilon$ expansion:
\[
\tilde{p}_{\frac{\pi}{4}}((\varepsilon,b)\,|\,(x,y),1-z) \underset{\varepsilon \to 0}{\sim} \varepsilon \cdot \frac{e^{-\frac{b^2 + x^2 + y^2}{2(1-z)}} \left(2x \sinh\left(\frac{b y}{1-z}\right) - 2y \sinh\left(\frac{b x}{1-z}\right)\right)}{\pi (1-z)^2}.
\]

Since the random initial value $a$ satisfies $a \sim dz^2$, it is small. Thus, we can expand the $\tilde{p}_{\frac{\pi}{2}}$ propagator around the point $(0,a\ll \sqrt{z})$:
\[
\tilde{p}_{\frac{\pi}{2}}((x, y)\,|\,(0, a), z) \underset{a^2 \ll z}{\sim} -a^2 \cdot \frac{(x^2 - y^2) e^{-\frac{x^2 + y^2}{2z}}}{2\pi z^3}.
\]

\subsection{Final result}
After a lengthy but physically appealing reformulation of $\Phi$ into an analytically tractable form using the Brownian web, and once the relevant constrained propagators are derived, the integrals in Eq.~\eqref{phi-tsaw-propag} can be computed exactly using \textit{Mathematica}. This yields a closed-form result—exact and surprisingly simple—valid for all class (III) processes:

\begin{keyboxedeq}[$\Phi$ function for class (III) processes]
\begin{equation}
    \label{phi-tsaw}
    \Phi(z) = \frac{ \sqrt{z(1 - z)} + (2z - 1)\, \arcsin\left( \sqrt{z} \right) }{\pi z}.
\end{equation}
\end{keyboxedeq}
This result reveals key features that are inaccessible from naive computations, and calls for several comments. 
\begin{itemize}
    \item As seen in Fig.~\ref{fig:phi-func}, the $\Phi$ function for class (III) increases with $z$, which is striking given that the TSAW is superdiffusive. This is the opposite behavior of superdiffusive fBM, highlighting subtle and nontrivial effects of memory in space exploration—effects that go beyond what is captured by standard scaling.
    
    \item The universal value $\Phi(1) = \phi = \frac{1}{2}$ is directly recovered from the exact expression \eqref{phi-tsaw}.

    \item As shown in \eqref{phi-expr}, the function $\Phi$ satisfies the key symmetry \eqref{phi-sym}, which we identified as characteristic of processes whose splitting probability takes the form of an incomplete Beta function.

    \item Most importantly, this represents—to our knowledge—the first explicit computation of an aged observable for the TSAW, and more broadly for any class (III) process. In the case of SIRWs, such observables had only been derived for the SATW, and only within this thesis (e.g., Eq.~\eqref{qm-aged-s0-deriv}).
    The exact result \eqref{phi-tsaw} thus serves as a reference point for aging observables in strongly self-interacting walks and demonstrates that explicit analytical treatment of such rich non-Markovian dynamics is achievable, though technically and conceptually involved. It also establishes a conceptual bridge between space exploration by SIRWs and fundamental probabilistic objects, such as the Brownian web used in this context.
\end{itemize}

\cleardoublepage
\chapter*{Conclusion}
\addcontentsline{toc}{chapter}{Global Conclusion}

This thesis set out to explore a fundamental question: how does memory reshape the way random walkers explore space? Across a diverse range of models and techniques, we have uncovered new insights into how non-Markovian dynamics—those in which the past directly affects the future—lead to rich, often unexpected behaviors in space exploration.

In the first part, we analyzed Locally Activated Random Walks (LARWs), where a walker’s motion depends on the time spent in specific locations. Despite their simple definition, these models display complex phenomena such as dynamical trapping, aging, and non-Gaussian spreading. We identified key conditions under which the walker becomes confined, discovered a dynamical phase transition driven by an interplay between activation and relaxation, and obtained explicit analytical results in several regimes.

The second part focused on Self-Interacting Random Walks (SIRWs), where the walker’s own history directly influences its future steps. By adapting advanced probabilistic tools—particularly Ray–Knight theory—we derived exact results for central observables: splitting probabilities, persistence exponents, first-passage times, number of visited sites, and even the influence of the walker's age on its future evolution. These results, including the first exact expressions for aged space exploration observables for the saturating SATW model, establish a strong analytical foundation for processes previously understood in the physics community mostly through heuristics and numerical simulations.

Finally, in the third part, we stepped back to develop a unified perspective on space exploration in non-Markovian systems. By introducing the concept of flip probabilities—measuring the walker's tendency to reverse direction of visitation—we uncovered universal patterns across all non-Markovian processes. Central to this theory is the function $\Phi(z)$, which encodes how the shape of the explored domain governs the probability of flips. We computed this function exactly in several important cases, including fractional Brownian motion and, notably, the true self-avoiding walk (TSAW). To the best of our knowledge, this result provides the first explicit expression for an aged observable in a non-saturating, strongly self-interacting walk, extending our own analytical results limited to the saturating SATW model.

Beyond their individual significance, our results collectively highlight a deeper message: even in systems governed by memory, where intuition often fails and exact results are rare, universal structures can emerge. Whether through the concept of flips, the interplay between activation and motion, or the fine structure of aged dynamics, our work opens analytical pathways to understand complexity in systems driven by their past.

This thesis thus lays the groundwork for future investigations into memory-driven dynamics—across physics, probability, and potentially, the life sciences—and shows that with the right observables and tools, the invisible influence of history can be made not only visible, but also quantifiable.

%% file: Annexe1.tex
%!TEX root = Manuscrit.tex
\section{Time-changed RWs}
An important reference model in the theory of stochastic processes, closely related to LARWs, is the time-changed continuous-time random walk (CTRW). In this model, a RWer $x(t)$ evolves on a graph $G$ and jumps from a site $s$ to a different site $s' \ne s$ with a \emph{deterministic}, time-dependent rate $1/\tau(t)$. This contrasts with the standard CTRW, where waiting times are typically independent and identically distributed.

Despite this explicit time dependence, one can define a renormalized time in which the walk becomes homogeneous. Specifically, introduce the change of variable:
\begin{equation}
\label{renorm-time}
    \tilde{t}(t) = G(t) \equiv \int_0^t \frac{ds}{\tau(s)}.
\end{equation}
The distribution $P(x,t) = \mathbb{P}(x(t) = x)$ satisfies a linear evolution equation:
\begin{equation}
    \partial_t P(x,t) = \frac{1}{\tau(t)} \mathcal{L} P(x,t),
\end{equation}
where $\mathcal{L}$ is a time-independent operator acting on the spatial variable $x$ (e.g., the graph Laplacian on $G$).
Because $\tau(t) > 0$, the function $G: t \mapsto \tilde{t}(t)$ is strictly increasing and invertible on the interval $[0, \tilde{T}_f)$, where we define
\begin{equation}
    \tilde{T}_f \equiv \int_0^\infty \frac{ds}{\tau(s)}.
\end{equation}
This upper bound $\tilde{T}_f$ plays a crucial role, as it separates cases where the renormalized time remains finite (the \emph{trapped} regime, where $x(t)$ is eventually trapped at a site) from those where it diverges (the \emph{free} regime, where $x(t)$ is never trapped). It thus provides a simple and robust criterion to distinguish between qualitatively different dynamical behaviors.
We now define the time-renormalized process:
\begin{equation}
\label{def-renorm-ctrw}
    y(\tilde{t}) \equiv x(t), \quad \text{with } t = G^{-1}(\tilde{t}).
\end{equation}
Since $P(y,\tilde{t}) = P(x,t)$ and $\partial_t = \tau(t)\, \partial_{\tilde{t}}$, the evolution equation becomes
\begin{equation}
    \partial_{\tilde{t}} P(y,\tilde{t}) = \mathcal{L} P(y,\tilde{t}),
\end{equation}
which shows that $\{y(\tilde{t})\}_{0 \leq \tilde{t} < \tilde{T}_f}$ is a CTRW with constant unit waiting time.

In particular, the long-time behavior of $x(t)$ is entirely governed by the behavior of the renormalized clock $\tilde{t}(t)$ and the value of $\tilde{T}_f$, which controls whether the process reaches a finite endpoint in renormalized time (trapping) or not (diffusion).

\section{A Collection of Functionals of Bessel processes and Brownian motions}
In this section, we derive explicit expressions for certain functionals of BESQ processes and Brownian motion that are required in our computations of relevant observables for SIRWs. To the best of our knowledge, these results do not appear in the existing literature, including in comprehensive references such as~\cite{borodinHandbookBrownianMotion2002}, which motivates their presentation here.

\subsection{Functionals of the BESQ process}
We consider a $\mathrm{BESQ}_\delta$ process $(X_t)_{t \geq 0}$ starting from $X_0 = a$, and compute the joint Laplace-transformed distributions of the following quantities:

\begin{enumerate}
    \item The joint distribution of the area under the process $\int_0^t X_u \, du$, and the position $X_t$ at time $t$, under reflecting boundary conditions at the origin:
    \begin{equation}
        \label{pos-area-refl}
        \mathbb{E}^a_{\delta,\mathrm{refl}} \left[ e^{-s \int_0^t X_u \, du} \,;\, X_t = b \right] 
        = \int_{\substack{X_0 = a,\, X_t = b \\ X \text{ reflected at } 0}} \mathcal{D}X \, e^{-s \int_0^t X_u \, du}.
    \end{equation}

    \item The joint distribution of the area $\int_0^t X_u \, du$ and the first-passage time $T_0$ to the origin:
    \begin{equation}
        \label{fpt-area-abs}
        \mathbb{E}^a_\delta \left[ e^{-s \int_0^t X_u \, du} \,;\, T_0 = t \right] 
        = \int_{\substack{X_0 = a,\, T_0 = t}} \mathcal{D}X \, e^{-s \int_0^t X_u \, du}.
    \end{equation}
\end{enumerate}
In both Eqs.~\eqref{pos-area-refl} and~\eqref{fpt-area-abs}, the path integral measure $\mathcal{D}X$ corresponds to that of a $\mathrm{BESQ}_\delta$ process. Accordingly, the trajectories are weighted according to the law of a $\mathrm{BESQ}_\delta$ process over the interval $[0,t]$. In Eq.~\eqref{pos-area-refl}, the integration is carried out over all paths that reflect at the origin and reach $X_t = b$ at time $t$. In contrast, Eq.~\eqref{fpt-area-abs} involves integration over all positive trajectories that are absorbed at zero, with the first-passage to zero $T_0$ occurring exactly at time $t$. \par 
Fortunately, the first functional~\eqref{pos-area-refl} can be computed using a result from~\cite{borodinHandbookBrownianMotion2002} (formula 1.9.7, p.~384). Defining $\mu := \sqrt{2s}$, the Laplace transform of the joint distribution of the area and the endpoint of a $\mathrm{BESQ}_\delta$ process reflected at zero is given by:
\begin{keyboxedeq}[Joint distribution of area and position of BESQ]
\begin{equation}
\label{laplace-area-endpoint}
\mathbb{E}^a_{\delta,\mathrm{refl}}\left[e^{-s \int_0^t X_u \, du} \,;\, X_t = b\right] 
= \frac{\mu  \left(\frac{b}{a}\right)^{\frac{\delta}{4} - \frac{1}{2}} 
e^{-\frac{1}{2} \mu (a + b) \coth(\mu t)} 
I_{\frac{\delta}{2} - 1} \left( \frac{\mu \sqrt{ab}}{\sinh(\mu t)} \right)}{2 \sinh(\mu t)}
\equiv \hat{P}_\delta(b \mid a, t, s),
\end{equation}
\end{keyboxedeq}
where $I_\nu$ denotes the modified Bessel function of the first kind. In the important case $a=0$, this becomes 
\begin{equation}
    \mathbb{E}^0_{\delta,\mathrm{refl}}\left[e^{-s \int_0^t X_u \, du} \,;\, X_t = b\right] = \frac{e^{\frac{-\mu b}{2} \coth (\mu  t)} \left(\frac{\mu  b}{2 \sinh (\mu  t)}\right)^{\delta /2}}{b \Gamma \left(\frac{\delta }{2}\right)}
\end{equation}

The second functional~\eqref{fpt-area-abs}, however, does not appear explicitly in the literature. We are only interested in the case $\delta < 2$, as for $\delta \geq 2$, the $\mathrm{BESQ}_\delta$ process almost surely never hits the origin~\cite{going-jaeschkeSurveyGeneralizationsBessel2003}.
To evaluate this functional, we invoke a classical time-reversal (or \emph{duality}) argument~\cite{lawlerNotesBessel}. Suppose the dimension satisfies $0 < \delta < 2$. In this regime, the $\mathrm{BESQ}_\delta$ process hits zero almost surely. In contrast, $4 - \delta > 2$, so the $\mathrm{BESQ}_{4 - \delta}$ process almost surely avoids zero. A key duality then holds: a $\mathrm{BESQ}_\delta$ process starting from $a > 0$ restricted (\emph{not} conditioned, see \eqref{duality-besq}) to reaching $X_t = b$ without being absorbed before time $t$ (i.e., restricted to the event $\{T_0 > t\}$, where $T_0$ denotes the hitting time of zero) has the same law as a time-reversed $\mathrm{BESQ}_{4 - \delta}$ process started from $b$, restricted to reaching $X_t = a$. More precisely, the following probabilities are equal:
\begin{equation}
    \label{duality-besq}
    \mathbb{P}_{\delta,\text{abs}}^a(X_t = b)\, db = \mathbb{P}_\delta^a(X_t = b,\, T_0 > t)\, db = \mathbb{P}_{4 - \delta}^b(X_t = a)\, da.
\end{equation}
In fact, this duality \eqref{duality-besq} extends beyond the one-point marginal to arbitrary functionals of the process. In particular, we obtain the identity:
\begin{equation}
    \label{time-rev-functional}
    \mathbb{E}_{\delta,\mathrm{abs}}^a\left[e^{-s \int_0^t X_u\, du} \,;\, X_t = b\right] 
    = \mathbb{E}_{4 - \delta,\mathrm{refl}}^b\left[e^{-s \int_0^t X_u\, du} \,;\, X_t = a\right].
\end{equation}

To obtain the second functional~\eqref{fpt-area-abs}, we now isolate the contribution of trajectories that hit zero for the first time in the infinitesimal time window $[t, t + dt]$. Using standard path integral reasoning, this gives:
\begin{equation}
\mathbb{E}^a_{\delta}\left[e^{-s \int_0^t X_u \, du} \,;\, T_0 = t\right] dt 
= \lim_{b \to 0} \mathbb{E}^a_{\delta,\mathrm{abs}}\left[e^{-s \int_0^t X_u \, du} \,;\, X_t = b\right] db.
\end{equation}

To proceed, we need to determine how the BESQ$_\delta$ process behaves as it approaches the origin. Recall that the $\mathrm{BESQ}_\delta$ process satisfies the stochastic differential equation:
\begin{equation}
dX_t = \delta\, dt + 2\sqrt{X_t}\, dB_t.
\end{equation}
When $X_t = b$ is small, the stochastic term becomes negligible compared to the deterministic drift. Thus, to leading order, $db \approx \delta\, dt$. Inserting this into the previous expression, we obtain:
\begin{equation}
\mathbb{E}^a_{\delta}\left[e^{-s \int_0^t X_u \, du} \,;\, T_0 = t\right] 
= \delta \lim_{b \to 0} \mathbb{E}^a_{\delta,\mathrm{abs}}\left[e^{-s \int_0^t X_u \, du} \,;\, X_t = b\right].
\end{equation}

Finally, applying the time-reversal identity~\eqref{time-rev-functional} and evaluating the limit using the expression~\eqref{laplace-area-endpoint}, we obtain the joint Laplace transform of the area and the first-passage time for a BESQ process of dimension $2 - \delta$:
\begin{keyboxedeq}[Joint distribution of area and FPT to $0$ of BESQ]
\begin{equation}
\label{joint-area-fpt}
\mathbb{E}^{a}_{2 - \delta}\left[e^{-s \int_0^t X_u \, du} \,;\, T_0 = t\right] 
= \frac{\mu\, e^{-\frac{1}{2} \mu a \coth(\mu t)} 
\left( \frac{\mu a}{2 \sinh(\mu t)} \right)^{\delta/2}}{
\Gamma\left( \frac{\delta}{2} \right) \sinh(\mu t)}
\equiv \hat{F}_{2 - \delta}(0 \mid a, t, s),
\end{equation}
\end{keyboxedeq}
where, as before, $\mu = \sqrt{2s}$. To the best of our knowledge, the only formula in the literature resembling~\eqref{joint-area-fpt} is equation (2.9.3) on p.~409 of~\cite{borodinHandbookBrownianMotion2002}. However, that expression is given in the Laplace domain with respect to $t$, and no explicit inversion is provided. Moreover, the Laplace transform appears analytically intractable to invert in closed form. We verified numerically, using Laplace inversion, that our explicit result~\eqref{joint-area-fpt} agrees with the Laplace-inverted form of the expression given in~\cite{borodinHandbookBrownianMotion2002}.

\subsection{Functionals of Brownian motion}
We now present expressions analogous to~\eqref{pos-area-refl} and~\eqref{fpt-area-abs}, but for standard Brownian motion (BM). Specifically, we consider the following two functionals:
\begin{align}
\label{pos-area-refl-bm}
\mathbb{E}^a_{\mathrm{BM,\,refl}} \left[ e^{-\gamma \int_0^t |X_u| \, du} \,;\, |X_t| = b \right] 
&= \int_{X_0 = a,\, X_t = b} \mathcal{D}X \, e^{-\gamma \int_0^t |X_u| \, du}, \\
\label{fpt-area-abs-bm}
\mathbb{E}^a_{\mathrm{BM}} \left[ e^{-\gamma \int_0^t X_u \, du} \,;\, T_0 = t \right] 
&= \int_{X_0 = a,\, T_0 = t} \mathcal{D}X \, e^{-\gamma \int_0^t X_u \, du},
\end{align}
where $\mathcal{D}X$ denotes the Wiener measure. In~\eqref{pos-area-refl-bm}, the process is reflected at the origin, while in~\eqref{fpt-area-abs-bm}, the process is absorbed upon first hitting zero.

To evaluate these functionals, we introduce an exponentially distributed random time $\tau \sim \mathrm{Exp}(\lambda)$, independent of the Brownian path. Formula 1.8.5 in~\cite{borodinHandbookBrownianMotion2002} provides the following expression:
\begin{multline}
\label{bm-refl-laplace}
\mathbb{E}^0_{\mathrm{BM,\,refl}} \left[ e^{-\gamma \int_0^\tau |X_u| \, du} \,;\, |X_\tau| \in db \right] 
= \lambda \int_0^\infty e^{-\lambda t} \mathbb{E}^0_{\mathrm{BM,\,refl}} \left[ e^{-\gamma \int_0^t |X_u| \, du} \,;\, |X_t| = b \right] dt \, db \\
= -2\lambda \cdot \frac{\mathrm{Ai}\left(2^{1/3} \gamma^{-2/3}(\lambda + \gamma b)\right)}{(2\gamma)^{1/3} \mathrm{Ai}'\left(2^{1/3} \gamma^{-2/3} \lambda\right)} db,
\end{multline}
where $\mathrm{Ai}(x)$ is the Airy function of the first kind. Inverting the Laplace transform in $\lambda \to t$ yields the first functional for deterministic time:
\begin{keyboxedeq}[Joint distribution of area and position for reflected BM]
\begin{equation}
\label{bm-refl-fixed-t}
\mathbb{E}^0_{\mathrm{BM}} \left[ e^{-\gamma \int_0^t |X_u| \, du} \,;\, |X_t| = z \right] 
= - (2\gamma)^{1/3} \sum_{k=1}^\infty e^{(\gamma^2/2)^{1/3} t \alpha_k'} \cdot \frac{\mathrm{Ai}\left( \alpha_k' + (2\gamma)^{1/3} z \right)}{\mathrm{Ai}(\alpha_k') \alpha_k'},
\end{equation}
\end{keyboxedeq}
where $\{\alpha_k'\}$ are the (negative) zeros of $\mathrm{Ai}'(x)$.

To compute the second functional~\eqref{fpt-area-abs-bm}, we again employ a time-reversal argument: a Brownian motion starting at $a$, ending at $b$, and avoiding the origin has the same law as a $\mathrm{BES}_3$ process moving backwards in time from $b$ to $a$. More precisely, we have, similarly to \eqref{duality-besq}:
\begin{equation}
\mathbb{E}^{b}_{R_s \sim \mathrm{BES}_3} \left[ e^{-\gamma \int_0^\tau R_s \, ds} \,;\, R_\tau = z \right] dz 
= \mathbb{E}^{z}_{B_s \sim \mathrm{BM}_{\mathrm{abs}}} \left[ e^{-\gamma \int_0^\tau B_s \, ds} \,;\, B_\tau = b \right] db.
\end{equation}

Using again formula 1.8.5 of~\cite{borodinHandbookBrownianMotion2002}, p.~446,\footnote{We note a misprint in the original reference \cite{borodinHandbookBrownianMotion2002}: an extraneous factor $2^{-2/3}$ appears in the formula, which must be removed to recover the correct normalization as $\gamma \to 0$.}
\begin{equation}
\mathbb{E}^0_{R_s \sim \mathrm{BES}_3} \left[ e^{-\gamma \int_0^\tau R_s \, ds} \,;\, R_\tau \in dz \right] 
= 2\lambda z \cdot \frac{\mathrm{Ai}\left(2^{1/3} \gamma^{-2/3}(\lambda + \gamma z)\right)}{\mathrm{Ai}\left(2^{1/3} \gamma^{-2/3} \lambda\right)} dz.
\end{equation}

To relate this to the first-passage-time functional, we use the correspondence $dt = 2b \, db$. This gives the double Laplace transform in the limit $b\to 0$:
\begin{equation}
\label{double-lpt-area-bm}
\mathbb{E}^z_{\mathrm{BM}} \left[ e^{-\gamma \int_0^{T_0} X_s \, ds - \lambda T_0} \right] 
= \frac{\mathrm{Ai}\left( 2^{1/3} \gamma^{-2/3} (z \gamma + \lambda) \right)}{\mathrm{Ai}\left( 2^{1/3} \gamma^{-2/3} \lambda \right)}.
\end{equation}

Taking the limit $\lambda \to 0$ of \eqref{double-lpt-area-bm}, we recover the known Laplace transform of the area under Brownian motion up to its first hitting time of the origin (see~\cite{borodinHandbookBrownianMotion2002}, p.~210):
\begin{equation}
\mathbb{E}^z_{\mathrm{BM}} \left[ e^{-\gamma \int_0^{T_0} X_s \, ds} \right] 
= 3^{2/3} \Gamma\left( \frac{2}{3} \right) \cdot \mathrm{Ai}\left( 2^{1/3} z \cdot \gamma^{1/3} \right).
\end{equation}
On the other hand, taking the limit $\gamma \to 0$ of \eqref{double-lpt-area-bm} recovers the Laplace transform of the FPT density to $0$:
\begin{equation}
    \mathbb{E}^z_{\mathrm{BM}} \left[ e^{-\lambda T_0} \right] = e^{-z \sqrt{2\lambda}}.
\end{equation}

Inverting the Laplace transform $\lambda \mapsto t$ in~\eqref{double-lpt-area-bm} yields the joint distribution of the area and first-passage time:
\begin{keyboxedeq}[Joint distribution of area and FPT to $0$ of BM]
\begin{equation}
\label{bm-fpt-area-inverted}
\mathbb{E}^z_{\mathrm{BM}} \left[ e^{-\gamma \int_0^t X_s \, ds} \,;\, T_0 = t \right] 
= \frac{\gamma^{2/3}}{2^{1/3}} \sum_{k=1}^\infty e^{(\gamma^2/2)^{1/3} t \alpha_k} \cdot \frac{\mathrm{Ai}\left( \alpha_k + (2\gamma)^{1/3} z \right)}{\mathrm{Ai}'(\alpha_k)},
\end{equation}
\end{keyboxedeq}
where $\{\alpha_k\}$ are the (negative) zeros of $\mathrm{Ai}(x)$.

\paragraph{Remark.} The following orthogonality relation for Airy functions (the Airy kernel) is useful \cite{tracyLevelspacingDistributions}:
\begin{equation}
\label{airykernel}
\int_0^\infty \mathrm{Ai}(t + x)\, \mathrm{Ai}(t + y)\, dt 
= \mathbb{A}i(x, y) := 
\begin{cases}
\displaystyle \frac{\mathrm{Ai}'(x) \mathrm{Ai}(y) - \mathrm{Ai}'(y) \mathrm{Ai}(x)}{y - x}, & x \neq y, \\[1em]
\displaystyle \mathrm{Ai}'(x)^2 - x \mathrm{Ai}(x)^2, & x = y.
\end{cases}
\end{equation}
In particular, for $x = \alpha_k$ and $y = \alpha_j'$, this gives:
\begin{equation}
\int_0^\infty \mathrm{Ai}(t + \alpha_k)\, \mathrm{Ai}(t + \alpha_j')\, dt 
= \frac{\mathrm{Ai}'(\alpha_k) \mathrm{Ai}(\alpha_j')}{\alpha_j' - \alpha_k}.
\end{equation}

\section{Some formulae involving sums of zeros of Airy and Airy'}

We begin with an identity that follows from the Weierstrass factorization of the Airy function \cite{olverAsymptoticsSpecial}:
\begin{equation}
    \label{weierstrass-airy}
    \frac{\mathrm{Ai}'(z)}{\mathrm{Ai}(z)} - \frac{\mathrm{Ai}'(0)}{\mathrm{Ai}(0)} = \sum_{j=1}^\infty \left[ \frac{1}{z - \alpha_j} + \frac{1}{\alpha_j} \right],
\end{equation}
where $\alpha_j$ denotes the $j$-th (negative) zero of the Airy function $\mathrm{Ai}(z)$.

Evaluating this expression at $z = \alpha_i'$ where $\alpha_i'$ is the $i$-th zero of $\mathrm{Ai}'(z)$ yields
\begin{equation}
    \label{sum-mixed-zerosai-aip}
    -\frac{\mathrm{Ai}'(0)}{\alpha_i'\mathrm{Ai}(0)} = \sum_{j=1}^\infty \frac{1}{(\alpha_i' - \alpha_j)\alpha_j}.
\end{equation}
From \eqref{weierstrass-airy}, we can obtain explicit expressions for the sums of negative powers of Airy zeros. Indeed, deriving \eqref{weierstrass-airy} $(n-1)$ times with respect to $z$, then plugging in $z=0$ yields, for $n\geq 2$
\begin{equation}
    \label{sum-airyzeros}
    \boxed{
    \sum_{j=1}^\infty \frac{1}{\alpha_j^n} = -\frac{1}{(n-1)!} \partial_z^{n-1} \left(\frac{\mathrm{Ai}'(z)}{\mathrm{Ai}(z)} \right)_{z=0}.
    }
\end{equation}
For instance, plugging in $n=2$ in the above yields 
\begin{equation}
    \sum_{j=1}^\infty \frac{1}{\alpha_j^2} = \frac{3^{2/3} \Gamma \left(\frac{2}{3}\right)^2}{\Gamma \left(\frac{1}{3}\right)^2}.
\end{equation}

% Summing over $i$ and rewriting, we see that the normalization condition for the Laplace-transformed FPT density $\hat{F}_k(s)$ given in Eq.~\eqref{fpt-density-tsaw} becomes:
% \begin{equation}
%     \label{sumairy1}
%     \sum_{i,j=1}^\infty \frac{1}{(\alpha_i' - \alpha_j) \alpha_j \alpha_i'} = -\frac{\mathrm{Ai}'(0)}{\mathrm{Ai}(0)} \sum_{i=1}^\infty \frac{1}{\alpha_i'^2}.
% \end{equation}

To obtain similar identities for the sums of negative powers of the zeros of the derivative of the Airy function, we consider the Weierstrass product representation for $\mathrm{Ai}'(z)$ and the differential identity $\mathrm{Ai}''(z) = z\,\mathrm{Ai}(z)$. We obtain :
\begin{equation}
    \label{weierstrass-airyprime}
    \frac{\mathrm{Ai}''(z)}{\mathrm{Ai}'(z)} - \frac{\mathrm{Ai}''(0)}{\mathrm{Ai}'(0)} 
    = \sum_{j=1}^\infty \left[ \frac{1}{z - \alpha_j'} + \frac{1}{\alpha_j'} \right] = \frac{z\,\mathrm{Ai}(z)}{\mathrm{Ai}'(z)}.
\end{equation}
% we evaluate at $z = \varepsilon$ and expand as $\varepsilon \to 0$. Extracting the coefficient of the linear term in $\varepsilon$ gives:
% \begin{equation}
%     \sum_{i=1}^\infty \frac{1}{\alpha_i'^2} = -\frac{1}{\frac{\mathrm{Ai}'(0)}{\mathrm{Ai}(0)}},
% \end{equation}
% which confirms the normalization:
% \[
% \sum_{i,j=1}^\infty \frac{1}{(\alpha_i' - \alpha_j) \alpha_j \alpha_i'} = 1.
% \]
Deriving \eqref{weierstrass-airyprime} with respect to $z$ and plugging $z=0$ at the end, we obtain 
\begin{equation}
    \label{sum-airyprimezeros}
    \boxed{
    \sum_{i=1}^\infty \frac{1}{(\alpha_i')^n} = -\frac{1}{(n-1)!} \partial_z^{n-1} \left(z\frac{\mathrm{Ai}(z)}{\mathrm{Ai'}(z)} \right)_{z=0}.
    }
\end{equation}
For instance, plugging in $n=3$ in \eqref{sum-airyprimezeros} yields the identity
\begin{equation}
    \label{sum-alphaip3}
    \sum_{i=1}^\infty \frac{1}{(\alpha_i')^3} = -1.
\end{equation}
Plugging in a zero of the Airy function $z = \alpha_j$ in \eqref{weierstrass-airyprime} yields another curious identity
\begin{equation}
    \label{vanishing-sum-airy}
    \sum_{i=1}^\infty \frac{1}{\alpha_i'(\alpha_i'-\alpha_j)} = 0.
\end{equation}
Similar identities can be obtained by first deriving \eqref{weierstrass-airyprime} (resp. \eqref{weierstrass-airy}) with respect to $z$, then plugging in a zero of the Airy function (resp. of its derivative). For example, for all $j\geq 1$, we have 
\begin{equation}
    \label{sum-squares-airy}
    \sum_{i=1}^\infty \frac{1}{(\alpha_i'-\alpha_j)^2} = -\alpha_j, \quad \sum_{i=1}^\infty \frac{1}{(\alpha_i'-\alpha_j)^3} = 1,
\end{equation}
while for all $i \geq 1$, we have the counterparts
\begin{equation}
    \label{sum-squares-airy-2}
    \sum_{j=1}^\infty \frac{1}{(\alpha_i'-\alpha_j)^2} = -\alpha_i', \quad \sum_{j=1}^\infty \frac{1}{(\alpha_i'-\alpha_j)^3} = \frac{1}{2}.
\end{equation}

\section{Distribution of the number of sites visited along certain rays for the simple RW on the star}
\subsection{Single-ray span $N_1(t)$}
In this section we compute the distribution of the number $N_1(t)$ of visited sites for a simple RW on the star $\mathcal{S}_n$, along the first ray $R_1$ (which can be any ray by symmetry). This has been computed in \cite{csakiLimitTheorems} using other methods, we recover it here using our Ray--Knight analysis. We begin by writing 
\begin{align}
    \mathcal{L}_{t \to s} \, \mathbb{P}(N_1(t) \geq N) 
    &= \frac{1}{s} \int_0^\infty \hat{q}_1(N, k_2, \dots, k_n) \, dk_2 \cdots dk_n \\
    &= \frac{(n-1)!}{s \sinh(N \mu)} \int_0^\infty 
    \frac{dx_2 \cdots dx_n}{\sinh^2(x_2) \cdots \sinh^2(x_n)} 
    \left( \coth(N \mu) + \coth(x_2) + \cdots + \coth(x_n) \right)^{-n} 
     \\
    &= \frac{(n-1)!}{s \sinh(N \mu)} \int_0^\infty 
    \left( n-1 + \coth(N \mu) + u_2 + \cdots + u_n \right)^{-n}
    \, du_2 \cdots du_n \\
    &= \frac{(n-1)!}{s \sinh(N \mu)} \int_0^\infty 
    \left( n-1 + \coth(N \mu) + S \right)^{-n} 
    \frac{S^{n-2}}{(n-2)!} \, dS \\
    &= \frac{(n-1)!}{s \sinh(N \mu)} \cdot \frac{1}{(n-2)!} \int_0^\infty 
    \frac{S^{n-2}}{(n-1+\coth(N \mu)+S)^n} \, dS \\
    &= \frac{(n-1)!}{s \sinh(N \mu)} \cdot \frac{1}{(n-1)!} \cdot \frac{1}{n-1+\coth(N \mu)} \\
    &= \frac{1}{s} \cdot \frac{1}{(n-1)\sinh(N \sqrt{2s}) + \cosh(N \sqrt{2s})} \\ 
    &= \frac{2}{s n} \sum_{k=0}^{\infty} \left(1-\frac{2}{n} \right)^k e^{-(2k+1)\sqrt{2s}N},
    \end{align}
where we developed in a geometric series for the last identity. Inverting the Laplace transform term-by-term, we recover the identity in \cite{csakiLimitTheorems}
\begin{boxedeq}
\begin{equation}
    \label{dist-oneray-star}
    \boxed{
    \mathbb{P}(N_1(t) \geq N) = \frac{2}{n}\sum_{k=0}^{\infty} \left(1-\frac{2}{n} \right)^k \erfc{\frac{(2k+1)N}{\sqrt{2t}}}.
    }
\end{equation}
\end{boxedeq}

\subsection{Two-ray distribution}

In a similar way, we compute the two-ray distribution $\mathbb{P}(N_1(t) \geq m_1, N_2(t) \geq m_2)$ by distinguishing whether the RW hits (and goes beyond) $m_2$ first before reaching $m_1$ or vice-versa. Integrating now over the $(n-3)$-dimensional simplex $\sum_{i=3}^n u_i = S$ of volume $\frac{S^{n-3}}{(n-3)!}$, we obtain
\begin{align}
    \label{jointdist-2ray-star}
    &\mathcal{L}_{t \to s} \, \mathbb{P}(N_1(t) \geq m_1, N_2(t) \geq m_2) = \\ 
    &\quad \frac{1}{s} \left( \int_{m_2 \mu}^\infty \frac{(n-2+\coth(m_1\mu)+\coth(u_2))^{-2}}{\sinh(m_1 \mu) \sinh^2(u_2)} \, du_2 + \int_{m_1 \mu}^\infty \frac{(n-2+\coth(m_2\mu)+\coth(u_1))^{-2}}{\sinh^2(u_1) \sinh(m_2 \mu)} \, du_1 \right) \\
    &= \frac{1}{s} \left( \int_{1}^{\coth(m_2 \mu)} \frac{(n-2+\coth(m_1\mu)+v_2)^{-2}}{\sinh(m_1 \mu)} \, dv_2 + \int_1^{\coth(m_1 \mu)} \frac{(n-2+\coth(m_2\mu)+v_1)^{-2}}{\sinh(m_2 \mu)} \, dv_1 \right) \\
    &= \frac{1}{s} \left( \frac{1}{\cosh(\mu m_1) + (n-1)\sinh(\mu m_1)} - \frac{1}{\cosh(\mu m_1) + (\coth(\mu m_2) + n-2)\sinh(\mu m_1)} + (m_1 \leftrightarrow m_2) \right).
\end{align}
We recover the (Laplace transform of the) single-ray distribution \eqref{dist-oneray-star} for $m_1$ when plugging $m_2=0$ in \eqref{jointdist-2ray-star}. Also, note that for $n=2$, i.e., the $1d$ simple RW, \eqref{jointdist-2ray-star} recovers the joint distribution of the min and max of Brownian motion \cite{borodinHandbookBrownianMotion2002}: 
\[
    \mathcal{L}_{t \to s} \, \mathbb{P}(N_1(t) \geq m_1, N_2(t) \geq m_2) = e^{-(m_1+m_2)\mu} \frac{\cosh(\frac{\mu (m_1-m_2)}{2})}{s \cosh(\frac{\mu(m_1+m_2)}{2})}.
\]
\par
To compute the correlations $\langle N_1(t) N_2(t) \rangle$, we integrate \eqref{jointdist-2ray-star} and directly inverse the Laplace transform:
\begin{align}
    \label{corr-rays-star}
    \langle N_1(t) N_2(t) \rangle &= \int_0^\infty \mathbb{P}(N_1(t) \geq m_1, N_2(t) \geq m_2) \, dm_1 \, dm_2 \\
    &= t \int_0^\infty \left( \frac{1}{\cosh(m_1) + (n-1)\sinh(m_1)} - \frac{1}{\cosh(m_1) + (\coth(m_2) + n-2)\sinh(m_1)} \right) \, dm_1 \, dm_2 \\
    &= t \int_1^\infty \frac{\log\left( \frac{n-1+u}{2} \right)}{\sqrt{u^2-1} (n-3+u)(n-1+u)} \, du,
\end{align}
where the second identity was obtained using Mathematica. This integral is nontrivial and is computed in the next section as $(n-1)! \cdot \mathcal{B}_n^{(2)}(\alpha=1)$.

\section{The variance of the number of visited sites for the simple RW on the star}
In this section we explicitly compute the integrals $\mathcal{B}_n^{(1)}(\alpha=1),\mathcal{B}_n^{(2)}(\alpha=1)$ that appear in \eqref{secmom-noft-star}. 
\begin{figure}
    \centering
    \begin{subfigure}[t]{0.48\textwidth}
        \includegraphics[width=\textwidth]{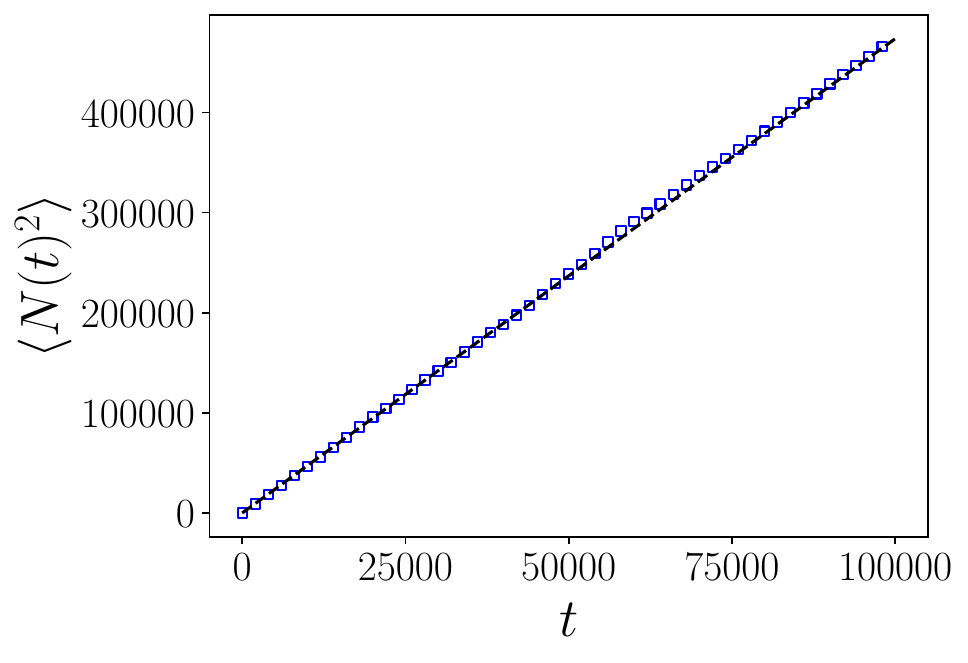}
        \caption{Star with $n=5$ rays.}
        \label{fig:meansquared-noft-15}
    \end{subfigure}
    \hfill
    \begin{subfigure}[t]{0.48\textwidth}
        \includegraphics[width=\textwidth]{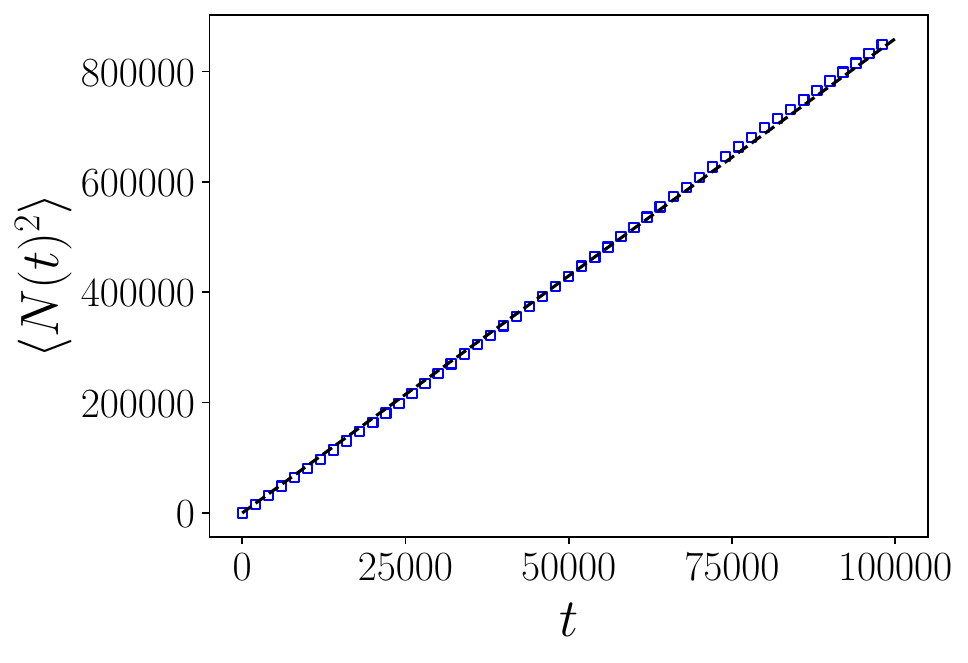}
        \caption{Star with $n=15$ rays.}
        \label{fig:meansquared-noft-n15}
    \end{subfigure}
    \caption{Numerical verification of the exact expression \eqref{secmom-noft-star} (black dashed line) for the second moment $\langle N(t)^2 \rangle$ for the SATW$_\alpha$ on a star graph.}
    \label{fig:meansquared-noft-star}
\end{figure}

\subsection{The first integral}
Here, we compute the simpler integral 
\begin{align}
    \label{app-bn1}
    \mathcal{B}_n^{(1)}(\alpha=1) &= \int_0^\infty \dots \int_0^\infty 
    \frac{\sinh^{-1}(\frac{1}{\sqrt{v_1(2+v_1)}})}{\sqrt{v_1(2 + v_1)}} \cdot 
    \frac{1}
    {(n + \sum_{i=1}^n v_i)^{n}} \, dv_1 \dots dv_n \\ &= \int_0^\infty dv_1 \frac{\sinh^{-1}(\frac{1}{\sqrt{v_1(2+v_1)}})}{\sqrt{v_1(2 + v_1)}} \int_0^\infty \frac{dS}{(n+v_1+S)^n} \frac{S^{n-2}}{(n-2)!} \\ &= \frac{1}{(n-1)!}\int_0^\infty dv_1 \frac{\sinh^{-1}(\frac{1}{\sqrt{v_1(2+v_1)}})}{\sqrt{v_1(2 + v_1)}(n+v_1)}.
\end{align}
Upon changing variables $v_1 = \coth(w)-1$, \eqref{app-bn1} can be computed \cite{gradshteynTableIntegrals}
\begin{equation}
    \boxed{
    \mathcal{B}_n^{(1)}(\alpha=1) = \frac{1}{(n-1)!} \int_0^\infty \frac{w}{\sinh(w)(n-1+\coth(w))} = \frac{1}{(n-1)!}\frac{2 \text{Li}_2\left(\sqrt{\frac{n-2}{n}}\right)-\frac{1}{2} \text{Li}_2\left(\frac{n-2}{n}\right)}{\sqrt{(n-2) n}}.
    }
\end{equation}
\subsection{The second integral}
Here, we compute the second, more difficult integral
\begin{align}
    \label{app-bn2}
    \mathcal{B}_n^{(2)}(\alpha=1) &= \int_0^\infty \dots \int_0^\infty 
    \frac{\sinh^{-1}(\frac{1}{\sqrt{v_2(2+v_2)}})}{\sqrt{v_1(2 + v_1)}} \cdot 
    \frac{1}
    {(n + \sum_{i=1}^n v_i)^{n}} \, dv_1 \dots dv_n \\ &= \int_0^\infty dv_1 \frac{1}{\sqrt{v_1(2 + v_1)}} \int_0^\infty dv_2 \sinh^{-1}(\frac{1}{\sqrt{v_2(2+v_2)}}) \int_0^\infty \frac{dS}{(n+v_1+v_2+S)^n} \frac{S^{n-3}}{(n-3)!} \\ &= \frac{1}{(n-1)!}\int_0^\infty dv_1 \frac{1}{\sqrt{v_1(2 + v_1)}} \int_0^\infty dv_2 \sinh^{-1}(\frac{1}{\sqrt{v_2(2+v_2)}}) \frac{1}{(n+v_1+v_2)^2} \\ &=  \frac{1}{(n-1)!}\int_0^\infty dx \frac{\log(\frac{n+x}{2})}{\sqrt{x(2 + x)}(n+x)(n-2+x)}.
\end{align}
This form is too complicated for Mathematica to handle. We will first use partial fractions decomposition 
\begin{equation}
    \frac{1}{(n+x)(n-2+x)} = \frac{1}{2(n-2+x)} - \frac{1}{2(n+x)}
\end{equation}
and write 
\begin{align}
    \mathcal{B}_n^{(2)}(\alpha=1) &= \frac{1}{(n-1)!}\left(\mathcal{A}^{(1)}_n + \mathcal{A}^{(1)}_n\right), \\ \mathcal{A}^{(1)}_n &= \int_0^\infty dx \frac{\log(\frac{n+x}{2})}{\sqrt{x(2 + x)}(n+x)}, \quad \mathcal{A}^{(2)}_n = \int_0^\infty dx \frac{\log(\frac{n+x}{2})}{\sqrt{x(2 + x)}(n+x-2)}. 
\end{align}
For both integrals, we perform the change of variables $x \mapsto \frac{2u^2}{1-u^2}$, which means $dx = \frac{4u}{(u^2-1)^2}du$ and yields 
\begin{equation}
    \mathcal{A}^{(1)}_n = \int_0^1 \frac{\log \left(\frac{(2-n) u^2+n}{2 \left(1-u^2\right)}\right)}{n+(2-n)u^2} du, \mathcal{A}^{(2)}_n = \int_0^1 \frac{\log \left(\frac{(2-n) u^2+n}{2 \left(1-u^2\right)}\right)}{(4-n) u^2+n-2} du.
\end{equation}
We are now ready to compute these integrals.

\subsubsection{First $\mathcal{A}$ integral}
We have 
\begin{equation}
    \label{app-a1}
    \mathcal{A}^{(1)}_n = \int_0^1 \frac{\log \left((2-n) u^2+n\right)}{n+(2-n)u^2} du - \int_0^1 \frac{\log \left(2(1-u)^2\right)}{n+(2-n)u^2} du.
\end{equation}
We have sufficiently reduced the complexity of the integrals that they can now be computed explicitly using Mathematica. After a lot of algebra, we find 
\begin{equation}
    \boxed{
    \begin{aligned}
    \mathcal{A}^{(1)}_n = \frac{1}{4\sqrt{(n-2)n}} \Bigg(
    & -2\, \text{Li}_2\left( \tfrac{1}{2} \left( \sqrt{\tfrac{n-2}{n}} + 1 \right) \right)
    + 2\, \text{Li}_2\left( \tfrac{1}{2} \left( 1 - \sqrt{\tfrac{n-2}{n}} \right) \right)
    \\
    & - 2\, \text{Li}_2\left(n - \sqrt{(n-2)n} - 1\right)
    + 2\, \text{Li}_2\left(n + \sqrt{(n-2)n} - 1\right)
    \\
    & + \log\left(\sqrt{n - 2} + \sqrt{n} \right)
    \left[
    \log\left(8\sqrt{n} - 8\sqrt{n - 2}\right)
    - 2 \log\left(1 - \sqrt{\tfrac{n - 2}{n}} \right)
    \right]
    \\
    & + 2 \left(
    \log\left( \tfrac{2(\sqrt{n - 2} + \sqrt{n})}{n - 2} \right) - i\pi
    \right)
    \log\left(n - \sqrt{(n - 2)n} - 1\right)
    \\
    & + \log\left( \sqrt{n} - \sqrt{n - 2} \right)
    \left[
    \log\left( \sqrt{n} - \sqrt{n - 2} \right)
    - 2 \log\left(2(n + \sqrt{(n - 2)n})\right)
    \right]
    \Bigg)
    \end{aligned}
    }
    \end{equation}

\subsubsection{Second $\mathcal{A}$ integral}
Likewise, we can write 
\begin{equation}
    \mathcal{A}^{(2)}_n = \int_0^1 \frac{\log \left((2-n) u^2+n\right)}{n-2+(4-n)u^2} du - \int_0^1 \frac{\log \left(2(1-u)^2\right)}{n-2+(4-n)u^2} du.
\end{equation}
Again, these can be computed by Mathematica. We obtain 
\begin{equation}
    \boxed{
    \begin{aligned}
    \mathcal{A}^{(2)}_n = \frac{1}{2\sqrt{(n-4)(n-2)}} \Bigg(
    & - \text{Li}_2\left(n - \sqrt{n^2 - 6n + 8} - 3\right)
    + \text{Li}_2\left(n + \sqrt{n^2 - 6n + 8} - 3\right)
    \\
    & + \left(
    \log\left(\sqrt{n - 2} - \sqrt{n - 4}\right)
    + \log\left(\sqrt{n - 2} + \sqrt{n - 4}\right)
    - \log(n - 4) - i\pi
    \right)
    \\
    & \quad \cdot \log\left(-\sqrt{n^2 - 6n + 8} + n - 3\right)
    - \log(2) \log\left(\sqrt{n^2 - 6n + 8} + n - 3\right)
    \\[4pt]
    & + \text{Li}_2\left(
    \frac{-n + \sqrt{(n - 4)(n - 2)} + 2}{-n + \sqrt{(n - 4)n} + 2}
    \right)
    \\
    & - \text{Li}_2\left(
    - \frac{n + \sqrt{(n - 4)(n - 2)} - 2}{-n + \sqrt{(n - 4)n} + 2}
    \right)
    \\
    & + \text{Li}_2\left(
    \frac{n - \sqrt{(n - 4)(n - 2)} - 2}{n + \sqrt{(n - 4)n} - 2}
    \right)
    \\
    & - \text{Li}_2\left(
    \frac{n + \sqrt{(n - 4)(n - 2)} - 2}{n + \sqrt{(n - 4)n} - 2}
    \right)
    \\
    & + 2\left(
    - \log(n - 4)
    + \log\left(n - \sqrt{(n - 4)n} - 2\right)
    + \log\left(n + \sqrt{(n - 4)n} - 2\right)
    - i\pi
    \right)
    \\
    & \quad \cdot \tanh^{-1}\left( \sqrt{\frac{n - 4}{n - 2}} \right)
    \Bigg)
    \end{aligned}
    }
    \end{equation}
    
We checked numerically that the sum of the two integrals $\mathcal{A}$ as obtained above give the integral $\mathcal{B}^{(2)}_n$.

\subsection{Putting all terms together}
Finally, putting everything together, we obtain the second moment of the number of visited sites on the star for the simple RW

\begin{equation}
    \label{secmom-noft-star}
    \boxed{
    \begin{aligned}
    \langle N^2(t) \rangle =\; & n t \Bigg(
    \frac{
    2\, \text{Li}_2\left( \sqrt{\frac{n-2}{n}} \right)
    - \frac{1}{2} \, \text{Li}_2\left( \frac{n-2}{n} \right)
    }{
    \sqrt{(n-2)n}
    }
    \\[4pt]
    & + \frac{n - 1}{2 \sqrt{n^2 - 6n + 8}} \Big[
    \text{Li}_2\left(
    \frac{-n + \sqrt{n^2 - 6n + 8} + 2}{-n + \sqrt{(n - 4)n} + 2}
    \right)
    \\
    &\quad - \text{Li}_2\left(
    -\frac{n + \sqrt{n^2 - 6n + 8} - 2}{-n + \sqrt{(n - 4)n} + 2}
    \right)
    + \text{Li}_2\left(
    \frac{n - \sqrt{n^2 - 6n + 8} - 2}{n + \sqrt{(n - 4)n} - 2}
    \right)
    \\
    &\quad - \text{Li}_2\left(
    \frac{n + \sqrt{n^2 - 6n + 8} - 2}{n + \sqrt{(n - 4)n} - 2}
    \right)
    - \text{Li}_2\left(n - \sqrt{n^2 - 6n + 8} - 3\right)
    \\
    &\quad + \text{Li}_2\left(n + \sqrt{n^2 - 6n + 8} - 3\right)
    + \left(
    \log(\sqrt{n - 2} - \sqrt{n - 4})
    + \log(\sqrt{n - 4} + \sqrt{n - 2})
    \right.
    \\
    &\quad\left.
    - \log(n - 4) - i\pi
    \right)
    \log\left(-\sqrt{n^2 - 6n + 8} + n - 3\right)
    \\
    &\quad - \log(2) \log\left(\sqrt{n^2 - 6n + 8} + n - 3\right)
    + 2\left(
    - \log(n - 4)
    + \log(n - \sqrt{(n - 4)n} - 2)
    \right.
    \\
    &\quad\left.
    + \log(n + \sqrt{(n - 4)n} - 2) - i\pi
    \right) \tanh^{-1}\left( \sqrt{\frac{n - 4}{n - 2}} \right)
    \Big]
    \\[4pt]
    & - \frac{1}{4 \sqrt{(n - 2)n}} \Big[
    -2\, \text{Li}_2\left( \frac{1}{2} \left( \sqrt{\frac{n - 2}{n}} + 1 \right) \right)
    + 2\, \text{Li}_2\left( \frac{1}{2} \left( 1 - \sqrt{\frac{n - 2}{n}} \right) \right)
    \\
    &\quad - 2\, \text{Li}_2\left(n - \sqrt{(n - 2)n} - 1\right)
    + 2\, \text{Li}_2\left(n + \sqrt{(n - 2)n} - 1\right)
    \\
    &\quad + \log\left(\sqrt{n - 2} + \sqrt{n}\right)
    \left[
    \log\left(8\sqrt{n} - 8\sqrt{n - 2}\right)
    - 2 \log\left(1 - \sqrt{\frac{n - 2}{n}}\right)
    \right]
    \\
    &\quad + 2\left(
    \log\left( \frac{2(\sqrt{n - 2} + \sqrt{n})}{n - 2} \right) - i\pi
    \right)
    \log\left(n - \sqrt{(n - 2)n} - 1\right)
    \\
    &\quad + \log\left(\sqrt{n} - \sqrt{n - 2}\right)
    \left[
    \log\left(\sqrt{n} - \sqrt{n - 2}\right)
    - 2 \log\left(2(n + \sqrt{(n - 2)n})\right)
    \right]
    \Big]
    \Bigg)
    \end{aligned}
    }
    \end{equation}
This lengthy expression~\eqref{secmom-noft-star}, although cumbersome, has the important advantage of being fully analytical and enables us to extract the nontrivial large-$n$ asymptotics. See FIG.~\ref{fig:meansquared-noft-star} for a check against numerical simulations. In particular, taking the limit $n \to 2$ in~\eqref{secmom-noft-star} recovers the known result \cite{hughes1995random} $\langle N^2(t) \rangle = 4\log(2)\, t$ for the $1d$ RW. Despite the appearance of $i\pi$ terms in~\eqref{secmom-noft-star}, the final expression is manifestly real.

\section{Antiderivative of the aged $q_{\pm}^{(2)}$ observables}
In this section, we show the antiderivative forms \eqref{qm-aged-s0-deriv},\eqref{qp-aged-s0-deriv} of the aged $q_{\pm}^{(2)}$ observables. We recall the definition of the $x_\pm$ variables \eqref{def-xp-xm-x}:
\begin{equation}
    x_- = \frac{m_2 - m_1}{k_1 + m_2}, \qquad
    x_+ = \frac{k_2 - k_1}{k_2 + m_1}.
\end{equation}
Then, we have the following Jacobians 
\begin{equation}
    \label{xpm-jacob}
    \frac{d x_+}{d k_2} = \frac{k_1+m_1}{(k_2+m_1)^2}, \qquad \frac{d x_-}{d m_2} = \frac{k_1+m_1}{(k_1+m_2)^2}.
\end{equation}

\subsection{Antiderivative of $q_{-}^{(2)}$}
Using \eqref{qm-aged-s0} and \eqref{xpm-jacob}, and using the variables $x_\pm$, we can write after some algebra
\begin{equation}
    \hat{q}_-^{(2)}(k_2,m_2, s=0 \mid k_1, m_1) dk_2 = \beta  (1-x_-)^{\alpha } \left(1-x_+\right)^{\beta -1} \, _2F_1(\alpha ,\beta +1;1;x_+ \cdot x_-) dx_+.
\end{equation}
Now, expand the ${}_2F_1(\alpha ,\beta +1;1;x_+ x_-)$ in series as
\begin{equation}
{}_2F_1(\alpha ,\beta +1;1;x_+ x_-) = \sum_{n=0}^\infty \frac{(\alpha)_n(1+\beta)_n}{n!^2} (x_+ x_-)^n.
\end{equation}
We will use the following Beta integral:
\begin{equation}
\label{betaint}
\int_0^{y} (1-u)^{a-1} u^{b-1} \, du = B(a,b) I_{y}(a,b).
\end{equation}
Then, the integral of $\hat{q}_-^{(2)}(k_2,m_2, s=0 \mid k_1, m_1)$ over $[k_1, k_2]$ is
\begin{equation}
\begin{aligned}
\label{proof-deriv-qm}
\int_{k_1}^{k_2} \hat{q}_-^{(2)}(k, m_2, s=0 \mid k_1, m_1) \, dk 
&= \int_0^{x_+} \beta (1 - x_-)^{\alpha} (1 - x_+)^{\beta - 1} \\
&\quad \times {}_2F_1(\alpha, \beta + 1; 1; x_+ x_-) \, dx_+ \\
&= \beta (1 - x_-)^{\alpha}\sum_{n=0}^\infty \frac{(\alpha)_n (1 + \beta)_n}{(n!)^2} x_-^n \, 
B(n + 1, \beta) \, I_{x_+}(n + 1, \beta).
\end{aligned}
\end{equation}
Simplifying the Gamma factors in the summand of \eqref{proof-deriv-qm}, we end up with \eqref{qm-aged-s0-deriv}. 

\subsection{Antiderivative of $q_{+}^{(2)}$}
In a similar spirit as above, and using \eqref{qp-aged-s0} and \eqref{xpm-jacob}, we can express the result in terms of the variables $x_\pm$. After some algebra, we obtain
\begin{equation}
\begin{aligned}
\hat{q}_+^{(2)}(k_2, m_2, s=0 \mid k_1, m_1) \, dm_2 
&= \alpha \beta \, x_+ (1 - x_-)^{\alpha - 1} (1 - x_+)^{\beta} \\
&\quad \times {}_2F_1(1 + \alpha, 1 + \beta; 2; x_+ x_-) \, dx_- \\
&\quad + (1 - x_+)^\beta \, \delta(x_-) \, dx_-.
\end{aligned}
\end{equation}
Then, the integral of $\hat{q}_+^{(2)}(k_2,m_2, s=0 \mid k_1, m_1)$ over $[m_1, m_2]$ is
\begin{equation}
    \label{proof-deriv-qp}
    \begin{aligned}
    \int_{m_1}^{m_2} \hat{q}_+^{(2)}(k_2, m, s=0 \mid k_1, m_1) \, dm
    &= \int_0^{x_-} \alpha \beta x_+ (1 - x_-)^{\alpha - 1} (1 - x_+)^{\beta} \\
    &\quad \times {}_2F_1(1 + \alpha, \beta + 1; 2; x_+ x_-) \, dx_- 
    + (1 - x_+)^\beta \\
    &= \alpha \beta (1 - x_+)^{\beta} \sum_{n=0}^\infty 
    \frac{(\alpha + 1)_n (1 + \beta)_n}{(n + 1)! \, n!} \, x_+^{n+1} \\
    &\quad \times B(n + 1, \alpha) \, I_{x_-}(n + 1, \alpha) 
    + (1 - x_+)^\beta.
    \end{aligned}
    \end{equation}
    
Simplifying the Gamma factors in the summand of \eqref{proof-deriv-qm}, we end up with \eqref{qm-aged-s0-deriv}. 

\section{Proof That Lévy Walks and the RAP Belong to Class (I)}

In this section, we show that for both Lévy Walks and the Random Acceleration Process (RAP), flips are independent events. This means that the probability of a flip does not depend on the eccentricity of the visited interval—in other words, the $\Phi$ function is constant.

Let \( X(t) \) denote the position of the walker and \( V(t) = X(t+1) - X(t) \) its speed at time \( t \). For the RAP, \( V(t) \) is a simple symmetric random walk; for Lévy Walks, \( V(t) = \pm 1 \) switches sign at waiting times drawn from a power-law distribution. For convenience, we assume \( V = 0 \) immediately before any change of sign.

A key property of both processes is that \( X(t) \) behaves as a Markov process between two zeros of the speed \( V(t) \).

Assume that the walker has visited \( n \) distinct sites and has eccentricity \( z = M/n \), where \( M \) is the maximum site visited. For both the RAP and Lévy Walks, the flip probability \( \tilde{\pi}_n(M) \sim \Phi(z)/n \) corresponds to the probability that:
\begin{itemize}
    \item the speed \( V \) is zero upon first reaching site \( M \), and
    \item the walker exits the visited interval \( [-(n-M), M] \) through the left endpoint \( -(n-M) \), starting with \( V = 0 \) at position \( M \).
\end{itemize}
Indeed, if \( V \neq 0 \) at \( M \), the walker immediately steps to \( M + 1 \), thereby exiting on the right.

Since the process is Markovian between zero-speed events, we can factor the flip probability as:
\begin{equation}
\label{class-i-indep}
\tilde{\pi}_n(M) = 
\mathbb{P}\left(V = 0 \text{ when first reaching } M \mid n \text{ sites visited} \right)
\cdot
\mathbb{P}\left( \text{exit through } -(n-M) \mid X_0 = M, V_0 = 0 \right).
\end{equation}

Both terms on the right-hand side of Eq.~\eqref{class-i-indep} are independent of \( M \), by translational invariance:

\begin{itemize}
    \item The second term depends only on the length of the interval, not on its eccentricity. Using translational invariance, we have:
    \[
    \mathbb{P}\left( \text{exit through } -(n-M) \mid X_0 = M, V_0 = 0 \right)
    = \mathbb{P}\left( \text{exit through } 0 \mid X_0 = n, V_0 = 0 \right).
    \]

    \item The first term is also independent of \( M \). The condition that the process visits \( n \) sites and has maximum \( M \) implies that it started at \( -(n - M) \), reached \( M \) without crossing \( -(n - M + 1) \), and was at zero speed at both endpoints. Using again translational invariance at zero speed, this probability is equal to:
    \[
    \mathbb{P}\left( V = 0 \text{ when first reaching } X = n \mid X_0 = 0, V_0 = 0, \text{ no crossing of } -1 \right).
    \]
\end{itemize}

Hence, both terms are independent of \( M \), and the flip probability is independent of eccentricity. This implies that \( \Phi(z) \equiv \phi \) is constant for both Lévy Walks and the RAP:
\[
    \Phi_{\text{(I)}}(z) \equiv \phi.
\]